\documentclass[english]{article}
\usepackage[T1]{fontenc}
\usepackage[utf8]{inputenc}
\usepackage{amsmath}
\usepackage{amssymb}

\makeatletter
\numberwithin{equation}{section}

\usepackage{jheppub}

\author[a,b]{Tamas Gombor}

\affiliation[a]{MTA-ELTE “Momentum” Integrable Quantum Dynamics Research Group, Department of Theoretical Physics, Eötvös
  Loránd University}
\affiliation[b]{Holographic QFT Group, HUN-REN Wigner Research Centre for Physics, Budapest, Hungary}

\emailAdd{gombort@caesar.elte.hu}

\abstract{We find closed formulas for the overlaps of Bethe eigenstates of $\mathfrak{gl}(N)$ symmetric spin chains and integrable boundary states.
We derive the general overlap formulas for $\mathfrak{gl}(M)\oplus\mathfrak{gl}(N-M)$ symmetric boundary states and give a well-established conjecture for the $\mathfrak{sp}(N)$ symmetric case.
Combining these results with the previously derived $\mathfrak{so}(N)$ symmetric formula, now we have the overlap functions for all integrable boundary states of the $\mathfrak{gl}(N)$ spin chains which are built from two-site states.
The calculations are independent from the representations of the quantum space therefore our formulas can be applied for the $SO(6)$ and the alternating $SU(4)$ spin chains which describe the scalar sectors of $\mathcal{N}=4$ super Yang-Mills and ABJM theories which are important application areas of our results.}


\makeatother

\usepackage{babel}
\begin{document}
\title{Exact overlaps for all integrable two-site boundary states of $\mathfrak{gl}(N)$
symmetric spin chains}
\maketitle

\section{Introduction}

In the last decade there has been growing interest in special overlaps
between particular spin-chain states called boundary states \cite{Ghoshal:1993tm,Piroli:2017sei}
and wavefunctions of integrable spin chains. These overlaps are important
in two parts of the theoretical physics. In statistical physics these
quantities appear in the context of non-equilibrium dynamics of the
integrable models and the overlaps play a central role in the study
of the quantum quenches \cite{Essler:2016ufo}. One of the main methods
for the investigation of the steady states is the so-called Quench
Action method \cite{Caux:2013ra}, where the knowledge of the exact
overlaps is an important input \cite{De_Nardis_2014,Pozsgay_2014,Wouters_2014,Piroli:2018ksf,Piroli:2018don,Rylands:2022gev}.

The other area of applications is the AdS/CFT correspondence. The
boundary states and the corresponding overlaps appear for various
setups: the overlaps describe one-point functions of the $\mathcal{N}=4$
super Yang-Mills and ABJM theories with domain wall defects \cite{deLeeuw:2015hxa,Buhl-Mortensen:2015gfd,Kristjansen:2021abc}
and the three-point functions of two determinant and one single trace
operators \cite{Jiang:2019xdz,Jiang:2019zig,Yang:2021hrl}, they also
relevant for the correlators between the ’t Hooft line and chiral
primaries of the $\mathcal{N}=4$ super Yang-Mills \cite{Kristjansen:2023ysz}
and for the correlation functions of a single trace operator and a
circular supersymmetric Wilson loop in ABJM theory \cite{Jiang:2023cdm}. 

In recent years, overlap functions have been determined for many boundary
states \cite{deLeeuw:2016umh,DeLeeuw:2018cal,deLeeuw:2019ebw,Pozsgay_2018,Gombor:2020auk,Gombor:2020kgu}.
In the most cases, these are only conjectures based on the observation
that the overlaps are proportional to the ratio of the so-called Gaudin
determinants. There exist exact proofs for the XXX and XXZ spin chains
\cite{Brockmann_2014,Brockmann_2014odd,Foda:2015nfk,Jiang:2020sdw},
but it is not clear how they can be extended to arbitrary representations
or nested systems. The first method which applied the algebraic Bethe
Ansatz for the derivations was published in \cite{Gombor:2021uxz}.
The method is based on the so-called $KT$-relation which allows us
to derive a recursion for the overlaps in a representation independent
way. This method was generalized for $\mathfrak{gl}(N)$ spin chains
in \cite{Gombor:2021hmj}. So far, this is the only precise proof
of boundary state overlaps in nested systems. This method was also
generalized for the proof of overlaps with crosscap states \cite{Gombor:2022deb}.
However, the results of \cite{Gombor:2021hmj} only include a subset
of possible boundary states. In this paper we extend the procedure
and determine the on-shell overlap formulas for all possible integrable
boundary states which are built from two-site states.

In \cite{Gombor:2021hmj} it was showed that we can distinguish between
untwisted and twisted boundary states. For untwisted or twisted boundary
states the Bethe roots have achiral or chiral pair structure. The
possible residual symmetries $\mathfrak{h}$ for the untwisted states
can be $\mathfrak{h}=\mathfrak{gl}(M)\otimes\mathfrak{gl}(N-M)$ and
for the twisted states they are $\mathfrak{h}=\mathfrak{so}(N)$ or
$\mathfrak{h}=\mathfrak{sp}(N)$. The derivations in \cite{Gombor:2021hmj}
could be applied for the $\mathfrak{h}=\mathfrak{gl}(\left\lfloor \frac{N}{2}\right\rfloor )\otimes\mathfrak{gl}(\left\lceil \frac{N}{2}\right\rceil )$
and $\mathfrak{h}=\mathfrak{so}(N)$ symmetric cases. In this paper
we extend the results for the remaining symmetry classes. We now list
the possible overlap functions for the $\mathfrak{gl}(N)$ spin chains.
For $\mathfrak{h}=\mathfrak{gl}(M)\otimes\mathfrak{gl}(N-M)$ the
on-shell overlaps are
\begin{equation}
\begin{split}\frac{\langle\Psi|\bar{u}\rangle}{\sqrt{\langle\bar{u}|\bar{u}\rangle}} & =\left[\prod_{\nu=1}^{M}\mathfrak{b}_{\nu}^{r_{\nu-1}-r_{\nu}}\right]\frac{Q_{M}(\mathfrak{a})}{\sqrt{\bar{Q}_{\frac{N}{2}}(0)Q_{\frac{N}{2}}(\frac{i}{2})}}\times\sqrt{\frac{\det G^{+}}{\det G^{-}}},\quad\text{where }N\text{ is even,}\\
\frac{\langle\Psi|\bar{u}\rangle}{\sqrt{\langle\bar{u}|\bar{u}\rangle}} & =\left[\prod_{\nu=1}^{M}\mathfrak{b}_{\nu}^{r_{\nu-1}-r_{\nu}}\right]\frac{Q_{M}(\mathfrak{a})}{\sqrt{Q_{\frac{N-1}{2}}(-\frac{i}{4})Q_{\frac{N-1}{2}}(\frac{i}{2})}}\times\sqrt{\frac{\det G^{+}}{\det G^{-}}},\quad\text{ where }N\text{ is odd.}
\end{split}
\label{eq:utw}
\end{equation}
For $\mathfrak{h}=\mathfrak{so}(N)$:
\begin{equation}
\frac{\langle\Psi|\bar{u}\rangle}{\sqrt{\langle\bar{u}|\bar{u}\rangle}}=\left[\prod_{\nu=1}^{N}y_{\nu}^{r_{\nu-1}-r_{\nu}}\right]\left[\prod_{\nu=1}^{N-1}\sqrt{\frac{Q_{\nu}(0)}{Q_{\nu}(\frac{i}{2})}}\right]\times\sqrt{\frac{\det G^{+}}{\det G^{-}}},\label{eq:twso}
\end{equation}
and for $\mathfrak{h}=\mathfrak{sp}(N)$:
\begin{equation}
\frac{\langle\Psi|\bar{u}\rangle}{\sqrt{\langle\bar{u}|\bar{u}\rangle}}=\left[\prod_{k=1}^{N/2}x_{k}^{r_{2k-2}-r_{2k}}\right]\sqrt{\frac{\prod_{k=1}^{N/2-1}Q_{2k}(0)Q_{2k}(i/2)}{\prod_{k=1}^{N/2}\bar{Q}_{2k-1}(0)Q_{2k-1}(i/2)}}\times\sqrt{\frac{\det G^{+}}{\det G^{-}}}.\label{eq:twsp}
\end{equation}
The $\mathfrak{a},\mathfrak{b}_{\nu},x_{k},y_{\nu}$ are parameters
of the boundary state, the $Q_{\nu}$-s are the usual $Q$-functions,
$r_{\nu}$-s are the number of Bethe roots and $\det G^{\pm}$ are
the Gaudin-like determinants. The proof for (\ref{eq:utw}) when $M=\left\lfloor \frac{N}{2}\right\rfloor $
and for (\ref{eq:twso}) can be found in \cite{Gombor:2021hmj}. In
this paper we extend the derivation of (\ref{eq:utw}) for general
$M$. We also give strong arguments for the correctness of (\ref{eq:twsp}),
but leave the precise proof to a later work. 

The formulas can be applied for any representations of $\mathfrak{gl}(N)$
therefore they give the overlaps of the $SO(6)$ and the alternating
$SU(4)$ spin chains which describe the scalar sectors of the $\mathcal{N}=4$
SYM and the ABJM theories. We show that our formulas agree with the
conjectures of the overlaps appeared in \cite{Yang:2021hrl,Kristjansen:2023ysz,Jiang:2023cdm}.

The paper is organized as follows. In section \ref{sec:Definitions}
we briefly recall the definitions of the $\mathfrak{gl}(N)$ symmetric
spin chains. In section \ref{sec:Integrable-final-states} we show
the properties the untwisted $KT$-relation and derive a sum formula
for the off-shell overlaps which gives the proof of the on-shell formula
(\ref{eq:utw}). In section \ref{sec:Egzact-overlaps-for} we continue
with the twisted $KT$-relation and derive the corresponding sum formula
for the off-shell overlaps of the $\mathfrak{sp}(N)$ symmetric boundary
states. We show that the sum formula has the same embedding rules
as the Gaudin-like determinants which allows us to conjecture (\ref{eq:twsp}).
In section \ref{sec:Applications-for-AdS/CFT} we apply our on-shell
formulas for the states which appeared earlier in the AdS/CFT correspondence.

\section{Definitions\label{sec:Definitions}}

In this section we review the definitions of the $\mathfrak{gl}(N)$
symmetric spin chains. Let us start with our conventions for $\mathfrak{gl}(N)$
algebra and its representations. Let \textbf{$E_{i,j}$}-s be the
$N\times N$ unit matrices which have components $(E_{i,j})_{a,b}=\delta_{i,a}\delta_{j,b}$
and they are satisfy the $\mathfrak{gl}(N)$ Lie-algebra:
\begin{equation}
[E_{i,j},E_{k,l}]=\delta_{j,k}E_{i,l}-\delta_{i,l}E_{k,j}.
\end{equation}
For the $N$-tuples $\Lambda=(\Lambda_{1},\dots,\Lambda_{N})$ we
can define a highest weight representation $\mathcal{V}^{\Lambda}$.
Let $E_{i,j}^{\Lambda}\in\mathrm{End}(\mathcal{V}^{\Lambda})$ and
$|0^{\Lambda}\rangle\in\mathcal{V}^{\Lambda}$ be the corresponding
generators and highest weight state for which
\begin{equation}
\begin{split}E_{i,j}^{\Lambda}|0^{\Lambda}\rangle & =\Lambda_{i}|0^{\Lambda}\rangle,\qquad\text{for }i=1,\dots,N,\\
E_{i,j}^{\Lambda}|0^{\Lambda}\rangle & =0,\qquad\qquad\text{for }1\leq i<j\leq N.
\end{split}
\end{equation}

Let us continue with the definition of the corresponding Yangian algebra
$Y(N)$ \cite{Molev:1994rs}. We use the notations of \cite{Hutsalyuk:2017tcx,Hutsalyuk:2017way,Liashyk:2018egk,Hutsalyuk:2020dlw,Gombor:2021hmj}.
The Yangian algebra $Y(N)$ is generated by the monodromy matrix $T(u)=\sum_{i,j=1}^{N}E_{i,j}\otimes T_{i,j}(u)\in\mathrm{End}(\mathbb{C}^{N})\otimes Y(N)$
which satisfies the usual $RTT$-relation
\begin{equation}
R_{12}(u-v)T_{1}(u)T_{2}(v)=T_{2}(v)T_{1}(u)R_{12}(u-v),\label{eq:RTT}
\end{equation}
where we use the $\mathfrak{gl}(N)$ R-matrix
\begin{equation}
R(u)=u\mathbf{1}+c\mathbf{P},
\end{equation}
where $c$ is a constant and $\mathbf{I}$, $\mathbf{P}$ are the
identity and the permutation operators in the vector space $\mathbb{C}^{N}\otimes\mathbb{C}^{N}$.
We can define representations of the Yangian $Y(N)$ on a quantum
space $\mathcal{H}$. A representation is highest weight if there
exists a unique pseudo-vacuum $|0\rangle\in\mathcal{H}$ such that
\begin{equation}
\begin{split}T_{i,i}|0\rangle & =\lambda_{i}(u)|0\rangle,\qquad\text{for }i=1,\dots,N,\\
T_{j,i}|0\rangle & =0,\qquad\qquad\text{for }1\leq i<j\leq N.
\end{split}
\end{equation}
The $\lambda_{i}(u)$-s are the vacuum eigenvalues. The irreducible
representations of $\mathfrak{gl}(N)$ can be generalized for representations
of $Y(N)$. We can define the matrices (Lax-operators)
\begin{equation}
L^{\Lambda}(u)=\mathbf{1}+\frac{c}{u}\sum_{i,j=1}^{N}E_{i,j}\otimes E_{j,i}^{\Lambda}\in\mathrm{End}(\mathbb{C}^{N})\otimes\mathrm{End}(\mathcal{V}^{\Lambda}),\label{eq:Lax}
\end{equation}
which are solutions of the $RTT$-relation. 

Let us consider the following tensor product quantum space $\mathcal{H}=\mathcal{H}^{(1)}\otimes\mathcal{H}^{(2)}$
and define monodromy matrices on each sub-spaces $T^{(i)}(u)\in\mathrm{End}(\mathbb{C}^{N})\otimes\mathcal{H}^{(i)}$
for $i=1,2$. We can define a monodromy matrix (which satisfy the
$RTT$-relation) on the tensor product space as
\begin{equation}
T_{i,j}(u)=\sum_{a=1}^{N}T_{a,j}^{(1)}(u)\otimes T_{i,a}^{(2)}(u).\label{eq:coprod}
\end{equation}
The consequence of this co-product property is that we can build
more general monodromy matrices using the elementary ones (\ref{eq:Lax}):

\begin{equation}
T_{0}(u)=L_{0,J}^{\Lambda^{(J)}}(u-\xi_{J})\dots L_{0,1}^{\Lambda^{(1)}}(u-\xi_{1}),\label{eq:genmod}
\end{equation}
where $\xi_{j}\in\mathbb{C}$ are the inhomogeneities. We can series
expand the monodromy matrix around $u=\infty$ as
\begin{equation}
T_{i,j}(u)=\delta_{i,j}+\frac{c}{u}\Delta(E_{j,i})+\mathcal{O}(u^{-2}),\label{eq:seriesMon}
\end{equation}
where we introduced the co-product of the $\mathfrak{gl}(N)$ generators
\begin{equation}
\Delta(E_{i,j})=\sum_{k}\left(E_{i,j}^{\Lambda^{(k)}}\right)_{k},
\end{equation}
where the operators $\left(E_{i,j}^{\Lambda^{(k)}}\right)_{k}$ act
non-trivially only on the site $k$. For this monodromy matrix (\ref{eq:genmod})
the pseudo vacuum is
\begin{equation}
|0\rangle=|0^{\Lambda^{(1)}}\rangle\otimes\dots\otimes|0^{\Lambda^{(J)}}\rangle,
\end{equation}
for which
\begin{equation}
\Delta(E_{i,i})|0\rangle=\mathbf{\boldsymbol{\Lambda}}_{i}|0\rangle,
\end{equation}
where $\mathbf{\boldsymbol{\Lambda}}_{i}=\sum_{j=1}^{J}\Lambda_{i}^{(j)}$
are the $\mathfrak{gl}(N)$ weights of the pseudo-vacuum. 

We can also define the transfer matrix
\begin{equation}
\mathcal{T}(u)=\mathrm{tr}T(u),
\end{equation}
which gives commuting quantities
\begin{equation}
[\mathcal{T}(u),\mathcal{T}(v)]=0.
\end{equation}
For a given sets of complex numbers $\bar{t}^{\mu}=\{t_{k}^{\mu}\}_{k=1}^{r_{k}}$,
$\mu=1,\dots,N-1$, following \cite{Hutsalyuk:2017tcx}, one can define
off-shell Bethe vectors
\begin{equation}
\mathbb{B}(\bar{t})\equiv\mathbb{B}(\bar{t}^{1},\dots,\bar{t}^{N-1}).
\end{equation}
The recursion for the definition of the off-shell Bethe vectors can
be found in the appendix \ref{sec:Off-shell-Bethe-vectors}. The Bethe
vector is on-shell if the Bethe roots $\bar{t}^{\mu}$ satisfies the
Bethe Ansatz equations
\begin{equation}
\alpha_{\mu}(t_{k}^{\mu}):=\frac{\lambda_{\mu}(t_{k}^{\mu})}{\lambda_{\mu+1}(t_{k}^{\mu})}=\frac{f(t_{k}^{\mu},\bar{t}_{k}^{\mu})}{f(\bar{t}_{k}^{\mu},t_{k}^{\mu})}\frac{f(\bar{t}^{\mu+1},t_{k}^{\mu})}{f(t_{k}^{\mu},\bar{t}^{\mu-1})},\label{eq:BE}
\end{equation}
where we used the following notations
\begin{equation}
\begin{split}g(u,v) & =\frac{c}{u-v},\qquad h(u,v)=\frac{f(u,v)}{g(u,v)},\\
f(u,v) & =1+g(u,v)=\frac{u-v+c}{u-v},\qquad\bar{t}_{k}^{\mu}=\bar{t}^{\mu}\backslash t_{k}^{\mu},\\
f(u,\bar{t}^{i}) & =\prod_{k=1}^{r_{i}}f(u,t_{k}^{i}),\quad f(\bar{t}^{i},u)=\prod_{k=1}^{r_{i}}f(t_{k}^{i},u),\quad f(\bar{t}^{i},\bar{t}^{j})=\prod_{k=1}^{r_{i}}f(t_{k}^{i},\bar{t}^{j}).
\end{split}
\end{equation}
The on-shell Bethe vectors are eigenvectors of the transfer matrix
\begin{equation}
\mathcal{T}(u)\mathbb{B}(\bar{t})=\tau(u|\bar{t})\mathbb{B}(\bar{t}),
\end{equation}
with the eigenvalue
\begin{equation}
\tau(u|\bar{t})=\sum_{i=1}^{N}\lambda_{i}(u)f(\bar{t}^{i},u)f(u,\bar{t}^{i-1}),\label{eq:eig}
\end{equation}
where $r_{0}=r_{N}=0$.

One can also define the left eigenvectors of the transfer matrix
\begin{equation}
\mathbb{C}(\bar{t})\mathcal{T}(u)=\tau(u|\bar{t})\mathbb{C}(\bar{t}),
\end{equation}
and the square of the norm of the on-shell Bethe states satisfies
the Gaudin hypothesis \cite{Hutsalyuk:2017way}
\begin{equation}
\mathbb{C}(\bar{t})\mathbb{B}(\bar{t})=\frac{\prod_{\nu=1}^{N-1}\prod_{k\neq l}f(t_{l}^{\nu},t_{k}^{\nu})}{\prod_{\nu=1}^{N-2}f(\bar{t}^{\nu+1},\bar{t}^{\nu})}\det G,\label{eq:norm}
\end{equation}
where $G$ is the Gaudin matrix given by
\begin{equation}
G_{j,k}^{(\mu,\nu)}=-c\frac{\partial\log\Phi_{j}^{(\mu)}}{\partial t_{k}^{\nu}},\label{eq:GM}
\end{equation}
where we defined the expressions
\begin{equation}
\Phi_{k}^{(\mu)}=\alpha_{\mu}(t_{k}^{\mu})\frac{f(\bar{t}_{k}^{\mu},t_{k}^{\mu})}{f(t_{k}^{\mu},\bar{t}_{k}^{\mu})}\frac{f(t_{k}^{\mu},\bar{t}^{\mu-1})}{f(\bar{t}^{\mu+1},t_{k}^{\mu})}.\label{eq:Phidef}
\end{equation}

We can also define an other monodromy matrix which satisfies the same
$RTT$-algebra \cite{Liashyk:2018egk}. This transfer matrix can be
obtained from one of the quantum minors as
\begin{align}
\widehat{T}_{N+1-j,N+1-i}(u) & =(-1)^{i+j}t_{1,\dots,\hat{i},\dots,N}^{1,\dots,\hat{j},\dots,N}(u-c)\mathrm{qdet}(T(u))^{-1},\\
t_{b_{1},b_{2},\dots,b_{m}}^{a_{1},a_{2},\dots,a_{m}}(u) & =\sum_{p}\mathrm{sgn}(p)T_{a,b_{p(1)}}(u)T_{a,b_{p(2)}}(u-c)\dots T_{a,b_{p(m)}}(u-(m-1)c),\label{eq:Qminors}\\
\mathrm{qdet}(T(u)) & =t_{1,2,\dots,N}^{1,2,\dots,N}(u).
\end{align}
Here $\hat{i}$ an\^{d} $\hat{j}$ mean that the corresponding indices
are omitted. We call $\widehat{T}$ as twisted monodromy matrix. The
twisted monodromy matrix $\widehat{T}$ is also a highest weight representation
of $Y(N)$, i.e.,
\begin{equation}
\begin{split}\widehat{T}_{i,i}\left|0\right\rangle  & =\hat{\lambda}_{i}(u)\left|0\right\rangle ,\qquad\text{for }i=1,\dots,N,\\
\widehat{T}_{j,i}\left|0\right\rangle  & =0,\qquad\qquad\text{for }1\leq i<j\leq N,
\end{split}
\end{equation}
where the pseudo-vacuum eigenvalues are
\begin{equation}
\hat{\lambda}_{i}(u)=\frac{1}{\lambda_{N-i+1}(u-(N-i)c)}\prod_{k=1}^{N-i}\frac{\lambda_{k}(u-kc)}{\lambda_{k}(u-(k-1)c)}.\label{eq:lambdah}
\end{equation}
The ratios of the vacuum eigenvalues have the following form
\begin{equation}
\hat{\alpha}_{i}(u)=\frac{\hat{\lambda}_{i}(u)}{\hat{\lambda}_{i+1}(u)}=\alpha_{N-i}(u-(N-i)c).\label{eq:idAlph}
\end{equation}
The twisted monodromy matrix $\widehat{T}$ is similar to the inverse
of the original monodromy matrix $T$:
\begin{equation}
V\widehat{T}^{t}(u)VT(u)=1,\label{eq:invemon}
\end{equation}
where $V$ is an off-diagonal $N\times N$ matrix of the auxiliary
space with the components $V_{i,j}=\delta_{i,N+1-j}$ and the superscript
$t$ is the transposition in the auxiliary space, i.e. $\left[\widehat{T}^{t}(u)\right]_{i,j}=\widehat{T}_{j,i}(u)$.
Applying this equation to the $RTT$-relation we obtain the $R\widehat{T}T$-relation
\begin{equation}
\bar{R}_{1,2}(u-v)\widehat{T}_{1}(u)T_{2}(v)=T_{2}(v)\widehat{T}_{1}(u)\bar{R}_{1,2}(u-v),\label{eq:RTbT}
\end{equation}
where we used the crossed $R$-matrix
\begin{equation}
\bar{R}_{1,2}(u)=V_{2}R_{1,2}^{t_{2}}(-u)V_{2}.
\end{equation}
One can also define the corresponding twisted transfer matrix as
\begin{equation}
\widehat{\mathcal{T}}(u)=\mathrm{tr}\widehat{T}(u)=\sum_{i=1}^{N}\widehat{T}_{i,i}(u).
\end{equation}
From the $R\widehat{T}T$-relation (\ref{eq:RTbT}) on can derive
that the original and twisted monodromy matrices are commuting
\begin{equation}
[\mathcal{T}(u),\widehat{\mathcal{T}}(v)]=0,
\end{equation}
therefore they have common eigenvectors. Let $\hat{\mathbb{B}}(\bar{t})$
be the off-shell Bethe vector generated from $\widehat{T}_{i,j}$.
In \cite{Liashyk:2018egk} the connection between the Bethe vectors
$\mathbb{B}(\bar{t})$ and $\hat{\mathbb{B}}(\bar{t})$ was determined
\begin{equation}
\hat{\mathbb{B}}(\bar{t})=(-1)^{\#\bar{t}}\left(\prod_{s=1}^{N-2}f(\bar{t}^{s+1},\bar{t}^{s})\right)^{-1}\mathbb{B}(\mu(\bar{t})),\label{eq:connBhatB}
\end{equation}
where
\begin{equation}
\mu(\bar{t})=\{\bar{t}^{N-1}-c,\bar{t}^{N-2}-2c,\dots,\bar{t}^{1}-(N-1)c\}.
\end{equation}
The eigenvalue of the twisted transfer matrix
\begin{equation}
\widehat{\mathcal{T}}(u)\mathbb{B}(\bar{t})=\hat{\tau}(u|\bar{t})\mathbb{B}(\bar{t})
\end{equation}
has the following form
\begin{equation}
\hat{\tau}(u|\bar{t})=\sum_{i=1}^{N}\hat{\lambda}_{i}(u)f(\bar{t}^{N-i}+(N-i)c,u)f(u,\bar{t}^{N-i+1}+(N-i+1)c).\label{eq:tweig}
\end{equation}

From the co-product form of the monodromy matrix (\ref{eq:genmod}),
we can obtain the co-product form of twisted monodromy matrix using
(\ref{eq:invemon})
\begin{equation}
\widehat{T}_{0}(u)=\widehat{L}_{0,J}^{\Lambda^{(J)}}(u-\xi_{J})\dots\widehat{L}_{0,1}^{\Lambda^{(1)}}(u-\xi_{1}),
\end{equation}
where
\begin{equation}
\widehat{L}_{0,1}^{\Lambda}(u)=V_{0}\left(\left(\widehat{L}_{0,1}^{\Lambda}(u)\right)^{-1}\right)^{t_{0}}V_{0}.\label{eq:crossedLax}
\end{equation}
For rectangular Young diagrams where $\Lambda_{j}=s$ for $j\leq a$,
$\Lambda_{j}=0$ for $j\geq a$ let us introduce the notation $L^{(s,a)}(u)$
for the Lax-operators. For these representations the Lax-operators
satisfy the unitarity property
\begin{equation}
L^{(s,a)}(u)L^{(s,a)}(-u-c(s-a))=\frac{(u+cs)(u-ca)}{u(u+c(s-a))}\mathbf{1},
\end{equation}
therefore
\begin{equation}
\widehat{L}_{0,1}^{(s,a)}(u)=\frac{u(u+c(s-a))}{(u+cs)(u-ca)}V_{0}\left(L_{0,1}^{(s,a)}(-u-c(s-a))\right)^{t_{0}}V_{0}.
\end{equation}

\section{Exact overlaps for untwisted boundary states\label{sec:Integrable-final-states}}

In this section we review the so-called $KT$-relation which serves
as the defining relation for the integrable boundary states \cite{Gombor:2021hmj}.
One can introduce the $KT$-relation in two different ways, and we
analyze the untwisted and the twisted $KT$-relations, separately.
The main advantage of these relations is that we can replace the creation
operators in the Bethe vectors with the annihilation ones, which opens
the way to calculate overlaps between Bethe and boundary states by
recursion. In this section we concentrate on the untwisted case. In
\cite{Gombor:2021hmj} the formulas for on-shell overlaps were derived
for a subset of integrable boundary states. In this section we review
these results and generalize them for all integrable boundary states.

\subsection{Definition of the untwisted integrable states}

The untwisted integrable boundary states $\langle\Psi|$ are defined
by the following untwisted $KT$-relation
\begin{equation}
K_{0}(u)\langle\Psi|T_{0}(u)=\langle\Psi|T_{0}(-u)K_{0}(u),\label{eq:KT_Utw}
\end{equation}
where $K(u)$ is an invertible $N\times N$ matrix of the auxiliary
space \footnote{Since the defining equation of the monodromy matrix ($RTT$-equation
(\ref{eq:RTT})) is homogeneous on $T$, the renormalized monodromy
matrix $\tilde{T}(z)=\tilde{\lambda}(z)T(z)$ is also a solution of
the $RTT$-relation. However this renormalized monodromy matrix satisfies
only a renormalized $KT$-relation therefore using the definition
(\ref{eq:KT_Utw}) we partially fixed the normalization freedom to
the symmetric functions $\tilde{\lambda}(-z)=\tilde{\lambda}(z)$.}. We define the twisted $KT$-relation in the next section. In \cite{Gombor:2021hmj}
it was showed that the consistency of the definition (\ref{eq:KT_Utw})
requires the reflection equation for the $K$-matrix
\begin{equation}
R_{1,2}(u-v)K_{1}(-u)R_{1,2}(u+v)K_{2}(-v)=K_{2}(-v)R_{1,2}(u+v)K_{1}(-u)R_{1,2}(u-v).\label{eq:refleq-2}
\end{equation}

In the following we analyze the $KT$-relation in more detail. We
investigate the properties of the possible $K$-matrices, show the
relation between $K$-matrices and boundary states and also show the
consequences of the $KT$-relation for the on-shell overlaps.

\subsubsection{Integrable $K$-matrices and their regular forms}

The most general solution of the reflection equation (\ref{eq:refleq-2})
is well known \cite{Arnaudon:2004sd}: 
\begin{equation}
K(u)=\frac{\mathfrak{a}}{u}\mathbf{1}+\mathcal{U},\label{eq:Kdef_UTw}
\end{equation}
where $\mathcal{U}$ is an $N\times N$ involution matrix, i.e.,
\begin{equation}
\mathcal{U}^{2}=\mathbf{1}.
\end{equation}
The constant $\mathfrak{a}\in\mathbb{C}$ is a free parameter of the
$K$-matrix. We note that there is another type of solutions of the
reflection equation for which $\mathcal{U}^{2}=0$, and we call them
singular solutions.

Since the monodromy matrix has $\mathfrak{gl}(N)$ symmetry
\begin{equation}
T_{0}(u)=G_{0}\Delta(G)T_{0}(u)G_{0}^{-1}\Delta(G^{-1}),
\end{equation}
(where $G\in GL(N)$ and $\Delta(G)\in\mathrm{End}(\mathcal{H})$
is the co-product of $G$) we can obtain transformed $K$-matrices
and boundary states 
\begin{equation}
K^{G}(u)=G_{0}^{-1}K_{0}(u)G_{0},\qquad\langle\Psi^{G}|=\langle\Psi|\Delta(G),
\end{equation}
which also satisfy the untwisted $KT$-relation:
\begin{equation}
K^{G}(u)\langle\Psi^{G}|T_{0}(u)=\langle\Psi^{G}|T_{0}(-u)K^{G}(u).
\end{equation}
One can diagonalize the involution $\mathcal{U}$ as
\begin{equation}
G^{-1}\mathcal{U}G=\mathrm{diag}(\underbrace{-1,\dots,-1}_{M},\underbrace{+1,\dots,+1}_{N-M}),\label{eq:sign}
\end{equation}
where the numbers of $-1$-s and $1$ are $M$ and $N-M$, respectively.
We say that the matrix $\mathcal{U}$ has signature $(M,N-M)$ and
we call it type $(N,M)$ involution. We can see that the $K$-matrix
commutes with a $\mathfrak{gl}(M)\oplus\mathfrak{gl}(N-M)$ subalgebra
of the original algebra $\mathfrak{gl}(N)$ therefore we call this
$K$-matrix and the corresponding boundary state $\langle\Psi|$ the
$\mathfrak{gl}(M)\oplus\mathfrak{gl}(N-M)$ symmetric $K$-matrix
and boundary state. Without loss of generality we can assume that
$M\leq\frac{N}{2}$. We also call the K-matrix and boundary states
with signature (\ref{eq:sign}) as type $(N,M)$ K-matrix and boundary
states. 

At the end of the day, we are interested in on-shell overlaps $\langle\Psi|\mathbb{B}(\bar{t})$.
We know that the on-shell Bethe states $\mathbb{B}(\bar{t})$ are
highest weight states, i.e.,
\begin{equation}
\Delta(E_{i,j})\mathbb{B}(\bar{t})=0,\quad\text{for }i<j.
\end{equation}
Applying this property we obtain that the on-shell overlaps are invariant
under $GL(N)$ transformations corresponding these generators
\begin{equation}
\langle\Psi^{G}|\mathbb{B}(\bar{t})=\langle\Psi|\Delta(G)\mathbb{B}(\bar{t})=\langle\Psi|\mathbb{B}(\bar{t}),
\end{equation}
where $G=\exp(\varphi E_{i,j})$ for $i<j$. In the following let
us try to apply such transformations, which leave the on-shell overlap
invariant, to obtain a fixed form for different $K$-matrices and
involutions $\mathcal{U}$ with the same type.

Let us define $GL(N)$ transformations which contains only generators
$E_{i,j}$ where $i<j$ (rising operators):
\begin{align}
G & =\exp(\sum_{c=2}^{N-1}\varphi_{c}E_{1,c})\exp(\sum_{c=2}^{N-1}\phi_{c}E_{c,N})\exp(\Phi E_{1,N}),
\end{align}
where the angles are fixed by
\begin{equation}
\varphi_{c}=-\frac{\mathcal{U}_{N,c}}{\mathcal{U}_{N,1}},\quad\phi_{c}=\frac{\mathcal{U}_{c,1}}{\mathcal{U}_{N,1}},\quad\Phi=-\frac{\mathcal{U}_{N,N}-s}{\mathcal{U}_{N,1}},\quad s=\pm1.
\end{equation}
We saw that the transformation $\langle\Psi|\Delta(G)$ does not change
the on-shell overlaps. The corresponding transformed $\mathcal{U}$-matrix
is
\begin{equation}
\mathcal{U}^{(2)}=G^{-1}\mathcal{U}G,
\end{equation}
which has the following components
\begin{align}
\mathcal{U}_{1,1}^{(2)} & =-s, & \mathcal{U}_{1,b}^{(2)} & =0, & \mathcal{U}_{1,N}^{(2)} & =0,\nonumber \\
\mathcal{U}_{a,1}^{(2)} & =0, & \mathcal{U}_{a,b}^{(2)} & =\mathcal{U}_{a,b}-\frac{\mathcal{U}_{a,1}\mathcal{U}_{N,b}}{\mathcal{U}_{N,1}}, & \mathcal{U}_{a,N}^{(2)} & =0,\label{eq:reg2}\\
\mathcal{U}_{N,1}^{(2)} & =\mathcal{U}_{N,1}, & \mathcal{U}_{N,b}^{(2)} & =0, & \mathcal{U}_{N,N}^{(2)} & =s,\nonumber 
\end{align}
where $a,b=2,\dots,N-1$. Without limiting the generality, let us
choose the convention $s=+1$. We can see that the new matrix $\mathcal{U}^{(2)}$
has a block diagonal form and the matrix $\left\{ \mathcal{U}_{a,b}^{(2)}\right\} _{a,b=2}^{N-1}$
in the $N-2$ dimensional invariant subspace is a type $(N-2,M-1)$
involution matrix. Applying the analog transformation on the $N-2$
dimensional invariant subspace, we obtain an $N-4$ dimensional invariant
subspace. Reaping this method, we obtain a series of matrices
\begin{equation}
\mathcal{U}^{(k+1)}=\left(G^{(k)}\right)^{-1}\mathcal{U}^{(k)}G^{(k)},\quad\mathcal{U}^{(1)}=\mathcal{U},
\end{equation}
where the $GL(N)$ transformations are
\begin{equation}
G^{(k)}=\exp(\sum_{c=k+1}^{N-k}\varphi_{c}^{(k)}E_{k,c})\exp(\sum_{c=k+1}^{N-k}\phi_{c}^{(k)}E_{c,N+1-k})\exp(\Phi^{(k)}E_{k,N+1-k}),\label{eq:GLN}
\end{equation}
with
\begin{equation}
\varphi_{c}^{(k)}=-\frac{\mathcal{U}_{N+1-k,c}^{(k)}}{\mathcal{U}_{N+1-k,k}^{(k)}},\quad\phi_{c}^{(k)}=\frac{\mathcal{U}_{c,k}^{(k)}}{\mathcal{U}_{N+1-k,k}^{(k)}},\quad\Phi^{(k)}=-\frac{\mathcal{U}_{N+1-k,N+1-k}^{(k)}-1}{\mathcal{U}_{N+1-k,1}^{(k)}}.
\end{equation}
The components of the matrices $\mathcal{U}^{(k)}$ have the following
recursion
\begin{align}
\mathcal{U}_{\alpha,\beta}^{(k+1)} & =-\delta_{\alpha,\beta}, & \mathcal{U}_{\alpha,b}^{(k+1)} & =0, & \mathcal{U}_{\alpha,\bar{\beta}}^{(2)} & =0,\nonumber \\
\mathcal{U}_{a,\beta}^{(k+1)} & =0, & \mathcal{U}_{a,b}^{(k+1)} & =\mathcal{U}_{a,b}^{(k)}-\frac{\mathcal{U}_{a,k}^{(k)}\mathcal{U}_{N+1-k,b}^{(k)}}{\mathcal{U}_{N+1-k,k}^{(k)}}, & \mathcal{U}_{a,\bar{\beta}}^{(2)} & =0,\\
\mathcal{U}_{\bar{\alpha},\beta}^{(k+1)} & =\mathcal{U}_{\bar{\alpha},\beta}^{(\beta)}\delta_{N+1-\bar{\alpha},\beta}, & \mathcal{U}_{\bar{\alpha},b}^{(k+1)} & =0, & \mathcal{U}_{\bar{\alpha},\bar{\beta}}^{(2)} & =+\delta_{\bar{\alpha},\bar{\beta}},\nonumber 
\end{align}
where $\alpha,\beta=1,\dots,k$, $a,b=k+1,\dots,N-k$, $\bar{\alpha},\bar{\beta}=N-k+1,\dots,N$.
We can see that the matrix $\left\{ \mathcal{U}_{a,b}^{(k+1)}\right\} _{a,b=k+1}^{N-k}$
in the $N-2k$ dimensional invariant subspace is a type $(N-2k,M-k)$
involution matrix. We can use this definition as long as $\mathcal{U}_{N+1-k,k}^{(k)}\neq0$,
which is equivalent to that the $N-2k+2$ dimensional matrix $\left\{ \mathcal{U}_{a,b}^{(k)}\right\} _{a,b=k}^{N+1-k}$
is not a type $(N-2k+2,0)$ involution matrix which is the identity,
i.e.  for the type $(N,M)$ involution the recursion stops at the
$k=M+1$ step and we have the involution matrix
\begin{align}
\mathcal{U}_{\alpha,\beta}^{(M+1)} & =-\delta_{\alpha,\beta}, & \mathcal{U}_{\alpha,b}^{(M+1)} & =0, & \mathcal{U}_{\alpha,\bar{\beta}}^{(M+1)} & =0,\nonumber \\
\mathcal{U}_{a,\beta}^{(M+1)} & =0, & \mathcal{U}_{a,b}^{(M+1)} & =+\delta_{a,b}, & \mathcal{U}_{a,\bar{\beta}}^{(M+1)} & =0,\\
\mathcal{U}_{\bar{\alpha},\beta}^{(M+1)} & =\mathfrak{b}_{\beta}\delta_{N+1-\bar{\alpha},\beta}, & \mathcal{U}_{\bar{\alpha},b}^{(M+1)} & =0, & \mathcal{U}_{\bar{\alpha},\bar{\beta}}^{(M+1)} & =+\delta_{\bar{\alpha},\bar{\beta}},\nonumber 
\end{align}
where $\alpha,\beta=1,\dots,M$, $a,b=M+1,\dots,N-M$, $\bar{\alpha},\bar{\beta}=N-M+1,\dots,N$
and 
\begin{equation}
\mathfrak{b}_{\beta}=\mathcal{U}_{N+1-\beta,\beta}^{(\beta)}.\label{eq:bdef}
\end{equation}
Hereafter, we call the matrix $\mathcal{U}^{(M+1)}$ (and the corresponding
$K$-matrix $K^{(M+1)}(z)=\frac{\mathfrak{a}}{z}\mathbf{1}+\mathcal{U}^{(M+1)}$
and boundary state $\langle\Psi^{(M+1)}|$) the regular form of the
type $(N,M)$ involution $\mathcal{U}$ (and K-matrix $K(z)$ and
boundary state $\langle\Psi|$). Since the $GL(N)$ transformations
(\ref{eq:GLN}) contain only rising operators, the on-shell overlaps
with the regular and the original boundary states are the same
\begin{equation}
\langle\Psi^{(M+1)}|\mathbb{B}(\bar{t})=\langle\Psi|\prod_{k=1}^{M}\Delta(G^{(k)})\mathbb{B}(\bar{t})=\langle\Psi|\mathbb{B}(\bar{t}).
\end{equation}
During the calculation of the overlaps, we concentrate on $K$-matrices
with regular forms. 

\subsubsection{Solutions of the untwisted $KT$-relations}

During the derivations of the overlaps, we do not need to specify
$\langle\Psi|$, it is enough to require the constraint (\ref{eq:KT_Utw}).
In the following we demonstrate that there exist non-trivial boundary
states $\langle\Psi|$ for all $K$-matrix and we show the connections
to the solutions of the reflection equations in different representations. 

We can build integrable boundary states using the co-product property
(\ref{eq:coprod}). For the tensor product quantum space $\mathcal{H}=\mathcal{H}^{(1)}\otimes\mathcal{H}^{(2)}$,
if the $\langle\Psi^{(1)}|\in\mathcal{H}^{(1)}$ and $\langle\Psi^{(2)}|\in\mathcal{H}^{(2)}$
are integrable boundary states with the same $K$-matrix $K(u)$,
i.e., the $KT$-relations are satisfied
\begin{equation}
K_{0}(u)\langle\Psi^{(i)}|T_{0}^{(i)}(u)=\langle\Psi^{(i)}|T_{0}^{(i)}(-u)K_{0}(u),\quad\text{for }i=1,2,
\end{equation}
then the tensor product state
\begin{equation}
\langle\Psi|=\langle\Psi^{(1)}|\otimes\langle\Psi^{(2)}|\in\mathcal{H}
\end{equation}
satisfies the $KT$-relation (\ref{eq:KT_Utw}) with the same $K$-matrix
$K(u)$ (see \cite{Gombor:2021hmj}). The consequence of this co-product
property is that we can build integrable boundary states as tensor
products of integrable two-site states. 

\paragraph{Elementary two-site states for the defining representations}

The simplest example is that where the quantum space is the tensor
product of the defining and its contra-gradient representations for
which the monodromy matrix is
\begin{equation}
T_{0}(u)=\bar{L}_{0,2}(u+\theta)L_{0,1}(u-\theta),\label{eq:elemUTwmon}
\end{equation}
where $L(u)$ is Lax operator (\ref{eq:Lax}) for $\Lambda=(1,0,\dots,0)$
and $\bar{L}(u)$ is Lax operator for $\Lambda=(0,\dots,0,-1)$, i.e.
\begin{equation}
L(u)=\mathbf{1}+\frac{c}{u}\sum_{i,j=1}^{N}E_{i,j}\otimes E_{j,i},\qquad\bar{L}(u)=\mathbf{1}-\frac{c}{u}\sum_{i,j=1}^{N}E_{i,j}\otimes E_{N+1-i,N+1-j}.
\end{equation}
For this ''elementary'' monodromy matrix the KT-relation reads as
\begin{equation}
K_{0}(u)\langle\psi(\theta)|\bar{L}_{0,2}(u+\theta)L_{0,1}(u-\theta)=\langle\psi(\theta)|\bar{L}_{0,2}(-u+\theta)L_{0,1}(-u-\theta)K_{0}(u).\label{eq:elemUtwKT}
\end{equation}
The ''elementary'' two-site state $\langle\psi(\theta)|$ acts on
the sites $1,2$. We can use the natural basis vectors $e_{i}$ in
$\mathbb{C}^{N}$ for which $E_{i,j}e_{k}=\delta_{j,k}e_{i}$. The
''elementary'' two-site state can be expressed as $\langle\psi(\theta)|=\sum\psi_{i,j}(\theta)\left(e_{i}\right)^{t}\otimes\left(e_{j}\right)^{t}$.
Using the equivalent matrix form of the two site state $\psi(\theta)=\sum\psi_{j,i}(\theta)E_{i,j}$
the $KT$-relation reads (\ref{eq:elemUtwKT}) as
\begin{equation}
K_{0}(u)\bar{L}_{0,1}(u+\theta)^{t_{1}}\psi_{1}(\theta)L_{0,1}(u-\theta)=\bar{L}_{0,1}(-u+\theta)^{t_{1}}\psi_{1}(\theta)L_{0,1}(-u-\theta)K_{0}(u).
\end{equation}
Since
\begin{align}
\bar{L}_{0,1}(u)^{t_{1}} & =\mathbf{1}-\frac{c}{u}\sum_{i,j=1}^{N}\left(E_{i,j}\right)_{0}\otimes\left(E_{N+1-i,N+1-j}\right)_{1}^{t}=\mathbf{1}-\frac{c}{u}\sum_{i,j=1}^{N}\left(E_{i,j}\right)_{0}\otimes\left(E_{N+1-j,N+1-i}\right)_{1}\nonumber \\
 & =V_{1}L_{0,1}(-u)V_{1},
\end{align}
the $KT$-relation simplifies as
\begin{equation}
K_{0}(u)L_{0,1}(-u-\theta)V_{1}\psi_{1}(\theta)L_{0,1}(u-\theta)=L_{0,1}(u-\theta)V_{1}\psi_{1}(\theta)L_{0,1}(-u-\theta)K_{0}(u).
\end{equation}
Since the permutation operator has the form $\mathbf{P}=\sum_{i,j}E_{i,j}\otimes E_{j,i}$
the Lax operator is just the $R$-matrix $L(u)=\frac{1}{u}R(u)$ therefore
the $KT$-relation is simplified as
\begin{equation}
K_{0}(u)R_{0,1}(-u-\theta)V_{1}\psi_{1}(\theta)R_{0,1}(u-\theta)=R_{0,1}(u-\theta)V_{1}\psi_{1}(\theta)R_{0,1}(-u-\theta)K_{0}(u).
\end{equation}
We just obtained the original reflection equation (\ref{eq:refleq-2})
therefore the two-site satisfies the ''elementary'' KT-relation (\ref{eq:elemUtwKT})
if
\begin{equation}
V\psi(\theta)=K(\theta)\quad\to\quad\psi_{j,i}(\theta)=K_{N+1-i,j}(\theta).\label{eq:elemTwoSite}
\end{equation}

\paragraph{Elementary two-site states for general representations}

We can generalize the two-site states and the corresponding $KT$-relation
for any representation $\Lambda$ as
\begin{equation}
K_{0}(u)\langle\psi^{\Lambda}(\theta)|L_{0,2}^{\bar{\Lambda}}(u+\theta)L_{0,1}^{\Lambda}(u-\theta)=\langle\psi^{\Lambda}(\theta)|L_{0,2}^{\bar{\Lambda}}(-u+\theta)L_{0,1}^{\Lambda}(-u-\theta)K_{0}(u),\label{eq:elementKT}
\end{equation}
where $\Lambda=(\Lambda_{1},\dots,\Lambda_{N})$ and $\Lambda=(-\Lambda_{N},\dots,-\Lambda_{1})$.
For representations $\Lambda$ (even for infinite dimensional ones)
for which there is a non-trivial solution of the above equation, we
can build general integrable two-site states as
\begin{equation}
\langle\Psi|=\langle\psi^{\Lambda_{1}}(\theta_{1})|\otimes\langle\psi^{\Lambda_{2}}(\theta_{2})|\otimes\dots\otimes\langle\psi^{\Lambda_{J/2}}(\theta_{J/2})|.
\end{equation}
In the following we demonstrate that there exist two-site states for
any finite dimensional irreps.

At first, let us choose a basis $e_{i}^{\Lambda}$ in $\mathcal{V}^{\Lambda}$
and their co-vectors are $\left(e_{i}^{\Lambda}\right)^{t}$. We can
express the two-site state as $\langle\psi^{\Lambda}(\theta)|=\sum\psi_{i,j}(\theta)\left(e_{i}^{\Lambda}\right)^{t}\otimes\left(e_{j}^{\Lambda}\right)^{t}$.
Using the equivalent matrix form of the two-site state $\psi^{\Lambda}(\theta)=\sum\psi_{j,i}^{\Lambda}(\theta)e_{i}^{\Lambda}\otimes\left(e_{j}^{\Lambda}\right)^{t}$,
the $KT$-relation (\ref{eq:elementKT}) reads as
\begin{equation}
K_{0}(u)L_{0,1}^{\bar{\Lambda}}(u+\theta)^{t_{1}}\psi_{1}^{\Lambda}(\theta)L_{0,1}^{\Lambda}(u-\theta)=L_{0,1}^{\bar{\Lambda}}(-u+\theta)^{t_{1}}\psi_{1}^{\Lambda}(\theta)L_{0,1}^{\Lambda}(-u-\theta)K_{0}(u).\label{eq:kkk}
\end{equation}
Let us look at the Lax operator $L^{\bar{\Lambda}}$ in more detail.
It contains the highest weight representation $E_{i,j}^{\bar{\Lambda}}$.
Let us define the following sets of operators
\begin{equation}
\bar{E}_{i,j}^{\Lambda}=-E_{N+1-j,N+1-i}^{\Lambda},
\end{equation}
which is also a highest weight irrep of $\mathfrak{gl}(N)$ and the
highest weights are $(-\Lambda_{N},\dots,-\Lambda_{1})$ therefore
we can choose the convention for the generators $E_{i,j}^{\bar{\Lambda}}$
as $E_{i,j}^{\bar{\Lambda}}=-E_{N+1-j,N+1-i}^{\Lambda}$. We also
choose the convention $\left(E_{i,j}^{\Lambda}\right)^{t}=E_{j,i}^{\Lambda}$
therefore 
\begin{equation}
L_{0,1}^{\bar{\Lambda}}(u)^{t_{1}}=\mathbf{1}-\frac{c}{u}\sum_{i,j=1}^{N}\left(E_{i,j}\right)_{0}\otimes\left(E_{N+1-j,N+1-i}^{\Lambda}\right)_{1}.
\end{equation}
The set of operators $\tilde{E}_{i,j}^{\Lambda}=E_{N+1-i,N+1-j}^{\Lambda}$
is also an irrep of $\mathfrak{gl}(N)$ but the state $|0^{\Lambda}\rangle$
is now a lowest weight state i.e. $\tilde{E}_{i,j}|0^{\Lambda}\rangle=0$
for $i>j$ and the lowest weights are $(\Lambda_{N},\dots,\Lambda_{1})$.
Since every finite dimensional irreps are highest weight reps therefore
there exists a highest weight state $|\tilde{0}^{\Lambda}\rangle$
with highest weights $(\Lambda_{1},\dots,\Lambda_{N})$. Since the
highest weight irreps are unique there exists a similarity transformation
$V^{\Lambda}$ for which $\tilde{E}_{i,j}^{\Lambda}=V^{\Lambda}E_{i,j}^{\Lambda}V^{\Lambda}$,
i.e.,
\begin{equation}
L_{0,1}^{\bar{\Lambda}}(u)^{t_{1}}=\mathbf{1}-\frac{c}{u}\sum_{i,j=1}^{N}\left(E_{i,j}\right)_{0}\otimes\left(V^{\Lambda}E_{j,i}^{\Lambda}V^{\Lambda}\right)_{1}=V_{1}^{\Lambda}L_{0,1}^{\Lambda}(-u)V_{1}^{\Lambda}.
\end{equation}
Substituting back to the equation (\ref{eq:kkk}), we obtain that
\begin{equation}
K_{0}(u)L_{0,1}^{\Lambda}(-u-\theta)V_{1}^{\Lambda}\psi_{1}^{\Lambda}(\theta)L_{0,1}^{\Lambda}(u-\theta)=L_{0,1}^{\Lambda}(u-\theta)V_{1}^{\Lambda}\psi_{1}^{\Lambda}(\theta)L_{0,1}^{\Lambda}(-u-\theta)K_{0}(u),
\end{equation}
which is equivalent to the reflection equation 
\begin{equation}
K_{0}(-u)L_{0,1}^{\Lambda}(\theta+u)K_{1}^{\Lambda}(-\theta)L_{0,1}^{\Lambda}(\theta-u)=L_{0,1}^{\Lambda}(\theta-u)K_{1}^{\Lambda}(-\theta)L_{0,1}^{\Lambda}(\theta+u)K_{0}(-u),
\end{equation}
where we used the notation
\begin{equation}
K_{1}^{\Lambda}(\theta)=V_{1}^{\Lambda}\psi_{1}^{\Lambda}(\theta)\quad\to\quad\psi_{j,i}^{\Lambda}(\theta)=\sum_{k}V_{i,k}^{\Lambda}K_{k,j}^{\Lambda}(\theta).
\end{equation}
Since the reflection equation has non-trivial solution for any finite
dimensional representation $\Lambda$ (e.g. it can be obtained from
the original K-matrix $K(u)$ using the fusion procedure \cite{fusion-open-chains})
we just showed that the two-site states exist for any finite dimensional
irreps. However, it is worth to note that the original equation (\ref{eq:elementKT})
also has infinite dimensional solutions and our later derivations
are independent from the quantum space therefore the results are valid
even for the infinite dimensional integrable two-site states. In section
\ref{subsec:Boundary-state-with} we show an example for infinite
dimensional boundary state.

\paragraph{Symmetric properties of the vacuum eigenvalues}

We showed that we can build general integrable states as
\begin{equation}
\langle\Psi|=\langle\psi^{\Lambda^{(1)}}(\theta_{1})|\otimes\langle\psi^{\Lambda^{(2)}}(\theta_{2})|\otimes\dots\otimes\langle\psi^{\Lambda^{(J)}}(\theta_{J})|,\label{eq:Psidef}
\end{equation}
where the monodromy matrices are defined as
\begin{equation}
T_{0}(u)=L_{0,2J}^{\bar{\Lambda}^{(J)}}(u+\theta_{J})L_{0,2J-1}^{\Lambda^{(J)}}(u-\theta_{J})\dots L_{0,2}^{\bar{\Lambda}^{(2)}}(u+\theta_{1})L_{0,1}^{\Lambda^{(1)}}(u-\theta_{1}).
\end{equation}
We can see that untwisted final states exist only for alternating
spin chains where the representation of site $2j$ is the conjugate
representation of site $2j-1$. For these alternating chain the pseudo-vacuum
eigenvalues are
\begin{equation}
\lambda_{i}(u)=\prod_{k=1}^{J}\frac{u-\theta_{k}+c\Lambda_{i}^{(k)}}{u-\theta_{k}}\frac{u+\theta_{k}+c\bar{\Lambda}_{i}^{(k)}}{u+\theta_{k}}=\prod_{k=1}^{J}\frac{u-\theta_{k}+c\Lambda_{i}^{(k)}}{u-\theta_{k}}\frac{u+\theta_{k}-c\Lambda_{N+1-i}^{(k)}}{u+\theta_{k}},
\end{equation}
therefore we have the following properties
\begin{equation}
\lambda_{i}(u)=\lambda_{N+1-i}(-u),\qquad\alpha_{i}(u)=\frac{1}{\alpha_{N-i}(-u)}.
\end{equation}

\subsubsection{Pair structure}

In this paper our goal is to calculate the overlaps between boundary
states and on-shell Bethe states: $\langle\Psi|\mathbb{B}(\bar{t})$.
There is an important property of these on-shell overlaps. From the
$KT$-relation (\ref{eq:KT_Utw}) we can easily show that
\begin{equation}
\langle\Psi|\left(\mathcal{T}(u)-\mathcal{T}(-u)\right)=0.
\end{equation}
Applying it on an on-shell Bethe vector we obtain that
\begin{equation}
\left(\tau(u|\bar{t})-\tau(-u|\bar{t})\right)\langle\Psi|\mathbb{B}(\bar{t})=0,
\end{equation}
therefore the non-vanishing on-shell overlaps ($\langle\Psi|\mathbb{B}(\bar{t})\neq0$)
require that 
\begin{equation}
\tau(u|\bar{t})=\tau(-u|\bar{t}).
\end{equation}
It can be shown that it is equivalent to that the Bethe roots have
achiral pair structure $\bar{t}^{N-\nu}=-\bar{t}^{\nu}$ i.e.,
\begin{equation}
\bar{t}=\pi^{a}(\bar{t}),\qquad\pi^{a}(\bar{t})=\{-\bar{t}^{N-1},-\bar{t}^{N-2},\dots,-\bar{t}^{1}\}.
\end{equation}

We introduce some notations. For odd $N$ or even $N$ where $r_{\frac{N}{2}}$
is even, we can decompose the set of Bethe roots as $\bar{t}=\bar{t}^{+}\cup\bar{t}^{-}$,
where $\bar{t}^{\pm}=\left\{ \bar{t}^{\pm,\nu}\right\} _{\nu=1}^{\left\lfloor \frac{N}{2}\right\rfloor }$
and $\bar{t}^{+,\nu}=\bar{t}^{\nu}$, $\bar{t}^{-,\nu}=\bar{t}^{N-\nu}$
for $\nu<\frac{N}{2}$ and $\bar{t}^{+,\frac{N}{2}}=\left\{ t_{k}^{\frac{N}{2}}\right\} _{k=1}^{r_{\frac{N}{2}}/2}$,
$\bar{t}^{-,\frac{N}{2}}=\left\{ t_{k}^{\frac{N}{2}}\right\} _{k=r_{\frac{N}{2}}/2+1}^{r_{\frac{N}{2}}}$.
In the pair structure limit $\bar{t}^{-}=-\bar{t}^{+}$. For even
$N$ where $r_{\frac{N}{2}}$ is odd, we can decompose the set of
Bethe roots as $\bar{t}=\bar{t}^{+}\cup\bar{t}^{-}\cup\bar{t}^{0}$,
where $\bar{t}^{\pm}=\left\{ \bar{t}^{\pm,\nu}\right\} _{\nu=1}^{\frac{N}{2}}$
and $\bar{t}^{+,\nu}=\bar{t}^{\nu}$, $\bar{t}^{-,\nu}=\bar{t}^{N-\nu}$
for $\nu<\frac{N}{2}$ and $\bar{t}^{+,\frac{N}{2}}=\left\{ t_{k}^{\frac{N}{2}}\right\} _{k=1}^{\frac{r_{\frac{N}{2}}-1}{2}}$,
$\bar{t}^{-,\frac{N}{2}}=\left\{ t_{k}^{\frac{N}{2}}\right\} _{k=\frac{r_{\frac{N}{2}}+1}{2}}^{r_{\frac{N}{2}}-1}$,
$\bar{t}^{0}=\left\{ t_{r_{N}}^{\frac{N}{2}}\right\} $. In the pair
structure limit $\bar{t}^{-}\to-\bar{t}^{+}$ and $\bar{t}^{0}\to\{0\}$.

We will show that the on-shell overlaps are proportional to the Gaudin-like
determinant $\det G^{+}$. When $\bar{t}^{0}=\emptyset$, we have
the decomposition $\bar{t}=\bar{t}^{+}\cup\bar{t}^{-}$ and we can
also make the corresponding decomposition of the set $\bar{\Phi}=\{\Phi_{k}^{\mu}\}_{\mu,k}$
(which are defined by (\ref{eq:Phidef})) as $\bar{\Phi}=\bar{\Phi}^{+}\cup\bar{\Phi}^{-}$.
Due to this decomposition, the original Gaudin matrix (\ref{eq:GM})
has the following block form
\begin{equation}
G=\left(\begin{array}{cc}
A^{++} & A^{+-}\\
A^{-+} & A^{--}
\end{array}\right),
\end{equation}
where we defined the following matrices
\begin{equation}
\begin{split}A_{j,k}^{++,(\mu,\nu)} & =-c\frac{\partial}{\partial t_{k}^{+,\nu}}\log\Phi_{j}^{+,(\mu)}\Biggr|_{\bar{t}^{-}=-\bar{t}^{+}},\qquad A_{j,k}^{+-,(\mu,\nu)}=-c\frac{\partial}{\partial t_{k}^{-,\nu}}\log\Phi_{j}^{+,(\mu)}\Biggr|_{\bar{t}^{-}=-\bar{t}^{+}},\\
A_{j,k}^{-+,(\mu,\nu)} & =-c\frac{\partial}{\partial t_{k}^{+,\nu}}\log\Phi_{j}^{-,(\mu)}\Biggr|_{\bar{t}^{-}=-\bar{t}^{+}},\qquad A_{j,k}^{--,(\mu,\nu)}=-c\frac{\partial}{\partial t_{k}^{-,\nu}}\log\Phi_{j}^{-,(\mu)}\Biggr|_{\bar{t}^{-}=-\bar{t}^{+}},
\end{split}
\end{equation}
It is easy to show that $A^{++}=A^{--}$ and $A^{+-}=A^{+-}$. Using
these identities, one can show that the original Gaudin determinant
factorizes in the pair structure limit as
\begin{equation}
\det G=\det G^{+}\det G^{-},\label{eq:Gfact}
\end{equation}
where 
\begin{equation}
\det G^{\pm}=\det\left(A^{++}\pm A^{+-}\right).
\end{equation}
For odd $r_{N/2}$, i.e, when $\bar{t}^{0}=\left\{ t_{r_{N/2}}^{N/2}\right\} $,
we have the decomposition $\bar{t}=\bar{t}^{+}\cup\bar{t}^{0}\cup\bar{t}^{-}$
and $\bar{\Phi}=\bar{\Phi}^{+}\cup\bar{\Phi}^{0}\cup\bar{\Phi}^{-}$.
Due to this decomposition, the original Gaudin matrix (\ref{eq:GM})
has the following block form
\begin{equation}
G=\left(\begin{array}{ccc}
A^{++} & A^{+0} & A^{+-}\\
A^{0+} & A^{00} & A^{0-}\\
A^{-+} & A^{-0} & A^{--}
\end{array}\right),
\end{equation}
where we defined the following matrices
\begin{equation}
\begin{split}A_{j}^{+0,(\mu)} & =-c\frac{\partial}{\partial t_{r_{N/2}}^{N/2}}\log\Phi_{j}^{+,(\mu)},\qquad A_{j}^{-0,(\mu)}=-c\frac{\partial}{\partial t_{r_{N/2}}^{N/2}}\log\Phi_{j}^{-,(\mu)},\\
A_{k}^{0+,(\nu)} & =-c\frac{\partial}{\partial t_{k}^{+,\nu}}\log\Phi_{r_{N/2}}^{(N/2)},\qquad A_{k}^{0-,(\nu)}=-c\frac{\partial}{\partial t_{k}^{-,\nu}}\log\Phi_{r_{N/2}}^{(N/2)},\\
A^{00} & =-c\frac{\partial}{\partial t_{r_{N/2}}^{N/2}}\log\Phi_{r_{N/2}}^{(N/2)}.
\end{split}
\label{eq:A0}
\end{equation}
In the rhs we took the pair structure limit $\bar{t}^{-,\nu}\to-\bar{t}^{+,\nu},\bar{t}^{0}\to\{0\}$
after the derivation. It is easy to show that 
\begin{equation}
\begin{split}A^{++} & =A^{--},\qquad A^{+-}=A^{-+},\\
A^{0+} & =A^{0-},\qquad A^{0-}=A^{0+},
\end{split}
\end{equation}
therefore the original Gaudin determinant factorizes also as (\ref{eq:Gfact})
with the definitions
\begin{equation}
\det G^{+}=\left|\begin{array}{cc}
A^{++}+A^{+-} & A^{+0}\\
A^{0+} & \frac{1}{2}A^{00}
\end{array}\right|,\quad\det G^{-}=2\left|A^{++}-A^{+-}\right|.\label{eq:degGodd}
\end{equation}

\subsection{Overlaps for the type $(N,\left\lfloor \frac{N}{2}\right\rfloor )$,
$(N,0)$ and singular K-matrices}

In this subsection we review the results of \cite{Gombor:2021hmj}
for the untwisted overlaps. In \cite{Gombor:2021hmj} the on-shell
overlap formulas were proved for the type $(N,\left\lfloor \frac{N}{2}\right\rfloor )$
boundary states. In the next subsection we use this result to prove
the on-shell overlap formulas for all untwisted boundary states.

\subsubsection{On-shell overlaps for type $(N,\left\lfloor \frac{N}{2}\right\rfloor )$
K-matrices}

 Let us choose the type $(N,\left\lfloor \frac{N}{2}\right\rfloor )$
$K$-matrix with its regular form
\begin{equation}
K(z)=\sum_{k=1}^{\left\lfloor \frac{N}{2}\right\rfloor }\frac{\mathfrak{a}-z}{z}E_{k,k}+\sum_{k=\left\lfloor \frac{N}{2}\right\rfloor +1}^{N}\frac{\mathfrak{a}+z}{z}E_{k,k}+\sum_{k=1}^{\left\lfloor \frac{N}{2}\right\rfloor }\mathfrak{b}_{k}E_{N+1-k,k}.\label{eq:typeN2}
\end{equation}
 In \cite{Gombor:2021hmj} it was proved that the normalized on-shell
overlap has the following simple form
\[
\frac{\langle\Psi|\mathbb{B}(\bar{t})}{\sqrt{\mathbb{C}(\bar{t})\mathbb{B}(\bar{t})}}=\left(-2\frac{\mathfrak{a}}{c}\right)^{\mathbf{r}^{0}}\prod_{\nu=1}^{\frac{N}{2}}\mathfrak{b}_{\nu}^{r_{\nu-1}-r_{\nu}}\mathbb{F}^{\left\lfloor \frac{N}{2}\right\rfloor }(\bar{t}^{+,\left\lfloor \frac{N}{2}\right\rfloor })\sqrt{\frac{\det G^{+}}{\det G^{-}}},
\]
where we introduced a quantum number $\mathbf{r}^{0}=\#\bar{t}^{0}$
and the one-particle overlap functions $\mathbb{F}^{\left\lfloor \frac{N}{2}\right\rfloor }$
read as
\begin{equation}
\begin{split}\mathbb{F}^{(\frac{N-1}{2})}(z) & =\frac{-\mathfrak{a}-z}{\sqrt{(-z)(c/2-z)}},\\
\mathbb{F}^{(\frac{N}{2})}(u) & =\frac{(\mathfrak{a}-z)(\mathfrak{a}+z)}{\sqrt{-z^{2}(c/2-z)(c/2+z)}}.
\end{split}
\label{eq:FK2}
\end{equation}

In the applications it is common to use the $c=i$ convention and
redefine the Bethe roots as $u_{k}^{(\mu)}:=t_{k}^{\mu}+\frac{i}{2}\mu+x$
for which the Bethe equations read as
\[
\alpha_{\mu}(u_{k}^{(\mu)}-\frac{i}{2}\mu-x)=\prod_{l\neq k}^{r_{\mu}}\frac{u_{k}^{(\mu)}-u_{l}^{(\mu)}+i}{u_{k}^{(\mu)}-u_{l}^{(\mu)}-i}\prod_{l=1}^{r_{\mu+1}}\frac{u_{k}^{(\mu)}-u_{l}^{(\mu+1)}-\frac{i}{2}}{u_{k}^{(\mu)}-u_{l}^{(\mu+1)}+\frac{i}{2}}\prod_{l=1}^{r_{\mu+1}}\frac{u_{k}^{(\mu)}-u_{l}^{(\mu-1)}-\frac{i}{2}}{u_{k}^{(\mu)}-u_{l}^{(\mu-1)}+\frac{i}{2}}.
\]
For $x=-i\frac{N}{4}$ the pair structure limit $t_{l}^{N-\mu}=-t_{k}^{\mu}$
becomes $u_{l}^{(N-\mu)}=-u_{k}^{(\mu)}$. Introducing the $Q$-functions
\begin{equation}
Q_{\mu}(u)=\prod_{k=1}^{r_{\mu}}(u-u_{k}^{\mu}),\label{eq:Qfunv}
\end{equation}
and 
\begin{equation}
\bar{Q}_{\mu}(u)=\begin{cases}
Q_{\mu}(u), & \text{if }0\notin\bar{u}^{(\mu)},\\
\frac{1}{u}Q_{\mu}(u), & \text{if }0\in\bar{u}^{(\mu)},
\end{cases}
\end{equation}
 the overlaps simplify as
\begin{equation}
\frac{\langle\Psi|\mathbb{B}(\bar{t})}{\sqrt{\mathbb{C}(\bar{t})\mathbb{B}(\bar{t})}}=\left[\prod_{\nu=1}^{\frac{N}{2}}\mathfrak{b}_{\nu}^{r_{\nu-1}-r_{\nu}}\right]\times\frac{Q_{\frac{N}{2}}(\mathfrak{a})}{\sqrt{\bar{Q}_{\frac{N}{2}}(0)Q_{\frac{N}{2}}(\frac{i}{2})}}\times\sqrt{\frac{\det G^{+}}{\det G^{-}}},
\end{equation}
for even $N$ and
\begin{equation}
\frac{\langle\Psi|\mathbb{B}(\bar{t})}{\sqrt{\mathbb{C}(\bar{t})\mathbb{B}(\bar{t})}}=\left[\prod_{\nu=1}^{\frac{N-1}{2}}\mathfrak{b}_{\nu}^{r_{\nu-1}-r_{\nu}}\right]\times\frac{Q_{\frac{N-1}{2}}(-\mathfrak{a}-\frac{i}{4})}{\sqrt{Q_{\frac{N-1}{2}}(\frac{i}{4})Q_{\frac{N-1}{2}}(-\frac{i}{4})}}\times\sqrt{\frac{\det G^{+}}{\det G^{-}}},
\end{equation}
for odd $N$.

\subsubsection{On-shell overlaps for the singular K-matrices}

The result of \cite{Gombor:2021hmj} can be also applied for singular
K-matrices. For simplicity, let us choose the singular $K$-matrix
in the regular form
\begin{equation}
K(z)=\frac{1}{z}\mathbf{1}+\sum_{k=1}^{\left\lfloor \frac{N}{2}\right\rfloor }\mathfrak{b}_{k}E_{N+1-k,k}.
\end{equation}
You can see that this singular $K$-matrix differs from the type $(N,\left\lfloor \frac{N}{2}\right\rfloor )$
(\ref{eq:typeN2}) only in its diagonal elements. The corresponding
on-shell overlaps read as
\begin{equation}
\frac{\langle\Psi|\mathbb{B}(\bar{t})}{\sqrt{\mathbb{C}(\bar{t})\mathbb{B}(\bar{t})}}=\left[\prod_{\nu=1}^{\left\lfloor \frac{N}{2}\right\rfloor }\mathfrak{b}_{\nu}^{r_{\nu-1}-r_{\nu}}\right]\mathbb{F}_{s}^{(\left\lfloor \frac{N}{2}\right\rfloor )}(\bar{t}^{+,\left\lfloor \frac{N}{2}\right\rfloor })\sqrt{\frac{\det G^{+}}{\det G^{-}}},
\end{equation}
where
\begin{equation}
\begin{split}\mathbb{F}_{s}^{(\frac{N-1}{2})}(z) & =\frac{1}{\sqrt{(-z)(c/2-z)}},\\
\mathbb{F}_{s}^{(\frac{N}{2})}(u) & =\frac{1}{\sqrt{-z^{2}(c/2-z)(c/2+z)}}.
\end{split}
\label{eq:FK2-2}
\end{equation}

In the next section we will use these overlap formulas in the limit
$\mathfrak{b}_{s}\to0$ for $s>M$. The limit of this formula is not
well defined and the reason is that we fixed the normalization of
the final state $\langle\Psi|$ as $\langle\Psi|0\rangle=1$ but in
$\mathfrak{b}_{s}\to0$ limit this cannot be done. Let us choose the
normalization in a $\mathfrak{b}_{s}$ dependent way:
\begin{equation}
\langle\Psi|0\rangle=A(\mathfrak{b}_{M+1},\dots,\mathfrak{b}_{\left\lfloor \frac{N}{2}\right\rfloor }),
\end{equation}
where $A$ is a function of $\mathfrak{b}$-s and it can also depend
on the quantum space i.e. the $\mathfrak{\ensuremath{gl}}(N)$ weights
of the pseudo-vacuum: $\mathbf{\boldsymbol{\Lambda}}=(\mathbf{\boldsymbol{\Lambda}}_{1},\dots,\mathbf{\boldsymbol{\Lambda}}_{N})$.
Using this normalization the overlap formula is modified as
\begin{equation}
\frac{\langle\Psi|\mathbb{B}(\bar{t})}{\sqrt{\mathbb{C}(\bar{t})\mathbb{B}(\bar{t})}}=A(\mathfrak{b}_{M+1},\dots,\mathfrak{b}_{\left\lfloor \frac{N}{2}\right\rfloor })\left[\prod_{\nu=1}^{\left\lfloor \frac{N}{2}\right\rfloor }\mathfrak{b}_{\nu}^{r_{\nu-1}-r_{\nu}}\right]\mathbb{F}_{s}^{(\left\lfloor \frac{N}{2}\right\rfloor )}(\bar{t}^{+,\left\lfloor \frac{N}{2}\right\rfloor })\sqrt{\frac{\det G^{+}}{\det G^{-}}}.\label{eq:renormSingOv}
\end{equation}
There is also a novelty in the $KT$-relation in this limit. Applying
the limit $\mathfrak{b}_{s}\to0$ for $s>M$ for the untwisted $KT$-relation,
we obtain that
\begin{equation}
\langle\Psi|T_{i,j}(u)=\langle\Psi|T_{i,j}(-u),\label{eq:KTsing}
\end{equation}
for $M<i,j<N+1-M$. Using the asymptotic expansion of the monodromy
matrices (\ref{eq:seriesMon}) in the $KT$-relation (\ref{eq:KTsing})
we obtain that
\begin{equation}
\langle\Psi|\Delta(E_{i,j})=0,\quad\text{for }M<i,j<N+1-M,
\end{equation}
therefore the final state $\langle\Psi|$ is a singlet for a $\mathfrak{gl}(N-M)$
subalgebra. To obtain non-vanishing overlaps, the Bethe states also
have to be singlets for this subalgebra. The $\mathfrak{gl}(N)$ weights
of the Bethe states are 
\begin{equation}
\Delta(E_{k,k})\mathbb{B}(\bar{t})=\left(\mathbf{\boldsymbol{\Lambda}}_{k}+r_{k-1}-r_{k}\right)\mathbb{B}(\bar{t}),
\end{equation}
therefore the non-vanishing overlaps require that
\begin{equation}
\mathbf{\boldsymbol{\Lambda}}_{k}=r_{k}-r_{k-1},\quad\text{for }M<k<N+1-M,\label{eq:selRule-2}
\end{equation}
i.e., we can fix the quantum numbers as
\begin{equation}
r_{k}=r_{M}+\sum_{l=M+1}^{k}\mathbf{\boldsymbol{\Lambda}}_{l},\quad\text{for }k=M+1,\dots,\left\lfloor \frac{N}{2}\right\rfloor ,
\end{equation}
in the limit $\mathfrak{b}_{s}\to0$ for $s>M$. Returning to the
on-shell formula, let us choose the normalization as
\begin{equation}
A(\mathfrak{b}_{M+1},\dots,\mathfrak{b}_{\left\lfloor \frac{N}{2}\right\rfloor })=\prod_{\nu=M+1}^{\left\lfloor \frac{N}{2}\right\rfloor }\mathfrak{b}_{\nu}^{\mathbf{\boldsymbol{\Lambda}}_{\nu}}.
\end{equation}
Substituting back to (\ref{eq:renormSingOv}), the renormalized overlap
formula reads as
\begin{equation}
\frac{\langle\Psi|\mathbb{B}(\bar{t})}{\sqrt{\mathbb{C}(\bar{t})\mathbb{B}(\bar{t})}}=\left[\prod_{\nu=1}^{M}\mathfrak{b}_{\nu}^{r_{\nu-1}-r_{\nu}}\right]\left[\prod_{\nu=M+1}^{\left\lfloor \frac{N}{2}\right\rfloor }\mathfrak{b}_{\nu}^{\mathbf{\boldsymbol{\Lambda}}_{\nu}+r_{\nu-1}-r_{\nu}}\right]\mathbb{F}_{s}^{(\left\lfloor \frac{N}{2}\right\rfloor )}(\bar{t}^{+,\left\lfloor \frac{N}{2}\right\rfloor })\sqrt{\frac{\det G^{+}}{\det G^{-}}}.
\end{equation}
We showed that the non-vanishing overlap have the selection rule (\ref{eq:selRule-2})
therefore the overlap formula simplifies as
\begin{equation}
\frac{\langle\Psi|\mathbb{B}(\bar{t})}{\sqrt{\mathbb{C}(\bar{t})\mathbb{B}(\bar{t})}}=\left[\prod_{\nu=1}^{M}\mathfrak{b}_{\nu}^{r_{\nu-1}-r_{\nu}}\right]\mathbb{F}_{s}^{(\left\lfloor \frac{N}{2}\right\rfloor )}(\bar{t}^{+,\left\lfloor \frac{N}{2}\right\rfloor })\sqrt{\frac{\det G^{+}}{\det G^{-}}},\label{eq:ovSing}
\end{equation}
which has a well defined limit.

\subsubsection{Sum formula for $M=0$\label{subsec:Sum-formula-forM0}}

For the $M=0$ case, where the $K$-matrix is just the identity $K(u)=\mathbf{1}$,
the $KT$-relation simplifies as
\begin{equation}
\langle\Psi|T_{i,j}(u)=\langle\Psi|T_{i,j}(-u),\quad\text{for }i,j=1,\dots,N.\label{eq:KTM0}
\end{equation}
We can see that we obtain the $M=0$ case from any $M$ in the $\mathfrak{a}\to\infty$
limit. In \cite{Gombor:2021hmj} we also derived a sum formula for
the off-shell overlaps of the type $(N,\left\lfloor \frac{N}{2}\right\rfloor )$
boundary states and taking the $\mathfrak{a}\to\infty$ limit we can
also obtain the sum formula for the off-shell overlaps of the $M=0$
case. See the details in appendix \ref{sec:Untwisted-off-shell-overlap}. 

A special property appears in the $M=0$ case. Using the asymptotic
expansion of the monodromy matrices (\ref{eq:seriesMon}) in the $KT$-relation
we obtain that
\begin{equation}
\langle\Psi|\Delta(E_{i,j})=0,\quad\text{for }i,j=1,\dots,N,
\end{equation}
therefore the boundary state $\langle\Psi|$ is a sum of singlet states.
Repeating the arguments of the previous sub-subsection, we obtain
that the Bethe states with non-vanishing overlaps have to be $\mathfrak{gl}(N)$
singlets, i.e. we have the selection rules
\begin{equation}
\mathbf{\boldsymbol{\Lambda}}_{k}=r_{k}-r_{k-1},
\end{equation}
i.e.,
\begin{equation}
r_{k}=r_{k}^{\mathbf{\boldsymbol{\Lambda}}}:=\sum_{l=1}^{k}\mathbf{\boldsymbol{\Lambda}}_{l}.\label{eq:M0SR}
\end{equation}
Since the functions $\alpha$-s depend on $\mathbf{\boldsymbol{\Lambda}}$,
the sum formula does not give off-shell overlap for arbitrarily $\alpha$-s
only for the $\alpha$-s which correspond to correct $\mathbf{\boldsymbol{\Lambda}}_{k}$-s.
If we first fix the quantum numbers $r_{k}$ then we have constrains
on the representation of the quantum space therefore the only remaining
freedom is to choose the limited number of inhomogeneities $\theta_{i}$-s
independently. The consequence is that the $\alpha_{\nu}(\bar{t}_{k}^{\nu})$-s
are not algebraically independent variables anymore.

Anyway, for proper $\alpha$-s we have a sum formula for the $(N,0)$
type off-shell overlaps:
\begin{equation}
\langle\Psi|\mathbb{B}(\bar{t})=S_{\bar{\alpha}}^{0}(\bar{t})=\sum_{\mathrm{part}}\frac{\prod_{\nu=1}^{N-1}f(\bar{t}_{\textsc{ii}}^{\nu},\bar{t}_{\textsc{i}}^{\nu})}{\prod_{\nu=1}^{N-2}f(\bar{t}_{\textsc{ii}}^{\nu+1},\bar{t}_{\textsc{i}}^{\nu})}\bar{Z}^{0}(\pi^{a}(\bar{t}_{\textsc{i}}))\bar{Z}^{0}(\bar{t}_{\textsc{ii}})\prod_{\nu=1}^{N-1}\alpha_{\nu}(\bar{t}_{\textsc{i}}^{\nu}),\label{eq:sumM=0000A7}
\end{equation}
where the recursion for the highest coefficients (HC) $\bar{Z}^{0}$
is given in (\ref{eq:rec_UTW0}).

\subsection{Recursion for the off-shell type $(N,M)$ overlaps\label{subsec:Reqursion-and-pair-general-M}}

In this section we show a recursive method for evaluation of the
off-shell overlaps where the $K$-matrix has type $(N,M)$ regular
form 
\begin{equation}
K(z)=\sum_{k=1}^{M}\mathfrak{b}_{k}E_{N+1-k,k}+\sum_{k=1}^{N}K_{k,k}(z)E_{k,k}.\label{eq:Ktemp}
\end{equation}
Choosing the diagonal part as
\begin{equation}
K_{k,k}(z)=\begin{cases}
\frac{\mathfrak{a}-z}{z}, & k=1,\dots,M,\\
\frac{\mathfrak{a}+z}{z}, & k=M+1,\dots,N,
\end{cases}
\end{equation}
we obtain the type $(N,M)$ $K$-matrix and choosing
\begin{equation}
K_{k,k}(z)=\frac{1}{z},\quad k=1,\dots,N,
\end{equation}
we obtain a singular $K$-matrix with the type $(N,M)$ form. The
calculations below can be used for both cases.

Our method starts with the recurrence relation of the off-shell Bethe
vectors (\ref{eq:rec1})
\begin{equation}
\mathbb{B}(\{z,\bar{t}^{1}\},\left\{ \bar{t}^{k}\right\} _{k=2}^{N-1})=\sum_{j=2}^{N}T_{1,j}(z)\sum_{\mathrm{part}(\bar{t})}(\dots)\mathbb{B}(\bar{t}^{1},\left\{ \bar{t}_{\textsc{ii}}^{k}\right\} _{k=2}^{j-1},\left\{ \bar{t}^{k}\right\} _{k=j}^{N-1}).
\end{equation}
Now we only want to sketch the method therefore we concentrate on
the operator content of the recurrence relation (\ref{eq:rec1}),
the explicit form of the numerical coefficients are not important,
we use the notation $(\dots)$ for them. We also concentrate on changes
of the quantum numbers $r_{k}$ therefore we introduce the shorthanded
notation 
\begin{equation}
\mathbb{B}(\bar{t})\to\mathbb{B}^{r_{1},r_{2},\dots,r_{N-1}},
\end{equation}
where $r_{j}=\#\bar{t}^{j}$. Using this notation, the recurrence
relation can be written as
\begin{equation}
\mathbb{B}^{r_{1}+1,r_{2},\dots,r_{N-1}}=\sum_{j=2}^{N}T_{1,j}\sum(\dots)\mathbb{B}^{r_{1},\tilde{r}_{2}\dots,\tilde{r}_{N-1}},\label{eq:rrr-1}
\end{equation}
where $\tilde{r}_{k}\leq r_{k}$. We also use the action formula (\ref{eq:act})
\begin{equation}
T_{i,j}\mathbb{B}^{r_{1},r_{2},\dots,r_{N-1}}=\begin{cases}
\sum(\dots)\mathbb{B}^{r_{1},\dots r_{i}+1,r_{i+1}+1,\dots,r_{j-1}+1,r_{j},\dots r_{N-1}}, & i\leq j,\\
\sum(\dots)\mathbb{B}^{r_{1},\dots r_{j}-1,r_{j+1}-1,\dots,r_{i-1}-1,r_{i},\dots r_{N-1}}, & i>j.
\end{cases}\label{eq:act0-2}
\end{equation}
We can see that the operators $T_{i,j}$ for $i<j$ \emph{increase}
the quantum numbers therefore they are the \emph{creation operators}
and analogously, the operators $T_{i,j}$ for $i>j$ decrease the
quantum numbers therefore they are the \emph{annihilation} operators.
We also use the $(N,j)$ components of the $KT$-relation

\begin{align}
\langle\Psi|T_{1,j}(u) & =\frac{\mathfrak{b}_{j}}{\mathfrak{b}_{1}}\langle\Psi|T_{N,N+1-j}(-u)+\frac{K_{j,j}(u)}{\mathfrak{b}_{1}}\langle\Psi|T_{N,j}(-u)-\frac{K_{N,N}(u)}{\mathfrak{b}_{1}}\langle\Psi|T_{N,j}(u),\text{ for }j\leq M,\\
\langle\Psi|T_{1,j}(u) & =\frac{K_{j,j}(u)}{\mathfrak{b}_{1}}\langle\Psi|T_{N,j}(-u)-\frac{K_{N,N}(u)}{\mathfrak{b}_{1}}\langle\Psi|T_{N,j}(u),\text{ for }j>M.
\end{align}
We can see that using this relation we can change the creation operators
$T_{1,j}$ in (\ref{eq:rrr-1}) to annihilation operators $T_{N,N+1-j},T_{N,j}$.
 Using the recurrence relation, the $KT$-relation and the action
formula, we obtain a recurrence equation for the off-shell overlap:
\begin{align}
\langle\Psi|\mathbb{B}^{r_{1},r_{2},\dots,r_{N-1}} & =\sum_{j=2}^{M}\sum(\dots)\langle\Psi|\mathbb{B}^{r_{1}-1,\dots,r_{j-1}-1,r_{j},\dots,r_{N-j},r_{N-j+1}-1,\dots,r_{N-1}-1}+\nonumber \\
 & +\sum(\dots)\langle\Psi|\mathbb{B}^{r_{1}-1,\dots,r_{N-1}-1}.
\end{align}
We can see that this recursion decreases $r_{1}$ and $r_{N-1}$ by
$1$ therefore the non-vanishing off-shell overlaps requires 
\begin{equation}
r_{N-1}=r_{1}.
\end{equation}
Using this recursion one can eliminate all $\bar{t}^{1}$ and $\bar{t}^{N-1}$
which leads to a $\mathfrak{gl}(N-2)$ overlap
\begin{equation}
\langle\Psi|\mathbb{B}^{r_{1},r_{2},\dots,r_{N-1}}=\sum_{\mathrm{part}}(\dots)\langle\Psi|\mathbb{B}^{0,\tilde{r}_{2},\dots,\tilde{r}_{N-2},0}.
\end{equation}
We can see that this recursion leaves the differences $r_{k+1}-r_{k}=\tilde{r}_{k+1}-\tilde{r}_{k}$
invariant for $k=M,M+1,\dots N-M$. The Bethe states $\mathbb{B}^{0,\tilde{r}_{2},\dots,\tilde{r}_{N-2},0}$
in the $\mathfrak{gl}(N-2)$ subsector are generated by $\left\{ T_{a,b}\right\} _{a,b=2}^{N-1}$.
Since $K_{N,a}=K_{a,N}=K_{1,a}=K_{a,1}=0$ for $a=2,\dots,N-1$, the
$KT$-relation is closed in this subsector, i.e.
\begin{equation}
\sum_{c=2}^{N-1}K_{a,c}(u)\langle\Psi|T_{c,b}(u)=\sum_{c=2}^{N-1}\langle\Psi|T_{a,c}(-u)K_{c,b}(u),
\end{equation}
for $a,b=2,\dots,N-1$. The $K$-matrix $\{K_{a,b}(u)\}_{a,b=2}^{N-1}$
of this subsector has type $(N-2,M-1)$ form. 

Repeating the recursion we obtain type $(N-2M,0)$ overlaps as
\begin{equation}
\langle\Psi|\mathbb{B}^{r_{1},r_{2},\dots,r_{N-1}}=\sum_{\mathrm{part}}(\dots)\langle\Psi|\mathbb{B}^{0,\dots,0,\tilde{r}_{M+1},\dots,\tilde{r}_{N-M-1},0,\dots,0}.
\end{equation}
Since the differences are invariant we can express $\tilde{r}_{k}$-s
with $r_{k}$-s as 
\begin{equation}
\tilde{r}_{k}=r_{k}-r_{M},\quad\text{for }k=M+1,\dots N-M-1.
\end{equation}
Since the type $(N-2M,0)$ $K$-matrix requires the selection rule
(\ref{eq:M0SR}), for non-vanishing overlaps we obtain that 
\begin{equation}
\tilde{r}_{k}=r_{k}-r_{M}=\sum_{k=M+1}^{s}\mathbf{\boldsymbol{\Lambda}}_{k},\quad\text{for }s=M+1,\dots,N-M-1.\label{eq:selRule}
\end{equation}
we can also invert this formula
\begin{equation}
\mathbf{\boldsymbol{\Lambda}}_{s}=r_{s}-r_{s-1},\quad\text{for }s=M+1,\dots,\left\lfloor \frac{N}{2}\right\rfloor .\label{eq:sr}
\end{equation}

\subsection{Sum formula and the on-shell limit}

In this section we show the sum formula of the off-shell overlap for
the $K$-matrices with the form (\ref{eq:Ktemp}). This sum formula
can be applied simultaneously for a type $(N,M)$ K-matrix and a singular
K-matrix for which the on-shell overlap was already derived (\ref{eq:ovSing}).
We show that the two off-shell formulas are proportional to each other
therefore specifying the proportionality factor and using the on-shell
formula of the singular case, we also obtain the on-shell formula
for the type $(N,M)$ boundary states.

For these types of $K$-matrices the off-shell overlaps have the following
sum formula
\begin{align}
\mathcal{S}_{\bar{\alpha}}(\bar{t}) & :=\langle\Psi|\mathbb{B}(\bar{t})=\sum\frac{\prod_{\nu=1}^{N-1}f(\bar{t}_{\textsc{ii}}^{\nu},\bar{t}_{\textsc{i}}^{\nu})}{\prod_{\nu=1}^{N-2}f(\bar{t}_{\textsc{ii}}^{\nu+1},\bar{t}_{\textsc{i}}^{\nu})}\frac{\prod_{s=M+1}^{N-M-1}f(\bar{t}_{\textsc{ii}}^{s},\bar{t}_{\textsc{iii}}^{s})f(\bar{t}_{\textsc{iii}}^{s},\bar{t}_{\textsc{i}}^{s})}{\prod_{s=M+1}^{N-M-1}f(\bar{t}_{\textsc{ii}}^{s+1},\bar{t}_{\textsc{iii}}^{s})f(\bar{t}_{\textsc{iii}}^{s},\bar{t}_{\textsc{i}}^{s-1})}\times\nonumber \\
 & \times\mathcal{Z}(\bar{t}_{\textsc{i}})\bar{\mathcal{Z}}(\bar{t}_{\textsc{ii}})S_{\{\alpha^{s}\}_{s=M+1}^{N-M-1}}^{0}(\{\bar{t}_{\textsc{iii}}^{s}\}_{s=M+1}^{N-M-1})\prod_{\nu=1}^{N-1}\alpha_{\nu}(\bar{t}_{\textsc{i}}^{\nu}),\label{eq:sumUTw}
\end{align}
where sum goes through the partitions $\text{\ensuremath{\bar{t}}}=\bar{t}_{\textsc{i}}\cup\bar{t}_{\textsc{ii}}\cup\bar{t}_{\textsc{iii}}$
where $\#\bar{t}_{\textsc{i}}^{N-s}=\#\bar{t}_{\textsc{i}}^{s}$,
$\#\bar{t}_{\textsc{ii}}^{N-s}=\#\bar{t}_{\textsc{ii}}^{s}$ , $\#\bar{t}_{\textsc{iii}}^{N-s}=\#\bar{t}_{\textsc{iii}}^{s}$,
$\#\bar{t}_{\textsc{i}}^{M}=\#\bar{t}_{\textsc{i}}^{M+1}=\dots=\#\bar{t}_{\textsc{i}}^{N-M}$,
$\#\bar{t}_{\textsc{ii}}^{M}=\#\bar{t}_{\textsc{ii}}^{M+1}=\dots=\#\bar{t}_{\textsc{ii}}^{N-M}$
and $\#\bar{t}_{\textsc{iii}}^{1}=\#\bar{t}_{\textsc{iii}}^{2}=\dots=\#\bar{t}_{\textsc{iii}}^{M}=0$,
$\bar{t}_{\textsc{iii}}^{N-M}=\dots=\#\bar{t}_{\textsc{iii}}^{N-1}=0$.
The sum formula contains an overlap function $S^{0}$ of a $\mathfrak{gl}(N-2M)$
spin chain with the identity $K$-matrix i.e. the type $(N-2M,0)$
overlaps which were previously calculated, see (\ref{eq:sumM=0000A7}).
The proof of the sum formula can be found in appendix \ref{sec:Proof-of-the}.

In appendix \ref{sec:Reqursion-for-the} we show that the HC-s $\mathcal{Z}$
and $\bar{\mathcal{Z}}$ are proportional to universal HC-s which
are independent from the diagonal element of the $K$-matrix: 
\begin{equation}
\begin{split}\mathcal{Z}(\bar{t}) & =\mathcal{G}(\bar{t}^{M})\mathcal{Z}^{0}(\bar{t}),\\
\bar{\mathcal{Z}}(\bar{t}) & =\mathcal{G}(\bar{t}^{M})\bar{\mathcal{Z}}^{0}(\bar{t}),
\end{split}
\label{eq:propo}
\end{equation}
where
\begin{equation}
\mathcal{G}(z)=\frac{1}{g(-z,z)}\frac{K_{N,N}(z)}{\mathfrak{b}_{1}}.\label{eq:Gfun}
\end{equation}
The universal HC-s $\mathcal{Z}^{0}$ and $\bar{\mathcal{Z}}^{0}$
are given by recurrence equations (\ref{eq:reqZ0}) and (\ref{eq:reqZb0}). 

Substituting back (\ref{eq:propo}) to the sum formula (\ref{eq:sumUTw})
we obtain that
\begin{align}
\mathcal{S}_{\bar{\alpha}}(\bar{t}) & =\mathcal{G}(\bar{t}^{M})\mathcal{S}_{\bar{\alpha}}^{0}(\bar{t}),\label{eq:fac}
\end{align}
where we introduced the universal off-shell overlap
\begin{align}
\mathcal{S}_{\bar{\alpha}}^{0}(\bar{t}) & =\sum\frac{\prod_{\nu=1}^{N-1}f(\bar{t}_{\textsc{ii}}^{\nu},\bar{t}_{\textsc{i}}^{\nu})}{\prod_{\nu=1}^{N-2}f(\bar{t}_{\textsc{ii}}^{\nu+1},\bar{t}_{\textsc{i}}^{\nu})}\frac{\prod_{s=M+1}^{N-M-1}f(\bar{t}_{\textsc{ii}}^{s},\bar{t}_{\textsc{iii}}^{s})f(\bar{t}_{\textsc{iii}}^{s},\bar{t}_{\textsc{i}}^{s})}{\prod_{s=M+1}^{N-M-1}f(\bar{t}_{\textsc{ii}}^{s+1},\bar{t}_{\textsc{iii}}^{s})f(\bar{t}_{\textsc{iii}}^{s},\bar{t}_{\textsc{i}}^{s-1})}\times\nonumber \\
 & \times\mathcal{Z}^{0}(\bar{t}_{\textsc{i}})\bar{\mathcal{Z}}^{0}(\bar{t}_{\textsc{ii}})S_{\{\alpha^{s}\}_{s=M+1}^{N-M-1}}^{0}(\{\bar{t}_{\textsc{iii}}^{s}\}_{s=M+1}^{N-M-1})\prod_{\nu=1}^{N-1}\alpha_{\nu}(\bar{t}_{\textsc{i}}^{\nu}),
\end{align}
which does not depend the diagonal part of the $K$-matrix. In the
following, let $\mathcal{S}_{\bar{\alpha}}^{M}(\bar{t})$ and $\mathcal{S}_{\bar{\alpha}}^{s}(\bar{t})$
be the off-shell overlaps of the type $(N,M)$ and the singular $K$-matrices
i.e.
\begin{align}
K_{1,1}(u) & =\dots=K_{M,M}(u)=\frac{\mathfrak{a}-z}{z},\;K_{M+1,M+1}(u)=\dots=K_{N,N}(u)=\frac{\mathfrak{a}+z}{z},\text{ for }\mathcal{S}_{\bar{\alpha}}^{M},\\
K_{1,1}(u) & =\dots=K_{M,M}(u)=K_{M+1,M+1}(u)=\dots=K_{N,N}(u)=\frac{1}{z},\text{ for }\mathcal{S}_{\bar{\alpha}}^{s},
\end{align}
Using (\ref{eq:fac}) and (\ref{eq:Gfun}) we have
\begin{equation}
\mathcal{S}_{\bar{\alpha}}^{M}(\bar{t})=\left[\prod_{k=1}^{r_{M}}\left(\mathfrak{a}+t_{k}^{M}\right)\right]\mathcal{S}_{\bar{\alpha}}^{s}(\bar{t}).
\end{equation}

In the previous section we already proved the normalized on-shell
overlap for the singular $K$-matrix (\ref{eq:ovSing}). Substituting
back, we obtain the normalized on-shell overlap for the type $(N,M)$
K-matrices
\begin{equation}
\frac{\langle\Psi^{M}|\mathbb{B}(\bar{t})}{\sqrt{\mathbb{C}(\bar{t})\mathbb{B}(\bar{t})}}=\left[\prod_{\nu=1}^{M}\mathfrak{b}_{\nu}^{r_{\nu-1}-r_{\nu}}\right]\prod_{k=1}^{r_{M}}\left(-\mathfrak{a}-t_{k}^{M}\right)\mathbb{F}_{0}^{(\left\lfloor \frac{N}{2}\right\rfloor )}(\bar{t}^{+,\left\lfloor \frac{N}{2}\right\rfloor })\sqrt{\frac{\det G^{+}}{\det G^{-}}}.
\end{equation}
For $c=i$ convention, using the $Q$-functions (\ref{eq:Qfunv})
the overlaps read as
\begin{equation}
\frac{\langle\Psi^{M}|\mathbb{B}(\bar{t})}{\sqrt{\mathbb{C}(\bar{t})\mathbb{B}(\bar{t})}}=\left[\prod_{\nu=1}^{M}\mathfrak{b}_{\nu}^{r_{\nu-1}-r_{\nu}}\right]\frac{Q_{M}(-\mathfrak{a}+\frac{i}{2}\left(M-\frac{N}{2}\right))}{\sqrt{\bar{Q}_{\frac{N}{2}}(0)\bar{Q}_{\frac{N}{2}}(\frac{i}{2})}}\times\sqrt{\frac{\det G^{+}}{\det G^{-}}},\label{eq:onShellUtw}
\end{equation}
for even $N$ and 
\begin{equation}
\frac{\langle\Psi^{M}|\mathbb{B}(\bar{t})}{\sqrt{\mathbb{C}(\bar{t})\mathbb{B}(\bar{t})}}=\left[\prod_{\nu=1}^{M}\mathfrak{b}_{\nu}^{r_{\nu-1}-r_{\nu}}\right]\frac{Q_{M}(-\mathfrak{a}+\frac{i}{2}\left(M-\frac{N}{2}\right))}{\sqrt{Q_{\frac{N-1}{2}}(-\frac{i}{4})Q_{\frac{N-1}{2}}(\frac{i}{4})}}\times\sqrt{\frac{\det G^{+}}{\det G^{-}}},\label{eq:onShellUtw2}
\end{equation}
for odd $N$.

\section{Exact overlaps for twisted boundary states\label{sec:Egzact-overlaps-for}}

In this section we investigate the overlap formulas for twisted boundary
states. In \cite{Gombor:2021hmj} we saw that there are two classes
of twisted boundary states, the $\mathfrak{so}(N)$ and the $\mathfrak{sp}(N)$
symmetric ones. In \cite{Gombor:2021hmj} the $\mathfrak{so}(N)$
on-shell overlaps were already derived and now we calculate the remaining
overlaps for the $\mathfrak{sp}(N)$ symmetric case. At first, we
specify the twisted integrable boundary states and the corresponding
$K$-matrices which satisfy the twisted $KT$-relation. Using this
relation we show a recursive method which allows the calculation of
the off-shell overlaps. After that we show a sum formula for the off-shell
overlaps. The sum formula has useful properties that are the same
as the embedding rules of the Gaudin-like determinants. These allow
us to assume the on-shell overlap formula, but further work is needed
for a precise proof, which we will postpone to later.

\subsection{Definition of the twisted integrable states}

Let us consider the boundary states $\langle\Psi|$ which satisfy
the twisted $KT$-relation
\begin{equation}
K_{0}(u)\langle\Psi|T_{0}(u)=\lambda_{0}(u)\langle\Psi|\widehat{T}_{0}(-u)K_{0}(u),\label{eq:KT_Utw-1}
\end{equation}
where $K(u)$ is an invertible $N\times N$ matrix of the auxiliary
space. In \cite{Gombor:2021hmj} it was showed that the consistency
of the definition (\ref{eq:KT_Utw-1}) requires the twisted reflection
equation for the $K$-matrix
\begin{equation}
R_{1,2}(u-v)K_{1}(-u)\bar{R}_{1,2}(u+v)K_{2}(-v)=K_{2}(-v)\bar{R}_{1,2}(u+v)K_{1}(-u)R_{1,2}(u-v).
\end{equation}
We note that for $N=2$ the twisted monodromy matrix is equivalent
to the original one (see equation (\ref{eq:gl2Equiv}))
\begin{equation}
\widehat{T}_{0}(u)=\frac{\hat{\lambda}_{2}(u)}{\lambda_{2}(u-\frac{c}{2})}\epsilon_{0}^{-1}T_{0}(u-\frac{c}{2})\epsilon_{0}=\frac{\lambda_{1}(u)}{\lambda_{2}(u-\frac{c}{2})}\sigma_{0}^{-1}T_{0}(u-\frac{c}{2})\sigma_{0},
\end{equation}
where we used (\ref{eq:lambdah}) and
\begin{equation}
\sigma=\left(\begin{array}{cc}
1 & 0\\
0 & -1
\end{array}\right).
\end{equation}
Substituting back to the twisted $KT$-relation, we obtain that
\begin{equation}
\left(\sigma_{0}K_{0}(u)\right)\langle\Psi|T_{0}(u)=\frac{\lambda_{0}(u)\lambda_{1}(-u)}{\lambda_{2}(-u-\frac{c}{2})}\langle\Psi|T_{0}(-u-\frac{c}{2})\left(\sigma_{0}K_{0}(u)\right).
\end{equation}
Choosing the normalization for which $\lambda_{0}(u)=\frac{\lambda_{2}(-u-\frac{c}{2})}{\lambda_{1}(-u)}$,
the shifted version of the monodromy matrix
\begin{equation}
\tilde{T}_{0}(u)=T_{0}(u-\frac{c}{2})
\end{equation}
satisfies the untwisted $KT$-relation
\begin{equation}
\tilde{K}_{0}(u)\langle\Psi|\tilde{T}_{0}(u)=\lambda_{0}(u)\langle\Psi|\tilde{T}_{0}(-u)\tilde{K}_{0}(u),
\end{equation}
where
\begin{equation}
\tilde{K}(u)=\sigma K(u-\frac{c}{2}).
\end{equation}
We just showed that the untwisted and twisted $KT$-relations are
equivalent for $N=2$. In the following we concentrate on $N>2$.

\subsubsection{Integrable $K$-matrices and their regular forms}

For $N>2$ the most general solution of this equation is well known
\cite{Arnaudon:2004sd}: 
\begin{equation}
K(u)=K=V\mathcal{U},\label{eq:Kdef_UTw-1}
\end{equation}
where $K$ and $\mathcal{U}$ are $N\times N$ matrices with the following
constraint
\begin{equation}
K^{t}=\pm VKV,\qquad\mathcal{U}^{t}=\pm\mathcal{U}.
\end{equation}
Since the monodromy matrices have $\mathfrak{gl}(N)$ symmetry
\begin{align}
T_{0}(u) & =G_{0}\Delta(G)T_{0}(u)G_{0}^{-1}\Delta(G^{-1}),\\
\widehat{T}_{0}(u) & =\left(VG_{0}^{-1}V\right)^{t_{0}}\Delta(G)T_{0}(u)(VG_{0}V)^{t_{0}}\Delta(G^{-1}),
\end{align}
(where $G\in GL(N)$) we can obtain transformed $K$-matrices and
boundary states as
\begin{equation}
K^{G}=VG^{t}VKG,\qquad,\mathcal{U}^{G}=G^{t}\mathcal{U}G\qquad\langle\Psi^{G}|=\langle\Psi|\Delta(G)\label{eq:Ktraf}
\end{equation}
which also satisfy the twisted $KT$-relation 
\begin{equation}
K^{G}\langle\Psi^{G}|T_{0}(u)=\lambda_{0}(u)\langle\Psi^{G}|\widehat{T}_{0}(-u)K^{G}.
\end{equation}
We can determine the subgroups which leave the $K$-matrices (and
equivalently the boundary states) invariant. Rearranging (\ref{eq:Ktraf}),
the defining equation for the subgroups is
\begin{equation}
\mathcal{U}=G^{t}\mathcal{U}G,\label{eq:sign-1}
\end{equation}
therefore $G\in SO(N)$ for $\mathcal{U}^{t}=+\mathcal{U}$ and $G\in Sp(N)$
for $\mathcal{U}^{t}=-\mathcal{U}$, thus we call these $K$-matrices
and the corresponding boundary states the $\mathfrak{so}(N)$ and
$\mathfrak{sp}(N)$ symmetric $K$-matrices and boundary states, respectively.
In \cite{Gombor:2021hmj} it was demonstrated that the twisted $KT$-relation
is very efficient tool to calculate the overlaps $\langle\Psi|\mathbb{B}(\bar{t})$
for $\mathfrak{so}(N)$ symmetric $K$-matrices. In the following
we concentrate on the $\mathfrak{sp}(N)$ symmetric case for which
the $N$ is always even.

At the end of the day, we are interested in on-shell overlaps $\langle\Psi|\mathbb{B}(\bar{t})$.
We know that the on-shell Bethe states $\mathbb{B}(\bar{t})$ are
highest weight states, therefor the transformations $G=\exp(\varphi E_{i,j})$
leave the on-shell overlap invariant for $i<j$ (rising operators).
In the following let us try to apply such transformations, which leave
the on-shell overlap invariant, to obtain more simple $K$- and $\mathcal{U}$-matrices
for the $\mathfrak{sp}(N)$ symmetric case.

Let us define $GL(N)$ transformations which contain only rising operators
as
\begin{align}
G & =\exp(\sum_{c=3}^{N}\varphi_{c}E_{1,c}+\sum_{c=3}^{N}\phi_{c}E_{2,c}),
\end{align}
where
\begin{equation}
\varphi_{c}=\frac{\mathcal{U}_{c,2}}{\mathcal{U}_{2,1}},\quad\phi_{c}=-\frac{\mathcal{U}_{c,1}}{\mathcal{U}_{2,1}}.
\end{equation}
We saw that this transformation does not change the on-shell overlaps.
The transformed $\mathcal{U}$-matrix is
\begin{equation}
\mathcal{U}^{(2)}=G^{t}\mathcal{U}G,
\end{equation}
which remains anti-symmetric and the components read as
\begin{equation}
\begin{split}\mathcal{U}_{2,1}^{(2)} & =\mathcal{U}_{2,1},\\
\mathcal{U}_{a,1}^{(2)} & =\mathcal{U}_{a,1}+\frac{\mathcal{U}_{a,1}\mathcal{U}_{1,2}}{\mathcal{U}_{2,1}}=0,\\
\mathcal{U}_{a,2}^{(2)} & =\mathcal{U}_{a,2}-\frac{\mathcal{U}_{a,2}\mathcal{U}_{2,1}}{\mathcal{U}_{2,1}}=0,\\
\mathcal{U}_{a,b}^{(2)} & =\mathcal{U}_{a,b}+\frac{\mathcal{U}_{a,1}\mathcal{U}_{b,2}}{\mathcal{U}_{2,1}}-\frac{\mathcal{U}_{a,2}\mathcal{U}_{b,1}}{\mathcal{U}_{2,1}},
\end{split}
\end{equation}
where $a,b=3,\dots,N$. We can see that the new matrix $\mathcal{U}^{(2)}$
has block-diagonal form and we can repeat the method on the $N-2$
dimensional invariant subspace. Reaping this method we obtain the
series of matrices
\begin{equation}
\mathcal{U}^{(k+1)}=\left(G^{(k)}\right)^{t}\mathcal{U}^{(k)}G^{(k)},\quad\mathcal{U}^{(1)}=\mathcal{U},
\end{equation}
where the $GL(N)$ transformations are
\begin{equation}
G^{(k)}=\exp(\sum_{c=2k+1}^{N}\varphi_{c}^{(k)}E_{2k-1,c}+\sum_{c=2k+1}^{N}\phi_{c}^{(k)}E_{2k,c}),\label{eq:GLN-1}
\end{equation}
with
\begin{equation}
\varphi_{c}^{(k)}=\frac{\mathcal{U}_{c,2k}^{(k)}}{\mathcal{U}_{2k,2k-1}^{(k)}},\quad\phi_{c}^{(k)}=-\frac{\mathcal{U}_{c,2k-1}^{(k)}}{\mathcal{U}_{2k,2k-1}^{(k)}}.
\end{equation}
The matrices have block diagonal forms
\begin{equation}
\mathcal{U}^{(k+1)}=\sum_{j=1}^{k}\mathcal{U}_{2j,2j-1}^{(j)}(E_{2j,2j-1}-E_{2j-1,2j})+\sum_{a,b=2k+1}^{N}(\mathcal{U}_{a,b}^{(k)}+\frac{\mathcal{U}_{a,2k-1}^{(k)}\mathcal{U}_{b,2k}^{(k)}}{\mathcal{U}_{2k,2k-1}^{(k)}}-\frac{\mathcal{U}_{a,2k}^{(k)}\mathcal{U}_{b,2k-1}^{(k)}}{\mathcal{U}_{2k,2k-1}^{(k)}})E_{a,b}.
\end{equation}
The recursion stops at the step $N/2$ and we have the involution
matrix
\begin{equation}
\mathcal{U}^{(\frac{N}{2})}=\sum_{j=1}^{\frac{N}{2}}x_{j}(E_{2j,2j-1}-E_{2j-1,2j}),
\end{equation}
where
\begin{equation}
x_{k}=\mathcal{U}_{2k,2k-1}^{(k)}.
\end{equation}
The corresponding $K$-matrix is
\begin{equation}
K^{(\frac{N}{2})}=V\mathcal{U}^{(\frac{N}{2})}=\sum_{j=1}^{\frac{N}{2}}x_{j}(E_{N-2j+1,2j-1}-E_{N-2j+2,2j}).\label{eq:regTwK}
\end{equation}
Hereafter, we call the matrix $\mathcal{U}^{(\frac{N}{2})}$ (and
the corresponding $K$-matrix and final state $\langle\Psi^{(\frac{N}{2})}|$)
the regular form of the original $\mathcal{U}$ (and K-matrix and
final state $\langle\Psi|$). Since the $GL(N)$ transformations (\ref{eq:GLN-1})
contain only rising operators, the on-shell overlaps with the regular
and the original final states are the same
\begin{equation}
\langle\Psi^{(\frac{N}{2})}|\mathbb{B}(\bar{t})=\langle\Psi|\prod_{k=1}^{\frac{N}{2}-1}\Delta(G^{(k)})\mathbb{B}(\bar{t})=\langle\Psi|\mathbb{B}(\bar{t}).
\end{equation}
During the calculation of the overlaps we will concentrate on the
regular $K$-matrices (\ref{eq:regTwK}). 

\subsubsection{Solutions of the untwisted $KT$-relations}

Similarly as for the untwisted case, we can build twisted integrable
final states using the co-product property (\ref{eq:coprod}). For
the tensor product quantum space $\mathcal{H}=\mathcal{H}^{(1)}\otimes\mathcal{H}^{(2)}$,
if the $\langle\Psi^{(1)}|\in\mathcal{H}^{(1)}$ and $\langle\Psi^{(2)}|\in\mathcal{H}^{(2)}$
are twisted integrable final states with the same $K$-matrix then
the tensor product state
\begin{equation}
\langle\Psi|=\langle\Psi^{(1)}|\otimes\langle\Psi^{(2)}|\in\mathcal{H}
\end{equation}
satisfies the twisted $KT$-relation (\ref{eq:KT_Utw-1}) with the
same $K$-matrix (see \cite{Gombor:2021hmj}). 

\paragraph{Elementary two-site states}

The consequence is that we can build integrable final states as a
tensor product of ''elementary'' two-site states where the quantum
space is the tensor product of two rectangular representations
\begin{equation}
T_{0}(u)=L_{0,2}^{(s,a)}(u+\theta-c(s-a))L_{0,1}^{(s,a)}(u-\theta),\qquad\widehat{T}_{0}(u)=\widehat{L}_{0,2}^{(s,a)}(u+\theta-c(s-a))\widehat{L}_{0,1}^{(s,a)}(u-\theta),
\end{equation}
where the twisted Lax-operators are
\begin{equation}
\begin{split}\widehat{L}_{0,1}^{(s,a)}(u) & =\frac{u(u+c(s-a))}{(u+cs)(u-ca)}\bar{L}_{0,1}^{(s,a)}(u+c(s-a)),\\
\bar{L}_{0,1}^{(s,a)}(u) & =V_{0}\left(L_{0,1}^{(s,a)}(-u)\right)^{t_{0}}V_{0}.
\end{split}
\end{equation}
Using this identity, the twisted $KT$-relation reads as
\begin{multline}
K_{0}\langle\psi^{(s,a)}(\theta)|L_{0,2}^{(s,a)}(u+\theta-c(s-a))L_{0,1}^{(s,a)}(u-\theta)=\\
\lambda_{0}(u)\langle\psi^{(s,a)}(\theta)|\widehat{L}_{0,2}^{(s,a)}(-u+\theta-c(s-a))\widehat{L}_{0,1}^{(s,a)}(-u-\theta)K_{0},
\end{multline}
which simplifies as
\begin{multline}
K_{0}\langle\psi^{(s,a)}(\theta)|L_{0,2}^{(s,a)}(u+\theta-c(s-a))L_{0,1}^{(s,a)}(u-\theta)=\\
\langle\psi^{(s,a)}(\theta)|\bar{L}_{0,2}^{(s,a)}(-u+\theta)\bar{L}_{0,1}^{(s,a)}(-u-\theta+c(s-a))K_{0},\label{eq:twistedtimeKT}
\end{multline}
where we defined the $\lambda_{0}$ as
\begin{equation}
\lambda_{0}(u)=\frac{(u^{2}-(\theta-cs)^{2})(u^{2}-(\theta+ca)^{2})}{(u^{2}-\theta^{2})(u^{2}-(\theta-c(s-a))^{2})}.
\end{equation}
The ''elementary'' two-site state acts on the sites $1,2$. Let us
choose a basis $e_{i}^{(s,a)}$ in $\mathcal{V}^{(s,a)}$ and their
co-vectors are $\left(e_{i}^{(s,a)}\right)^{t}$ for which $\langle\psi^{(s,a)}(\theta)|=\sum\psi_{i,j}(\theta)\left(e_{i}^{(s,a)}\right)^{t}\otimes\left(e_{j}^{(s,a)}\right)^{t}$.
Using the equivalent matrix form of the two-site state $\psi^{(s,a)}(\theta)=\sum\psi_{j,i}(\theta)E_{i,j}^{(s,a)}$,
the $KT$-relation reads as
\begin{multline}
K_{0}L_{0,1}^{(s,a)}(u+\theta-c(s-a))^{t_{1}}\psi_{1}^{(s,a)}(\theta)L_{0,1}^{(s,a)}(u-\theta)=\\
\bar{L}_{0,1}^{(s,a)}(-u+\theta)^{t_{1}}\psi_{1}^{(s,a)}(\theta)\bar{L}_{0,1}^{(s,a)}(-u-\theta+c(s-a))K_{0}.
\end{multline}
Let us look at the Lax operator $L^{\bar{\Lambda}}$ in more details.
It contains the highest weight representation $E_{i,j}^{\bar{\Lambda}}$.
Let us define the following sets of operators
\begin{equation}
\bar{E}_{i,j}^{\Lambda}=-E_{N+1-j,N+1-i}^{\Lambda},
\end{equation}
which is also a highest weight irrep of $\mathfrak{gl}(N)$ and the
highest weights are $(-\Lambda_{N},\dots,-\Lambda_{1})$ therefore
we can choose the convention for the generators $E_{i,j}^{\bar{\Lambda}}$
as $E_{i,j}^{\bar{\Lambda}}=-E_{N+1-j,N+1-i}^{\Lambda}$. We also
choose the convention $\left(E_{i,j}^{\Lambda}\right)^{t}=E_{j,i}^{\Lambda}$
therefore 
\begin{equation}
\begin{split}\bar{L}_{0,1}^{(s,a)}(u)^{t_{1}} & =\mathbf{1}-\frac{c}{u}\sum_{i,j=1}^{N}\left(E_{i,j}\right)_{0}\otimes\left(E_{N+1-j,N+1-i}^{(s,a)}\right)_{1},\\
L_{0,1}^{(s,a)}(u)^{t_{1}} & =\mathbf{1}+\frac{c}{u}\sum_{i,j=1}^{N}\left(E_{i,j}\right)_{0}\otimes\left(E_{i,j}^{(s,a)}\right)_{1}.
\end{split}
\end{equation}
The set of operators $\tilde{E}_{i,j}^{\Lambda}=E_{N+1-i,N+1-j}^{\Lambda}$
is also an irrep of $\mathfrak{gl}(N)$ but the state $|0^{\Lambda}\rangle$
is now a lowest weight state i.e. $\tilde{E}_{i,j}|0^{\Lambda}\rangle=0$
for $i>j$ and the lowest weights are $(\Lambda_{N},\dots,\Lambda_{1})$.
Since every finite dimensional irreps are highest weight reps therefore
there exist a highest weight state $|\tilde{0}^{\Lambda}\rangle$
for which the highest weights are $(\Lambda_{1},\dots,\Lambda_{N})$.
Since the highest weight irreps are unique there exists a similarity
transformation $V^{(s,a)}$ for which $\tilde{E}_{i,j}^{(s,a)}=V^{(s,a)}E_{i,j}^{(s,a)}V^{(s,a)}$
i.e.
\begin{equation}
\begin{split}\bar{L}_{0,1}^{(s,a)}(u)^{t_{1}} & =\mathbf{1}-\frac{c}{u}\sum_{i,j=1}^{N}\left(E_{i,j}\right)_{0}\otimes\left(V^{(s,a)}E_{j,i}^{(s,a)}V^{(s,a)}\right)_{1}=V_{1}^{(s,a)}L_{0,1}^{(s,a)}(-u)V_{1}^{(s,a)},\\
L_{0,1}^{(s,a)}(u)^{t_{1}} & =\mathbf{1}+\frac{c}{u}\sum_{i,j=1}^{N}\left(E_{N+1-i,N+1-j}\right)_{0}\otimes\left(V^{(s,a)}E_{i,j}^{(s,a)}V^{(s,a)}\right)_{1}=V_{1}^{(s,a)}\bar{L}_{0,1}^{(s,a)}(-u)V_{1}^{(s,a)},
\end{split}
\end{equation}
Substituting back, the $KT$-relation simplifies as ($v_{1}=-\theta+c\frac{s-a}{2}$,$v_{2}=-u+c\frac{s-a}{2}$)
\begin{equation}
K_{0}\bar{L}_{0,1}^{(s,a)}(v_{1}+v_{2})K_{1}^{(s,a)}(-v_{1})L_{0,1}^{(s,a)}(v_{1}-v_{2})=L_{0,1}^{(s,a)}(v_{1}-v_{2})K_{1}^{(s,a)}(-v_{1})\bar{L}_{0,1}^{(s,a)}(v_{1}+v_{2})K_{0},
\end{equation}
where we used the notation
\begin{equation}
K_{1}^{(s,a)}(v)=V_{1}^{(s,a)}\psi_{1}^{(s,a)}(v-c\frac{s-a}{2}),\quad\to\quad\psi_{j,i}^{(s,a)}(v)=\sum_{k}V_{i,k}^{(s,a)}K_{k,j}^{(s,a)}(v+c\frac{s-a}{2}).
\end{equation}
We just obtained the twisted reflection equation for the rectangular
representation $(s,a)$. Since the reflection equation has solution
for any rectangular representation (e.g. it can be obtained from the
original K-matrix using the fusion procedure) we just showed that
the ''elementary'' two-site state exists for any rectangular representation.
However, it is worth to note that the original equation (\ref{eq:KT_Utw-1})
also has infinite dimensional solutions and our later derivations
are independent from the quantum space therefore the results are valid
even for the infinite dimensional integrable two-site states. 

\paragraph{Symmetric properties of the vacuum eigenvalues}

In summary, we can build general integrable states as
\begin{equation}
\langle\Psi|=\langle\psi^{(s_{1},a_{1})}(\theta_{1})|\otimes\dots\otimes\langle\psi^{(s_{J},a_{J})}(\theta_{J})|,\label{eq:Psidef-2}
\end{equation}
where the monodromy matrices are defined as
\begin{equation}
T_{0}(u)=L_{0,2J}^{(s_{J},a_{J})}(u+\theta_{J}-c(s_{J}-a_{J}))L_{0,2J-1}^{(s_{J},a_{J})}(u-\theta_{J})\dots L_{0,2}^{(s_{1},a_{1})}(u+\theta-c(s_{1}-a_{1}))L_{0,1}^{(s_{1},a_{1})}(u-\theta_{1}).
\end{equation}
We can see that twisted final states exist for spin chains where the
representation of site $2j$ is the same as $2j-1$. For these chains
the pseudo-vacuum eigenvalues have the following properties
\begin{equation}
\lambda_{k}(u)=\lambda_{0}(u)\hat{\lambda}_{N+1-k}(-u),\qquad\alpha_{k}(u)=\frac{1}{\alpha_{k}(-u-kc)}.\label{eq:twistedalpha}
\end{equation}

\paragraph{Elementary one-site states}

For the $\mathfrak{sp}(N)$ symmetric $K$-matrices there exists an
other type of ''elementary'' integrable state which are one-site states.
In the following we show an example where the quantum space is a rectangular
representation for which $s=1,a=2$. The monodromy matrix is
\begin{equation}
T_{0}(z)=L_{0,1}^{(1,2)}(z+c/2)=\mathbf{1}+\frac{c}{z+c/2}\left(E_{i,j}\right)_{0}\otimes\left(E_{j,i}^{(1,2)}\right)_{1}.\label{eq:oneSiteMon}
\end{equation}
We show that there exists a non-trivial solution of the twisted $KT$-relation
\begin{equation}
K_{0}\langle\psi|T_{0}(z)=\lambda_{0}(z)\langle\psi|\widehat{T}_{0}(-z)K_{0}.\label{eq:elemTwKT}
\end{equation}
We know that the representation $s=1,a=2$ can be obtained from the
tensor product of the defining representation by antisymmetrization.
If $e_{j}$-s are the basis vectors of the defining representation
(i.e. $E_{i,j}e_{k}=\delta_{j,k}e_{i}$) then let $e_{(i,j)}$ for
$1\leq i<j\leq N$ be the basis vectors of the representation $s=1,a=2$
with the identification $e_{(i,j)}=e_{i}\otimes e_{j}-e_{j}\otimes e_{i}$.
The generators has matrix elements:
\begin{equation}
\left(E_{i,j}^{(1,2)}\right)_{(a_{1},b_{1}),(a_{2},b_{2})}=\delta_{a_{1},i}\delta_{a_{2},j}\delta_{b_{1},b_{2}}-\delta_{b_{1},i}\delta_{a_{2},j}\delta_{a_{1},b_{2}}-\delta_{a_{1},i}\delta_{b_{2},j}\delta_{b_{1},a_{2}}+\delta_{b_{1},i}\delta_{b_{2},j}\delta_{a_{1},a_{2}}.
\end{equation}
The twisted monodromy matrix reads as
\begin{equation}
\widehat{T}_{0}(z)=\frac{1}{\lambda_{0}(z)}\left(\mathbf{1}-\frac{c}{z-c/2}\left(E_{i,j}\right)_{0}\otimes\left(E_{N+1-i,N+1-j}^{(1,2)}\right)_{1}\right).\label{eq:onesiteTb}
\end{equation}
For these chains the pseudo-vacuum eigenvalues are
\begin{align}
\lambda_{k}(u) & =\begin{cases}
\frac{z+3c/2}{z+c/2}, & \text{for }k=1,2,\\
1, & \text{for }k>2,
\end{cases}\\
\hat{\lambda}_{k}(u) & =\begin{cases}
\frac{1}{\lambda_{0}(z)}\frac{z-3c/2}{z-c/2}, & \text{for }k=N-1,N,\\
\frac{1}{\lambda_{0}(z)}, & \text{for }k>2.
\end{cases}
\end{align}
The pseudo-vacuum eigenvalues have the same properties as before (\ref{eq:twistedalpha}). 

We can fix the $\lambda_{0}$ from the defining equation of the twisted
monodromy matrix 
\begin{align}
T_{0}(z)V_{0}\widehat{T}_{0}^{t_{0}}(z)V_{0} & =\frac{1}{\lambda_{0}(z)}\left(\mathbf{1}+\frac{c}{z+c/2}\sum_{i,j}E_{i,j}\otimes E_{j,i}^{(1,2)}\right)\left(\mathbf{1}-\frac{c}{z-c/2}\sum_{k,l}E_{k,l}\otimes E_{l,k}^{(1,2)}\right)=\nonumber \\
 & =\frac{1}{\lambda_{0}(z)}\left(\mathbf{1}-\frac{c^{2}}{z^{2}-c^{2}/4}\sum_{i,j}E_{i,j}\otimes E_{j,i}^{(1,2)}-\frac{c^{2}}{z^{2}-c^{2}/4}\sum_{i,j,l}E_{i,l}\otimes E_{j,i}^{(1,2)}E_{l,j}^{(1,2)}\right).
\end{align}
Using the identity
\begin{equation}
\sum_{j=1}^{N}E_{j,i}^{(1,2)}E_{l,j}^{(1,2)}=2\delta_{i,l}\mathbf{1}-E_{l,i}^{(1,2)},
\end{equation}
we obtain that
\begin{equation}
T_{0}(z)V_{0}\widehat{T}_{0}^{t_{0}}(z)V_{0}=\frac{1}{\lambda_{0}(z)}\frac{z^{2}-9c^{2}/4}{z^{2}-c^{2}/4}\mathbf{1},
\end{equation}
therefore the $\lambda_{0}(z)$ function reads as:
\begin{equation}
\lambda_{0}(z)=\frac{z^{2}-9c^{2}/4}{z^{2}-c^{2}/4}.
\end{equation}
Fixing the $K$-matrix to the regular form (\ref{eq:regTwK}). Substituting
back to the $KT$-relation, we have four types of equations
\begin{align}
x_{a}\langle\psi|T_{2a,2b}(z) & =\lambda_{0}(z)\langle\psi|\widehat{T}_{N+2-2a,N+2-2b}(-z)x_{b},\\
-x_{a}\langle\psi|T_{2a,2b-1}(z) & =\lambda_{0}(z)\langle\psi|\widehat{T}_{N+2-2a,N+1-2b}(-z)x_{b},\\
-x_{a}\langle\psi|T_{2a-1,2b}(z) & =\lambda_{0}(z)\langle\psi|\widehat{T}_{N+1-2a,N+2-2b}(-z)x_{b},\\
x_{a}\langle\psi|T_{2a-1,2b-1}(z) & =\lambda_{0}(z)\langle\psi|\widehat{T}_{N+1-2a,N+1-2b}(-z)x_{b},
\end{align}
which are equivalent to
\begin{align}
x_{a}\langle\psi|E_{2b,2a}^{(1,2)} & =\langle\psi|E_{2a-1,2b-1}^{(1,2)}x_{b},\label{eq:AS1}\\
-x_{a}\langle\psi|E_{2b-1,2a}^{(1,2)} & =\langle\psi|E_{2a-1,2b}^{(1,2)}x_{b},\label{eq:AS2}\\
-x_{a}\langle\psi|E_{2b,2a-1}^{(1,2)} & =\langle\psi|E_{2a,2b-1}^{(1,2)}x_{b}.\label{eq:AS3}
\end{align}
For $a=b$ we have
\begin{align}
\langle\psi|\left(E_{2a,2a}^{(1,2)}-E_{2a-1,2a-1}^{(1,2)}\right) & =0.
\end{align}
The following ansatz for the one-site state solves the equation above:
\begin{equation}
\langle\psi|=\sum_{a}y_{a}e_{(2a-1,2a)}.
\end{equation}
Substituting back to the equation (\ref{eq:AS1}) for $a\neq b$ we
have
\begin{equation}
\langle\psi|\left(x_{a}E_{2b,2a}^{(1,2)}-x_{b}E_{2a-1,2b-1}^{(1,2)}\right)=\begin{cases}
\left(x_{a}y_{b}-x_{b}y_{a}\right)(2b-1,2a), & \text{for }b\leq a,\\
\left(x_{b}y_{a}-x_{a}y_{b}\right)(2a,2b-1), & \text{for }b>a.
\end{cases}
\end{equation}
These equations are satisfied iff
\begin{equation}
y_{a}=x_{a}.
\end{equation}
One can also show that the one-site state
\begin{equation}
\langle\psi|=\sum_{a}x_{a}e_{(2a-1,2a)}\label{eq:one-site}
\end{equation}
satisfies the remaining equations (\ref{eq:AS2}) and (\ref{eq:AS3}),
too. 

\subsubsection{Pair structure}

The on-shell overlaps have an important property. From the twisted
$KT$-relation we can easily show that
\begin{equation}
\langle\Psi|\left(\mathcal{T}(u)-\lambda_{0}(u)\widehat{\mathcal{T}}(-u)\right)=0.
\end{equation}
Applying it on an on-shell Bethe vector we obtain that
\begin{equation}
\left(\tau(u|\bar{t})-\lambda_{0}(u)\hat{\tau}(-u|\bar{t})\right)\langle\Psi|\mathbb{B}(\bar{t})=0,
\end{equation}
therefore the non-vanishing on-shell overlaps require that 
\begin{equation}
\tau(u|\bar{t})=\lambda_{0}(u)\hat{\tau}(-u|\bar{t}).
\end{equation}
It can be shown that it is equivalent to that the Bethe roots have
chiral pair structure $\bar{t}^{\nu}=-\bar{t}^{\nu}-\nu c$ i.e.,
\begin{equation}
\bar{t}=\pi^{c}(\bar{t}),\qquad\pi^{c}(\bar{t})=\{-\bar{t}^{1}-c,-\bar{t}^{2}-2c,\dots,-\bar{t}^{N-1}-(N-1)c\}.\label{eq:chiralPair}
\end{equation}

We introduce some notations. Let $\mathfrak{r}$ be a set for which
$\mathfrak{r}=\{\nu|r_{\nu}\text{ is odd}\}$. We can decompose the
set of Bethe roots as $\bar{t}=\bar{t}^{+}\cup\bar{t}^{0}\cup\bar{t}^{-}$,
where $\bar{t}^{\pm}=\left\{ \bar{t}^{\pm,\nu}\right\} _{\nu=1}^{N}$,
$\bar{t}^{0}=\left\{ t^{0,\nu}\right\} _{\nu\in\mathfrak{r}}$ for
$\bar{t}^{+,\nu}=\left\{ t_{k}^{\nu}\right\} _{k=1}^{\left\lfloor \frac{r_{\nu}}{2}\right\rfloor }$,
$\bar{t}^{-,\nu}=\left\{ t_{k}^{\nu}\right\} _{k=\left\lfloor \frac{r_{\nu}}{2}\right\rfloor +1}^{2\left\lfloor \frac{r_{\nu}}{2}\right\rfloor }$
and $t^{0,\nu}=t_{r_{\nu}}^{\nu}$. In the pair structure limit $\bar{t}^{-,\nu}=\pi^{c}(\bar{t}^{+,\nu})$
and $t^{0,\nu}=-c\nu/2$.  

In the following, we argue that the on-shell overlaps are proportional
to the Gaudin-like determinant $\det G^{+}$. Due to the decomposition
$\bar{t}=\bar{t}^{+}\cup\bar{t}^{0}\cup\bar{t}^{-}$ and $\bar{\Phi}=\bar{\Phi}^{+}\cup\bar{\Phi}^{0}\cup\bar{\Phi}^{-}$,
the original Gaudin matrix has the following block form
\begin{equation}
G=\left(\begin{array}{ccc}
A^{++} & A^{+0} & A^{+-}\\
A^{0+} & A^{00} & A^{0-}\\
A^{-+} & A^{-0} & A^{--}
\end{array}\right),
\end{equation}
where we defined the following matrices
\begin{equation}
\begin{split}A_{j,k}^{++,(\mu,\nu)} & =-c\frac{\partial}{\partial t_{k}^{+,\nu}}\log\Phi_{j}^{+,(\mu)},\qquad A_{j,k}^{+-,(\mu,\nu)}=-c\frac{\partial}{\partial t_{k}^{-,\nu}}\log\Phi_{j}^{+,(\mu)},\\
A_{j,k}^{-+,(\mu,\nu)} & =-c\frac{\partial}{\partial t_{k}^{-,\nu}}\log\Phi_{j}^{+,(\mu)},\qquad A_{j,k}^{--,(\mu,\nu)}=-c\frac{\partial}{\partial t_{k}^{-,\nu}}\log\Phi_{j}^{-,(\mu)},\\
A_{j}^{+0,(\mu,\nu)} & =-c\frac{\partial}{\partial t_{r_{\nu}}^{\nu}}\log\Phi_{j}^{+,(\mu)},\qquad A_{j}^{-0,(\mu,\nu)}=-c\frac{\partial}{\partial t_{r_{\nu}}^{\nu}}\log\Phi_{j}^{-,(\mu)},\\
A_{k}^{0+,(\mu,\nu)} & =-c\frac{\partial}{\partial t_{k}^{+,\nu}}\log\Phi^{0,(\mu)},\qquad A_{k}^{0-,(\mu,\nu)}=-c\frac{\partial}{\partial t_{k}^{-,\nu}}\log\Phi^{0,(\mu)},\\
A^{00,(\mu,\nu)} & =-c\frac{\partial}{\partial t_{r_{\nu}}^{\nu}}\log\Phi^{0,(\mu)}.
\end{split}
\label{eq:A0-1}
\end{equation}
In the rhs we took the pair structure limit after the derivation.
It is easy to show that 
\begin{equation}
\begin{split}A^{++} & =A^{--},\qquad A^{+-}=A^{-+},\\
A^{0+} & =A^{0-},\qquad A^{0-}=A^{0+},
\end{split}
\end{equation}
therefore the original Gaudin determinant factorize as 
\begin{equation}
\det G=\det G^{+}\det G^{-},
\end{equation}
 with the definitions
\begin{equation}
\det G^{+}=\left|\begin{array}{cc}
A^{++}+A^{+-} & A^{+0}\\
A^{0+} & \frac{1}{2}A^{00}
\end{array}\right|,\quad\det G^{-}=2^{\#\mathfrak{r}}\left|A^{++}-A^{+-}\right|,\label{eq:degGodd-1}
\end{equation}
where $\mathbf{r}^{0}=\#\bar{t}^{0}=\#\mathfrak{r}$. 

The Gaudin-like determinant $\det G^{+}$ depends on the Bethe roots
$t_{k}^{+,\mu}$ and the derivatives $X_{k}^{+,\mu}$, $X^{0,\mu}$
which are defined as
\begin{equation}
X_{k}^{+,\mu}=-c\frac{\partial}{\partial z}\log\alpha(z)\Biggr|_{z=t_{k}^{+,\mu}},\quad X^{0,\mu}=-\frac{c}{2}\frac{\partial}{\partial z}\log\alpha(z)\Biggr|_{z=-\mu c/2}.
\end{equation}
 Let us introduce the number of $X_{k}^{+,\mu}$ (which is equal to
the number of $t_{k}^{+,\mu}$) as $\mathbf{r}^{+}=\sum_{\nu=1}^{N}\#\bar{t}^{+,\mu}=\sum_{\nu=1}^{N}\left\lfloor \frac{r_{\nu}}{2}\right\rfloor $.
The set of functions $\mathbf{F}^{(\mathbf{r}^{+},\mathbf{r}^{0})}(\bar{X}^{+}|\bar{X}^{0}|\bar{t}^{+})$
equals to the Gaudin-like determinant, i.e., $\mathbf{F}^{(\mathbf{r}^{+},\mathbf{r}^{0})}(\bar{X}^{+}|\bar{X}^{0}|\bar{t}^{+})=\det G^{+}$
iff it satisfies the following Korepin criteria: 
\begin{enumerate}
\item \label{enum:prop1-1-2}The function $\mathbf{F}^{(\mathbf{r}^{+},\mathbf{r}^{0})}(\bar{X}^{+}|\bar{X}^{0}|\bar{t}^{+})$
is symmetric over the replacement of the pairs $(X_{j}^{+,\mu},t_{j}^{+,\mu})\leftrightarrow(X_{k}^{+,\mu},t_{k}^{+,\mu})$.
\item \label{enum:prop2-1-2}It is linear function of each $X_{j}^{+,\mu}$
and $X^{0,\mu}$.
\item \label{enum:prop3-1-2}$\mathbf{F}^{(1,0)}(X_{1}^{+,\nu}|\emptyset|t_{1}^{+,\nu})=X_{1}^{+,\nu}$
and $\mathbf{F}^{(0,1)}(\emptyset|X^{0,\nu}|\emptyset)=X^{0,\nu}$.
\item \label{enum:prop4-1-2}The coefficient of $X_{j}^{+,\mu}$ is given
by the function $\mathbf{F}^{(\mathbf{r}^{+}-1,\mathbf{r}^{0})}$
with modified parameters 
\begin{equation}
\frac{\partial\mathbf{F}^{(\mathbf{r}^{+},\mathbf{r}^{0})}(\bar{X}^{+}|\bar{X}^{0}|\bar{t}^{+})}{\partial X_{j}^{+,\mu}}=\mathbf{F}^{(\mathbf{r}^{+}-1,\mathbf{r}^{0})}(\bar{X}^{+,mod}\backslash X_{j}^{+,\mu,mod}|\bar{X}^{0,mod}|\bar{t}^{+}\backslash t_{j}^{+,\mu}),\label{eq:derivFX-1-2}
\end{equation}
where the original variables should be replaced by
\begin{equation}
\begin{split}X_{k}^{+,\nu,mod} & =X_{k}^{+,\nu}-c\frac{d}{du}\log\beta_{\nu}(u|t_{j}^{+,\mu})\Biggr|_{u=t_{k}^{+,\nu}}.\\
X^{0,\nu,mod} & =X^{0,\nu}-c\frac{d}{du}\log\beta_{\nu}(u|t_{j}^{+,\mu})\Biggr|_{u=-\frac{\nu c}{2}}.
\end{split}
\end{equation}
The coefficient of $X^{0,\mu}$ is given by the function $\mathbf{F}^{(\mathbf{r}^{+},\mathbf{r}^{0}-1)}$
with modified parameters 
\begin{equation}
\frac{\partial\mathbf{F}^{(\mathbf{r}^{+},\mathbf{r}^{0})}(\bar{X}^{+}|\bar{X}^{0}|\bar{t}^{+})}{\partial X^{0,\mu}}=\mathbf{F}^{(\mathbf{r}^{+},\mathbf{r}^{0}-1)}(\bar{X}^{+,mod}|\bar{X}^{0,mod}\backslash X^{0,\mu,mod},\bar{t}^{+}),\label{eq:derivFX-1-2-1}
\end{equation}
where the original variables should be replaced by
\begin{equation}
\begin{split}X_{k}^{+,\nu,mod} & =X_{k}^{+,\nu}-c\frac{d}{du}\log\gamma_{\nu}(u)\Biggr|_{u=t_{k}^{+,\nu}},\\
X^{0,\nu,mod} & =X^{0,\nu}-c\frac{d}{du}\log\gamma_{\nu}(u)\Biggr|_{u=-\frac{\nu c}{2}}.
\end{split}
\end{equation}
\item \label{enum:prop5-1-2}$\mathbf{F}^{(\mathbf{r}^{+})}(\bar{X}^{+}|\bar{X}^{0}|\bar{t}^{+})=0$,
if all $X_{j}^{+,\mu}=0$ and $X^{0,\mu}=0$.
\end{enumerate}
The definitions of the $\beta$-s are
\begin{equation}
\beta_{\nu}(u|t_{j}^{+,\mu})=\begin{cases}
\frac{f(t_{j}^{+,\mu},u)}{f(u,t_{j}^{+,\mu})}\frac{f(-t_{j}^{+,\mu}-\mu c,u)}{f(u,-t_{j}^{+,\mu}-\mu c)} & \text{for }\nu=\mu,\\
\frac{1}{f(t_{j}^{+,\mu},u)}\frac{1}{f(-t_{j}^{+,\mu}-\mu c,u)} & \text{for }\nu=\mu-1,\\
f(u,t_{j}^{+,\mu})f(u,-t_{j}^{+,\mu}-\mu c) & \text{for }\nu=\mu+1,\\
1 & \text{for }\nu\neq\mu-1,\mu,\mu+1.
\end{cases}
\end{equation}
The $\gamma$ functions are defined as
\begin{equation}
\gamma_{\nu}(u)=\begin{cases}
\frac{f(-\mu c/2,u)}{f(u,-\mu c/2)} & \text{for }\nu=\mu,\\
\frac{1}{f(-\mu c/2,u)} & \text{for }\nu=\mu-1,\\
f(u,-\mu c/2) & \text{for }\nu=\mu+1,\\
1 & \text{for }\nu\neq\mu-1,\mu,\mu+1.
\end{cases}
\end{equation}

\subsection{Recursion for the off-shell overlaps\label{subsec:Reqursion-for-the-TW}}

In \cite{Gombor:2021hmj} a recursion was introduced for the off-shell
overlaps of the $\mathfrak{so}(N)$ symmetric $K$-matrices. The recursion
assumed that $K_{N,1}\neq0$ which could have been prescribed in the
$\mathfrak{so}(N)$ case but for the $\mathfrak{sp}(N)$ symmetric
$K$-matrix we have $K_{i,j}=-K_{N+1-j,N+1-i}$ therefore $K_{N,1}=0$.
It means that the recursion of \cite{Gombor:2021hmj} cannot be used
here in the twisted case. Now we use an alternative recursion for
the Bethe states
\begin{align}
\mathbb{B}(\bar{t}^{1},\{z,\bar{t}^{2}\},\left\{ \bar{t}^{k}\right\} _{k=2}^{N-1}) & =\sum_{j=3}^{N}T_{2,j}(z)\sum_{\mathrm{part}(\bar{t})}(\dots)\mathbb{B}(\bar{t}^{1},\bar{t}^{2},\left\{ \bar{t}_{\textsc{ii}}^{k}\right\} _{k=3}^{j-1},\left\{ \bar{t}^{k}\right\} _{k=j}^{N-1})+\nonumber \\
 & +\sum_{j=3}^{N}T_{1,j}(z)\sum_{\mathrm{part}(\bar{t})}(\dots)\mathbb{B}(\bar{t}_{\textsc{ii}}^{1},\bar{t}^{2},\left\{ \bar{t}_{\textsc{ii}}^{k}\right\} _{k=3}^{j-1},\left\{ \bar{t}^{k}\right\} _{k=j}^{N-1}),
\end{align}
where the dots denote some coefficients which can be found in appendix
\ref{sec:A-new-recurrance}. Using the simplified notation we have
\begin{equation}
\mathbb{B}^{r_{1},r_{2}+1,r_{3},\dots,r_{N-1}}=\sum_{j=3}^{N}T_{2,j}\sum_{\mathrm{part}}(\dots)\mathbb{B}^{r_{1},r_{2},\tilde{r}_{2}\dots,\tilde{r}_{N-1}}+\sum_{j=3}^{N}T_{1,j}\sum_{\mathrm{part}}(\dots)\mathbb{B}^{r_{1}-1,r_{2},\tilde{r}_{2}\dots,\tilde{r}_{N-1}},\label{eq:reqTW}
\end{equation}
where $\tilde{r}_{j}\leq r_{j}$. In the following we show that the
twisted $KT$-relation can be used to replace the creation operators
in $\mathbb{B}(\bar{t})$ to annihilation ones therefore we can obtain
a recursion for the off-shell overlaps.  We use the regular form
of the $\mathfrak{sp}(N)$ symmetric K-matrix (\ref{eq:regTwK}).
We need the following components of the twisted $KT$-relation
\begin{equation}
\begin{split}K_{N,2}\langle\Psi|T_{2,2a-1}(z) & =\lambda_{0}(u)\langle\Psi|\widehat{T}_{N,N-2a+1}(-z)K_{N-2a+1,2a-1},\\
K_{N,2}\langle\Psi|T_{2,2a}(z) & =\lambda_{0}(u)\langle\Psi|\widehat{T}_{N,N-2a+2}(-z)K_{N-2a+2,2a},\\
K_{N-1,2}\langle\Psi|T_{1,2a-1}(z) & =\lambda_{0}(u)\langle\Psi|\widehat{T}_{N-1,N-2a+1}(-z)K_{N-2a+1,2a-1},\\
K_{N-1,2}\langle\Psi|T_{2,2a}(z) & =\lambda_{0}(u)\langle\Psi|\widehat{T}_{N-1,N-2a+2}(-z)K_{N-2a+2,2a}.
\end{split}
\end{equation}
Using the recurrence relation (\ref{eq:reqTW}) and the $KT$-relation
we obtain that
\begin{align}
\langle\Psi|\mathbb{B}(\bar{t}^{1},\{z,\bar{t}^{2}\},\{\bar{t}^{s}\}_{s=3}^{N-1})= & \sum_{a=2}^{N/2}\sum_{\mathrm{part}(\bar{t})}\langle\Psi|\widehat{T}_{N,N-2a+1}(-z)\mathbb{B}(\bar{t}^{1},\bar{t}^{2},\{\bar{t}_{\textsc{ii}}^{s}\}_{s=3}^{2a-2},\{\bar{t}^{s}\}_{s=2a-1}^{N-1})(\dots)\nonumber \\
+ & \sum_{a=2}^{N/2}\sum_{\mathrm{part}(\bar{t})}\langle\Psi|\widehat{T}_{N,N-2a+2}(-z)\mathbb{B}(\bar{t}^{1},\bar{t}^{2},\{\bar{t}_{\textsc{ii}}^{s}\}_{s=3}^{2a-1},\{\bar{t}^{s}\}_{s=2a}^{N-1})(\dots)\label{eq:recTW-1}\\
+ & \sum_{a=2}^{N/2}\sum_{\mathrm{part}(\bar{t})}\langle\Psi|\widehat{T}_{N-1,N-2a+1}(-z)\mathbb{B}(\bar{t}_{\textsc{ii}}^{1},\bar{t}^{2},\{\bar{t}_{\textsc{ii}}^{s}\}_{s=3}^{2a-2},\{\bar{t}^{s}\}_{s=2a-1}^{N-1})(\dots),\nonumber \\
+ & \sum_{a=2}^{N/2}\sum_{\mathrm{part}(\bar{t})}\langle\Psi|\widehat{T}_{N-1,N-2a+2}(-z)\mathbb{B}(\bar{t}_{\textsc{ii}}^{1},\bar{t}^{2},\{\bar{t}_{\textsc{ii}}^{s}\}_{s=3}^{2a-1},\{\bar{t}^{s}\}_{s=2a}^{N-1})(\dots).\nonumber 
\end{align}
Applying the action formula (\ref{eq:actTw}) with the simplified
notations
\begin{equation}
\begin{split}\widehat{T}_{i,j}\mathbb{B}^{r_{1},r_{2},\dots,r_{N-1}} & =\begin{cases}
\sum_{\mathrm{part}}(\dots)\mathbb{B}^{r_{1},\dots r_{N-j+1}+1,r_{N-j+2}+1,\dots,r_{N-i}+1,r_{N-i+1},\dots r_{N-1}}, & i\leq j,\\
\sum_{\mathrm{part}}(\dots)\mathbb{B}^{r_{1},\dots r_{N-i+1}-1,r_{N-i+2}-1,\dots,r_{N-j}-1,r_{N-j+1},\dots r_{N-1}}, & i>j.
\end{cases}\end{split}
\end{equation}
the recurrence equation (\ref{eq:recTW-1}) for the off-shell overlap
reads as
\begin{align}
\langle\Psi|\mathbb{B}^{r_{1},r_{2},\dots,r_{N-1}} & =\sum_{a=1}^{N/2}\sum_{\mathrm{part}}(\dots)\langle\Psi|\mathbb{B}^{r_{1}-1,r_{2}-2,r_{3}-2,\dots,r_{2a-2}-2,r_{2a-1}-1,r_{2a},\dots,r_{N-1}}.
\end{align}
We can see that we just obtained a recursion which decrease the quantum
number $r_{2}$ by $2$. We also note that this recursion leaves the
combinations $2r_{2k-1}-r_{2k-2}-r_{2k}$ invariant for $k=1,\dots,N/2$.
Applying this recursion iteratively we can eliminate all of the second
Bethe roots as
\begin{equation}
\langle\Psi|\mathbb{B}^{r_{1},r_{2},\dots,r_{N-1}}=\sum_{\mathrm{part}}(\dots)\langle\Psi|\mathbb{B}^{\tilde{r}_{1},0,\tilde{r}_{3}\dots,\tilde{r}_{N-1}}.
\end{equation}
Since recursion decrease the quantum number $r_{2}$ by 2, we obtain
non-vanishing overlap only when $r_{2}$ is even. The Bethe states
with $r_{2}=0$ are generated by the operators $T_{\alpha,\beta}$
and $T_{a,b}$ for $\alpha,\beta\leq2$ and $a,b\geq3$. The $\left\{ T_{\alpha,\beta}\right\} _{\alpha,\beta\leq2}$
and $\left\{ T_{a,b}\right\} _{a,b\geq3}$ are closed $Y(2)$ and
$Y(N-2)$ subalgebras of the original Yangian $Y(N)$, furthermore,
they are commuting subalgebras $[T_{\alpha,\beta}(u),T_{a,b}(v)]=0$.
We recall that the operators $\left\{ \widehat{T}_{\bar{\alpha},\bar{\beta}}\right\} _{\bar{\alpha},\bar{\beta}\geq N-1}$
or $\left\{ \widehat{T}_{\bar{a},\bar{b}}\right\} _{\bar{a},\bar{b}\leq N-2}$
generate the same subalgebras as $\left\{ T_{\alpha,\beta}\right\} _{\alpha,\beta\leq2}$
or $\left\{ T_{a,b}\right\} _{a,b\geq3}$ , respectively. In summary,
we have traced back the off-shell overlaps of $\mathfrak{gl}(N)$
spin chain to $\mathfrak{gl}(2)\oplus\mathfrak{gl}(N-2)$ spin chain,
i.e.
\begin{equation}
\langle\Psi|\mathbb{B}^{\tilde{r}_{1},0,\tilde{r}_{3}\dots,\tilde{r}_{N-1}}=\left(\langle\Psi'|\mathbb{B}^{\tilde{r}_{1}}\right)\times\left(\langle\Psi''|\mathbb{B}^{\tilde{r}_{3}\dots,\tilde{r}_{N-1}}\right),
\end{equation}
where the boundary states $\langle\Psi'|$, $\langle\Psi''|$ correspond
to the $K$-matrices of the $\mathfrak{gl}(2)$ and $\mathfrak{gl}(N-2)$
subsectors. Repeating the recursion, the overlaps can be expressed
as
\begin{equation}
\langle\Psi|\mathbb{B}^{r_{1},r_{2},\dots,r_{N-1}}=\sum_{\mathrm{part}}(\dots)\langle\Psi|\mathbb{B}^{\tilde{r}_{1},0,\tilde{r}_{3},0,\tilde{r}_{5},\dots,0,\tilde{r}_{N-1}}=\sum_{\mathrm{part}}(\dots)\prod_{a=1}^{N/2}\langle\Psi'|\mathbb{B}^{\tilde{r}_{2a-1}},
\end{equation}
which is a product of $\mathfrak{gl}(2)$ overlaps. The boundary states
$\langle\Psi'|$ of these $\mathfrak{gl}(2)$ subsectors correspond
to the $K$-matrix
\begin{equation}
K=\left(\begin{array}{cc}
1 & 0\\
0 & -1
\end{array}\right).
\end{equation}
We showed that the twisted and untwisted $KT$-relations are equivalent
for the $\mathfrak{gl}(2)$ spin chains and the $K$-matrix in the
untwisted convention reads as
\begin{equation}
\tilde{K}=\sigma K=\left(\begin{array}{cc}
1 & 0\\
0 & 1
\end{array}\right),
\end{equation}
which is a type $(2,0)$ $K$-matrix. We saw that only the $\mathfrak{gl}(2)$
singlets have non-vanishing overlaps for these boundary states therefore
we have the selection rules $\tilde{r}_{2k-1}=\frac{\mathbf{\boldsymbol{\Lambda}}_{2k-1}-\mathbf{\boldsymbol{\Lambda}}_{2k}}{2}$.
Since the combinations
\begin{equation}
\tilde{r}_{2k-1}=r_{2k-1}-\frac{r_{2k-2}+r_{2k}}{2},\quad\text{for }k=1,\dots,N/2,
\end{equation}
are invariant under the recursion we can express $\tilde{r}_{k}$-s
with $r_{k}$-s, therefore the non-vanishing overlaps require that
\begin{equation}
r_{2k-1}-\frac{r_{2k-2}+r_{2k}}{2}=\frac{\mathbf{\boldsymbol{\Lambda}}_{2k-1}-\mathbf{\boldsymbol{\Lambda}}_{2k}}{2},\quad\text{for }k=1,\dots,N/2.\label{eq:selRule-1}
\end{equation}

\subsection{Sum formula and the pair structure limit\label{subsec:Sum-formulaTW}}

The recursion described in the previous subsection provides a systematic
way to compute the overlaps. It turns out that combining this method
with the co-product property of the Bethe states and the boundary
states $\langle\Psi|$, we can derive the following sum formula for
the off-shell overlaps: 
\begin{align}
\mathcal{S}_{\bar{\alpha}}(\bar{t}) & :=\langle\Psi|\mathbb{B}(\bar{t})=\sum\frac{\prod_{\nu=1}^{N-1}f(\bar{t}_{\textsc{ii}}^{\nu},\bar{t}_{\textsc{i}}^{\nu})}{\prod_{\nu=1}^{N-2}f(\bar{t}_{\textsc{ii}}^{\nu+1},\bar{t}_{\textsc{i}}^{\nu})}\frac{\prod_{a=1}^{N/2}f(\bar{t}_{\textsc{ii}}^{2a-1},\bar{t}_{\textsc{iii}}^{2a-1})f(\bar{t}_{\textsc{iii}}^{2a-1},\bar{t}_{\textsc{i}}^{2a-1})}{\prod_{a=1}^{N/2-1}f(\bar{t}_{\textsc{ii}}^{2a},\bar{t}_{\textsc{iii}}^{2a-1})f(\bar{t}_{\textsc{iii}}^{2a+1},\bar{t}_{\textsc{i}}^{2a})}\times\nonumber \\
 & \times\mathcal{Z}(\bar{t}_{\textsc{i}})\bar{\mathcal{Z}}(\bar{t}_{\textsc{ii}})\prod_{a=1}^{N/2}S_{\alpha_{2a-1}}^{(2a-1)}(\bar{t}_{\textsc{iii}}^{2a-1})\prod_{\nu=1}^{N-1}\alpha_{\nu}(\bar{t}_{\textsc{i}}^{\nu}),\label{eq:sumTW}
\end{align}
where sum goes through the partitions $\text{\ensuremath{\bar{t}}}=\bar{t}_{\textsc{i}}\cup\bar{t}_{\textsc{ii}}\cup\bar{t}_{\textsc{iii}}$
where $\#\bar{t}_{\textsc{iii}}^{2a}=0$ and $\#\bar{t}_{\textsc{i}}^{2a},\#\bar{t}_{\textsc{ii}}^{2a}$
are even for $a=1,\dots,N/2-1$; and $\#\bar{t}_{\textsc{i}}^{2a-1}=\frac{\#\bar{t}_{\textsc{i}}^{2a-2}-\#\bar{t}_{\textsc{i}}^{2a}}{2}$,
$\#\bar{t}_{\textsc{ii}}^{2a-1}=\frac{\#\bar{t}_{\textsc{ii}}^{2a-2}-\#\bar{t}_{\textsc{ii}}^{2a}}{2}$,
$\#\bar{t}_{\textsc{iii}}^{2a-1}=\frac{\mathbf{\boldsymbol{\Lambda}}_{2a-1}-\mathbf{\boldsymbol{\Lambda}}_{2a}}{2}$
for $a=1,\dots,N/2$. The sum formula contains the overlap functions
$S_{\alpha_{2a-1}}^{(2a-1)}$ of the $\mathfrak{gl}(2)$ spin chain
\begin{equation}
\begin{split}S_{\alpha}^{(s)}(\bar{t}) & :=S_{\alpha}(\bar{t}+c\frac{s}{2})\Biggr|_{\alpha_{2a-1}(z)\to\alpha_{2a-1}(z-c\frac{s}{2})},\\
S_{\alpha}(\bar{t}) & =\sum f(\bar{t}_{\textsc{ii}}^{\nu},\bar{t}_{\textsc{i}}^{\nu})Z^{0}(\bar{t}_{\textsc{i}})Z^{0}(-\bar{t}_{\textsc{ii}})\alpha(\bar{t}_{\textsc{i}}),
\end{split}
\end{equation}
where
\begin{equation}
Z^{0}(\bar{t})=\kappa(\bar{t})\prod_{k<l}f(-t_{k},t_{l}),\qquad\kappa(z)=\frac{1}{z}.
\end{equation}
The proof of the sum formula can be found in appendix \ref{subsec:Sum-formula-Twisted-case}.
The HC-s $\mathcal{Z},\bar{\mathcal{Z}}$ are determined by the recursions
(\ref{eq:recZTw}) and (\ref{eq:recZbTw}). There is also a connection
between the HC-s
\begin{equation}
\mathcal{Z}(\bar{t})=\frac{1}{\prod_{s=1}^{N-2}f(\bar{t}^{s+1},\bar{t}^{s})}\bar{\mathcal{Z}}(\pi^{c}(\bar{t}))).
\end{equation}

Since the on-shell overlaps are non-vanishing only when the Bethe
roots have chiral pair structure (\ref{eq:chiralPair}), we need to
take the pair structure limit $t_{l}^{s}\to-t_{k}^{s}-sc$ of the
off-shell overlap. It turns out that the HC-s have poles in this limit.
The recurrence relations of HC-s can be used to derive the residue
of these poles at the pair structure limit $t_{l}^{s}\to-t_{k}^{s}-sc$.
For even $s$ i.e. in the limit $t_{l}^{2a}\to-t_{k}^{2a}-2ac$ we
obtain that
\begin{align}
\bar{\mathcal{Z}}(\bar{t}) & \to\frac{c}{t_{l}^{2a}+t_{k}^{2a}+2ac}\left[\frac{x_{a+1}}{x_{a}}h(t_{k}^{2a},t_{l}^{2a})h(t_{l}^{2a},t_{k}^{2a})\right]\frac{f(\bar{\tau}^{2a},t_{k}^{2a})f(\bar{\tau}^{2a},t_{l}^{2a})}{f(\bar{\tau}^{2a+1},t_{k}^{2a})f(\bar{\tau}^{2a+1},t_{l}^{2a})}\times\nonumber \\
 & \times\sum_{\mathrm{part}(\bar{\tau}^{2a-1},\bar{\tau}^{2a+1})}\bar{\mathcal{Z}}(\bar{\tau}_{\mathrm{i}})\frac{f(\bar{\tau}_{\mathrm{i}}^{2a-1},\bar{\tau}_{\mathrm{iii}}^{2a-1})}{f(\bar{\tau}^{2a},\bar{\tau}_{\mathrm{iii}}^{2a-1})}\frac{f(\bar{\tau}_{\mathrm{i}}^{2a+1},\bar{\tau}_{\mathrm{iii}}^{2a+1})}{f(\bar{\tau}^{2a+2},\bar{\tau}_{\mathrm{iii}}^{2a+1})}\times\label{eq:poleZb-1}\\
 & \times\left[\frac{1}{h(t_{k}^{2a},\bar{\tau}_{\mathrm{iii}}^{2a-1})h(t_{l}^{2a},\bar{\tau}_{\mathrm{iii}}^{2a-1})}\right]\left[g(\bar{\tau}_{\mathrm{iii}}^{2a+1},t_{k}^{2a})g(\bar{\tau}_{\mathrm{iii}}^{2a+1},t_{l}^{2a})\right]+reg.\nonumber 
\end{align}
where $\bar{\tau}=\bar{t}\backslash\{t_{k}^{2a},t_{l}^{2a}\}$. The
summation goes thought the partitions $\bar{\tau}^{s}=\bar{\tau}_{\mathrm{i}}^{s}\cup\bar{\tau}_{\mathrm{iii}}^{s}$
where $\#\bar{\tau}_{\mathrm{iii}}^{s}=\delta_{s,2a-1}+\delta_{s,2a+1}$.
For odd $s$ i.e. in the limit $t_{l}^{2a-1}\to-t_{k}^{2a-1}-(2a-1)c$
the HC-s are regular. These properties are derived in the appendix
\ref{sec:Poles-of-the}.

There is an other difference between the even and odd $s$. For $s=2b$
there is no extra selection rule for the $\alpha$-functions $\alpha_{2b}(z)$
i.e. it can contain arbitrary number of parameters (inhomogeneities)
therefore we can handle the expressions $t_{k}^{2b}$, $\alpha_{2b}(t_{k}^{2b})$
and also the derivatives $\alpha_{2b}'(t_{k}^{2b})$ (or equivalently
$X_{k}^{2b}=-c\alpha'(t_{k}^{2b})/\alpha(t_{k}^{2b})$) as independent
variables. On the other hand we have extra selection rules for $s=2b-1$
since the numbers of Bethe roots restrict the possible quantum spaces,
such as (\ref{eq:selRule-1})
\begin{equation}
n_{b}:=\frac{\mathbf{\boldsymbol{\Lambda}}_{2b-1}-\mathbf{\boldsymbol{\Lambda}}_{2b}}{2}=r_{2b-1}-\frac{r_{2b-2}+r_{2b}}{2}.\label{eq:selection}
\end{equation}
Since the $\alpha$-functions $\alpha_{2b-1}(z)$ reads as
\begin{equation}
\alpha_{2b-1}(z)=\prod_{j=1}^{m_{2b-1}}\frac{z-\theta_{j}^{(2b-1)}+s_{j}^{(2b-1)}c}{z-\theta_{j}^{(2b-1)}}\frac{z+c(2b-1)+\theta_{j}^{(2b-1)}}{z+c(2b-1)+\theta_{j}^{(2b-1)}-s_{j}^{(2b-1)}c},
\end{equation}
where $m_{2b-1}$ is the number of rectangular representation $(s,2b-1)$
in the quantum space and
\begin{equation}
\frac{\mathbf{\boldsymbol{\Lambda}}_{2b-1}-\mathbf{\boldsymbol{\Lambda}}_{2b}}{2}=n_{b}=\sum_{j=1}^{m_{2b-1}}s_{j}^{(2b-1)}.
\end{equation}
For a fixed $n_{b}$ the maximum number of inhomogeneities $\theta_{j}^{(2b-1)}$
appear when the all $s_{j}^{(2b-1)}=1$ therefore
\begin{equation}
\alpha_{2b-1}(z)=\prod_{j=1}^{n_{b}}\frac{z-\theta_{j}^{(2b-1)}+c}{z-\theta_{j}^{(2b-1)}}\frac{z+c(2b-1)+\theta_{j}^{(2b-1)}}{z+c(2b-1)+\theta_{j}^{(2b-1)}-c}.
\end{equation}
For $n_{b}\geq2$ we have at least two independent parameters $\theta_{1}^{(2b-1)},\theta_{2}^{(2b-1)}$
in the function $\alpha_{2b-1}(z)$. Taking a Bethe root $t_{k}^{2b-1}$
for a fixed $k$, we can handle the expressions $t_{k}^{2b-1}$, $\alpha_{2b-1}(t_{k}^{2b-1})$
and also the derivatives $\alpha_{2b-1}'(t_{k}^{2b-1})$ as independent
variables however the other expressions $\alpha_{2b-1}(t_{l}^{2b-1})$
and $\alpha_{2b-1}'(t_{l}^{2b-1})$ are not independent any more and
for $n_{b}=2$ we can express them as a function of $t_{k}^{2b-1}$,
$\alpha_{2b-1}(t_{k}^{2b-1})$ and $\alpha_{2b-1}'(t_{k}^{2b-1})$. 

It turns out that taking the $t_{l}^{s}\to-t_{k}^{s}-sc$ limit (for
$s=2b-1$ we assume that $n_{b}\geq2$ and we handle the expressions
$t_{k}^{2b-1}$, $\alpha_{2b-1}(t_{k}^{2b-1})$ and $X_{k}^{2b-1}$
as independent variables) of the sum formula we obtain that the $X_{k}^{s}$
dependence of the off-shell overlap is
\begin{equation}
\lim_{t_{l}^{s}\to-t_{k}^{s}-sc}S_{\bar{\alpha}}(\bar{t})=X_{k}^{s}\times F^{(s)}(t_{k}^{s})\frac{f(\bar{\tau}^{s},t_{k}^{s})f(\bar{\tau}^{s}t_{l}^{s})}{f(\bar{\tau}^{s+1},t_{k}^{s})f(\bar{\tau}^{s+1},t_{l}^{s})}S_{\bar{\alpha}^{mod}}(\bar{\tau})+\tilde{S},
\end{equation}
where $\bar{\tau}=\bar{t}\backslash\{t_{k}^{s},t_{l}^{s}\}$ and $\tilde{S}$
is independent from $X_{k}^{s}$ . We also used the following functions:
\begin{equation}
\begin{split}F^{(2b-1)}(z) & =\frac{c^{2}}{4}g(z,-z-(2b-1)c)g(-z-(2b-1)c,z),\\
F^{(2b)}(z) & =\frac{x_{b+1}}{x_{b}}\frac{c^{2}}{4}h(z,-z-2bc)h(-z-2bc,z),
\end{split}
\end{equation}
and the modified $\alpha$-s are 
\begin{equation}
\begin{split}\alpha_{s-1}^{mod}(z) & =\alpha_{s-1}(z)\frac{1}{f(t_{k}^{s},z)f(-t_{k}^{s}-sc,z)},\\
\alpha_{s}^{mod}(z) & =\alpha_{s}(z)\frac{f(t_{k}^{s},z)f(-t_{k}^{s}-sc,z)}{f(z,t_{k}^{s})f(z,-t_{k}^{s}-sc)},\\
\alpha_{s+1}^{mod}(z) & =\alpha_{s+1}(z)f(z,t_{k}^{s})f(z,-t_{k}^{s}-sc),\\
\alpha_{\nu}^{mod}(z) & =\alpha_{\nu}(z),\quad\text{for }\nu\neq s-1,s,s+1.
\end{split}
\end{equation}
The derivation can be found in appendix \ref{sec:Pair-structure-limit}.

We saw that for non-vanishing overlaps the quantum numbers $r_{2b}$
are even, however the quantum numbers $r_{2b-1}$ can be odd, for
which the limit $t_{k}^{2b-1}\to-(b-1/2)c$ is relevant. It turns
out that taking the $t_{k}^{2b-1}\to-(b-1/2)c$ limit (we assume that
$n_{b}\geq2$ and we handle the expression $X^{0,2b-1}=-c\alpha'(-(b-1/2)c)$
as independent variable) of the sum formula, we obtain that the $X^{0,2b-1}$
dependence of the off-shell overlap is
\begin{equation}
\lim_{t_{k}^{2b-1}\to-(b-1/2)c}S_{\bar{\alpha}}(\bar{t})=X^{0,2b-1}\left(-\frac{2}{c}\right)\frac{f(\bar{\tau}^{s},-(b-1/2)c)}{f(\bar{\tau}^{s+1},-(b-1/2)c)}S_{\bar{\alpha}^{mod}}(\bar{\tau})+\tilde{S},
\end{equation}
where $\bar{\tau}=\bar{t}\backslash\{t_{k}^{2a-1}\}$ and the modified
$\alpha$-s are 
\begin{equation}
\begin{split}\alpha_{2b-2}^{mod}(z) & =\alpha_{2b-2}(z)\frac{1}{f(-(b-1/2)c,z)},\\
\alpha_{2b-1}^{mod}(z) & =\alpha_{2b-1}(z)\frac{f(-(b-1/2)c,z)}{f(z,-(b-1/2)c)},\\
\alpha_{2b}^{mod}(z) & =\alpha_{2b}(z)f(z,-(b-1/2)c),\\
\alpha_{\nu}^{mod}(z) & =\alpha_{\nu}(z),\quad\text{for }\nu\neq2b-2,2b-1,2b+1.
\end{split}
\end{equation}
The derivation can be found in appendix \ref{subsec:Pair-structure-limit0}.

\subsection{On-shell limit}

Let us renormalize the overlap function as
\begin{align}
\mathbf{N}_{\bar{\alpha}}(\bar{t}) & =\left(-\frac{c}{2}\right)^{\mathbf{r}^{0}}\frac{1}{\prod_{\nu=1}^{N-1}F^{(\nu)}(\bar{t}^{+,\nu})\prod_{k\neq l}f(t_{l}^{+,\nu},t_{k}^{+,\nu})\prod_{k<l}f(t_{l}^{+,\nu},-t_{k}^{+,\nu}-\nu c)f(-t_{k}^{+,\nu}-\nu c,t_{l}^{+,\nu})}\times\nonumber \\
 & \times\frac{1}{\prod_{\nu\in\mathfrak{r}}\prod_{k}f(-\nu c/2,t_{k}^{+,\nu})f(t_{k}^{+,\nu},-\nu c/2)}\mathcal{S}_{\bar{\alpha}}(\bar{t}).
\end{align}
We can easily show that 
\begin{equation}
\lim_{t_{l}^{s}\to-t_{k}^{s}-sc}\mathbf{N}_{\bar{\alpha}}(\bar{t})=X_{k}^{s}\mathbf{N}_{\bar{\alpha}^{mod}}(\bar{\tau})+\tilde{\mathbf{N}},\label{eq:reqq}
\end{equation}
and
\begin{equation}
\lim_{t_{k}^{2b-1}\to-(b-1/2)c}\mathbf{N}_{\bar{\alpha}}(\bar{t})=X^{0,2b-1}\mathbf{N}_{\bar{\alpha}^{mod}}(\bar{\tau})+\tilde{\mathbf{N}},\label{eq:reqq2}
\end{equation}
which agrees with the embedding of the Gaudin-like determinant (\ref{eq:derivFX-1-2}),
(\ref{eq:derivFX-1-2-1}). However it is not enough for the proof
our conjecture. We can see that the Gaudin-like determinant is defined
for every set of quantum numbers $\mathbf{r}^{+},\mathbf{r}^{0}$
however the overlap function is defined only for $r_{2b-1}\geq\frac{r_{2b-2}+r_{2b}}{2}$
therefore we should use more complicated initial condition then property
(\ref{enum:prop3-1-2}). Another problem is that we have constraints
for the functions $\alpha_{2b-1}(z)$ and their most general form
is
\begin{equation}
\alpha_{2b-1}(z)=\mathcal{A}_{2b-1}(z|\bar{\theta}^{2b-1})=\prod_{j=1}^{n_{b}}\frac{z-\theta_{j}^{2b-1}+c}{z-\theta_{j}^{2b-1}}\frac{z+c(2b-1)+\theta_{j}^{2b-1}}{z+c(2b-1)+\theta_{j}^{2b-1}-c},
\end{equation}
which contain only $n_{b}=r_{2b-1}-\frac{r_{2b-2}+r_{2b}}{2}$ number
of free variable $\theta_{j}^{2b-1}$ for $j=1,\dots,n_{b}$ therefore
the expressions $\alpha'(t_{k}^{2b-1})$ are not independent, i.e.,
the pair structure limit can not handle as a function of the variables
$X_{k}^{+,\nu}$. We can only define functions 
\begin{equation}
\hat{\mathbf{N}}^{+}(\bar{t}^{+}|\{\bar{X}^{+,2a}\}_{a=1}^{\frac{N}{2}-1}|\{\bar{\theta}^{2a-1}\}_{a=1}^{\frac{N}{2}})=\lim_{\alpha_{2b}(t_{k}^{2b})\to\mathcal{F}_{k}^{2b}}\left[\lim_{\bar{t}^{-}\to\pi^{c}(\bar{t}^{+})}\mathbf{N}_{\bar{\alpha}}(\bar{t})\right],\label{eq:Nb}
\end{equation}
where the second limit is the formal on-shell limit with replacement
\begin{equation}
\alpha_{2b}(t_{k}^{2b})\to\mathcal{F}_{k}^{2b}:=\frac{f(t_{k}^{\mu},\bar{t}_{k}^{\mu})}{f(\bar{t}_{k}^{\mu},t_{k}^{\mu})}\frac{f(\bar{t}^{\mu+1},t_{k}^{\mu})}{f(t_{k}^{\mu},\bar{t}^{\mu-1})}.
\end{equation}
The expression $\hat{\mathbf{N}}^{+}$ (\ref{eq:Nb}) depends only
on the ''half'' of the $X$ variables $X_{k}^{+,2b}$ and the inhomogeneities
$\theta_{j}^{2a-1}$. Our conjecture is

\begin{equation}
\hat{\mathbf{N}}^{+}(\bar{t}^{+}|\{\bar{X}^{+,2a}\}_{a=1}^{\frac{N}{2}-1}|\{\bar{\theta}^{2a-1}\}_{a=1}^{\frac{N}{2}})=\mathbf{F}^{(\mathbf{r}^{+})}(\bar{X}^{+},\bar{t}^{+})\Biggr|_{X_{k}^{2b-1}\to-c\partial_{z}\log\mathcal{A}_{2b-1}(z|\bar{\theta}^{2b-1})\Biggr|_{z\to t_{k}^{2b-1}}},\label{eq:Conj}
\end{equation}
for on-shell Bethe roots. We postpone the precise proof to a later
work. Nevertheless, we believe that the proof of the recursion rules
(\ref{eq:reqq}) and (\ref{eq:reqq2}) sufficiently establishes the
correctness of the statement. 

Applying our conjecture, the on-shell overlap can be written as
\begin{align}
\langle\Psi|\mathbb{B}(\bar{t}) & =\left(-\frac{2}{c}\right)^{\mathbf{r}^{0}}\prod_{\nu=1}^{N-1}\left[F^{(\nu)}(\bar{t}^{+,\nu})\prod_{k\neq l}f(t_{l}^{+,\nu},t_{k}^{+,\nu})\prod_{k<l}f(t_{l}^{+,\nu},-t_{k}^{+,\nu}-\nu c)f(-t_{k}^{+,\nu}-\nu c,t_{l}^{+,\nu})\right]\times\nonumber \\
 & \times\prod_{\nu\in\mathfrak{r}}\left[\prod_{k}f(-\nu c/2,t_{k}^{+,\nu})f(t_{k}^{+,\nu},-\nu c/2)\right]\det G^{+},
\end{align}
and the normalized on-shell overlap has the following simple form
\begin{equation}
\frac{\langle\Psi|\mathbb{B}(\bar{t})}{\sqrt{\mathbb{C}(\bar{t})\mathbb{B}(\bar{t})}}=\left(-\frac{2}{c}\right)^{\mathbf{r}^{0}}\prod_{\nu=1}^{N-1}\mathbb{F}^{(\nu)}(\bar{t}^{+,\nu})\sqrt{\frac{\det G^{+}}{\det G^{-}}}.
\end{equation}
The one-particle overlap functions $\mathbb{F}^{(\nu)}$ read as
\begin{equation}
\mathbb{F}^{(\nu)}(z)=\frac{F^{(\nu)}(z)}{\sqrt{f(z,-z-\nu c)f(-z-\nu c,z)}},
\end{equation}
therefore
\begin{equation}
\begin{split}\mathbb{F}^{(2b)}(z) & =\left(\frac{x_{b+1}}{x_{b}}\right)\sqrt{(z+bc)^{2}\left(c^{2}/4-(z+bc)^{2}\right)},\\
\mathbb{F}^{(2k-1)}(z) & =\frac{1}{\sqrt{(z+\frac{2b-1}{2}c)^{2}\left(c^{2}/4-(z+(b-1/2)c)^{2}\right)}}.
\end{split}
\end{equation}

For $c=i$ convention, using the $Q$-functions the overlaps simplifies
as
\begin{equation}
\frac{\langle\Psi|\mathbb{B}(\bar{t})}{\sqrt{\mathbb{C}(\bar{t})\mathbb{B}(\bar{t})}}=\left[\prod_{k=1}^{N/2}x_{k}^{r_{2k-2}-r_{2k}}\right]\sqrt{\frac{\prod_{k=1}^{N/2-1}Q_{2k}(0)Q_{2k}(i/2)}{\prod_{k=1}^{N/2}\bar{Q}_{2k-1}(0)Q_{2k-1}(i/2)}}\times\sqrt{\frac{\det G^{+}}{\det G^{-}}}.\label{eq:twOnShell}
\end{equation}

\section{Applications for AdS/CFT correspondence\label{sec:Applications-for-AdS/CFT}}

In this section we apply our formulas for states which are relevant
for the AdS/CFT correspondence. In the $\mathcal{N}=4$ SYM and the
ABJM theories the scalar sectors can be described by $\mathfrak{gl}(4)$
spin chain where the quantum spaces are the tensor product of the
six dimensional $\Lambda=(1,1,0,0)$ representation and the alternating
tensor product of the defining $\Lambda=(1,0,0,0)$ and its contra-gradient
$\Lambda=(0,0,0,-1)$ representation. For the untwisted case the possible
residual symmetries are $\mathfrak{gl}(2)\oplus\mathfrak{gl}(2)$
or $\mathfrak{gl}(3)$ and for twisted case the possible residual
symmetries are $\mathfrak{so}(4)$ and $\mathfrak{sp}(4)$. The $\mathfrak{gl}(2)\oplus\mathfrak{gl}(2)$
and $\mathfrak{so}(4)$ symmetric boundary states were handled in
the previous paper \cite{Gombor:2021hmj}. Now we take the remaining
$\mathfrak{gl}(3)$ and $\mathfrak{sp}(4)$ symmetric boundary states
which appeared earlier in the context of defect AdS/CFT correspondence.
We show that the previously conjectured formulas agree with our general
result. We would like to emphasize that in the untwisted case our
formula is precisely proven.

\subsection{$\mathfrak{sp}(4)$ symmetric boundary state}

In this section, we reconstruct the overlap formula of \cite{deLeeuw:2019ebw,Kristjansen:2023ysz}
for the $\mathfrak{sp}(4)\cong\mathfrak{so}(5)$ symmetric boundary
state of the scalar sector of the $\mathcal{N}=4$ SYM which can be
described by an $\mathfrak{so}(6)$ spin chain. Now we take a homogeneous
spin chain with the representations $s_{k}=1$, $r_{k}=2$ of $\mathfrak{gl}(4)$
which is isomorphic to the defining representation of the $\mathfrak{so}(6)$
algebra. The monodromy matrix is 
\begin{equation}
T_{a}(u)=L_{a,J}^{(1,2)}(u+c/2)\dots L_{a,1}^{(1,2)}(u+c/2).
\end{equation}
This monodromy matrix is built from Lax-operators (\ref{eq:oneSiteMon})
for which the one-site state (\ref{eq:one-site}) satisfies the elementary
twisted $KT$-relation (\ref{eq:elemTwKT}) therefore the homogeneous
tensor product states satisfy
\begin{equation}
\langle\Psi|=\langle\psi|^{\otimes J},
\end{equation}
 the twisted $KT$-relation. We can use the notation $(Z,Y,X,\bar{X},\bar{Y},\bar{Z})$
for the basis vectors of the six dimensional representation as
\begin{equation}
\begin{split}Z & \to(1,1,0,0)\equiv e_{(1,2)},\qquad\bar{Z}\to(0,0,1,1)\equiv e_{(3,4)},\\
Y & \to(1,0,1,0)\equiv e_{(1,3)},\qquad\bar{Y}\to(0,1,0,1)\equiv e_{(2,4)},\\
X & \to(1,0,0,1)\equiv e_{(1,4)},\qquad\bar{X}\to(0,1,1,0)\equiv e_{(2,3)}.
\end{split}
\end{equation}

Using these notations the elementary state reads as
\begin{equation}
\langle\psi|=Z+\bar{Z},
\end{equation}
for $x_{1}=x_{2}=1$ which agrees with the $\mathfrak{sp}(4)\cong\mathfrak{so}(5)$
symmetric boundary states in \cite{deLeeuw:2019ebw}. This state is
also relevant for defect CFT defined by a ’t Hooft line embedded in
the $\mathcal{N}=4$ SYM \cite{Kristjansen:2023ysz}. We can apply
our on-shell overlap formula (\ref{eq:twOnShell}) as
\begin{equation}
\frac{\mathbb{C}(\bar{t})|\Psi\rangle\langle\Psi|\mathbb{B}(\bar{t})}{\mathbb{C}(\bar{t})\mathbb{B}(\bar{t})}=\frac{Q_{2}(0)Q_{2}(i/2)}{Q_{1}(0)Q_{1}(i/2)Q_{3}(0)Q_{3}(i/2)}\frac{\det G^{+}}{\det G^{-}}.
\end{equation}
We can see that this result is in complete agreement with the conjecture
of \cite{deLeeuw:2019ebw}.

\subsection{Boundary states in the ABJM theory}

In this section, we reconstruct the overlap formula of \cite{Yang:2021hrl}
and \cite{Jiang:2023cdm} for the scalar sector of the ABJM theory
which can be described by an alternating $\mathfrak{gl}(4)$ spin
chain. In \cite{Yang:2021hrl} a $\mathfrak{gl}(3)$ symmetric boundary
state appeared in the context of three-point function of two-determinant
and one single trace operator. In \cite{Jiang:2023cdm} a $\mathfrak{gl}(2)\oplus\mathfrak{gl}(2)$
and a $\mathfrak{gl}(3)$ symmetric boundary state corresponds to
correlation functions of single trace operators and a circular supersymmetric
Wilson loop. 

Now we take a alternating spin chain with the defining representations
$\Lambda=(1,0,0,0)$ and its contra-gradient $\Lambda=(0,0,0,-1)$
one. The monodromy matrix is 
\begin{equation}
T_{a}(u)=\bar{L}_{a,2J-1}(u+c)L_{a,2J-1}(u-c)\dots\bar{L}_{a,2}(u+c)L_{a,1}(u-c).
\end{equation}
This monodromy matrix is built from the elementary monodromy matrices
(\ref{eq:elemUTwmon}) ($\theta=c)$, for which the two-site state
(\ref{eq:elemTwoSite}) satisfies the elementary untwisted $KT$-relation
(\ref{eq:elemUtwKT}) therefore the tensor product states 
\begin{equation}
\langle\Psi|=\langle\psi(c)|^{\otimes J}
\end{equation}
satisfy the untwisted $KT$-relation. We can use the notation $(Y^{1},Y^{2},Y^{3},Y^{4})$,
$(\bar{Y}_{1},\bar{Y}_{2},\bar{Y}_{3},\bar{Y}_{4})$ for the basis
vectors of the representations $\Lambda=(1,0,0,0)$ and its contra-gradient
$\Lambda=(0,0,0,-1)$, respectively, as
\begin{equation}
\begin{split}Y^{1} & \to(1,0,0,0),\qquad\bar{Y}_{1}\to(-1,0,0,0),\\
Y^{2} & \to(0,1,0,0),\qquad\bar{Y}_{2}\to(0,-1,0,0),\\
Y^{3} & \to(0,0,1,0),\qquad\bar{Y}_{3}\to(0,0,-1,0),\\
Y^{4} & \to(0,0,0,1),\qquad\bar{Y}_{4}\to(0,0,0,-1).
\end{split}
\end{equation}

\subsubsection{Three-point functions}

Let us choose a $\mathfrak{gl}(3)$ symmetric $K$-matrix as $\mathfrak{a}=-c$
and 
\begin{equation}
\mathcal{U}=\left(\begin{array}{cccc}
0 & 0 & 0 & -1\\
0 & 1 & 0 & 0\\
0 & 0 & 1 & 0\\
-1 & 0 & 0 & 0
\end{array}\right),
\end{equation}
for which the explicit form of the $K$-matrix is
\begin{equation}
K(u)=\left(\begin{array}{cccc}
\frac{-c}{u} & 0 & 0 & -1\\
0 & \frac{-c+u}{u} & 0 & 0\\
0 & 0 & \frac{-c+u}{u} & 0\\
-1 & 0 & 0 & \frac{-c}{u}
\end{array}\right).
\end{equation}
The two-site state (\ref{eq:elemTwoSite}) reads as
\begin{equation}
\langle\psi|=\sum_{A,B}K_{B,A}(c)Y^{A}\otimes\bar{Y}_{B}=-(Y^{1}+Y^{4})\otimes(\bar{Y}_{1}+\bar{Y}_{4}),
\end{equation}
which agrees with the boundary states of the maximal giant graviton
in \cite{Yang:2021hrl}. To apply our overlap formula we need to calculate
the regular form of the $\mathcal{U}$-matrix (\ref{eq:reg2}), which
is
\begin{equation}
\mathcal{U}^{(2)}=\left(\begin{array}{cccc}
-1 & 0 & 0 & 0\\
0 & 1 & 0 & 0\\
0 & 0 & 1 & 0\\
-1 & 0 & 0 & 1
\end{array}\right).
\end{equation}
The parameter $\mathfrak{b}_{1}$ can be calculated from (\ref{eq:bdef}),
i.e., 
\begin{equation}
\mathfrak{b}_{1}=-1.
\end{equation}
Now, we can apply our on-shell overlap formula (\ref{eq:onShellUtw})
as
\begin{equation}
\frac{\langle\Psi|\mathbb{B}(\bar{t})}{\sqrt{\mathbb{C}(\bar{t})\mathbb{B}(\bar{t})}}=(-1)^{r_{1}}\frac{Q_{1}(\frac{i}{2})}{\sqrt{Q_{2}(0)Q_{2}(\frac{i}{2})}}\times\sqrt{\frac{\det G^{+}}{\det G^{-}}}.
\end{equation}
We can see that this result is in complete agreement with the conjecture
of \cite{Yang:2021hrl}.

\subsubsection{Wilson loop one-point functions}

\paragraph{Overlap of 1/6 BPS Wilson loops}

Let us choose a $\mathfrak{gl}(2)$ symmetric $K$-matrix as $\mathfrak{a}=0$
and 
\begin{equation}
K(u)=\mathcal{U}=\left(\begin{array}{cccc}
-1 & 0 & 0 & 0\\
0 & -1 & 0 & 0\\
0 & 0 & 1 & 0\\
0 & 0 & 0 & 1
\end{array}\right).\label{eq:K16}
\end{equation}
The elementary two-site state (\ref{eq:elemTwoSite}) reads as
\begin{equation}
\langle\psi|=\lim_{u\to\infty}\sum_{A,B}K_{B,A}(c)Y^{A}\otimes\bar{Y}_{B}=-Y^{1}\otimes\bar{Y}_{1}-Y^{2}\otimes\bar{Y}_{2}+Y^{3}\otimes\bar{Y}_{3}+Y^{4}\otimes\bar{Y}_{4},\label{eq:psi16}
\end{equation}
which agrees with the boundary states of the of 1/6 BPS Wilson loops
in \cite{Jiang:2023cdm}. We derived the overlap for the regular form
$K$-matrix
\begin{equation}
K^{r}(u)=\frac{\mathfrak{a}}{z}+\left(\begin{array}{cccc}
-1 & 0 & 0 & 0\\
0 & -1 & 0 & 0\\
0 & \mathfrak{b}_{2} & 1 & 0\\
\mathfrak{b}_{1} & 0 & 0 & 1
\end{array}\right),
\end{equation}
therefore we can obtain the overlap corresponding to $K$-matrix (\ref{eq:K16})
by taking the $\mathfrak{b}_{1},\mathfrak{b}_{2},\mathfrak{a}\to0$
limit. However this limit of our overlap formula is not well defined
since we fixed the normalization of the boundary state as $\langle\Psi|0\rangle=1$,
however when $\mathfrak{b}_{1}=0$ the overlap with the pseudo-vacuum
is zero $\langle\Psi|0\rangle=0$. It can be seen from the explicit
form of the corresponding two-site state (given by (\ref{eq:elemTwoSite}))
as
\begin{equation}
\langle\psi^{r}|=\lim_{u\to\infty}\sum_{A,B}K_{B,A}^{r}(c)Y^{A}\otimes\bar{Y}_{B}=\mathfrak{b}_{1}Y^{1}\otimes\bar{Y}_{4}+\mathfrak{b}_{2}Y^{2}\otimes\bar{Y}_{3}-Y^{1}\otimes\bar{Y}_{1}-Y^{2}\otimes\bar{Y}_{2}+Y^{3}\otimes\bar{Y}_{3}+Y^{4}\otimes\bar{Y}_{4}.
\end{equation}
To obtain the two-site state (\ref{eq:psi16}) in the limit $\mathfrak{b}_{1},\mathfrak{b}_{2}\to0$,
we have to fix the normalization in a $\mathfrak{b}_{1}$ dependent
way: $\langle\Psi|0\rangle=\mathfrak{b}_{1}^{J}$. Using this normalization,
our overlap function for the regular $K$-matrix (\ref{eq:onShellUtw})
modifies as
\begin{equation}
\frac{\langle\Psi^{r}|\mathbb{B}(\bar{t})}{\sqrt{\mathbb{C}(\bar{t})\mathbb{B}(\bar{t})}}=\left[\mathfrak{b}_{1}^{J-r_{1}}\mathfrak{b}_{2}^{r_{1}-r_{2}}\right]\times\frac{Q_{2}(\mathfrak{a})}{\sqrt{Q_{2}(0)Q_{2}(\frac{i}{2})}}\times\sqrt{\frac{\det G^{+}}{\det G^{-}}}.
\end{equation}
Now we can take the limit. Since $J\geq r_{1}$ and $r_{1}\geq r_{2}$,
we obtain non-vanishing overlaps only when $J=r_{1}=r_{2}$ and the
final formula simplifies as
\begin{equation}
\frac{\langle\Psi|\mathbb{B}(\bar{t})}{\sqrt{\mathbb{C}(\bar{t})\mathbb{B}(\bar{t})}}=\lim_{\mathfrak{b}_{1},\mathfrak{b}_{2},\mathfrak{a}\to0}\frac{\langle\Psi^{r}|\mathbb{B}(\bar{t})}{\sqrt{\mathbb{C}(\bar{t})\mathbb{B}(\bar{t})}}=\sqrt{\frac{Q_{2}(0)}{Q_{2}(\frac{i}{2})}}\times\sqrt{\frac{\det G^{+}}{\det G^{-}}}.
\end{equation}
We can see that this result is in complete agreement with the conjecture
(4.29) in \cite{Jiang:2023cdm}.

\paragraph{Overlap of 1/2 BPS Wilson loops}

Let us choose a $\mathfrak{gl}(3)$ symmetric $K$-matrix as $\mathfrak{a}=0$
and 
\begin{equation}
K(u)=\mathcal{U}=\left(\begin{array}{cccc}
-1 & 0 & 0 & 0\\
0 & 1 & 0 & 0\\
0 & 0 & 1 & 0\\
0 & 0 & 0 & 1
\end{array}\right).\label{eq:K16-1}
\end{equation}
The elementary two-site state (\ref{eq:elemTwoSite}) reads as
\begin{equation}
\langle\psi|=\lim_{u\to\infty}\sum_{A,B}K_{B,A}(c)Y^{A}\otimes\bar{Y}_{B}=-Y^{1}\otimes\bar{Y}_{1}+Y^{2}\otimes\bar{Y}_{2}+Y^{3}\otimes\bar{Y}_{3}+Y^{4}\otimes\bar{Y}_{4},
\end{equation}
which agrees with the boundary states of the of 1/2 BPS Wilson loops
in \cite{Jiang:2023cdm}. We derived the overlap for the regular form
$K$-matrix
\begin{equation}
K^{r}(u)=\frac{\mathfrak{a}}{z}+\left(\begin{array}{cccc}
-1 & 0 & 0 & 0\\
0 & 1 & 0 & 0\\
0 & 0 & 1 & 0\\
\mathfrak{b}_{1} & 0 & 0 & 1
\end{array}\right),
\end{equation}
therefore we can obtain the overlap corresponding to $K$-matrix (\ref{eq:K16-1})
by taking the $\mathfrak{b}_{1},\mathfrak{a}\to0$ limit. Applying
the previous argument we need to fix the normalization as $\langle\Psi|0\rangle=\mathfrak{b}_{1}^{J}$.
Using this normalization, our overlap function for the regular $K$-matrix
(\ref{eq:onShellUtw}) modifies as
\begin{equation}
\frac{\langle\Psi^{r}|\mathbb{B}(\bar{t})}{\sqrt{\mathbb{C}(\bar{t})\mathbb{B}(\bar{t})}}=\mathfrak{b}_{1}^{J-r_{1}}\frac{Q_{1}(-\frac{i}{2})}{\sqrt{Q_{2}(0)Q_{2}(\frac{i}{2})}}\times\sqrt{\frac{\det G^{+}}{\det G^{-}}},
\end{equation}
where we have the selection rule $r_{1}=r_{2}$. Now we can take the
limit. Since $J\geq r_{1}$, we obtain non-vanishing overlaps only
when $J=r_{1}$ and the final formula simplifies as
\begin{equation}
\frac{\langle\Psi|\mathbb{B}(\bar{t})}{\sqrt{\mathbb{C}(\bar{t})\mathbb{B}(\bar{t})}}=\lim_{\mathfrak{b}_{1},\mathfrak{a}\to0}\frac{\langle\Psi^{r}|\mathbb{B}(\bar{t})}{\sqrt{\mathbb{C}(\bar{t})\mathbb{B}(\bar{t})}}=\frac{Q_{1}(-\frac{i}{2})}{\sqrt{Q_{2}(0)Q_{2}(\frac{i}{2})}}\times\sqrt{\frac{\det G^{+}}{\det G^{-}}}.
\end{equation}
We can see that this result is in complete agreement with the conjecture
(5.7) in \cite{Jiang:2023cdm}.

\subsection{Boundary state with infinite dimensional representation\label{subsec:Boundary-state-with}}

One of the main advantages of our results is that they are representation-independent,
i.e. they can be applied to infinite dimensional quantum spaces. Here
is a concrete example that is relevant for correlators between the
’t Hooft line embedded in $\mathcal{N}=4$ SYM theory and non-protected
bulk operators \cite{Kristjansen:2023ysz}. The $SL(2)$ sector can
be described by homogeneous $\mathfrak{gl}(2)$ spin chains where
the quantum spaces are tensor products of infinite dimensional representation
with highest weight $\Lambda=(-1/2,1/2)$. The monodromy matrix is
\begin{equation}
T_{0}(u)=L_{0,J}^{(-\frac{1}{2},\frac{1}{2})}(u)\dots L_{0,1}^{(-\frac{1}{2},\frac{1}{2})}(u),
\end{equation}
and the Lax-operator $L(u)$ has a usual form
\begin{equation}
L^{(-\frac{1}{2},\frac{1}{2})}(u)=\mathbf{1}+\frac{c}{u}\sum_{k,l=1}^{2}E_{k,l}\otimes E_{l,k}^{(-\frac{1}{2},\frac{1}{2})}=\sum_{k,l=1}^{2}E_{k,l}\otimes\ell_{k,l}(u).
\end{equation}
Let us choose the representation $E_{k,l}^{(-\frac{1}{2},\frac{1}{2})}$
as 
\begin{equation}
\begin{split}E_{1,1}^{(-\frac{1}{2},\frac{1}{2})}|n\rangle & =-(n+\frac{1}{2})|n\rangle,\\
E_{2,2}^{(-\frac{1}{2},\frac{1}{2})}|n\rangle & =(n+\frac{1}{2})|n\rangle,\\
E_{2,1}^{(-\frac{1}{2},\frac{1}{2})}|n\rangle & =-i(n+1)|n+1\rangle,\\
E_{1,2}^{(-\frac{1}{2},\frac{1}{2})}|n\rangle & =-in|n-1\rangle.
\end{split}
\end{equation}
The boundary state which describes correlators for the ’t Hooft line
embedded in $\mathcal{N}=4$ SYM theory can be written as
\begin{equation}
\langle\Psi|=\langle B|^{\otimes J},\label{eq:fullB}
\end{equation}
where the elementary one-site state is
\begin{equation}
\langle B|=\left(\sum_{n=0}^{\infty}(-1)^{n}P_{n}(\cos\theta)\langle n|\right).
\end{equation}
 Now we have to find the corresponding $K$-matrix of the one-site
state which satisfies the untwisted $KT$-relation\footnote{For the $\mathfrak{gl}(2)$ spin chains the untwisted and twisted
KT-relations are equivalent. For simplicity, we prefer to use the
former notation. }
\begin{equation}
K_{0}(u)\langle B|T_{0}(u)=\langle B|T_{0}(-u)K_{0}(u),
\end{equation}
for $J=1$. Let us parameterize the $K$-matrix in the usual way
\begin{equation}
K(u)=\frac{\mathfrak{a}}{u}\mathbf{1}+\mathcal{U},\qquad\mathcal{U}=\left(\begin{array}{cc}
\mathcal{U}_{1,1} & \mathcal{U}_{1,2}=\frac{1-\mathcal{U}_{1,1}^{2}}{\mathcal{U}_{2,1}}\\
\mathcal{U}_{2,1} & \mathcal{U}_{2,2}=-\mathcal{U}_{1,1}
\end{array}\right).\label{eq:Kansatz}
\end{equation}
In the auxiliary space we have four equations:
\begin{align}
K_{1,1}(u)\langle B|\ell_{1,1}(u)+K_{1,2}(u)\langle B|\ell_{2,1}(u) & =\langle B|\ell_{1,1}(-u)K_{1,1}(u)+\langle B|\ell_{1,2}(-u)K_{2,1}(u),\label{eq:first}\\
K_{1,1}(u)\langle B|\ell_{1,2}(u)+K_{1,2}(u)\langle B|\ell_{2,2}(u) & =\langle B|\ell_{1,1}(-u)K_{1,2}(u)+\langle B|\ell_{1,2}(-u)K_{2,2}(u),\label{eq:sec}\\
K_{2,1}(u)\langle B|\ell_{1,1}(u)+K_{2,2}(u)\langle B|\ell_{2,1}(u) & =\langle B|\ell_{2,1}(-u)K_{1,1}(u)+\langle B|\ell_{2,2}(-u)K_{2,1}(u),\label{eq:third}\\
K_{2,1}(u)\langle B|\ell_{1,2}(u)+K_{2,2}(u)\langle B|\ell_{2,2}(u) & =\langle B|\ell_{2,1}(-u)K_{1,2}(u)+\langle B|\ell_{2,2}(-u)K_{2,2}(u).\label{eq:forth}
\end{align}
Since
\begin{equation}
E_{1,1}^{(-\frac{1}{2},\frac{1}{2})}|n\rangle=-E_{2,2}^{(-\frac{1}{2},\frac{1}{2})}|n\rangle,
\end{equation}
for all $n$ we have the identity
\begin{equation}
\langle B|\ell_{2,2}(-u)=\langle B|\ell_{1,1}(u).
\end{equation}
Let us apply this identity in the third equation (\ref{eq:third}).
It simplifies as
\begin{equation}
K_{2,2}(u)\langle B|\ell_{2,1}(u)=\langle B|\ell_{2,1}(-u)K_{1,1}(u),\label{eq:elemKT2}
\end{equation}
i.e.,
\begin{equation}
\left(K_{2,2}(u)+K_{1,1}(u)\right)\langle B|E_{1,2}^{(-\frac{1}{2},\frac{1}{2})}=0.
\end{equation}
Since $\langle B|E_{1,2}^{(-\frac{1}{2},\frac{1}{2})}\neq0$, we obtain
that
\begin{equation}
K_{2,2}(u)=-K_{1,1}(u),
\end{equation}
therefore
\begin{equation}
\mathfrak{a}=0.
\end{equation}
Let us continue with the first equation (\ref{eq:first}). It simplifies
as
\begin{equation}
2\mathcal{U}_{1,1}\langle B|E_{1,1}^{(-\frac{1}{2},\frac{1}{2})}+\mathcal{U}_{1,2}\langle B|E_{1,2}^{(-\frac{1}{2},\frac{1}{2})}+\langle B|E_{2,1}^{(-\frac{1}{2},\frac{1}{2})}\mathcal{U}_{2,1}=0.
\end{equation}
Let us multiply with the state $|n\rangle$:
\begin{equation}
-2(n+\frac{1}{2})\mathcal{U}_{1,1}\langle B|n\rangle-in\mathcal{U}_{1,2}\langle B|n-1\rangle-i(n+1)\mathcal{U}_{2,1}\langle B|n+1\rangle=0,
\end{equation}
therefore we obtain that
\begin{equation}
\mathcal{U}_{2,1}(n+1)P_{n+1}(\cos\theta)=-i\mathcal{U}_{1,1}(2n+1)P_{n}(\cos\theta)-n\mathcal{U}_{1,2}P_{n-1}(\cos\theta).
\end{equation}
Using the defining relation of the Legendre polynomials
\begin{equation}
(n+1)P_{n+1}(\cos\theta)=(2n+1)\cos\theta P_{n}(\cos\theta)-nP_{n-1}(\cos\theta),
\end{equation}
we can fix the unknowns as
\begin{align}
\mathcal{U}_{1,2} & =\mathcal{U}_{2,1},\\
\mathcal{U}_{1,1} & =i\cos\theta\mathcal{U}_{2,1}.
\end{align}
The involution property of $\mathcal{U}$ fixes the normalization
freedom as
\begin{equation}
\mathcal{U}_{1,2}=\frac{1-\mathcal{U}_{1,1}^{2}}{\mathcal{U}_{2,1}}\to\mathcal{U}_{2,1}=\frac{1+\left(\cos\theta\mathcal{U}_{2,1}\right)^{2}}{\mathcal{U}_{2,1}}\to\mathcal{U}_{2,1}=\pm\frac{1}{\sin\theta}.
\end{equation}
Without loss of generality, choose the positive sign for which the
$\mathcal{U}$-matrix and $K$-matrix read as
\begin{equation}
K(z)=\mathcal{U}=\left(\begin{array}{cc}
i\cot\theta & \frac{1}{\sin\theta}\\
\frac{1}{\sin\theta} & -i\cot\theta
\end{array}\right).
\end{equation}
This K-matrix satisfies the remaining two equations (\ref{eq:sec})
and (\ref{eq:forth}) therefore the elementary $KT$-equation (\ref{eq:elemKT2})
is satisfied. Using the co-product property of $KT$-relation, the
tensor product boundary states (\ref{eq:fullB}) satisfy the $KT$-relation
for any $J$. The existence of the $K$-matrix proves the integrability
of the boundary state (\ref{eq:fullB}). To use our overlap formula,
we need to calculate the regular form of the $\mathcal{U}$-matrix
(\ref{eq:reg2}):
\begin{equation}
\mathcal{U}^{(2)}=\left(\begin{array}{cc}
-1 & 0\\
\frac{1}{\sin\theta} & 1
\end{array}\right).
\end{equation}
and the variable $\mathfrak{b}_{1}$ is defined by (\ref{eq:bdef}),
i.e.,
\begin{equation}
\mathfrak{b}_{1}=\frac{1}{\sin\theta}.
\end{equation}
Now we can apply our on-shell overlap formula (\ref{eq:onShellUtw})
\begin{equation}
\frac{\langle\Psi|\mathbb{B}(\bar{t})}{\sqrt{\mathbb{C}(\bar{t})\mathbb{B}(\bar{t})}}=\left(\sin\theta\right)^{r_{1}}\times\sqrt{\frac{Q_{1}(0)}{Q_{1}(\frac{i}{2})}}\times\sqrt{\frac{\det G^{+}}{\det G^{-}}}.
\end{equation}
We can see that this result is in complete agreement with the conjecture
of \cite{Kristjansen:2023ysz}. 

\subsection{Boundary state in the gluon subsector}

In the $\mathcal{N}=4$ SYM there is the gluonic spin-1 $\mathfrak{su}(2)$
subsector which is formed by the self-dual components of the field
strength. The one-loop mixing matrix of the gluon operators is the
Hamiltonian of an $SO(3)$ spin chain which is a homogeneous $\mathfrak{gl}(2)$
spin chain where the quantum space is a tensor product of the representation
$\Lambda=(1,-1)$.

We can define the monodromy matrix as
\begin{equation}
T_{0}(u)=L_{0,J}^{(1,-1)}(u)\dots L_{0,1}^{(1,-1)}(u),
\end{equation}
and the Lax-operator $L(u)$ has a usual form
\begin{equation}
L^{(1,-1)}(u)=\mathbf{1}+\frac{c}{u}\sum_{k,l=1}^{2}E_{k,l}\otimes E_{l,k}^{(1,-1)}=\sum_{k,l=1}^{2}E_{k,l}\otimes\ell_{k,l}(u).
\end{equation}
We can choose the convention as
\begin{equation}
\begin{split}E_{1,1}^{(1,-1)} & =S_{z},\quad E_{1,2}^{(1,-1)}=S_{x}+iS_{y},\\
E_{2,2}^{(1,-1)} & =-S_{z},\quad E_{2,1}^{(1,-1)}=S_{x}-iS_{y},
\end{split}
\end{equation}
where
\begin{equation}
S_{x}=\left(\begin{array}{ccc}
0 & 0 & 0\\
0 & 0 & i\\
0 & -i & 0
\end{array}\right),\quad S_{y}=\left(\begin{array}{ccc}
0 & 0 & -i\\
0 & 0 & 0\\
i & 0 & 0
\end{array}\right),\quad S_{z}=\left(\begin{array}{ccc}
0 & -i & 0\\
i & 0 & 0\\
0 & 0 & 0
\end{array}\right).
\end{equation}
The pseudo-vacuum is
\begin{equation}
|0^{(1,-1)}\rangle=\frac{1}{\sqrt{2}}\left(\begin{array}{c}
1\\
i\\
0
\end{array}\right).
\end{equation}
The boundary state which describes correlators for the ’t Hooft line
embedded in $\mathcal{N}=4$ SYM theory can be written as
\begin{equation}
\langle\Psi|=\langle B|^{\otimes J},\label{eq:fullB-1}
\end{equation}
where the elementary one-site state is
\begin{equation}
\langle B|=\left(\begin{array}{c}
x_{1}\\
x_{2}\\
x_{3}
\end{array}\right)^{T}=\mathbf{x}^{T}.
\end{equation}
Let us introduce the notations
\begin{equation}
z=x_{1}+ix_{2},\qquad\rho^{2}=x_{1}^{2}+x_{2}^{2}+x_{3}^{2}.
\end{equation}
The vacuum overlap is
\begin{equation}
\langle B|0\rangle=\frac{1}{2^{J/2}}z^{J}.
\end{equation}
Now we have to find the corresponding $K$-matrix of the one-site
state which satisfies the untwisted $KT$-relation
\begin{equation}
K_{0}(u)\langle B|T_{0}(u)=\langle B|T_{0}(-u)K_{0}(u),
\end{equation}
for $J=1$. The calculation is completely analogue as was in the previous
section. We use the same ansatz (\ref{eq:Kansatz}) for the $K$-matrix.
Solving the equations (\ref{eq:first})-(\ref{eq:forth}) we obtain
that
\begin{equation}
\mathfrak{a}=0,\quad\mathcal{U}_{1,1}=-\frac{x_{3}}{z}\mathcal{U}_{2,1},\quad\mathcal{U}_{2,2}=\frac{x_{3}}{z}\mathcal{U}_{2,1},\quad\mathcal{U}_{1,2}=\frac{x_{1}-ix_{2}}{z}\mathcal{U}_{2,1}.
\end{equation}
From the normalization $\mathcal{U}^{2}=\mathbf{1}$ we obtain that
\begin{equation}
\mathcal{U}_{2,1}=\frac{z}{\rho}.
\end{equation}
The existence of the $K$-matrix proves the integrability of the boundary
state (\ref{eq:fullB-1}). To use our overlap formula, we need to
calculate the regular form of the $\mathcal{U}$-matrix (\ref{eq:reg2}):
\begin{equation}
\mathcal{U}^{(2)}=\left(\begin{array}{cc}
-1 & 0\\
\frac{z}{\rho} & 1
\end{array}\right).
\end{equation}
and the variable $\mathfrak{b}_{1}$ is defined by (\ref{eq:bdef}),
i.e.,
\begin{equation}
\mathfrak{b}_{1}=\frac{z}{\rho}.
\end{equation}
Now we can apply our on-shell overlap formula (\ref{eq:onShellUtw})
\begin{equation}
\frac{\langle\Psi|\mathbb{B}(\bar{t})}{\sqrt{\mathbb{C}(\bar{t})\mathbb{B}(\bar{t})}}=\frac{1}{2^{J/2}}\frac{z^{J-r_{1}}}{\rho^{-r_{1}}}\times\sqrt{\frac{Q_{1}(0)}{Q_{1}(\frac{i}{2})}}\times\sqrt{\frac{\det G^{+}}{\det G^{-}}}.
\end{equation}
We can see that this result is in complete agreement with the conjecture
of \cite{Kristjansen:2023ysz}. 

\section{Conclusion}

In this paper we determined the on-shell overlap functions of the
$\mathfrak{gl}(N)$ symmetric spin chains for the $\mathfrak{sp}(N)$
and $\mathfrak{gl}(M)\oplus\mathfrak{gl}(N-M)$ symmetric boundary
states (see (\ref{eq:onShellUtw}),(\ref{eq:onShellUtw2}) and (\ref{eq:twOnShell})).
Combing them with the results of \cite{Gombor:2021hmj}, the on-shell
overlap formulas are available for every integrable boundary states
of the $\mathfrak{gl}(N)$ spin chains which are built from one- or
two-site states. We gave a complete proof for (\ref{eq:onShellUtw})
and (\ref{eq:onShellUtw2}), however the precise derivation of (\ref{eq:twOnShell})
is positioned to a later work.

It would be interesting to generalize our method for $K$-matrices
with extra boundary degrees of freedom, i.e., when the boundary states
are given by matrix product states (MPS) \cite{Pozsgay_2019}. This
generalization would be very important since these MPS appear in the
context of defect AdS/CFT correspondence \cite{Buhl-Mortensen:2015gfd,deLeeuw:2016umh,DeLeeuw:2018cal,deLeeuw:2019ebw,Kristjansen:2021abc,Gombor:2022aqj}.

Another interesting direction could be the generalization to supersymmetric
$\mathfrak{gl}(N|M)$ spin chains. Applying our methods one might
be able to find the overlap functions for every grading, which could
provide an alternative proof of the previously discovered fermionic
duality rules of the on-shell overlaps \cite{Kristjansen:2020vbe,Kristjansen:2021xno}.

\section*{Acknowledgments}

This paper was supported by the NKFIH grant PD142929 and the János
Bolyai Research Scholarship of the Hungarian Academy of Science.

\appendix

\section{Off-shell Bethe vectors\label{sec:Off-shell-Bethe-vectors}}

In this section we review the recurrence and action formulas of the
off-shell Bethe vectors which are used in our derivations of the overlaps.
These formulas can be found in \cite{Hutsalyuk:2017tcx,Hutsalyuk:2017way,Liashyk:2018egk,Hutsalyuk:2020dlw}.

The off-shell Bethe vectors can be calculated from the following sum
formula \cite{Hutsalyuk:2017tcx}

\begin{equation}
\mathbb{B}(\{z,\bar{t}^{1}\},\left\{ \bar{t}^{k}\right\} _{k=2}^{N-1})=\sum_{j=2}^{N}\frac{T_{1,j}(z)}{\lambda_{2}(z)}\sum_{\mathrm{part}(\bar{t})}\mathbb{B}(\bar{t}^{1},\left\{ \bar{t}_{\textsc{ii}}^{k}\right\} _{k=2}^{j-1},\left\{ \bar{t}^{k}\right\} _{k=j}^{N-1})\frac{\prod_{\nu=2}^{j-1}\alpha_{\nu}(\bar{t}_{\textsc{i}}^{\nu})g(\bar{t}_{\textsc{i}}^{\nu},\bar{t}_{\textsc{i}}^{\nu-1})f(\bar{t}_{\textsc{ii}}^{\nu},\bar{t}_{\textsc{i}}^{\nu})}{\prod_{\nu=1}^{j-1}f(\bar{t}^{\nu+1},\bar{t}_{\textsc{i}}^{\nu})},\label{eq:rec1}
\end{equation}
where the sum goes over all the possible partitions $\bar{t}^{\nu}=\bar{t}_{\textsc{i}}^{\nu}\cup\bar{t}_{\textsc{ii}}^{\nu}$
for $\nu=2,\dots,j-1$ where $\bar{t}_{\textsc{i}}^{\nu},\bar{t}_{\textsc{ii}}^{\nu}$
are disjoint subsets and $\#\bar{t}_{\textsc{i}}^{\nu}=1$. We set
by definition $\bar{t}_{\textsc{i}}^{1}=\{z\}$ and $\bar{t}^{N}=\emptyset$. 

There is another sum formula
\begin{equation}
\mathbb{B}(\left\{ \bar{t}^{k}\right\} _{k=1}^{N-2},\{z,\bar{t}^{N-1}\})=\sum_{j=1}^{N-1}\frac{T_{j,N}(z)}{\lambda_{N}(z)}\sum_{\mathrm{part}(\bar{t})}\mathbb{B}(\left\{ \bar{t}^{k}\right\} _{k=1}^{j-1},\left\{ \bar{t}_{\textsc{ii}}^{k}\right\} _{k=j}^{N-2},\bar{t}^{N-1})\frac{\prod_{\nu=j}^{N-2}g(\bar{t}_{\textsc{i}}^{\nu+1},\bar{t}_{\textsc{i}}^{\nu})f(\bar{t}_{\textsc{i}}^{\nu},\bar{t}_{\textsc{ii}}^{\nu})}{\prod_{\nu=j}^{N-1}f(\bar{t}_{\textsc{i}}^{\nu},\bar{t}^{\nu-1})},\label{eq:rec2-1}
\end{equation}
where the sum goes over all the possible partitions $\bar{t}^{\nu}=\bar{t}_{\textsc{i}}^{\nu}\cup\bar{t}_{\textsc{ii}}^{\nu}$
for $\nu=j,\dots,N-2$ where $\bar{t}_{\textsc{i}}^{\nu},\bar{t}_{\textsc{ii}}^{\nu}$
are disjoint subsets and $\#\bar{t}_{\textsc{i}}^{\nu}=1$.

We also use the following action formula for the off-shell Bethe vectors
\cite{Hutsalyuk:2020dlw}

\begin{multline}
T_{i,j}(z)\mathbb{B}(\bar{t})=\lambda_{N}(z)\sum_{\mathrm{part}(\bar{w})}\mathbb{B}(\bar{w}_{\textsc{ii}})\frac{\prod_{s=j}^{i-1}f(\bar{w}_{\textsc{i}}^{s},\bar{w}_{\textsc{iii}}^{s})}{\prod_{s=j}^{i-2}f(\bar{w}_{\textsc{i}}^{s+1},\bar{w}_{\textsc{iii}}^{s})}\times\\
\prod_{s=1}^{i-1}\frac{f(\bar{w}_{\textsc{i}}^{s},\bar{w}_{\textsc{ii}}^{s})}{h(\bar{w}_{\textsc{i}}^{s},\bar{w}_{\textsc{i}}^{s-1})f(\bar{w}_{\textsc{i}}^{s},\bar{w}_{\textsc{ii}}^{s-1})}\prod_{s=j}^{N-1}\frac{\alpha_{s}(\bar{w}_{\textsc{iii}}^{s})f(\bar{w}_{\textsc{ii}}^{s},\bar{w}_{\textsc{iii}}^{s})}{h(\bar{w}_{\textsc{iii}}^{s+1},\bar{w}_{\textsc{iii}}^{s})f(\bar{w}_{\textsc{ii}}^{s+1},\bar{w}_{\textsc{iii}}^{s})},\label{eq:act}
\end{multline}
where $\bar{w}^{\nu}=\{z,\bar{t}^{\nu}\}$. The sum goes over all
the partitions of $\bar{w}^{\nu}=\bar{w}_{\textsc{i}}^{\nu}\cup\bar{w}_{\textsc{ii}}^{\nu}\cup\bar{w}_{\textsc{iii}}^{\nu}$
where $\bar{w}_{\textsc{i}}^{\nu},\bar{w}_{\textsc{ii}}^{\nu},\bar{w}_{\textsc{iii}}^{\nu}$
are disjoint sets for a fixed $\nu$ and $\#\bar{w}_{\textsc{i}}^{\nu}=\Theta(i-1-\nu)$,
$\#\bar{w}_{\textsc{iii}}^{\nu}=\Theta(\nu-j)$. We also set $\bar{w}_{\textsc{i}}^{0}=\bar{w}_{\textsc{iii}}^{N}=\{z\}$
and $\bar{w}_{\textsc{ii}}^{0}=\bar{w}_{\textsc{iii}}^{0}=\bar{w}_{\textsc{i}}^{N}=\bar{w}_{\textsc{ii}}^{N}=\emptyset$.
We also used the unit step function $\Theta(k)$ which is defined
as $\Theta(k)=1$ for $k\geq0$ and $\Theta(k)=0$ for $k<0$.

We can see that the diagonal elements $T_{i,i}(u)$ do not change
the quantum numbers $r_{j}$. The creation operators i.e. $T_{i,j}(u)$
where $i<j$ increase the quantum numbers $r_{i},r_{i+1},\dots,r_{j-1}$
by one. The annihilation operators i.e. $T_{j,i}(u)$ where $i<j$
decrease the quantum numbers $r_{i},r_{i+1},\dots,r_{j-1}$ by one. 

We also use the co-product formula of the off-shell Bethe vectors.
Let $\mathcal{H}^{(1)},\mathcal{H}^{(2)}$ be two quantum spaces for
which $\mathcal{H}=\mathcal{H}^{(1)}\otimes\mathcal{H}^{(2)}$ and
the corresponding off-shell states are $\mathbb{B}^{(1)}(\bar{t}),\mathbb{B}^{(2)}(\bar{t})$.
The co-product formula reads as \cite{Hutsalyuk:2016srn}
\begin{equation}
\mathbb{B}(\bar{t})=\sum_{\mathrm{part}(\bar{t})}\frac{\prod_{\nu=1}^{N-1}\alpha_{\nu}^{(2)}(\bar{t}_{\mathrm{i}}^{\nu})f(\bar{t}_{\mathrm{ii}}^{\nu},\bar{t}_{\mathrm{i}}^{\nu})}{\prod_{\nu=1}^{N-2}f(\bar{t}_{\mathrm{ii}}^{\nu+1},\bar{t}_{\mathrm{i}}^{\nu})}\mathbb{\mathbb{B}}^{(1)}(\bar{t}_{\mathrm{i}})\otimes\mathbb{B}^{(2)}(\bar{t}_{\mathrm{ii}}),\label{eq:coproduct}
\end{equation}
where the sum goes over all the possible partitions $\bar{t}^{\nu}=\bar{t}_{\mathrm{i}}^{\nu}\cup\bar{t}_{\mathrm{ii}}^{\nu}$
where $\bar{t}_{\mathrm{i}}^{\nu},\bar{t}_{\mathrm{ii}}^{\nu}$ are
disjoint subsets.

We also need the action of the twisted monodromy matrix on the off-shell
Bethe vectors \cite{Gombor:2021hmj}.
\begin{multline}
\widehat{T}_{i,j}(z)\mathbb{B}(\bar{t})=(-1)^{i-j}\hat{\lambda}_{N}(z)\sum_{\mathrm{part}(\bar{w})}\mathbb{B}(\bar{w}_{\textsc{ii}})\frac{\prod_{s=2}^{N-1}f(\bar{t}^{s-1}-c,\bar{t}^{s})}{\prod_{s=2}^{N-1}f(\bar{w}_{\textsc{ii}}^{s-1}-c,\bar{w}_{\textsc{ii}}^{s})}\frac{\prod_{s=N-i+1}^{N-j}f(\bar{w}_{\textsc{i}}^{s},\bar{w}_{\textsc{iii}}^{s})}{\prod_{s=N-i+2}^{N-j}f(\bar{w}_{\textsc{i}}^{s-1}-c,\bar{w}_{\textsc{iii}}^{s})}\times\\
\prod_{s=N-i+1}^{N-1}\frac{f(\bar{w}_{\textsc{i}}^{s},\bar{w}_{\textsc{ii}}^{s})}{h(\bar{w}_{\textsc{i}}^{s},\bar{w}_{\textsc{i}}^{s+1}+c)f(\bar{w}_{\textsc{i}}^{s},\bar{w}_{\textsc{ii}}^{s+1}+c)}\prod_{s=1}^{N-j}\frac{\alpha_{s}(\bar{w}_{\textsc{iii}}^{s})f(\bar{w}_{\textsc{ii}}^{s},\bar{w}_{\textsc{iii}}^{s})}{h(\bar{w}_{\textsc{iii}}^{s-1}-c,\bar{w}_{\textsc{iii}}^{s})f(\bar{w}_{\textsc{ii}}^{s-1}-c,\bar{w}_{\textsc{iii}}^{s})},\label{eq:actTw}
\end{multline}
where $\bar{w}^{\nu}=\{z-sc,\bar{t}^{\nu}\}$. The sum goes over all
the partition of $\bar{w}^{\nu}=\bar{w}_{\textsc{i}}^{\nu}\cup\bar{w}_{\textsc{ii}}^{\nu}\cup\bar{w}_{\textsc{iii}}^{\nu}$
where $\bar{w}_{\textsc{i}}^{\nu},\bar{w}_{\textsc{ii}}^{\nu},\bar{w}_{\textsc{iii}}^{\nu}$
are disjoint sets for a fixed $\nu$ and $\#\bar{w}_{\textsc{i}}^{\nu}=\Theta(\nu-N-1+i)$,
$\#\bar{w}_{\textsc{iii}}^{\nu}=\Theta(N-j-\nu)$. We also set $\bar{w}_{\textsc{iii}}^{0}=\{z\}$,
$\bar{w}_{\textsc{i}}^{N}=\{z-Nc\}$ and $\bar{w}_{\textsc{i}}^{0}=\bar{w}_{\textsc{ii}}^{0}=\bar{w}_{\textsc{ii}}^{N}=\bar{w}_{\textsc{iii}}^{N}=\emptyset$.

There are $\mathfrak{gl}(2)$ subsectors which are spanned by the
Bethe vectors $\mathbb{B}(\emptyset^{\times k-1},\bar{t}^{k},\emptyset^{\times N-1-k})$
and the generators $\{T_{i,j}\}_{i,j=k}^{k+1}$ and $\{\widehat{T}_{i,j}\}_{i,j=N-k}^{N+1-k}$
leave this subspace invariant. We can compare the actions (\ref{eq:act})
and (\ref{eq:actTw}) on this subspace and we find the following identities
\begin{equation}
\begin{split}\frac{\widehat{T}_{N-k,N-k}(z)}{\hat{\lambda}_{N+1-k}(z)}\mathbb{B}(\emptyset^{\times k-1},\bar{t}^{k},\emptyset^{\times N-1-k}) & =\frac{T_{k,k}(z-kc)}{\lambda_{k+1}(z-kc)}\mathbb{B}(\emptyset^{\times k-1},\bar{t}^{k},\emptyset^{\times N-1-k}),\\
\frac{\widehat{T}_{N-k,N-k+1}(z)}{\hat{\lambda}_{N+1-k}(z)}\mathbb{B}(\emptyset^{\times k-1},\bar{t}^{k},\emptyset^{\times N-1-k}) & =-\frac{T_{k,k+1}(z-kc)}{\lambda_{k+1}(z-kc)}\mathbb{B}(\emptyset^{\times k-1},\bar{t}^{k},\emptyset^{\times N-1-k}),\\
\frac{\widehat{T}_{N-k+1,N-k}(z)}{\hat{\lambda}_{N+1-k}(z)}\mathbb{B}(\emptyset^{\times k-1},\bar{t}^{k},\emptyset^{\times N-1-k}) & =-\frac{T_{k+1,k}(z-kc)}{\lambda_{k+1}(z-kc)}\mathbb{B}(\emptyset^{\times k-1},\bar{t}^{k},\emptyset^{\times N-1-k}),\\
\frac{\widehat{T}_{N-k+1,N-k+1}(z)}{\hat{\lambda}_{N+1-k}(z)}\mathbb{B}(\emptyset^{\times k-1},\bar{t}^{k},\emptyset^{\times N-1-k}) & =\frac{T_{k+1,k+1}(z-kc)}{\lambda_{k+1}(z-kc)}\mathbb{B}(\emptyset^{\times k-1},\bar{t}^{k},\emptyset^{\times N-1-k}),
\end{split}
\end{equation}
therefore the original and the twisted monodromy matrices are equivalent
in the $\mathfrak{gl}(2)$ subsectors, i.e.,
\begin{multline}
\frac{1}{\hat{\lambda}_{N+1-k}(z)}\left(\begin{array}{cc}
\widehat{T}_{N-k,N-k}(z) & \widehat{T}_{N-k,N-k+1}(z)\\
\widehat{T}_{N-k+1,N-k}(z) & \widehat{T}_{N-k+1,N-k+1}(z)
\end{array}\right)\cong\\
\frac{1}{\lambda_{k+1}(z-kc)}\left(\begin{array}{cc}
T_{k,k}(z-kc) & -T_{k,k+1}(z-kc)\\
-T_{k+1,k}(z-kc) & T_{k+1,k+1}(z-kc)
\end{array}\right).\label{eq:gl2Equiv}
\end{multline}

\section{Untwisted off-shell overlap of $M=0$ case\label{sec:Untwisted-off-shell-overlap}}

Let us take the following singular $K$-matrix
\begin{equation}
K(u)=\frac{1}{u}\mathbf{1}+\mathcal{U}=\frac{1}{u}\mathbf{1}+\mathfrak{b}_{j}\sum_{j=1}^{\left\lfloor \frac{N}{2}\right\rfloor }E_{N+1-j,j}.\label{eq:singK}
\end{equation}
We can apply the result of \cite{Gombor:2021hmj} for the off-shell
overlaps. The off-shell overlap has the sum formula
\begin{equation}
S_{\bar{\alpha}}(\bar{t}|\bar{\mathfrak{b}})=A\sum_{\mathrm{part}}\frac{\prod_{\nu=1}^{N-1}f(\bar{t}_{\textsc{ii}}^{\nu},\bar{t}_{\textsc{i}}^{\nu})}{\prod_{\nu=1}^{N-2}f(\bar{t}_{\textsc{ii}}^{\nu+1},\bar{t}_{\textsc{i}}^{\nu})}\bar{Z}(\pi^{a}(\bar{t}_{\textsc{i}}))\bar{Z}(\bar{t}_{\textsc{ii}})\prod_{\nu=1}^{N-1}\alpha_{\nu}(\bar{t}_{\textsc{i}}^{\nu}),
\end{equation}
where we introduced a non-trivial vacuum overlap
\begin{equation}
A:=\langle\Psi|\Omega\rangle.
\end{equation}
The HC has the recursion relation (see \cite{Gombor:2021hmj})
\begin{multline}
\bar{Z}(\{z,\bar{t}^{1}\},\left\{ \bar{t}^{s}\right\} _{s=2}^{N-1})=\\
\frac{\mathfrak{b}_{2}}{\mathfrak{b}_{1}}\sum_{\mathrm{part}}\bar{Z}(\{\bar{\omega}_{\textsc{ii}}^{s}\}_{s=1}^{N-2},\bar{t}_{\textsc{ii}}^{N-1})\prod_{s=1}^{N-2}\frac{f(\bar{\omega}_{\textsc{i}}^{s},\bar{\omega}_{\textsc{ii}}^{s})}{h(\bar{\omega}_{\textsc{i}}^{s},\bar{\omega}_{\textsc{i}}^{s-1})f(\bar{\omega}_{\textsc{i}}^{s},\bar{\omega}_{\textsc{ii}}^{s-1})}\frac{f(\bar{t}_{\textsc{i}}^{N-1},\bar{t}_{\textsc{ii}}^{N-1})f(\bar{t}^{N-1},-z)}{h(\bar{t}_{\textsc{i}}^{N-1},\bar{\omega}_{\textsc{i}}^{N-2})f(\bar{t}_{\textsc{i}}^{N-1},\bar{\omega}_{\textsc{ii}}^{N-2})f(\bar{t}^{2},z)}-\\
-\frac{1}{\mathfrak{b}_{1}}\frac{1}{z}\sum_{\mathrm{part}}\bar{Z}(\bar{w}_{\textsc{ii}}^{1},\{\bar{t}_{\textsc{ii}}^{s}\}_{s=2}^{N-1})\frac{f(\bar{w}_{\textsc{i}}^{1},\bar{w}_{\textsc{ii}}^{1})}{h(\bar{w}_{\textsc{i}}^{1},z)}\frac{f(\bar{t}_{\textsc{i}}^{2},\bar{t}_{\textsc{ii}}^{2})}{h(\bar{t}_{\textsc{i}}^{2},\bar{w}_{\textsc{i}}^{1})f(\bar{t}_{\textsc{i}}^{2},\bar{w}_{\textsc{ii}}^{1})}\prod_{s=3}^{N-1}\frac{f(\bar{t}_{\textsc{i}}^{s},\bar{t}_{\textsc{ii}}^{s})}{h(\bar{t}_{\textsc{i}}^{s},\bar{t}_{\textsc{i}}^{s-1})f(\bar{t}_{\textsc{i}}^{s},\bar{t}_{\textsc{ii}}^{s-1})}.\label{eq:recursionZ_UTw1-1}
\end{multline}
Let us renormalize the HC as
\begin{equation}
\bar{Z}(\bar{t})=\frac{1}{\prod_{s=1}^{\frac{N}{2}}\mathfrak{b}_{s}^{r_{s}-r_{s-1}}}\bar{Z}^{0}(\bar{t}).
\end{equation}
Using the renormalized HC-s, the recursion will be independent on
parameters of the K-matrix
\begin{multline}
\bar{Z}^{0}(\{z,\bar{t}^{1}\},\left\{ \bar{t}^{s}\right\} _{s=2}^{N-1})=\\
\sum_{\mathrm{part}}\bar{Z}^{0}(\{\bar{\omega}_{\textsc{ii}}^{s}\}_{s=1}^{N-2},\bar{t}_{\textsc{ii}}^{N-1})\prod_{s=1}^{N-2}\frac{f(\bar{\omega}_{\textsc{i}}^{s},\bar{\omega}_{\textsc{ii}}^{s})}{h(\bar{\omega}_{\textsc{i}}^{s},\bar{\omega}_{\textsc{i}}^{s-1})f(\bar{\omega}_{\textsc{i}}^{s},\bar{\omega}_{\textsc{ii}}^{s-1})}\frac{f(\bar{t}_{\textsc{i}}^{N-1},\bar{t}_{\textsc{ii}}^{N-1})f(\bar{t}^{N-1},-z)}{h(\bar{t}_{\textsc{i}}^{N-1},\bar{\omega}_{\textsc{i}}^{N-2})f(\bar{t}_{\textsc{i}}^{N-1},\bar{\omega}_{\textsc{ii}}^{N-2})f(\bar{t}^{2},z)}-\\
-\frac{1}{z}\sum_{\mathrm{part}}\bar{Z}^{0}(\bar{w}_{\textsc{ii}}^{1},\{\bar{t}_{\textsc{ii}}^{s}\}_{s=2}^{N-1})\frac{f(\bar{w}_{\textsc{i}}^{1},\bar{w}_{\textsc{ii}}^{1})}{h(\bar{w}_{\textsc{i}}^{1},z)}\frac{f(\bar{t}_{\textsc{i}}^{2},\bar{t}_{\textsc{ii}}^{2})}{h(\bar{t}_{\textsc{i}}^{2},\bar{w}_{\textsc{i}}^{1})f(\bar{t}_{\textsc{i}}^{2},\bar{w}_{\textsc{ii}}^{1})}\prod_{s=3}^{N-1}\frac{f(\bar{t}_{\textsc{i}}^{s},\bar{t}_{\textsc{ii}}^{s})}{h(\bar{t}_{\textsc{i}}^{s},\bar{t}_{\textsc{i}}^{s-1})f(\bar{t}_{\textsc{i}}^{s},\bar{t}_{\textsc{ii}}^{s-1})}.\label{eq:rec_UTW0}
\end{multline}
The $\mathfrak{b}_{s}$ dependence appears only in the sum formula
as
\begin{equation}
S_{\bar{\alpha}}(\bar{t}|\bar{\mathfrak{b}})=\frac{A}{\prod_{s=1}^{\frac{N}{2}}\mathfrak{b}_{s}^{r_{s}-r_{s-1}}}S_{\bar{\alpha}}^{0}(\bar{t}),
\end{equation}
where we defined the $\mathfrak{b}_{s}$ independent quantity
\begin{equation}
S_{\bar{\alpha}}^{0}(\bar{t})=\sum_{\mathrm{part}}\frac{\prod_{\nu=1}^{N-1}f(\bar{t}_{\textsc{ii}}^{\nu},\bar{t}_{\textsc{i}}^{\nu})}{\prod_{\nu=1}^{N-2}f(\bar{t}_{\textsc{ii}}^{\nu+1},\bar{t}_{\textsc{i}}^{\nu})}\bar{Z}^{0}(\pi^{a}(\bar{t}_{\textsc{i}}))\bar{Z}^{0}(\bar{t}_{\textsc{ii}})\prod_{\nu=1}^{N-1}\alpha_{\nu}(\bar{t}_{\textsc{i}}^{\nu}).
\end{equation}
It is clear that we obtain the $M=0$ $K$-matrix in the limit $\mathfrak{b}_{j}\to0$.
We saw that the non-vanishing overlaps require the selection rules
$r_{k}=r_{k}^{\boldsymbol{\Lambda}}=\sum_{l=1}^{k}\mathbf{\boldsymbol{\Lambda}}_{k}$
(see (\ref{eq:M0SR})). For these quantum numbers, the limit $\mathfrak{b}_{j}\to0$
of the overlap is non-singular only if the vacuum expectation value
scales as $A\sim\prod_{s=1}^{\frac{N}{2}}\mathfrak{b}_{s}^{r_{s}^{\boldsymbol{\Lambda}}-r_{s-1}^{\boldsymbol{\Lambda}}}=\prod_{s=1}^{\frac{N}{2}}\mathfrak{b}_{s}^{\mathbf{\boldsymbol{\Lambda}}_{s}}$.
Let us choose the normalization of the two-site state as
\begin{equation}
A=\prod_{s=1}^{\frac{N}{2}}\mathfrak{b}_{s}^{\mathbf{\boldsymbol{\Lambda}}_{s}},\label{eq:normA}
\end{equation}
therefore the off-shell overlap reads as
\begin{equation}
S_{\bar{\alpha}}(\bar{t}|\bar{\mathfrak{b}})=\prod_{s=1}^{\frac{N}{2}}\mathfrak{b}_{s}^{\mathbf{\boldsymbol{\Lambda}}_{s}+r_{s-1}-r_{s}}S_{\bar{\alpha}}^{0}(\bar{t}).\label{eq:ovsing}
\end{equation}
 Now we can take the $\mathfrak{b}_{s}\to0$ limit as
\begin{equation}
S_{\bar{\alpha}}(\bar{t}|\bar{\mathfrak{b}}=\bar{0})\Biggr|_{\mathbf{r}=\mathbf{r}^{\boldsymbol{\Lambda}}}=S_{\bar{\alpha}}^{0}(\bar{t}).
\end{equation}
We just derived that the non-vanishing off-shell overlaps for $M=0$
is $S_{\bar{\alpha}}^{0}(\bar{t})$ with the recursion (\ref{eq:rec_UTW0}).

There is a significant simplification in the $N=2$ case. In this
case we have only one set of Bethe roots $\bar{t}\equiv\bar{t}^{1}$
and $\alpha$-function $\alpha\equiv\alpha_{1}$. The sum formula
simplifies as
\begin{equation}
S_{\alpha}^{0}(\bar{t})=\sum_{\mathrm{part}}f(\bar{t}_{\textsc{ii}},\bar{t}_{\textsc{i}})Z^{0}(\bar{t}_{\textsc{i}})Z^{0}(-\bar{t}_{\textsc{ii}})\alpha(\bar{t}_{\textsc{i}}),\label{eq:offov2}
\end{equation}
and we can give the HC in a closed form:
\begin{equation}
Z^{0}(\bar{t})=\kappa(\bar{t})\prod_{k<l}f(-t_{k},t_{l}),\qquad\kappa(z)=\frac{1}{z}.\label{eq:HC2}
\end{equation}
We also need some limits of the $\mathfrak{gl}(2)$ overlaps. The
HC-s have formal poles in the limits $t_{l}\to-t_{k}$ and $t_{k}\to0$.
For the first limit let us introduce the notation $\bar{\tau}=\bar{t}\backslash\{t_{k},t_{l}\}$.
Taking the limit $t_{l}\to-t_{k}$ we have
\begin{align}
 & S_{\alpha}^{0}(\bar{t})\to\\
 & g(-t_{k},t_{l})\alpha(t_{k})\alpha(t_{l})\frac{1}{-t_{k}^{2}}\sum_{\mathrm{part}(\bar{\tau})}f(\bar{\tau}_{\textsc{ii}},t_{k})f(\bar{\tau}_{\textsc{ii}},-t_{k})f(-t_{k},\bar{\tau}_{\textsc{i}})f(t_{k},\bar{\tau}_{\textsc{i}})f(\bar{\tau}_{\textsc{ii}},\bar{\tau}_{\textsc{i}})Z^{0}(\bar{\tau}_{\textsc{i}})Z^{0}(-\bar{\tau}_{\textsc{ii}})\alpha(\bar{\tau}_{\textsc{i}})+\nonumber \\
 & g(t_{k},-t_{l})\frac{1}{-t_{k}^{2}}\sum_{\mathrm{part}(\bar{\tau})}f(t_{k},\bar{\tau}_{\textsc{i}})f(-t_{k},\bar{\tau}_{\textsc{i}})f(\bar{\tau}_{\textsc{ii}},-t_{k})f(\bar{\tau}_{\textsc{ii}},t_{k})f(\bar{\tau}_{\textsc{ii}},\bar{\tau}_{\textsc{i}})Z^{0}(\bar{\tau}_{\textsc{i}})Z^{0}(-\bar{\tau}_{\textsc{ii}})\alpha(\bar{\tau}_{\textsc{i}})+reg.\nonumber 
\end{align}
After simplifications we have
\begin{align}
S_{\alpha}^{0}(\bar{t}) & \to g(-t_{k},t_{l})\left(\alpha(t_{k})\alpha(t_{l})-1\right)\times\\
 & \times\frac{1}{-t_{k}^{2}}f(\bar{\tau},t_{k})f(\bar{\tau},-t_{k})\sum_{\mathrm{part}(\bar{\tau})}f(\bar{\tau}_{\textsc{ii}},\bar{\tau}_{\textsc{i}})Z^{0}(\bar{\tau}_{\textsc{i}})Z^{0}(-\bar{\tau}_{\textsc{ii}})\left[\alpha(\bar{\tau}_{\textsc{i}})\frac{f(t_{k},\bar{\tau}_{\textsc{i}})f(-t_{k},\bar{\tau}_{\textsc{i}})}{f(\bar{\tau}_{\textsc{i}},t_{k})f(\bar{\tau}_{\textsc{i}},-t_{k})}\right]+reg.\nonumber 
\end{align}
Introducing the modified $\alpha$-s as
\begin{equation}
\alpha^{mod}(z)=\alpha(z)\frac{f(t_{k},z)f(-t_{k},z)}{f(z,t_{k})f(z,-t_{k})},
\end{equation}
and using the sum formula (\ref{eq:offov2}) we obtain that
\begin{equation}
S_{\alpha}^{0}(\bar{t})\to X_{k}\frac{1}{-t_{k}^{2}}f(\bar{\tau},t_{k})f(\bar{\tau},-t_{k})S_{\alpha^{mod}}^{0}(\bar{\tau})+\tilde{S},\label{eq:XdepGL2}
\end{equation}
where $\tilde{S}$ does not depend on $X_{k}$ which is defined as
\begin{equation}
\lim_{t_{l}\to-t_{k}}g(-t_{k},t_{l})\left(\alpha(t_{k})\alpha(t_{l})-1\right)=-c\frac{\alpha'(t_{k})}{\alpha(t_{k})}\equiv X_{k}.
\end{equation}

Taking the other limit $t_{k}\to0$ we have
\begin{align}
S_{\alpha}^{0}(\bar{t}) & \to\frac{1}{t_{k}}\alpha(t_{k})\sum_{\mathrm{part}(\bar{\tau})}f(\bar{\tau}_{\textsc{ii}},0)f(0,\bar{\tau}_{\textsc{i}})f(\bar{\tau}_{\textsc{ii}},\bar{\tau}_{\textsc{i}})Z^{0}(\bar{\tau}_{\textsc{i}})Z^{0}(-\bar{\tau}_{\textsc{ii}})\alpha(\bar{t}_{\textsc{i}})+\\
 & +\frac{1}{-t_{k}}\sum_{\mathrm{part}(\bar{\tau})}f(0,\bar{\tau}_{\textsc{i}})f(\bar{\tau}_{\textsc{ii}},0)f(\bar{\tau}_{\textsc{ii}},\bar{\tau}_{\textsc{i}})Z^{0}(\bar{\tau}_{\textsc{i}})Z^{0}(-\bar{\tau}_{\textsc{ii}})\alpha(\bar{t}_{\textsc{i}})+reg.,\nonumber 
\end{align}
where we introduced the set $\bar{\tau}=\bar{t}\backslash\{t_{k}\}$.
After simplifications we have
\begin{equation}
S_{\alpha}^{0}(\bar{t})\to\frac{1}{t_{k}}(\alpha(t_{k})-1)f(\bar{\tau},0)\sum_{\mathrm{part}(\bar{\tau})}f(\bar{\tau}_{\textsc{ii}},\bar{\tau}_{\textsc{i}})Z^{0}(\bar{\tau}_{\textsc{i}})Z^{0}(-\bar{\tau}_{\textsc{ii}})\left[\alpha(\bar{t}_{\textsc{i}})\frac{f(0,\bar{\tau}_{\textsc{i}})}{f(\bar{\tau}_{\textsc{i}},0)}\right]+reg.
\end{equation}
Introducing the modified $\alpha$-s as
\begin{equation}
\alpha^{mod}(z)=\alpha(z)\frac{f(0,z)}{f(z,0)},
\end{equation}
and using the sum formula (\ref{eq:offov2}) we obtain that
\begin{equation}
S_{\alpha}^{0}(\bar{t})\to X^{0}\left(-\frac{2}{c}\right)f(\bar{\tau},0)S_{\alpha^{mod}}^{0}(\bar{\tau})+\tilde{S},\label{eq:XdepGL2-1}
\end{equation}
where $\tilde{S}$ does not depend on $X_{k}$ which is defined as
\begin{equation}
\lim_{t_{k}\to0}\frac{1}{t_{k}}(\alpha(t_{k})-1)=\alpha'(0)\equiv\left(-\frac{2}{c}\right)X^{0}.
\end{equation}

\section{A new recurrence relations for the Bethe vectors\label{sec:A-new-recurrance}}

In section \ref{sec:Off-shell-Bethe-vectors} we show recurrence relations
for the off-shell Bethe vectors (\ref{eq:rec1}) and (\ref{eq:rec2-1})
which are recursions on the quantum numbers $r_{1}$ and $r_{N-1}$.
In this section we derive two other recurrence relations for the off-shell
Bethe vectors which decreases the quantum number $r_{2}$ and $r_{N-2}$.

\subsection*{Recursion in $r_{2}$}

In this subsection we derive the recurrence relation
\begin{align}
 & \mathbb{B}(\bar{t}^{1},\{z,\bar{t}^{2}\},\{\bar{t}^{s}\}_{s=3}^{N-1})\nonumber \\
 & =\sum_{j=3}^{N}\sum_{\mathrm{part}(\bar{t})}\frac{T_{2,j}(z)}{\lambda_{3}(z)}\mathbb{B}(\bar{t}^{1},\bar{t}^{2},\{\bar{t}_{\textsc{ii}}^{s}\}_{s=3}^{j-1},\{\bar{t}^{s}\}_{s=j}^{N-1})\frac{1}{f(z,\bar{t}^{1})}\frac{\prod_{s=3}^{j-1}\alpha_{s}(\bar{t}_{\textsc{i}}^{s})g(\bar{t}_{\textsc{i}}^{s},\bar{t}_{\textsc{i}}^{s-1})f(\bar{t}_{\textsc{ii}}^{s},\bar{t}_{\textsc{i}}^{s})}{\prod_{s=2}^{j-1}f(\bar{t}^{s+1},\bar{t}_{\textsc{i}}^{s})}\nonumber \\
 & +\sum_{j=3}^{N}\sum_{\mathrm{part}(\bar{t})}\frac{T_{1,j}(z)}{\lambda_{3}(z)}\mathbb{B}(\bar{t}_{\textsc{ii}}^{1},\bar{t}^{2},\{\bar{t}_{\textsc{ii}}^{s}\}_{s=3}^{j-1},\{\bar{t}^{s}\}_{s=j}^{N-1})\frac{g(z,\bar{t}_{\textsc{i}}^{1})f(\bar{t}_{\textsc{i}}^{1},\bar{t}_{\textsc{ii}}^{1})}{f(z,\bar{t}^{1})}\frac{\prod_{s=3}^{j-1}\alpha_{s}(\bar{t}_{\textsc{i}}^{s})g(\bar{t}_{\textsc{i}}^{s},\bar{t}_{\textsc{i}}^{s-1})f(\bar{t}_{\textsc{ii}}^{s},\bar{t}_{\textsc{i}}^{s})}{\prod_{s=2}^{j-1}f(\bar{t}^{s+1},\bar{t}_{\textsc{i}}^{s})},\label{eq:rect2}
\end{align}
where the sum goes over all the possible partitions $\bar{t}^{\nu}=\bar{t}_{\textsc{i}}^{\nu}\cup\bar{t}_{\textsc{ii}}^{\nu}$
for $\nu=1$ and $\nu=3,\dots,j-1$ where $\bar{t}_{\textsc{i}}^{\nu},\bar{t}_{\textsc{ii}}^{\nu}$
are disjoint subsets and $\#\bar{t}_{\textsc{i}}^{\nu}=1$. We set
by definition $\bar{t}_{\textsc{i}}^{2}=\{z\}$ and $\bar{t}^{N}=\emptyset$.

During the derivation we use the action formula (\ref{eq:act}). Applying
it for $i=2$ we obtain that
\begin{multline}
\frac{T_{2,j}(z)}{\lambda_{3}(z)}\mathbb{B}(\bar{t}^{1},\bar{t}^{2},\{\bar{t}_{\textsc{ii}}^{s}\}_{s=3}^{j-1},\{\bar{t}^{s}\}_{s=j}^{N-1})=\frac{1}{\prod_{s=3}^{N-1}\alpha_{s}(z)}\sum_{\mathrm{part}(\bar{w})}\mathbb{B}(\bar{w}_{\textsc{ii}}^{1},\{z,\bar{t}^{2}\},\{z,\bar{t}_{\textsc{ii}}^{s}\}_{s=3}^{j-1},\{\bar{w}_{\textsc{ii}}^{s}\}_{s=j}^{N-1})\times\\
\frac{f(\bar{w}_{\textsc{i}}^{1},\bar{w}_{\textsc{ii}}^{1})}{h(\bar{w}_{\textsc{i}}^{1},z)}\prod_{s=j}^{N-1}\frac{\alpha_{s}(\bar{w}_{\textsc{iii}}^{s})f(\bar{w}_{\textsc{ii}}^{s},\bar{w}_{\textsc{iii}}^{s})}{h(\bar{w}_{\textsc{iii}}^{s+1},\bar{w}_{\textsc{iii}}^{s})f(\bar{w}_{\textsc{ii}}^{s+1},\bar{w}_{\textsc{iii}}^{s})},
\end{multline}
where $\bar{w}^{\nu}=\{z,\bar{t}^{\nu}\}$ and the sum goes over all
the partitions of $\bar{w}^{1}=\bar{w}_{\textsc{i}}^{1}\cup\bar{w}_{\textsc{ii}}^{1}$
and $\bar{w}^{\nu}=\bar{w}_{\textsc{ii}}^{\nu}\cup\bar{w}_{\textsc{iii}}^{\nu}$
for $\nu=j,j+1,\dots,N-1$. We also set $\bar{w}_{\textsc{iii}}^{N}=\{z\}$
and $\bar{w}_{\textsc{ii}}^{N}=\emptyset$.

We can do the summation for the partitions of $\bar{w}^{1}$:
\begin{align}
 & \frac{T_{2,j}(z)}{\lambda_{3}(z)}\mathbb{B}(\bar{t}^{1},\bar{t}^{2},\{\bar{t}_{\textsc{ii}}^{s}\}_{s=3}^{j-1},\{\bar{t}^{s}\}_{s=j}^{N-1})=\frac{1}{\prod_{s=3}^{N-1}\alpha_{s}(z)}\times\label{eq:temp1}\\
 & \Biggl[\sum_{\mathrm{part}(\{\bar{w}^{s}\}_{s=j}^{N-1})}\mathbb{B}(\bar{t}^{1},\{z,\bar{t}^{2}\},\{z,\bar{t}_{\textsc{ii}}^{s}\}_{s=3}^{j-1},\{\bar{w}_{\textsc{ii}}^{s}\}_{s=j}^{N-1})f(z,\bar{t}^{1})\prod_{s=j}^{N-1}\frac{\alpha_{s}(\bar{w}_{\textsc{iii}}^{s})f(\bar{w}_{\textsc{ii}}^{s},\bar{w}_{\textsc{iii}}^{s})}{h(\bar{w}_{\textsc{iii}}^{s+1},\bar{w}_{\textsc{iii}}^{s})f(\bar{w}_{\textsc{ii}}^{s+1},\bar{w}_{\textsc{iii}}^{s})}+\nonumber \\
 & \sum_{\mathrm{part}(\bar{t}^{1},\{\bar{w}^{s}\}_{s=j}^{N-1})}\mathbb{B}(\{z,\bar{t}_{\textsc{ii}}^{1}\},\{z,\bar{t}^{2}\},\{z,\bar{t}_{\textsc{ii}}^{s}\}_{s=3}^{j-1},\{\bar{w}_{\textsc{ii}}^{s}\}_{s=j}^{N-1})g(\bar{t}_{\textsc{i}}^{1},z)f(\bar{t}_{\textsc{i}}^{1},\bar{t}_{\textsc{ii}}^{1})\prod_{s=j}^{N-1}\frac{\alpha_{s}(\bar{w}_{\textsc{iii}}^{s})f(\bar{w}_{\textsc{ii}}^{s},\bar{w}_{\textsc{iii}}^{s})}{h(\bar{w}_{\textsc{iii}}^{s+1},\bar{w}_{\textsc{iii}}^{s})f(\bar{w}_{\textsc{ii}}^{s+1},\bar{w}_{\textsc{iii}}^{s})}\Biggr].\nonumber 
\end{align}
In the second line the sum goes over all the partitions of $\bar{t}^{1}=\bar{t}_{\textsc{i}}^{1}\cup\bar{t}_{\textsc{ii}}^{1}$
and $\bar{w}^{\nu}=\bar{w}_{\textsc{ii}}^{\nu}\cup\bar{w}_{\textsc{iii}}^{\nu}$
where $\#\bar{t}_{\textsc{i}}^{1}=\#\bar{w}_{\textsc{iii}}^{\nu}=1$
for $\nu=j,j+1,\dots,N-1$. Let us concentrate on the summation of
the first line. We can do the summation for the partitions using that
if $\bar{w}_{\textsc{iii}}^{k}=\{z\}$ than the term $1/f(\bar{w}_{\textsc{ii}}^{k+1},\bar{w}_{\textsc{iii}}^{k})=0$
for the partitions where $z\in\bar{w}_{\textsc{ii}}^{k+1}$ therefore
if $\bar{w}_{\textsc{iii}}^{k}=\{z\}$ then $\bar{w}_{\textsc{iii}}^{k+1}=\{z\},\bar{w}_{\textsc{ii}}^{k+1}=\bar{t}^{k+1}$
for non-vanishing terms. We can repeat this argument and we obtain
that if $\bar{w}_{\textsc{iii}}^{k}=\{z\}$ then $\bar{w}_{\textsc{iii}}^{s}=\{z\},\bar{w}_{\textsc{ii}}^{s}=\bar{t}^{s}$
for $s>k$, i.e.,
\begin{align}
 & \frac{1}{\prod_{s=3}^{N-1}\alpha_{s}(z)}\sum_{\mathrm{part}(\bar{w})}\mathbb{B}(\bar{t}^{1},\{z,\bar{t}^{2}\},\{z,\bar{t}_{\textsc{ii}}^{s}\}_{s=3}^{j-1},\{\bar{w}_{\textsc{ii}}^{s}\}_{s=j}^{N-1})f(z,\bar{t}^{1})\prod_{s=j}^{N-1}\frac{\alpha_{s}(\bar{w}_{\textsc{iii}}^{s})f(\bar{w}_{\textsc{ii}}^{s},\bar{w}_{\textsc{iii}}^{s})}{h(\bar{w}_{\textsc{iii}}^{s+1},\bar{w}_{\textsc{iii}}^{s})f(\bar{w}_{\textsc{ii}}^{s+1},\bar{w}_{\textsc{iii}}^{s})}=\nonumber \\
 & \frac{1}{\prod_{s=3}^{j-1}\alpha_{s}(z)}\mathbb{B}(\bar{t}^{1},\{z,\bar{t}^{2}\},\{z,\bar{t}_{\textsc{ii}}^{s}\}_{s=3}^{j-1},\{\bar{t}^{s}\}_{s=j}^{N-1})f(z,\bar{t}^{1})f(\bar{t}^{j},z)+\\
 & \sum_{k=j+1}^{N}\frac{1}{\prod_{s=3}^{k-1}\alpha_{s}(z)}\sum_{\mathrm{part}(\{\bar{t}^{s}\}_{s=j}^{k-1})}\mathbb{B}(\bar{t}^{1},\{z,\bar{t}^{2}\},\{z,\bar{t}_{\textsc{ii}}^{s}\}_{s=3}^{k-1},\{\bar{t}^{s}\}_{s=k}^{N-1})f(z,\bar{t}^{1})f(\bar{t}^{k},z)\times\nonumber \\
 & \qquad\qquad\qquad\qquad\qquad\times\frac{\prod_{s=j}^{k-1}\alpha_{s}(\bar{t}_{\textsc{i}}^{s})f(\bar{t}_{\textsc{ii}}^{s},\bar{t}_{\textsc{i}}^{s})}{\prod_{s=j}^{k-2}h(\bar{t}_{\textsc{i}}^{s+1},\bar{t}_{\textsc{i}}^{s})f(\bar{t}_{\textsc{ii}}^{s+1},\bar{t}_{\textsc{i}}^{s})}\frac{g(z,\bar{t}_{\textsc{i}}^{k-1})}{f(\bar{t}^{k},\bar{t}_{\textsc{i}}^{k-1})}.\nonumber 
\end{align}
In the third line the sum goes over all the partitions of $\bar{t}^{\nu}=\bar{t}_{\textsc{i}}^{\nu}\cup\bar{t}_{\textsc{ii}}^{\nu}$
where $\#\bar{t}_{\textsc{i}}^{\nu}=1$ for $\nu=j,j+1,\dots,k-1$.
We can repeat the analogous calculation for the second line of (\ref{eq:temp1})
and we obtain that
\begin{align}
 & \frac{T_{2,j}(z)}{\lambda_{3}(z)}\mathbb{B}(\bar{t}^{1},\bar{t}^{2},\{\bar{t}_{\textsc{ii}}^{s}\}_{s=3}^{j-1},\{\bar{t}^{s}\}_{s=j}^{N-1})\nonumber \\
 & =\frac{1}{\prod_{s=3}^{j-1}\alpha_{s}(z)}\mathbb{B}(\bar{t}^{1},\{z,\bar{t}^{2}\},\{z,\bar{t}_{\textsc{ii}}^{s}\}_{s=3}^{j-1},\{\bar{t}^{s}\}_{s=j}^{N-1})f(z,\bar{t}^{1})f(\bar{t}^{j},z)+\nonumber \\
 & +\sum_{k=j+1}^{N}\frac{1}{\prod_{s=3}^{k-1}\alpha_{s}(z)}\sum_{\mathrm{part}(\{\bar{t}^{s}\}_{s=j}^{k-1})}\mathbb{B}(\bar{t}^{1},\{z,\bar{t}^{2}\},\{z,\bar{t}_{\textsc{ii}}^{s}\}_{s=3}^{k-1},\{\bar{t}^{s}\}_{s=k}^{N-1})f(z,\bar{t}^{1})f(\bar{t}^{k},z)\nonumber \\
 & \qquad\qquad\qquad\qquad\qquad\times\frac{\prod_{s=j}^{k-1}\alpha_{s}(\bar{t}_{\textsc{i}}^{s})f(\bar{t}_{\textsc{ii}}^{s},\bar{t}_{\textsc{i}}^{s})}{\prod_{s=j}^{k-2}h(\bar{t}_{\textsc{i}}^{s+1},\bar{t}_{\textsc{i}}^{s})f(\bar{t}_{\textsc{ii}}^{s+1},\bar{t}_{\textsc{i}}^{s})}\frac{g(z,\bar{t}_{\textsc{i}}^{k-1})}{f(\bar{t}^{k},\bar{t}_{\textsc{i}}^{k-1})}+\nonumber \\
 & +\frac{1}{\prod_{s=3}^{j-1}\alpha_{s}(z)}\sum_{\mathrm{part}(\bar{t}^{1})}\mathbb{B}(\{z,\bar{t}_{\textsc{ii}}^{1}\},\{z,\bar{t}^{2}\},\{z,\bar{t}_{\textsc{ii}}^{s}\}_{s=3}^{j-1},\{\bar{t}^{s}\}_{s=j}^{N-1})g(\bar{t}_{\textsc{i}}^{1},z)f(\bar{t}_{\textsc{i}}^{1},\bar{t}_{\textsc{ii}}^{1})f(\bar{t}^{j},z)+\label{eq:tt2}\\
 & +\sum_{k=j+1}^{N}\frac{1}{\prod_{s=3}^{k-1}\alpha_{s}(z)}\sum_{\mathrm{part}(\bar{t}^{1},\{\bar{t}^{s}\}_{s=j}^{k-1})}\mathbb{B}(\{z,\bar{t}_{\textsc{ii}}^{1}\},\{z,\bar{t}^{2}\},\{z,\bar{t}_{\textsc{ii}}^{s}\}_{s=3}^{k-1},\{\bar{t}^{s}\}_{s=k}^{N-1})g(\bar{t}_{\textsc{i}}^{1},z)f(\bar{t}_{\textsc{i}}^{1},\bar{t}_{\textsc{ii}}^{1})f(\bar{t}^{k},z)\times\nonumber \\
 & \qquad\qquad\qquad\qquad\qquad\times\frac{\prod_{s=j}^{k-1}\alpha_{s}(\bar{t}_{\textsc{i}}^{s})f(\bar{t}_{\textsc{ii}}^{s},\bar{t}_{\textsc{i}}^{s})}{\prod_{s=j}^{k-2}h(\bar{t}_{\textsc{i}}^{s+1},\bar{t}_{\textsc{i}}^{s})f(\bar{t}_{\textsc{ii}}^{s+1},\bar{t}_{\textsc{i}}^{s})}\frac{g(z,\bar{t}_{\textsc{i}}^{k-1})}{f(\bar{t}^{k},\bar{t}_{\textsc{i}}^{k-1})}.\nonumber 
\end{align}
We can repeat the analogous calculation for
\begin{align}
 & \frac{T_{1,j}(z)}{\lambda_{3}(z)}\mathbb{B}(\bar{t}_{\textsc{ii}}^{1},\bar{t}^{2},\{\bar{t}_{\textsc{ii}}^{s}\}_{s=3}^{j-1},\{\bar{t}^{s}\}_{s=j}^{N-1})\label{eq:tt1}\\
 & =\frac{1}{\prod_{s=3}^{j-1}\alpha_{s}(z)}\mathbb{B}(\{z,\bar{t}_{\textsc{ii}}^{1}\},\{z,\bar{t}^{2}\},\{z,\bar{t}_{\textsc{ii}}^{s}\}_{s=3}^{j-1},\{\bar{t}^{s}\}_{s=j}^{N-1})f(\bar{t}^{j},z)+\nonumber \\
 & +\sum_{k=j+1}^{N}\frac{1}{\prod_{s=3}^{k-1}\alpha_{s}(z)}\sum_{\mathrm{part}(\{\bar{t}^{s}\}_{s=j}^{k-1})}\mathbb{B}(\{z,\bar{t}_{\textsc{ii}}^{1}\},\{z,\bar{t}^{2}\},\{z,\bar{t}_{\textsc{ii}}^{s}\}_{s=3}^{k-1},\{\bar{t}^{s}\}_{s=k}^{N-1})f(\bar{t}^{k},z)\times\nonumber \\
 & \qquad\qquad\qquad\qquad\qquad\times\frac{\prod_{s=j}^{k-1}\alpha_{s}(\bar{t}_{\textsc{i}}^{s})f(\bar{t}_{\textsc{ii}}^{s},\bar{t}_{\textsc{i}}^{s})}{\prod_{s=j}^{k-2}h(\bar{t}_{\textsc{i}}^{s+1},\bar{t}_{\textsc{i}}^{s})f(\bar{t}_{\textsc{ii}}^{s+1},\bar{t}_{\textsc{i}}^{s})}\frac{g(z,\bar{t}_{\textsc{i}}^{k-1})}{f(\bar{t}^{k},\bar{t}_{\textsc{i}}^{k-1})}.\nonumber 
\end{align}
Let us introduce the following notation
\begin{align}
\mathcal{T}_{j}(z|\bar{t}^{1},\bar{t}^{2},\{\bar{t}_{\textsc{ii}}^{s}\}_{s=3}^{j-1},\{\bar{t}^{s}\}_{s=j}^{N-1}) & :=\frac{T_{2,j}(z)}{\lambda_{3}(z)}\mathbb{B}(\bar{t}^{1},\bar{t}^{2},\{\bar{t}_{\textsc{ii}}^{s}\}_{s=3}^{j-1},\{\bar{t}^{s}\}_{s=j}^{N-1})-\label{eq:defTj}\\
 & -\sum_{\mathrm{part}(\bar{t}^{1})}\frac{T_{1,j}(z)}{\lambda_{3}(z)}\mathbb{B}(\bar{t}_{\textsc{ii}}^{1},\bar{t}^{2},\{\bar{t}_{\textsc{ii}}^{s}\}_{s=3}^{j-1},\{\bar{t}^{s}\}_{s=j}^{N-1})g(\bar{t}_{\textsc{i}}^{1},z)f(\bar{t}_{\textsc{i}}^{1},\bar{t}_{\textsc{ii}}^{1}),\nonumber 
\end{align}
where the sum goes over all the partitions of $\bar{t}^{1}=\bar{t}_{\textsc{i}}^{1}\cup\bar{t}_{\textsc{ii}}^{1}$
for $\#\bar{t}_{\textsc{i}}^{1}=1$. Using the previous results (\ref{eq:tt2})
and (\ref{eq:tt1}) we obtain that
\begin{align}
 & \mathcal{T}_{j}(z|\bar{t}^{1},\bar{t}^{2},\{\bar{t}_{\textsc{ii}}^{s}\}_{s=3}^{j-1},\{\bar{t}^{s}\}_{s=j}^{N-1})\nonumber \\
 & =\frac{1}{\prod_{s=3}^{j-1}\alpha_{s}(z)}\mathbb{B}(\bar{t}^{1},\{z,\bar{t}^{2}\},\{z,\bar{t}_{\textsc{ii}}^{s}\}_{s=3}^{j-1},\{\bar{t}^{s}\}_{s=j}^{N-1})f(z,\bar{t}^{1})f(\bar{t}^{j},z)+\\
 & +\sum_{k=j+1}^{N}\frac{1}{\prod_{s=3}^{k-1}\alpha_{s}(z)}\sum_{\mathrm{part}(\{\bar{t}^{s}\}_{s=j}^{k-1})}\mathbb{B}(\bar{t}^{1},\{z,\bar{t}^{2}\},\{z,\bar{t}_{\textsc{ii}}^{s}\}_{s=3}^{k-1},\{\bar{t}^{s}\}_{s=k}^{N-1})f(z,\bar{t}^{1})f(\bar{t}^{k},z)\times\nonumber \\
 & \qquad\qquad\qquad\qquad\qquad\times\frac{\prod_{s=j}^{k-1}\alpha_{s}(\bar{t}_{\textsc{i}}^{s})f(\bar{t}_{\textsc{ii}}^{s},\bar{t}_{\textsc{i}}^{s})}{\prod_{s=j}^{k-2}h(\bar{t}_{\textsc{i}}^{s+1},\bar{t}_{\textsc{i}}^{s})f(\bar{t}_{\textsc{ii}}^{s+1},\bar{t}_{\textsc{i}}^{s})}\frac{g(z,\bar{t}_{\textsc{i}}^{k-1})}{f(\bar{t}^{k},\bar{t}_{\textsc{i}}^{k-1})}.\nonumber 
\end{align}
We can see that the $\mathcal{T}_{j}$-s are expressed with $\mathbb{B}(\bar{t}^{1},\{z,\bar{t}^{2}\},\{z,\bar{t}_{\textsc{ii}}^{s}\}_{s=3}^{j-1},\{\bar{t}^{s}\}_{s=j}^{N-1})$
for $j=3,\dots,N$. We can invert these equations and express $\mathbb{B}(\bar{t}^{1},\{z,\bar{t}^{2}\},\{\bar{t}^{s}\}_{s=3}^{N-1})$
with $\mathcal{T}_{j}$-s as
\begin{align}
 & \mathbb{B}(\bar{t}^{1},\{z,\bar{t}^{2}\},\{\bar{t}^{s}\}_{s=3}^{N-1})=\nonumber \\
 & \sum_{k=3}^{N}\sum_{\mathrm{part}(\bar{t})}\mathcal{T}_{k}(z,\bar{t}^{1},\bar{t}^{2},\{\bar{t}_{\textsc{ii}}^{s}\}_{s=3}^{k-1},\{\bar{t}^{s}\}_{s=k}^{N-1})\frac{1}{f(z,\bar{t}^{1})}\frac{\prod_{s=3}^{k-1}\alpha_{s}(\bar{t}_{\textsc{i}}^{s})g(\bar{t}_{\textsc{i}}^{s},\bar{t}_{\textsc{i}}^{s-1})f(\bar{t}_{\textsc{ii}}^{s},\bar{t}_{\textsc{i}}^{s})}{\prod_{s=3}^{k}f(\bar{t}^{s},\bar{t}_{\textsc{i}}^{s-1})},
\end{align}
where the sum goes over all the partitions of $\bar{t}^{\nu}=\bar{t}_{\textsc{i}}^{\nu}\cup\bar{t}_{\textsc{ii}}^{\nu}$
for $\nu=3,4,\dots,k-1$ and $\#\bar{t}_{\textsc{i}}^{\nu}=1$. Substituting
the definition of $\mathcal{T}_{j}$ (\ref{eq:defTj}) we obtain what
we wanted to prove (\ref{eq:rect2}).

\subsection*{Recursion in $r_{N-2}$}

In this subsection we derive an alternative recurrence relation
\begin{align}
 & \mathbb{B}(\{\bar{t}^{s}\}_{s=1}^{N-3},\{z,\bar{t}^{N-2}\},\bar{t}^{N-2})\nonumber \\
 & =\sum_{j=1}^{N-2}\sum_{\mathrm{part}(\bar{t})}\frac{T_{j,N-1}(z)}{\lambda_{N-1}(z)}\mathbb{B}(\{\bar{t}^{s}\}_{s=1}^{j-1},\{\bar{t}_{\textsc{ii}}^{s}\}_{s=j}^{N-3},\bar{t}^{N-2},\bar{t}^{N-1})\frac{\prod_{\nu=j}^{N-3}g(\bar{t}_{\textsc{i}}^{\nu+1},\bar{t}_{\textsc{i}}^{\nu})f(\bar{t}_{\textsc{i}}^{\nu},\bar{t}_{\textsc{ii}}^{\nu})}{\prod_{\nu=j}^{N-2}f(\bar{t}_{\textsc{i}}^{\nu},\bar{t}^{\nu-1})}\frac{1}{f(\bar{t}^{N-1},z)}+\nonumber \\
 & +\sum_{j=1}^{N-2}\sum_{\mathrm{part}(\bar{t})}\frac{T_{j,N}(z)}{\lambda_{N-1}(z)}\mathbb{B}(\{\bar{t}^{s}\}_{s=1}^{j-1},\{\bar{t}_{\textsc{ii}}^{s}\}_{s=j}^{N-3},\bar{t}^{N-2},\bar{t}_{\textsc{ii}}^{N-1})\frac{\prod_{\nu=j}^{N-3}g(\bar{t}_{\textsc{i}}^{\nu+1},\bar{t}_{\textsc{i}}^{\nu})f(\bar{t}_{\textsc{i}}^{\nu},\bar{t}_{\textsc{ii}}^{\nu})}{\prod_{\nu=j}^{N-2}f(\bar{t}_{\textsc{i}}^{\nu},\bar{t}^{\nu-1})}\times\label{eq:rect2-2}\\
 & \qquad\qquad\qquad\qquad\qquad\qquad\qquad\qquad\times\frac{\alpha_{N-1}(\bar{t}_{\textsc{i}}^{N-1})g(\bar{t}_{\textsc{i}}^{N-1},z)f(\bar{t}_{\textsc{ii}}^{N-1},\bar{t}_{\textsc{i}}^{N-1})}{f(\bar{t}^{N-1},z)},\nonumber 
\end{align}
where the sum goes over all the possible partitions $\bar{t}^{\nu}=\bar{t}_{\textsc{i}}^{\nu}\cup\bar{t}_{\textsc{ii}}^{\nu}$
for $\nu=N-1$ and $\nu=j,\dots,N-3$ where $\bar{t}_{\textsc{i}}^{\nu},\bar{t}_{\textsc{ii}}^{\nu}$
are disjoint subsets and $\#\bar{t}_{\textsc{i}}^{\nu}=1$. We set
by definition $\bar{t}_{\textsc{i}}^{N-2}=\{z\}$ and $\bar{t}^{0}=\emptyset$.

The derivation is analogous to the previous subsection. We start with
the action formulas (\ref{eq:act}):
\begin{align}
 & \frac{T_{j,N-1}(z)}{\lambda_{N-1}(z)}\mathbb{B}(\{\bar{t}^{s}\}_{s=1}^{j-1},\{\bar{t}_{\textsc{ii}}^{s}\}_{s=j}^{N-3},\bar{t}^{N-2},\bar{t}^{N-1})=\nonumber \\
 & =\frac{1}{\alpha_{N-1}(z)}\sum_{\mathrm{part}(\bar{w})}\mathbb{B}(\{\bar{w}_{\textsc{ii}}^{s}\}_{s=1}^{j-1},\{\bar{t}_{\textsc{ii}}^{s},z\}_{s=j}^{N-3},\{\bar{t}^{N-2},z\},\bar{w}_{\textsc{ii}}^{N-1})\times\\
 & \times\prod_{s=1}^{j-1}\frac{f(\bar{w}_{\textsc{i}}^{s},\bar{w}_{\textsc{ii}}^{s})}{h(\bar{w}_{\textsc{i}}^{s},\bar{w}_{\textsc{i}}^{s-1})f(\bar{w}_{\textsc{i}}^{s},\bar{w}_{\textsc{ii}}^{s-1})}\frac{\alpha_{N-1}(\bar{w}_{\textsc{iii}}^{N-1})f(\bar{w}_{\textsc{ii}}^{N-1},\bar{w}_{\textsc{iii}}^{N-1})}{h(z,\bar{w}_{\textsc{iii}}^{N-1})},\nonumber 
\end{align}
and
\begin{align}
 & \frac{T_{j,N}(z)}{\lambda_{N-1}(z)}\mathbb{B}(\{\bar{t}^{s}\}_{s=1}^{j-1},\{\bar{t}_{\textsc{ii}}^{s}\}_{s=j}^{N-3},\bar{t}^{N-2},\bar{t}_{\textsc{ii}}^{N-1})=\\
 & =\frac{1}{\alpha_{N-1}(z)}\sum_{\mathrm{part}(\bar{w})}\mathbb{B}(\{\bar{w}_{\textsc{ii}}^{s}\}_{s=1}^{j-1},\{\bar{t}_{\textsc{ii}}^{s},z\}_{s=j}^{N-3},\{\bar{t}^{N-2},z\},\{\bar{t}_{\textsc{ii}}^{N-1},z\})\prod_{s=1}^{j-1}\frac{f(\bar{w}_{\textsc{i}}^{s},\bar{w}_{\textsc{ii}}^{s})}{h(\bar{w}_{\textsc{i}}^{s},\bar{w}_{\textsc{i}}^{s-1})f(\bar{w}_{\textsc{i}}^{s},\bar{w}_{\textsc{ii}}^{s-1})},\nonumber 
\end{align}
where $\bar{w}^{\nu}=\{z,\bar{t}^{\nu}\}$ and the sum goes over all
the partitions of $\bar{w}^{N-1}=\bar{w}_{\textsc{ii}}^{N-1}\cup\bar{w}_{\textsc{iii}}^{N-1}$
and $\bar{w}^{\nu}=\bar{w}_{\textsc{i}}^{\nu}\cup\bar{w}_{\textsc{ii}}^{\nu}$
where $\#\bar{w}_{\textsc{iii}}^{N-1}=\#\bar{w}_{\textsc{i}}^{\nu}=1$
for $\nu=1,\dots,j-1$. We also set $\bar{w}_{\textsc{i}}^{1}=\{z\}$
and $\bar{w}_{\textsc{ii}}^{1}=\emptyset$.

We can do the summation for the partitions of $\bar{w}^{N-1}$:
\begin{align}
 & \frac{T_{j,N-1}(z)}{\lambda_{N-1}(z)}\mathbb{B}(\{\bar{t}^{s}\}_{s=1}^{j-1},\{\bar{t}_{\textsc{ii}}^{s}\}_{s=j}^{N-3},\bar{t}^{N-2},\bar{t}^{N-1})\nonumber \\
 & =\sum_{\mathrm{part}(\bar{w})}\mathbb{B}(\{\bar{w}_{\textsc{ii}}^{s}\}_{s=1}^{j-1},\{\bar{t}_{\textsc{ii}}^{s},z\}_{s=j}^{N-3},\{\bar{t}^{N-2},z\},\bar{t}^{N-1})\prod_{s=1}^{j-1}\frac{f(\bar{w}_{\textsc{i}}^{s},\bar{w}_{\textsc{ii}}^{s})}{h(\bar{w}_{\textsc{i}}^{s},\bar{w}_{\textsc{i}}^{s-1})f(\bar{w}_{\textsc{i}}^{s},\bar{w}_{\textsc{ii}}^{s-1})}f(\bar{t}^{N-1},z)\nonumber \\
 & +\frac{1}{\alpha_{N-1}(z)}\sum_{\mathrm{part}(\{\bar{w}\}_{1}^{j-1},\bar{t}^{N-1})}\mathbb{B}(\{\bar{w}_{\textsc{ii}}^{s}\}_{s=1}^{j-1},\{\bar{t}_{\textsc{ii}}^{s},z\}_{s=j}^{N-3},\{\bar{t}^{N-2},z\},\{\bar{t}_{\textsc{ii}}^{N-1},z\})\times\label{eq:temp1-2}\\
 & \qquad\qquad\qquad\times\prod_{s=1}^{j-1}\frac{f(\bar{w}_{\textsc{i}}^{s},\bar{w}_{\textsc{ii}}^{s})}{h(\bar{w}_{\textsc{i}}^{s},\bar{w}_{\textsc{i}}^{s-1})f(\bar{w}_{\textsc{i}}^{s},\bar{w}_{\textsc{ii}}^{s-1})}\alpha_{N-1}(\bar{t}_{\textsc{i}}^{N-1})g(z,\bar{t}_{\textsc{i}}^{N-1})f(\bar{t}_{\textsc{ii}}^{N-1},\bar{t}_{\textsc{i}}^{N-1}).\nonumber 
\end{align}
Let us concentrate on the summation of the first line. We can do the
summation using the same trick as in the previous subsection, i.e.,
if $\bar{w}_{\textsc{i}}^{k}=\{z\}$ then $\bar{w}_{\textsc{i}}^{s}=\{z\},\bar{w}_{\textsc{ii}}^{s}=\bar{t}^{s}$
for $s<k$:
\begin{align}
 & \sum_{\mathrm{part}(\bar{w})}\mathbb{B}(\{\bar{w}_{\textsc{ii}}^{s}\}_{s=1}^{j-1},\{\bar{t}_{\textsc{ii}}^{s},z\}_{s=j}^{N-3},\{\bar{t}^{N-2},z\},\bar{t}^{N-1})\prod_{s=1}^{j-1}\frac{f(\bar{w}_{\textsc{i}}^{s},\bar{w}_{\textsc{ii}}^{s})}{h(\bar{w}_{\textsc{i}}^{s},\bar{w}_{\textsc{i}}^{s-1})f(\bar{w}_{\textsc{i}}^{s},\bar{w}_{\textsc{ii}}^{s-1})}f(\bar{t}^{N-1},z)=\nonumber \\
 & =\mathbb{B}(\{\bar{t}^{s}\}_{s=1}^{j-1},\{\bar{t}_{\textsc{ii}}^{s},z\}_{s=j}^{N-3},\{\bar{t}^{N-2},z\},\bar{t}^{N-1})f(z,\bar{t}^{j-1})f(\bar{t}^{N-1},z)+\\
 & +\sum_{k=1}^{j-1}\sum_{\mathrm{part}(\{\bar{t}\}_{k}^{j-1})}\mathbb{B}(\{\bar{t}^{s}\}_{s=1}^{k-1},\{\bar{t}_{\textsc{ii}}^{s},z\}_{s=k}^{N-3},\{\bar{t}^{N-2},z\},\bar{t}^{N-1})f(z,\bar{t}^{k-1})f(\bar{t}^{N-1},z)\times\nonumber \\
 & \qquad\qquad\qquad\times\frac{g(\bar{t}_{\textsc{i}}^{k},z)f(\bar{t}_{\textsc{i}}^{k},\bar{t}_{\textsc{ii}}^{k})}{f(\bar{t}_{\textsc{i}}^{k},\bar{t}^{k-1})}\prod_{s=k+1}^{j-1}\frac{g(\bar{t}_{\textsc{i}}^{s},\bar{t}_{\textsc{i}}^{s-1})f(\bar{t}_{\textsc{i}}^{s},\bar{t}_{\textsc{ii}}^{s})}{f(\bar{t}_{\textsc{i}}^{s},\bar{t}^{s-1})},\nonumber 
\end{align}
where the sum goes over all the partitions of $\bar{t}^{\nu}=\bar{t}_{\textsc{i}}^{\nu}\cup\bar{t}_{\textsc{ii}}^{\nu}$
and $\#\bar{t}_{\textsc{i}}^{\nu}=1$ for $\nu=k,\dots,j-1$ . We
can repeat the analogous calculation for the second line of (\ref{eq:temp1-2})
and we obtain that
\begin{align}
 & \sum_{\mathrm{part}(\bar{w})}\mathbb{B}(\{\bar{w}_{\textsc{ii}}^{s}\}_{s=1}^{j-1},\{\bar{t}_{\textsc{ii}}^{s},z\}_{s=j}^{N-3},\{\bar{t}^{N-2},z\},\{\bar{t}_{\textsc{ii}}^{N-1},z\})\prod_{s=1}^{j-1}\frac{f(\bar{w}_{\textsc{i}}^{s},\bar{w}_{\textsc{ii}}^{s})}{h(\bar{w}_{\textsc{i}}^{s},\bar{w}_{\textsc{i}}^{s-1})f(\bar{w}_{\textsc{i}}^{s},\bar{w}_{\textsc{ii}}^{s-1})}=\nonumber \\
 & =\mathbb{B}(\{\bar{t}^{s}\}_{s=1}^{j-1},\{\bar{t}_{\textsc{ii}}^{s},z\}_{s=j}^{N-3},\{\bar{t}^{N-2},z\},\{\bar{t}_{\textsc{ii}}^{N-1},z\})f(z,\bar{t}^{j-1})+\\
 & +\sum_{k=1}^{j-1}\sum_{\mathrm{part}(\{\bar{t}\}_{k}^{j-1})}\mathbb{B}(\{\bar{t}^{s}\}_{s=1}^{k-1},\{\bar{t}_{\textsc{ii}}^{s},z\}_{s=k}^{N-3},\{\bar{t}^{N-2},z\},\{\bar{t}_{\textsc{ii}}^{N-1},z\})f(z,\bar{t}^{k-1})\times\nonumber \\
 & \qquad\qquad\qquad\times\frac{g(\bar{t}_{\textsc{i}}^{k},z)f(\bar{t}_{\textsc{i}}^{k},\bar{t}_{\textsc{ii}}^{k})}{f(\bar{t}_{\textsc{i}}^{k},\bar{t}^{k-1})}\prod_{s=k+1}^{j-1}\frac{g(\bar{t}_{\textsc{i}}^{s},\bar{t}_{\textsc{i}}^{s-1})f(\bar{t}_{\textsc{i}}^{s},\bar{t}_{\textsc{ii}}^{s})}{f(\bar{t}_{\textsc{i}}^{s},\bar{t}^{s-1})}.\nonumber 
\end{align}
Substituting back to (\ref{eq:temp1-2}) we obtain that
\begin{align}
 & \frac{T_{j,N-1}(z)}{\lambda_{N-1}(z)}\mathbb{B}(\{\bar{t}^{s}\}_{s=1}^{j-1},\{\bar{t}_{\textsc{ii}}^{s}\}_{s=j}^{N-3},\bar{t}^{N-2},\bar{t}^{N-1})=\nonumber \\
 & =\mathbb{B}(\{\bar{t}^{s}\}_{s=1}^{j-1},\{\bar{t}_{\textsc{ii}}^{s},z\}_{s=j}^{N-3},\{\bar{t}^{N-2},z\},\bar{t}^{N-1})f(z,\bar{t}^{j-1})f(\bar{t}^{N-1},z)+\nonumber \\
 & +\sum_{k=1}^{j-1}\sum_{\mathrm{part}(\{\bar{t}\}_{k}^{j-1})}\mathbb{B}(\{\bar{t}^{s}\}_{s=1}^{k-1},\{\bar{t}_{\textsc{ii}}^{s},z\}_{s=k}^{N-3},\{\bar{t}^{N-2},z\},\bar{t}^{N-1})\frac{f(z,\bar{t}^{k-1})}{1}f(\bar{t}^{N-1},z)\times\nonumber \\
 & \qquad\qquad\qquad\times\frac{g(\bar{t}_{\textsc{i}}^{k},z)f(\bar{t}_{\textsc{i}}^{k},\bar{t}_{\textsc{ii}}^{k})}{f(\bar{t}_{\textsc{i}}^{k},\bar{t}^{k-1})}\prod_{s=k+1}^{j-1}\frac{g(\bar{t}_{\textsc{i}}^{s},\bar{t}_{\textsc{i}}^{s-1})f(\bar{t}_{\textsc{i}}^{s},\bar{t}_{\textsc{ii}}^{s})}{f(\bar{t}_{\textsc{i}}^{s},\bar{t}^{s-1})}+\nonumber \\
 & +\frac{1}{\alpha_{N-1}(z)}\sum_{\mathrm{part}(\bar{t}^{N-1})}\Biggl[\mathbb{B}(\{\bar{t}^{s}\}_{s=1}^{j-1},\{\bar{t}_{\textsc{ii}}^{s},z\}_{s=j}^{N-3},\{\bar{t}^{N-2},z\},\{\bar{t}_{\textsc{ii}}^{N-1},z\})f(z,\bar{t}^{j-1})+\label{eq:tt2-1}\\
 & \qquad\qquad\qquad+\sum_{\mathrm{part}(\{\bar{t}\}_{k}^{j-1})}\mathbb{B}(\{\bar{t}^{s}\}_{s=1}^{k-1},\{\bar{t}_{\textsc{ii}}^{s},z\}_{s=k}^{N-3},\{\bar{t}^{N-2},z\},\{\bar{t}_{\textsc{ii}}^{N-1},z\})f(z,\bar{t}^{k-1})\times\nonumber \\
 & \qquad\qquad\qquad\times\frac{g(\bar{t}_{\textsc{i}}^{k},z)f(\bar{t}_{\textsc{i}}^{k},\bar{t}_{\textsc{ii}}^{k})}{f(\bar{t}_{\textsc{i}}^{k},\bar{t}^{k-1})}\prod_{s=k+1}^{j-1}\frac{g(\bar{t}_{\textsc{i}}^{s},\bar{t}_{\textsc{i}}^{s-1})f(\bar{t}_{\textsc{i}}^{s},\bar{t}_{\textsc{ii}}^{s})}{f(\bar{t}_{\textsc{i}}^{s},\bar{t}^{s-1})}\Biggr]\alpha_{N-1}(\bar{t}_{\textsc{i}}^{N-1})g(z,\bar{t}_{\textsc{i}}^{N-1})f(\bar{t}_{\textsc{ii}}^{N-1},\bar{t}_{\textsc{i}}^{N-1}).\nonumber 
\end{align}
We can repeat the analogous calculation for
\begin{align}
 & \frac{T_{j,N}(z)}{\lambda_{N-1}(z)}\mathbb{B}(\{\bar{t}^{s}\}_{s=1}^{j-1},\{\bar{t}_{\textsc{ii}}^{s}\}_{s=j}^{N-3},\bar{t}^{N-2},\bar{t}_{\textsc{ii}}^{N-1})=\label{eq:tt1-1}\\
 & =\frac{1}{\alpha_{N-1}(z)}\mathbb{B}(\{\bar{t}^{s}\}_{s=1}^{j-1},\{\bar{t}_{\textsc{ii}}^{s},z\}_{s=j}^{N-3},\{\bar{t}^{N-2},z\},\{\bar{t}_{\textsc{ii}}^{N-1},z\})f(z,\bar{t}^{j-1})+\nonumber \\
 & +\frac{1}{\alpha_{N-1}(z)}\sum_{k=1}^{j-1}\sum_{\mathrm{part}(\{\bar{t}\}_{k}^{j-1})}\mathbb{B}(\{\bar{t}^{s}\}_{s=1}^{k-1},\{\bar{t}_{\textsc{ii}}^{s},z\}_{s=k}^{N-3},\{\bar{t}^{N-2},z\},\{\bar{t}_{\textsc{ii}}^{N-1},z\})f(z,\bar{t}^{k-1})\times\nonumber \\
 & \qquad\qquad\qquad\times\frac{g(\bar{t}_{\textsc{i}}^{k},z)f(\bar{t}_{\textsc{i}}^{k},\bar{t}_{\textsc{ii}}^{k})}{f(\bar{t}_{\textsc{i}}^{k},\bar{t}^{k-1})}\prod_{s=k+1}^{j-1}\frac{g(\bar{t}_{\textsc{i}}^{s},\bar{t}_{\textsc{i}}^{s-1})f(\bar{t}_{\textsc{i}}^{s},\bar{t}_{\textsc{ii}}^{s})}{f(\bar{t}_{\textsc{i}}^{s},\bar{t}^{s-1})}.
\end{align}
Let us introduce the following notation
\begin{multline}
\mathcal{T}_{j}(z|\{\bar{t}^{s}\}_{s=1}^{j-1},\{\bar{t}_{\textsc{ii}}^{s}\}_{s=j}^{N-3},\bar{t}^{N-2},\bar{t}^{N-1}):=\frac{T_{j,N-1}(z)}{\lambda_{N-1}(z)}\mathbb{B}(\{\bar{t}^{s}\}_{s=1}^{j-1},\{\bar{t}_{\textsc{ii}}^{s}\}_{s=j}^{N-3},\bar{t}^{N-2},\bar{t}^{N-1})-\\
-\sum_{\mathrm{part}(\bar{t}^{N-1})}\frac{T_{j,N}(z)}{\lambda_{N-1}(z)}\mathbb{B}(\{\bar{t}^{s}\}_{s=1}^{j-1},\{\bar{t}_{\textsc{ii}}^{s}\}_{s=j}^{N-3},\bar{t}^{N-2},\bar{t}_{\textsc{ii}}^{N-1})\alpha_{N-1}(\bar{t}_{\textsc{i}}^{N-1})g(z,\bar{t}_{\textsc{i}}^{N-1})f(\bar{t}_{\textsc{ii}}^{N-1},\bar{t}_{\textsc{i}}^{N-1}),\label{eq:defTj-1}
\end{multline}
where the sum goes over all the partitions of $\bar{t}^{N-1}=\bar{t}_{\textsc{i}}^{N-1}\cup\bar{t}_{\textsc{ii}}^{N-1}$
for $\#\bar{t}_{\textsc{i}}^{N-1}=1$. Using the previous results
(\ref{eq:tt2-1}) and (\ref{eq:tt1-1}) we obtain that
\begin{align}
 & \mathcal{T}_{j}(z|\{\bar{t}^{s}\}_{s=1}^{j-1},\{\bar{t}_{\textsc{ii}}^{s}\}_{s=j}^{N-3},\bar{t}^{N-2},\bar{t}^{N-1})=\nonumber \\
 & =\mathbb{B}(\{\bar{t}^{s}\}_{s=1}^{j-1},\{\bar{t}_{\textsc{ii}}^{s},z\}_{s=j}^{N-3},\{\bar{t}^{N-2},z\},\bar{t}^{N-1})f(z,\bar{t}^{j-1})f(\bar{t}^{N-1},z)+\\
 & +\sum_{k=1}^{j-1}\sum_{\mathrm{part}(\{\bar{t}\}_{k}^{j-1})}\mathbb{B}(\{\bar{t}^{s}\}_{s=1}^{k-1},\{\bar{t}_{\textsc{ii}}^{s},z\}_{s=k}^{N-3},\{\bar{t}^{N-2},z\},\bar{t}^{N-1})\frac{f(z,\bar{t}^{k-1})}{1}f(\bar{t}^{N-1},z)\times\nonumber \\
 & \qquad\qquad\qquad\times\frac{g(\bar{t}_{\textsc{i}}^{k},z)f(\bar{t}_{\textsc{i}}^{k},\bar{t}_{\textsc{ii}}^{k})}{f(\bar{t}_{\textsc{i}}^{k},\bar{t}^{k-1})}\prod_{s=k+1}^{j-1}\frac{g(\bar{t}_{\textsc{i}}^{s},\bar{t}_{\textsc{i}}^{s-1})f(\bar{t}_{\textsc{i}}^{s},\bar{t}_{\textsc{ii}}^{s})}{f(\bar{t}_{\textsc{i}}^{s},\bar{t}^{s-1})}.\nonumber 
\end{align}
We can see that the $\mathcal{T}_{j}$-s are expressed with $\mathbb{B}(\bar{t}^{1},\{z,\bar{t}^{2}\},\{z,\bar{t}_{\textsc{ii}}^{s}\}_{s=3}^{j-1},\{\bar{t}^{s}\}_{s=j}^{N-1})$
for $j=1,\dots,N-2$. We can invert these equations and express $\mathbb{B}(\bar{t}^{1},\{z,\bar{t}^{2}\},\{\bar{t}^{s}\}_{s=3}^{N-1})$
with $\mathcal{T}_{j}$-s as
\begin{multline}
\mathbb{B}(\{\bar{t}^{s}\}_{s=1}^{N-3},\{\bar{t}^{N-2},z\},\bar{t}^{N-1}))=\\
\sum_{j=1}^{N-2}\sum_{\mathrm{part}(\bar{t})}\mathcal{T}_{j}(z|\{\bar{t}^{s}\}_{s=1}^{j-1},\{\bar{t}_{\textsc{ii}}^{s}\}_{s=j}^{N-3},\bar{t}^{N-2},\bar{t}^{N-1})\frac{\prod_{s=j}^{N-3}g(\bar{t}_{\textsc{i}}^{s+1},\bar{t}_{\textsc{i}}^{s})f(\bar{t}_{\textsc{ii}}^{s},\bar{t}_{\textsc{i}}^{s})}{\prod_{s=j}^{N-2}f(\bar{t}_{\textsc{i}}^{s},\bar{t}^{s-1})}\frac{1}{f(\bar{t}^{N-1},z)},
\end{multline}
where the sum goes over all the partitions of $\bar{t}^{s}=\bar{t}_{\textsc{i}}^{s}\cup\bar{t}_{\textsc{ii}}^{s}$
for $s=j,\dots,N-3$ and $\#\bar{t}_{\textsc{i}}^{\nu}=1$. Substituting
the definition of $\mathcal{T}_{j}$ (\ref{eq:defTj-1}) we obtain
what we wanted to prove (\ref{eq:rect2-2}).

\section{Proof of the sum formulas}

In this section we prove the sum formulas (\ref{eq:sumUTw}) and (\ref{eq:sumTW}).
We deal the untwisted and twisted cases separately.

\subsection{Untwisted case\label{sec:Proof-of-the}}

For the untwisted case, let us fix the $K$-matrix to the type $(N,M)$
regular form: 
\begin{equation}
K(z)=\sum_{k=1}^{M}\mathfrak{b}_{k}E_{N+1-k,k}+\sum_{k=1}^{N}K_{k,k}(z)E_{k,k}.\label{eq:regform}
\end{equation}
We do not need the explicit form of the non-zero matrix entries therefore
the result is valid also for the singular $K$-matrices with the form
(\ref{eq:regform}). In section \ref{subsec:Reqursion-and-pair-general-M}
we already saw that the variables $\alpha_{s}(t_{k}^{s})$ can be
handled as independent variables for $s=1,\dots,M$ and $s=N-M,\dots,N-1$.
Repeating the previous derivation of \cite{Gombor:2021hmj} (appendix
(C.2)) we can show that the off-shell overlap has the sum formula
\begin{equation}
\mathcal{S}_{\bar{\alpha}}(\bar{t})=\sum_{\mathrm{part}(\{\bar{t}^{s}\}_{s\in\mathfrak{s}^{+}})}\mathcal{W}_{\{\alpha_{s}\}_{s\in\mathfrak{s}^{-}}}(\{\bar{t}_{\textsc{i}}^{s}\}_{s\in\mathfrak{s}^{+}}|\{\bar{t}_{\textsc{ii}}^{s}\}_{s\in\mathfrak{s}^{+}}|\{\bar{t}^{s}\}_{s\in\mathfrak{s}^{-}})\prod_{s\in\mathfrak{s}^{+}}\alpha_{s}(\bar{t}_{\textsc{i}}^{s}),\label{eq:sumGF0}
\end{equation}
where we defined the sets $\mathfrak{s}^{+}=\{1,\dots,M\}\cup\{N-M,\dots,N-1\}$
and $\mathfrak{s}^{-}=\{M+1,\dots,N-M-1\}$. The coefficients $\mathcal{W}_{\{\alpha_{s}\}_{s\in\mathfrak{s}^{-}}}(\{\bar{t}_{\textsc{i}}^{s}\}_{s\in\mathfrak{s}^{+}}|\{\bar{t}_{\textsc{ii}}^{s}\}_{s\in\mathfrak{s}^{+}}|\{\bar{t}^{s}\}_{s\in\mathfrak{s}^{-}})$
do not depend on the $\alpha$-functions $\alpha_{s}$ for $s\in\mathfrak{s}^{+}$.
For completeness, we present this proof.

\subsubsection{Proof of (\ref{eq:sumGF0})}

We want to prove the sum formula (\ref{eq:sumGF0}) where the weights
$\mathcal{W}_{\{\alpha_{s}\}_{s\in\mathfrak{s}^{-}}}$ do not depend
on $\alpha_{s}$ for $s\in\mathfrak{s}^{+}$. In the derivation we
only care about $\alpha_{s}$ ($s\in\mathfrak{s}^{+}$) dependence
of the overlap therefore we use the notation $(\dots)$ for the $\alpha_{s}$
independent coefficients ($s\in\mathfrak{s}^{+}$). Using this notation
the sum formula takes the form
\begin{equation}
\mathcal{S}_{\bar{\alpha}}(\bar{t})=\sum_{\mathrm{part}(\bar{t})}\prod_{s\in\mathfrak{s}^{+}}\alpha_{s}(\bar{t}_{\textsc{i}}^{s})(\dots).\label{eq:sumrule_UTw}
\end{equation}
We prove this sum formula using induction on $M$. Let us start with
$M=1$. We use the recurrence formula (\ref{eq:rec1})
\begin{equation}
\mathbb{B}(\{z,\bar{t}^{1}\},\left\{ \bar{t}^{s}\right\} _{s=2}^{N-1})=\sum_{j=2}^{N}\frac{T_{1,j}(z)}{\lambda_{2}(z)}\sum_{\mathrm{part}(\bar{t})}\mathbb{B}(\bar{t}^{1},\left\{ \bar{t}_{\textsc{ii}}^{s}\right\} _{s=2}^{j-1},\left\{ \bar{t}^{s}\right\} _{s=j}^{N-1})\prod_{\nu=2}^{j-1}\alpha_{\nu}(t_{\textsc{i}}^{\nu})\times(\dots).
\end{equation}
We saw that, for non-vanishing overlap, the numbers of the first and
$(N-1)$-th type of Bethe roots are equal therefore we fix the quantum
numbers as $\#\bar{t}^{1}=r_{1}-1$ and $\#\bar{t}^{N-1}=r_{1}$.
Using the $KT$-relation (\ref{eq:KT_Utw}) we obtain that
\begin{align}
\langle\Psi|\mathbb{B}(\{z,\bar{t}^{1}\},\left\{ \bar{t}^{s}\right\} _{s=2}^{N-1}) & =\frac{1}{\lambda_{2}(z)}\sum_{j=2}^{N-1}\frac{K_{j,j}(z)}{\mathfrak{b}_{1}}\sum_{\mathrm{part}(\bar{t})}\langle\Psi|T_{N,j}(-z)\mathbb{B}(\bar{t}^{1},\left\{ \bar{t}_{\textsc{ii}}^{s}\right\} _{s=2}^{j-1},\left\{ \bar{t}^{s}\right\} _{s=j}^{N-1})(\dots)\nonumber \\
 & -\frac{1}{\lambda_{2}(z)}\frac{K_{N,N}(z)}{\mathfrak{b}_{1}}\sum_{j=2}^{N-1}\sum_{\mathrm{part}(\bar{t})}\langle\Psi|T_{N,j}(z)\mathbb{B}(\bar{t}^{1},\left\{ \bar{t}_{\textsc{ii}}^{s}\right\} _{s=2}^{j-1},\left\{ \bar{t}^{s}\right\} _{s=j}^{N-1})(\dots)\label{eq:RC1}\\
 & +\frac{1}{\lambda_{2}(z)}\frac{K_{N,N}(z)}{\mathfrak{b}_{1}}\sum_{\mathrm{part}(\bar{t})}\langle\Psi|T_{N,N}(-z)\mathbb{B}(\bar{t}^{1},\left\{ \bar{t}_{\textsc{ii}}^{s}\right\} _{s=2}^{N-1})\alpha_{N-1}(\bar{t}_{\textsc{i}}^{N-1})(\dots)\nonumber \\
 & -\frac{1}{\lambda_{2}(z)}\frac{K_{N,N}(z)}{\mathfrak{b}_{1}}\sum_{\mathrm{part}(\bar{t})}\langle\Psi|T_{N,N}(z)\mathbb{B}(\bar{t}^{1},\left\{ \bar{t}_{\textsc{ii}}^{s}\right\} _{s=2}^{N-1})\alpha_{N-1}(\bar{t}_{\textsc{i}}^{N-1})(\dots).\nonumber 
\end{align}
We can use the action formulas (\ref{eq:act})
\begin{equation}
\begin{split}\frac{T_{N,j}(-z)}{\lambda_{2}(z)}\mathbb{B}(\bar{t}) & =\alpha_{1}(z)\sum_{\mathrm{part}}\mathbb{B}(\bar{\omega}_{\textsc{ii}})\alpha_{N-1}(\bar{\omega}_{\textsc{iii}}^{N-1})\times(\dots),\\
\frac{T_{N,j}(z)}{\lambda_{2}(z)}\mathbb{B}(\bar{t}) & =\frac{1}{\alpha_{N-1}(z)}\sum_{\mathrm{part}}\mathbb{B}(\bar{w}_{\textsc{ii}})\alpha_{N-1}(\bar{w}_{\textsc{iii}}^{N-1})\times(\dots),
\end{split}
\label{eq:RC2}
\end{equation}
for $j>1$, where where $\bar{w}^{\nu}=\{z,\bar{t}^{\nu}\}$, $\bar{\omega}^{\nu}=\{-z,\bar{t}^{\nu}\}$
and $\bar{w}^{\nu}=\bar{w}_{\textsc{i}}^{\nu}\cup\bar{w}_{\textsc{ii}}^{\nu}\cup\bar{w}_{\textsc{iii}}^{\nu}$,
$\bar{\omega}^{\nu}=\bar{\omega}_{\textsc{i}}^{\nu}\cup\bar{\omega}_{\textsc{ii}}^{\nu}\cup\bar{\omega}_{\textsc{iii}}^{\nu}$.
We can see that in these formulas the $\alpha$-dependent terms are
$\alpha_{N-1}(t_{k}^{N-1})$ and $\alpha_{1}(z),\alpha_{N-1}(z)$
therefore combining (\ref{eq:RC1}) and (\ref{eq:RC2}) we obtain
the following recursion for the overlap
\begin{equation}
\langle\Psi|\mathbb{B}(\{z,\bar{t}^{1}\},\left\{ \bar{t}^{s}\right\} _{s=2}^{N-1})=\sum_{\mathrm{part}}\langle\Psi|\mathbb{B}(\bar{w}_{\textsc{ii}})\alpha_{N-1}(\bar{t}_{\textsc{i}}^{N-1})\times(\alpha_{1}(z),\alpha_{N-1}(z),\dots),
\end{equation}
where $\bar{w}_{\textsc{ii}}^{\nu}\subset\{z,-z,\bar{t}_{\textsc{ii}}^{\nu}\}$
and $\bar{t}=\bar{t}_{\textsc{i}}\cup\bar{t}_{\textsc{ii}}$ where
$\bar{w}_{\textsc{ii}}^{1}=\bar{w}_{\textsc{ii}}^{N-1}=r_{1}-1$.
Applying this recursion for the overlap we can eliminate the Bethe
roots $\bar{t}^{1}$ and $\bar{t}^{N-1}$ as
\begin{equation}
\langle\Psi|\mathbb{B}(\bar{t})=\sum_{\mathrm{part}}\langle\Psi|\mathbb{B}(\emptyset,\left\{ \bar{w}_{\textsc{ii}}^{k}\right\} _{k=2}^{N-2},\emptyset)\alpha_{N-1}(\bar{t}_{\textsc{i}}^{N-1})\times(\alpha_{1}(t_{k}^{1}),\alpha_{N-1}(t_{k}^{1}),\dots),
\end{equation}
where $\bar{w}_{\textsc{ii}}^{\nu}\subset\{\bar{t}^{1},-\bar{t}^{1},\bar{t}_{\textsc{ii}}^{\nu}\}$.
Since the type $(N-2,0)$ overlap does not depend on $\alpha_{1}$
or $\alpha_{N-1}$, we obtain that
\begin{equation}
\mathcal{S}_{\bar{\alpha}}(\bar{t})=\sum_{\mathrm{part}}\alpha_{N-1}(\bar{t}_{\textsc{i}}^{N-1})\times(\alpha_{1}(t_{k}^{1}),\alpha_{N-1}(t_{k}^{1}),\dots),\label{eq:prop1}
\end{equation}
for $M=1$. We can repeat this calculation for the other type of recursion
of the off-shell overlap (\ref{eq:rec2-1}):
\begin{equation}
\mathbb{B}(\left\{ \bar{t}^{k}\right\} _{k=1}^{N-2},\{z,\bar{t}^{N-1}\})=\sum_{j=1}^{N-1}\frac{T_{j,N}(z)}{\lambda_{N}(z)}\sum_{\mathrm{part}(\bar{t})}\mathbb{B}(\left\{ \bar{t}^{k}\right\} _{k=1}^{j-1},\left\{ \bar{t}_{\textsc{ii}}^{k}\right\} _{k=j}^{N-2},\bar{t}^{N-1})(\dots).
\end{equation}
Using the $KT$-relation (\ref{eq:KT_Utw}) we obtain that
\begin{equation}
\begin{split}\langle\Psi|\mathbb{B}(\left\{ \bar{t}^{k}\right\} _{k=1}^{N-2},\{z,\bar{t}^{N-1}\}) & =\frac{1}{\lambda_{N}(z)}\sum_{j=1}^{N-1}\frac{K_{j,j}(-z)}{\mathfrak{b}_{1}}\sum\langle\Psi|T_{j,1}(-z)\mathbb{B}(\left\{ \bar{t}^{k}\right\} _{k=1}^{j-1},\left\{ \bar{t}_{\textsc{ii}}^{k}\right\} _{k=j}^{N-2},\bar{t}^{N-1})(\dots)\\
 & -\frac{1}{\lambda_{N}(z)}\frac{K_{1,1}(-z)}{\mathfrak{b}_{1}}\sum_{j=1}^{N-1}\sum\langle\Psi|T_{j,1}(z)\mathbb{B}(\bar{t}^{1},\left\{ \bar{t}_{\textsc{ii}}^{s}\right\} _{s=2}^{j-1},\left\{ \bar{t}^{s}\right\} _{s=j}^{N-1})(\dots).
\end{split}
\end{equation}
We can use the action formulas (\ref{eq:act})
\begin{equation}
\begin{split}\frac{T_{j,1}(-z)}{\lambda_{2}(z)}\mathbb{B}(\bar{t}) & =\alpha_{1}(z)\alpha_{N-1}(z)\sum_{\mathrm{part}}\mathbb{B}(\bar{\omega}_{\textsc{ii}})\alpha_{1}(\bar{\omega}_{\textsc{iii}}^{1})\alpha_{N-1}(\bar{\omega}_{\textsc{iii}}^{N-1})\times(\dots),\\
\frac{T_{j,1}(z)}{\lambda_{2}(z)}\mathbb{B}(\bar{t}) & =\sum_{\mathrm{part}}\mathbb{B}(\bar{w}_{\textsc{ii}})\alpha_{1}(\bar{w}_{\textsc{iii}}^{1})\alpha_{N-1}(\bar{w}_{\textsc{iii}}^{N-1})\times(\dots).
\end{split}
\end{equation}
We can see that in these formulas the $\alpha$-dependent terms are
$\alpha_{1}(t_{k}^{1}),\alpha_{N-1}(t_{k}^{N-1})$ and $\alpha_{1}(z),\alpha_{N-1}(z)$
therefore combining the equations above we obtain the following recursion
for the overlap
\begin{equation}
\langle\Psi|\mathbb{B}(\{z,\bar{t}^{1}\},\left\{ \bar{t}^{s}\right\} _{s=2}^{N-1})=\sum_{\mathrm{part}}\langle\Psi|\mathbb{B}(\bar{w}_{\textsc{ii}})\alpha_{1}(\bar{t}_{\textsc{i}}^{1})\alpha_{N-1}(\bar{t}_{\textsc{i}}^{N-1})\times(\alpha_{1}(z),\alpha_{N-1}(z),\dots),
\end{equation}
where $\bar{w}_{\textsc{ii}}^{\nu}\subset\{z,-z,\bar{t}_{\textsc{ii}}^{\nu}\}$
and $\bar{t}=\bar{t}_{\textsc{i}}\cup\bar{t}_{\textsc{ii}}$ where
$\bar{w}_{\textsc{ii}}^{1}=\bar{w}_{\textsc{ii}}^{N-1}=r_{1}-1$.
Applying this recursion for the overlap we can eliminate the Bethe
roots $\bar{t}^{1}$ and $\bar{t}^{N-1}$ as
\begin{equation}
\langle\Psi|\mathbb{B}(\bar{t})=\sum_{\mathrm{part}}\langle\Psi|\mathbb{B}(\emptyset,\left\{ \bar{w}_{\textsc{ii}}^{k}\right\} _{k=2}^{N-2},\emptyset)\alpha_{1}(\bar{t}_{\textsc{i}}^{1})\times(\alpha_{1}(t_{k}^{N-1}),\alpha_{N-1}(t_{k}^{N-1}),\dots),
\end{equation}
where $\bar{w}_{\textsc{ii}}^{\nu}\subset\{\bar{t}^{1},-\bar{t}^{1},\bar{t}_{\textsc{ii}}^{\nu}\}$.
Since the type $(N-2,0)$ overlap does not depend on $\alpha_{1}$
or $\alpha_{N-1}$, we obtain that
\begin{equation}
\mathcal{S}_{\bar{\alpha}}(\bar{t})=\sum_{\mathrm{part}}\alpha_{1}(\bar{t}_{\textsc{i}}^{1})\times(\alpha_{1}(t_{k}^{N-1}),\alpha_{N-1}(t_{k}^{N-1}),\dots),\label{eq:prop2}
\end{equation}
for $M=1$. Combining the two properties (\ref{eq:prop1}) and (\ref{eq:prop2})
of the $M=1$ off-shell formula we obtain that
\begin{equation}
\mathcal{S}_{\bar{\alpha}}(\bar{t})=\sum_{\mathrm{part}}\alpha_{1}(\bar{t}_{\textsc{i}}^{1})\alpha_{N-1}(\bar{t}_{\textsc{i}}^{N-1})\times(\dots),\label{eq:finalform1}
\end{equation}
which proves (\ref{eq:sumGF0}) for type $M=1$. 

Let us continue with general $M$. Let us assume that the type $(N-2,M-1)$
overlaps have the sum formula (\ref{eq:sumGF0}). Let us turn to the
type $(N,M)$ overlaps. Combining the recurrence relation (\ref{eq:rec1})
with the $KT$-relation we obtain the equation
\begin{align}
 & \langle\Psi|\mathbb{B}(\{z,\bar{t}^{1}\},\left\{ \bar{t}^{s}\right\} _{s=2}^{N-1})=\nonumber \\
 & =\frac{1}{\lambda_{2}(z)}\sum_{j=2}^{M}\frac{\mathfrak{b}_{j}}{\mathfrak{b}_{1}}\sum\langle\Psi|T_{N,j}(-z)\mathbb{B}(\bar{t}^{1},\left\{ \bar{t}_{\textsc{ii}}^{s}\right\} _{s=2}^{j-1},\left\{ \bar{t}^{s}\right\} _{s=j}^{N-1})\prod_{s\in\mathfrak{s}^{+}\cap[2,\dots,j-1]}\alpha_{s}(\bar{t}_{\textsc{i}}^{s})(\dots)\\
 & +\frac{1}{\lambda_{2}(z)}\sum_{j=2}^{N-1}\frac{K_{j,j}(z)}{\mathfrak{b}_{1}}\sum\langle\Psi|T_{N,j}(-z)\mathbb{B}(\bar{t}^{1},\left\{ \bar{t}_{\textsc{ii}}^{s}\right\} _{s=2}^{j-1},\left\{ \bar{t}^{s}\right\} _{s=j}^{N-1})\prod_{s\in\mathfrak{s}^{+}\cap[2,\dots,j-1]}\alpha_{s}(\bar{t}_{\textsc{i}}^{s})(\dots)\nonumber \\
 & -\frac{1}{\lambda_{2}(z)}\frac{K_{N,N}(z)}{\mathfrak{b}_{1}}\sum_{j=2}^{N-1}\sum\langle\Psi|T_{N,j}(z)\mathbb{B}(\bar{t}^{1},\left\{ \bar{t}_{\textsc{ii}}^{s}\right\} _{s=2}^{j-1},\left\{ \bar{t}^{s}\right\} _{s=j}^{N-1})\prod_{s\in\mathfrak{s}^{+}\cap[2,\dots,j-1]}\alpha_{s}(\bar{t}_{\textsc{i}}^{s})(\dots).\nonumber 
\end{align}
We can also use the action formula (\ref{eq:act})
\begin{equation}
\begin{split}\frac{T_{N,j}(-z)}{\lambda_{2}(z)}\mathbb{B}(\bar{t}) & =\alpha_{1}(z)\sum_{\mathrm{part}}\mathbb{B}(\bar{\omega}_{\textsc{ii}})\prod_{s\in\mathfrak{s}^{+}\cap[j,\dots,N-1]}\alpha_{s}(\bar{t}_{\textsc{i}}^{s})\times(\dots),\\
\frac{T_{N,j}(z)}{\lambda_{2}(z)}\mathbb{B}(\bar{t}) & =\frac{1}{\prod_{s\in\mathfrak{s}^{+}\cap[j,\dots,N-1]}\alpha_{s}(z)}\sum_{\mathrm{part}}\mathbb{B}(\bar{w}_{\textsc{ii}})\prod_{s\in\mathfrak{s}^{+}\cap[j,\dots,N-1]}\alpha_{s}(\bar{w}_{\textsc{iii}}^{s})\times(\dots),
\end{split}
\end{equation}
for $j>1$. We can see that in these formulas the $\alpha$-dependent
terms are $\alpha_{s}(t_{k}^{s})$ for $s\in\mathfrak{s}^{+}\cap[2,\dots,N-1]$
and $\alpha_{s}(z)$ for $s\in\mathfrak{s}^{+}$ therefore combining
them we obtain the following recursion for the overlap
\begin{equation}
\langle\Psi|\mathbb{B}(\{z,\bar{t}^{1}\},\left\{ \bar{t}^{s}\right\} _{s=2}^{N-1})=\sum_{\mathrm{part}}\langle\Psi|\mathbb{B}(\bar{w}_{\textsc{ii}})\prod_{s\in\mathfrak{s}^{+}\cap[2,\dots,N-1]}\alpha_{s}(\bar{t}_{\textsc{i}}^{s})\times(\alpha_{s}(z),\dots),
\end{equation}
where $\bar{w}_{\textsc{ii}}^{\nu}\subset\{z,-z,\bar{t}_{\textsc{ii}}^{\nu}\}$
and $\bar{t}=\bar{t}_{\textsc{i}}\cup\bar{t}_{\textsc{ii}}$ where
$\bar{w}_{\textsc{ii}}^{1}=\bar{w}_{\textsc{ii}}^{N-1}=r_{1}-1$.
Applying this recursion for the overlap we can eliminate the Bethe
roots $\bar{t}^{1}$ and $\bar{t}^{N-1}$ as
\begin{equation}
\langle\Psi|\mathbb{B}(\bar{t})=\sum_{\mathrm{part}}\langle\Psi|\mathbb{B}(\emptyset,\left\{ \bar{w}_{\textsc{ii}}^{k}\right\} _{k=2}^{N-2},\emptyset)\prod_{s\in\mathfrak{s}^{+}\cap[2,\dots,N-1]}\alpha_{s}(\bar{t}_{\textsc{i}}^{s})\times(\alpha_{s}(t_{k}^{1}),\dots),
\end{equation}
where $\bar{w}_{\textsc{ii}}^{\nu}\subset\{\bar{t}^{1},-\bar{t}^{1},\bar{t}_{\textsc{ii}}^{\nu}\}$.
Since the type $(N-2,M-1)$ overlap satisfies the induction hypothesis,
we obtain that
\begin{equation}
\mathcal{S}_{\bar{\alpha}}(\bar{t})=\sum_{\mathrm{part}}\prod_{s\in\mathfrak{s}^{+}\cap[2,\dots,N-1]}\alpha_{s}(\bar{t}_{\textsc{i}}^{s})\times(\alpha_{s}(t_{k}^{1}),\dots).\label{eq:prop1-1}
\end{equation}
We can repeat this calculation for the other type of recursion of
the off-shell overlap (\ref{eq:rec2-1}). Combining it with the $KT$-relation
(\ref{eq:KT_Utw}) and the action formulas (\ref{eq:act}), we obtain
a sum formula where the $\alpha$-dependent terms are $\alpha_{s}(t_{k}^{s})$
and $\alpha_{s}(z)$ for $s\in\mathfrak{s}^{+}$ i.e.
\begin{equation}
\langle\Psi|\mathbb{B}(\{z,\bar{t}^{1}\},\left\{ \bar{t}^{s}\right\} _{s=2}^{N-1})=\sum_{\mathrm{part}}\langle\Psi|\mathbb{B}(\bar{w}_{\textsc{ii}})\prod_{s\in\mathfrak{s}^{+}}\alpha_{s}(\bar{t}_{\textsc{i}}^{s})\times(\alpha_{s}(z),\dots),
\end{equation}
where $\bar{w}_{\textsc{ii}}^{\nu}\subset\{z,-z,\bar{t}_{\textsc{ii}}^{\nu}\}$
and $\bar{t}=\bar{t}_{\textsc{i}}\cup\bar{t}_{\textsc{ii}}$ where
$\bar{w}_{\textsc{ii}}^{1}=\bar{w}_{\textsc{ii}}^{N-1}=r_{1}-1$.
Applying this recursion for the overlap we can eliminate the Bethe
roots $\bar{t}^{1}$ and $\bar{t}^{N-1}$ as
\begin{equation}
\langle\Psi|\mathbb{B}(\bar{t})=\sum_{\mathrm{part}}\langle\Psi|\mathbb{B}(\emptyset,\left\{ \bar{w}_{\textsc{ii}}^{k}\right\} _{k=2}^{N-2},\emptyset)\prod_{s\in\mathfrak{s}^{+}\cap[1,\dots,N-2]}\alpha_{s}(\bar{t}_{\textsc{i}}^{s})\times(\alpha_{s}(t_{k}^{N-1}),\dots),
\end{equation}
where $\bar{w}_{\textsc{ii}}^{\nu}\subset\{\bar{t}^{1},-\bar{t}^{1},\bar{t}_{\textsc{ii}}^{\nu}\}$.
Since the type $(N-2,M-1)$ overlap satisfies the induction hypothesis,
we obtain that
\begin{equation}
\mathcal{S}_{\bar{\alpha}}(\bar{t})=\sum_{\mathrm{part}}\prod_{s\in\mathfrak{s}^{+}\cap[1,\dots,N-2]}\alpha_{s}(\bar{t}_{\textsc{i}}^{s})\times(\alpha_{s}(t_{k}^{N-1}),\dots).\label{eq:prop2-1}
\end{equation}
Combining the two properties (\ref{eq:prop1-1}) and (\ref{eq:prop2-1})
we obtain that
\begin{equation}
\mathcal{S}_{\bar{\alpha}}(\bar{t})=\sum_{\mathrm{part}}\prod_{s\in\mathfrak{s}^{+}}\alpha_{s}(\bar{t}_{\textsc{i}}^{s})\times(\dots),\label{eq:finalform1-1}
\end{equation}
which proves (\ref{eq:sumGF0}) for general $M$. 

\subsubsection{Proof of (\ref{eq:sumUTw})}

Now we derive some useful identities for the undetermined coefficients
\begin{equation}
\mathcal{W}_{\{\alpha_{s}\}_{s\in\mathfrak{s}^{-}}}(\{\bar{t}_{\textsc{i}}^{s}\}_{s\in\mathfrak{s}^{+}}|\{\bar{t}_{\textsc{ii}}^{s}\}_{s\in\mathfrak{s}^{+}}|\{\bar{t}^{s}\}_{s\in\mathfrak{s}^{-}}),
\end{equation}
of the sum formula (\ref{eq:sumGF0}). Let us renormalize the overlap
formula as
\begin{equation}
\tilde{\mathcal{S}}_{\bar{\alpha}}(\bar{t})=\mathcal{S}_{\bar{\alpha}}(\bar{t})\prod_{s=1}^{N-1}\lambda_{s+1}(\bar{t}^{s}).
\end{equation}
For the renormalized overlap the sum formula (\ref{eq:sumGF0}) reads
as
\begin{equation}
\tilde{\mathcal{S}}_{\bar{\alpha}}(\bar{t})=\sum_{\mathrm{part}(\{\bar{t}^{s}\}_{s\in\mathfrak{s}^{+}})}\mathcal{W}_{\{\alpha_{s}\}_{s\in\mathfrak{s}^{-}}}(\{\bar{t}_{\textsc{i}}^{s}\}_{s\in\mathfrak{s}^{+}}|\{\bar{t}_{\textsc{ii}}^{s}\}_{s\in\mathfrak{s}^{+}}|\{\bar{t}^{s}\}_{s\in\mathfrak{s}^{-}})\prod_{s\in\mathfrak{s}^{+}}\lambda_{s}(\bar{t}_{\textsc{i}}^{s})\lambda_{s+1}(\bar{t}_{\textsc{ii}}^{s})\prod_{s\in\mathfrak{s}^{-}}\lambda_{s+1}(\bar{t}^{s}).\label{eq:sumt}
\end{equation}
We can us the co-product formula (\ref{eq:coproduct})
\begin{equation}
\tilde{\mathcal{S}}_{\bar{\alpha}}(\bar{t})=\sum_{\mathrm{part}(\bar{t})}\frac{\prod_{\nu=1}^{N-1}\lambda_{\nu}^{(2)}(\bar{t}_{\mathrm{i}}^{\nu})\lambda_{\nu+1}^{(1)}(\bar{t}_{\mathrm{ii}}^{\nu})f(\bar{t}_{\mathrm{ii}}^{\nu},\bar{t}_{\mathrm{i}}^{\nu})}{\prod_{\nu=1}^{N-2}f(\bar{t}_{\mathrm{ii}}^{\nu+1},\bar{t}_{\mathrm{i}}^{\nu})}\tilde{\mathcal{S}}_{\bar{\alpha}^{(1)}}(\bar{t}_{\mathrm{i}})\tilde{\mathcal{S}}_{\bar{\alpha}^{(2)}}(\bar{t}_{\mathrm{ii}}).\label{eq:sp}
\end{equation}
Let us choose a fixed partition $\bar{t}^{\nu}=\bar{t}_{\mathrm{i}}^{\nu}\cup\bar{t}_{\mathrm{ii}}^{\nu}$
for every $\nu\in\mathfrak{s}^{+}$. Let us choose the vacuum eigenvalues
$\lambda_{\nu}(z)$ as
\begin{equation}
\begin{split}\lambda_{\nu}^{(2)}(t_{k}^{\nu}) & =0,\quad\text{where }t_{k}^{\nu}\in\bar{t}_{\mathrm{ii}}^{\nu},\text{ for }\nu\in\mathfrak{s}^{+},\\
\lambda_{\nu+1}^{(1)}(t_{k}^{\nu}) & =0,\quad\text{where }t_{k}^{\nu}\in\bar{t}_{\mathrm{i}}^{\nu},\text{ for }\nu\in\mathfrak{s}^{+}.
\end{split}
\end{equation}
Using these conditions in the sum formula (\ref{eq:sumt}) we obtain
that
\begin{equation}
\tilde{\mathcal{S}}_{\bar{\alpha}}(\bar{t})=\mathcal{W}_{\{\alpha_{s}\}_{s\in\mathfrak{s}^{-}}}(\{\bar{t}_{\mathrm{i}}^{s}\}_{s\in\mathfrak{s}^{+}}|\{\bar{t}_{\mathrm{ii}}^{s}\}_{s\in\mathfrak{s}^{+}}|\{\bar{t}^{s}\}_{s\in\mathfrak{s}^{-}})\prod_{s\in\mathfrak{s}^{+}}\lambda_{s}(\bar{t}_{\mathrm{i}}^{s})\lambda_{s+1}(\bar{t}_{\mathrm{ii}}^{s})\prod_{s\in\mathfrak{s}^{-}}\lambda_{s+1}(\bar{t}^{s}).
\end{equation}
Substituting to the co-product formula (\ref{eq:sp}) we obtain that
\begin{equation}
\tilde{\mathcal{S}}_{\bar{\alpha}}(\bar{t})=\sum_{\mathrm{part}(\{\bar{t}^{s}\}_{s\in\mathfrak{s}^{-}})}\frac{\prod_{\nu=1}^{N-1}f(\bar{t}_{\mathrm{ii}}^{\nu},\bar{t}_{\mathrm{i}}^{\nu})}{\prod_{\nu=1}^{N-2}f(\bar{t}_{\mathrm{ii}}^{\nu+1},\bar{t}_{\mathrm{i}}^{\nu})}\tilde{\mathcal{S}}_{\bar{\alpha}^{(1)}}(\bar{t}_{\mathrm{i}})\tilde{\mathcal{S}}_{\bar{\alpha}^{(2)}}(\bar{t}_{\mathrm{ii}})\prod_{\nu=1}^{N-1}\lambda_{\nu}^{(2)}(\bar{t}_{\mathrm{i}}^{\nu})\lambda_{\nu+1}^{(1)}(\bar{t}_{\mathrm{ii}}^{\nu}),
\end{equation}
where the sum goes through to the partitions $\bar{t}^{\nu}=\bar{t}_{\mathrm{i}}^{\nu}\cup\bar{t}_{\mathrm{ii}}^{\nu}$
for $\nu\in\mathfrak{s}^{-}$ and the partitions for $\nu\in\mathfrak{s}^{+}$
are fixed. The overlaps are 
\begin{align}
\tilde{\mathcal{S}}_{\bar{\alpha}^{(1)}}(\bar{t}_{\mathrm{i}}) & =\mathcal{W}_{\{\alpha_{s}^{(1)}\}_{s\in\mathfrak{s}^{-}}}(\{\bar{t}_{\mathrm{i}}^{s}\}_{s\in\mathfrak{s}^{+}}|\emptyset|\{\bar{t}_{\mathrm{i}}^{s}\}_{s\in\mathfrak{s}^{-}})\prod_{s\in\mathfrak{s}^{+}}\lambda_{s}^{(1)}(\bar{t}_{\mathrm{i}}^{s})\prod_{s\in\mathfrak{s}^{-}}\lambda_{s+1}^{(1)}(\bar{t}_{\mathrm{i}}^{s}),\\
\tilde{\mathcal{S}}_{\bar{\alpha}^{(2)}}(\bar{t}_{\mathrm{ii}}) & =\mathcal{W}_{\{\alpha_{s}^{(2)}\}_{s\in\mathfrak{s}^{-}}}(\emptyset|\{\bar{t}_{\mathrm{ii}}^{s}\}_{s\in\mathfrak{s}^{+}}|\{\bar{t}_{\mathrm{ii}}^{s}\}_{s\in\mathfrak{s}^{-}})\prod_{s\in\mathfrak{s}^{+}}\lambda_{s+1}^{(2)}(\bar{t}_{\mathrm{ii}}^{s})\prod_{s\in\mathfrak{s}^{-}}\lambda_{s+1}^{(2)}(\bar{t}_{\mathrm{ii}}^{s}).
\end{align}
Substituting back we obtain that
\begin{multline}
\mathcal{W}_{\{\alpha_{s}\}_{s\in\mathfrak{s}^{-}}}(\{\bar{t}_{\mathrm{i}}^{s}\}_{s\in\mathfrak{s}^{+}}|\{\bar{t}_{\mathrm{ii}}^{s}\}_{s\in\mathfrak{s}^{+}}|\{\bar{t}^{s}\}_{s\in\mathfrak{s}^{-}})=\sum_{\mathrm{part}(\{\bar{t}^{s}\}_{s\in\mathfrak{s}^{-}})}\frac{\prod_{\nu=1}^{N-1}f(\bar{t}_{\mathrm{ii}}^{\nu},\bar{t}_{\mathrm{i}}^{\nu})}{\prod_{\nu=1}^{N-2}f(\bar{t}_{\mathrm{ii}}^{\nu+1},\bar{t}_{\mathrm{i}}^{\nu})}\times\\
\times\mathcal{W}_{\{\alpha_{s}^{(1)}\}_{s\in\mathfrak{s}^{-}}}(\{\bar{t}_{\mathrm{i}}^{s}\}_{s\in\mathfrak{s}^{+}}|\emptyset|\{\bar{t}_{\mathrm{i}}^{s}\}_{s\in\mathfrak{s}^{-}})\mathcal{W}_{\{\alpha_{s}^{(2)}\}_{s\in\mathfrak{s}^{-}}}(\emptyset|\{\bar{t}_{\mathrm{ii}}^{s}\}_{s\in\mathfrak{s}^{+}}|\{\bar{t}_{\mathrm{ii}}^{s}\}_{s\in\mathfrak{s}^{-}})\prod_{s\in\mathfrak{s}^{-}}\alpha_{s}^{(2)}(\bar{t}_{\mathrm{i}}^{s}).
\end{multline}
Let us choose the quantum space $\mathcal{H}^{(1)}$ as $\alpha_{s}^{(1)}(z)=1$
for $s\in\mathfrak{s}^{-}$. The formula above simplifies as
\begin{multline}
\mathcal{W}_{\{\alpha_{s}\}_{s\in\mathfrak{s}^{-}}}(\{\bar{t}_{\mathrm{i}}^{s}\}_{s\in\mathfrak{s}^{+}}|\{\bar{t}_{\mathrm{ii}}^{s}\}_{s\in\mathfrak{s}^{+}}|\{\bar{t}^{s}\}_{s\in\mathfrak{s}^{-}})=\\
\sum_{\mathrm{part}(\{\bar{t}^{s}\}_{s\in\mathfrak{s}^{-}})}\frac{\prod_{\nu=1}^{N-1}f(\bar{t}_{\mathrm{ii}}^{\nu},\bar{t}_{\mathrm{i}}^{\nu})}{\prod_{\nu=1}^{N-2}f(\bar{t}_{\mathrm{ii}}^{\nu+1},\bar{t}_{\mathrm{i}}^{\nu})}\mathcal{Z}(\bar{t}_{\mathrm{i}})\mathcal{W}_{\{\alpha_{s}\}_{s\in\mathfrak{s}^{-}}}(\emptyset|\{\bar{t}_{\mathrm{ii}}^{s}\}_{s\in\mathfrak{s}^{+}}|\{\bar{t}_{\mathrm{ii}}^{s}\}_{s\in\mathfrak{s}^{-}})\prod_{s\in\mathfrak{s}^{-}}\alpha_{s}(\bar{t}_{\mathrm{i}}^{s}),\label{eq:WW1}
\end{multline}
where we introduced the HC as 
\begin{equation}
\mathcal{Z}(\bar{t}):=\mathcal{W}_{\{\alpha_{s}\}_{s\in\mathfrak{s}^{-}}}(\{\bar{t}^{s}\}_{s\in\mathfrak{s}^{+}}|\emptyset|\{\bar{t}^{s}\}_{s\in\mathfrak{s}^{-}})\Biggr|_{\alpha_{s}(z)=1}.
\end{equation}
We can also choose the other quantum space $\mathcal{H}^{(2)}$ as
$\alpha_{s}^{(2)}(z)=1$ for which
\begin{multline}
\mathcal{W}_{\{\alpha_{s}\}_{s\in\mathfrak{s}^{-}}}(\{\bar{t}_{\mathrm{i}}^{s}\}_{s\in\mathfrak{s}^{+}}|\{\bar{t}_{\mathrm{ii}}^{s}\}_{s\in\mathfrak{s}^{+}}|\{\bar{t}^{s}\}_{s\in\mathfrak{s}^{-}})=\\
\sum_{\mathrm{part}(\{\bar{t}^{s}\}_{s\in\mathfrak{s}^{-}})}\frac{\prod_{\nu=1}^{N-1}f(\bar{t}_{\mathrm{ii}}^{\nu},\bar{t}_{\mathrm{i}}^{\nu})}{\prod_{\nu=1}^{N-2}f(\bar{t}_{\mathrm{ii}}^{\nu+1},\bar{t}_{\mathrm{i}}^{\nu})}\mathcal{W}_{\{\alpha_{s}\}_{s\in\mathfrak{s}^{-}}}(\{\bar{t}_{\mathrm{i}}^{s}\}_{s\in\mathfrak{s}^{+}}|\emptyset|\{\bar{t}_{\mathrm{i}}^{s}\}_{s\in\mathfrak{s}^{-}})\bar{\mathcal{Z}}(\bar{t}_{\mathrm{ii}}^{s}),\label{eq:secf}
\end{multline}
where we introduced the other HC as 
\begin{equation}
\bar{\mathcal{Z}}(\bar{t}):=\mathcal{W}_{\{\alpha_{s}\}_{s\in\mathfrak{s}^{-}}}(\emptyset|\{\bar{t}^{s}\}_{s\in\mathfrak{s}^{+}}|\{\bar{t}^{s}\}_{s\in\mathfrak{s}^{-}})\Biggr|_{\alpha_{s}(z)=1}.
\end{equation}
Substituting $\bar{t}_{\mathrm{i}}^{s}=\emptyset$ for $s\in\mathfrak{s}^{+}$
to the second formula (\ref{eq:secf}) we obtain that
\begin{equation}
\mathcal{W}_{\{\alpha_{s}\}_{s\in\mathfrak{s}^{-}}}(\emptyset|\{\bar{t}_{\mathrm{ii}}^{s}\}_{s\in\mathfrak{s}^{+}}|\{\bar{t}^{s}\}_{s\in\mathfrak{s}^{-}})=\sum_{\mathrm{part}(\{\bar{t}^{s}\}_{s\in\mathfrak{s}^{-}})}\frac{\prod_{s=M+1}^{N-M-1}f(\bar{t}_{\mathrm{ii}}^{s},\bar{t}_{\mathrm{i}}^{s})}{\prod_{s=M+1}^{N-M-1}f(\bar{t}_{\mathrm{ii}}^{s+1},\bar{t}_{\mathrm{i}}^{s})}S_{\{\alpha_{s}\}_{s\in\mathfrak{s}^{-}}}^{0}(\{\bar{t}_{\mathrm{i}}^{s}\}_{s\in\mathfrak{s}^{-}})\bar{\mathcal{Z}}(\bar{t}_{\mathrm{ii}}^{s}),
\end{equation}
where we used that 
\begin{equation}
\mathcal{W}_{\{\alpha_{s}\}_{s\in\mathfrak{s}^{-}}}(\emptyset||\emptyset|\{\bar{t}^{s}\}_{s\in\mathfrak{s}^{-}})=\langle\Psi|\mathbb{B}(\emptyset,\dots,\emptyset,\bar{t}^{M+1},\dots,\bar{t}^{N-M-1},\emptyset,\dots,\emptyset)=S_{\{\alpha_{s}\}_{s\in\mathfrak{s}^{-}}}^{0}(\{\bar{t}_{\mathrm{i}}^{s}\}_{s\in\mathfrak{s}^{-}}).
\end{equation}
We can substitute to (\ref{eq:WW1}):
\begin{multline}
\mathcal{W}_{\{\alpha_{s}\}_{s\in\mathfrak{s}^{-}}}(\{\bar{t}_{\mathrm{i}}^{s}\}_{s\in\mathfrak{s}^{+}}|\{\bar{t}_{\mathrm{ii}}^{s}\}_{s\in\mathfrak{s}^{+}}|\{\bar{t}^{s}\}_{s\in\mathfrak{s}^{-}})=\sum_{\mathrm{part}(\{\bar{t}^{s}\}_{s\in\mathfrak{s}^{-}})}\frac{\prod_{\nu=1}^{N-1}f(\bar{t}_{\mathrm{ii}}^{\nu},\bar{t}_{\mathrm{i}}^{\nu})}{\prod_{\nu=1}^{N-2}f(\bar{t}_{\mathrm{ii}}^{\nu+1},\bar{t}_{\mathrm{i}}^{\nu})}\times\\
\frac{\prod_{s=M+1}^{N-M-1}f(\bar{t}_{\mathrm{iii}}^{s},\bar{t}_{\mathrm{i}}^{s})f(\bar{t}_{\mathrm{ii}}^{s},\bar{t}_{\mathrm{iii}}^{s})}{\prod_{s=M+1}^{N-M-1}f(\bar{t}_{\mathrm{iii}}^{s},\bar{t}_{\mathrm{i}}^{s-1})f(\bar{t}_{\mathrm{ii}}^{s+1},\bar{t}_{\mathrm{iii}}^{s})}\mathcal{Z}(\bar{t}_{\mathrm{i}})\bar{\mathcal{Z}}(\bar{t}_{\mathrm{ii}}^{s})S_{\{\alpha_{s}\}_{s\in\mathfrak{s}^{-}}}^{0}(\{\bar{t}_{\mathrm{iii}}^{s}\}_{s\in\mathfrak{s}^{-}})\prod_{s\in\mathfrak{s}^{-}}\alpha_{s}(\bar{t}_{\mathrm{i}}^{s}).
\end{multline}
Substituting back to the original sum formula (\ref{eq:sumGF0}),
we obtain that
\begin{align}
\mathcal{S}_{\bar{\alpha}}(\bar{t}) & =\sum_{\mathrm{part}(\bar{t})}\frac{\prod_{\nu=1}^{N-1}f(\bar{t}_{\mathrm{ii}}^{\nu},\bar{t}_{\mathrm{i}}^{\nu})}{\prod_{\nu=1}^{N-2}f(\bar{t}_{\mathrm{ii}}^{\nu+1},\bar{t}_{\mathrm{i}}^{\nu})}\frac{\prod_{s=M+1}^{N-M-1}f(\bar{t}_{\mathrm{iii}}^{s},\bar{t}_{\mathrm{i}}^{s})f(\bar{t}_{\mathrm{ii}}^{s},\bar{t}_{\mathrm{iii}}^{s})}{\prod_{s=M+1}^{N-M-1}f(\bar{t}_{\mathrm{iii}}^{s},\bar{t}_{\mathrm{i}}^{s-1})f(\bar{t}_{\mathrm{ii}}^{s+1},\bar{t}_{\mathrm{iii}}^{s})}\times\nonumber \\
 & \times\mathcal{Z}(\bar{t}_{\mathrm{i}})\bar{\mathcal{Z}}(\bar{t}_{\mathrm{ii}}^{s})S_{\{\alpha_{s}\}_{s\in\mathfrak{s}^{-}}}^{0}(\{\bar{t}_{\mathrm{iii}}^{s}\}_{s\in\mathfrak{s}^{-}})\prod_{s=1}^{N-1}\alpha_{s}(\bar{t}_{\mathrm{i}}^{s}),
\end{align}
which is what we wanted to prove (\ref{eq:sumUTw}).

\subsection{Twisted case\label{subsec:Sum-formula-Twisted-case}}

For the twisted case let us fix the $K$-matrix to the regular form
\[
K=\sum_{a=1}^{N/2}x_{a}E_{N+1-2a,2a-1}-x_{a}E_{N+2-2a,2a}.
\]
In section \ref{subsec:Reqursion-for-the-TW} we already saw that
the variables $\alpha_{2a}(t_{k}^{2a})$ can be handled as independent
variables for $a=1,\dots,\frac{N}{2}-1$. Using a similar derivation
of \cite{Gombor:2021hmj} (appendix (C.1)) we can show that the overlap
has the sum formula
\begin{equation}
\mathcal{S}_{\bar{\alpha}}(\bar{t})=\sum_{\mathrm{part}(\{\bar{t}^{2a}\}_{a=1}^{N/2-1})}\mathcal{W}_{\{\alpha_{2a-1}\}_{a=1}^{N/2}}(\{\bar{t}_{\textsc{i}}^{2a}\}_{a=1}^{N/2-1}|\{\bar{t}_{\textsc{ii}}^{2a}\}_{a=1}^{N/2-1}|\{\bar{t}^{2a-1}\}_{a=1}^{N/2})\prod_{a=1}^{N/2-1}\alpha_{2a}(\bar{t}_{\textsc{i}}^{2a}).\label{eq:sumTWW}
\end{equation}
For completeness, we present this proof.

\subsubsection{Proof of (\ref{eq:sumTWW})}

We start with the proof of the sum formula (\ref{eq:sumTWW}) where
the weights $\mathcal{W}_{\{\alpha_{2a-1}\}_{a=1}^{N/2}}$ do not
depend on $\alpha_{2a}$ for $a=1,\dots,\frac{N}{2}-1$. In the derivation
we only care about $\alpha_{2a}$ dependence of the overlap therefore
we use the notation $(\dots)$ for the $\alpha_{2a}$ independent
coefficients. Using this notation the sum formula takes the form
\begin{equation}
\mathcal{S}_{\bar{\alpha}}(\bar{t})=\sum_{\mathrm{part}(\bar{t})}\prod_{a=1}^{N/2-1}\alpha_{2a}(\bar{t}_{\textsc{i}}^{2a})(\dots).\label{eq:sumrule_Tw}
\end{equation}
We prove this sum formula using induction on $N$. Let us start with
$N=4$. Using the recurrence formula (\ref{eq:rect2})
\begin{align}
\mathbb{B}(\bar{t}^{1},\{z,\bar{t}^{2}\},\bar{t}^{3}) & =\sum_{j=3}^{4}\sum_{\mathrm{part}(\bar{t})}\frac{T_{2,j}(z)}{\lambda_{3}(z)}\mathbb{B}(\bar{t}^{1},\bar{t}^{2},\{\bar{t}_{\textsc{ii}}^{s}\}_{s=3}^{j-1},\{\bar{t}^{s}\}_{s=j}^{3})(\dots)\nonumber \\
 & +\sum_{j=3}^{4}\sum_{\mathrm{part}(\bar{t})}\frac{T_{1,j}(z)}{\lambda_{3}(z)}\mathbb{B}(\bar{t}_{\textsc{ii}}^{1},\bar{t}^{2},\{\bar{t}_{\textsc{ii}}^{s}\}_{s=3}^{j-1},\{\bar{t}^{s}\}_{s=j}^{3})(\dots),
\end{align}
and the $KT$-relation
\begin{equation}
\begin{split}T_{2,3}(z) & =-\frac{x_{2}}{x_{1}}\lambda_{0}(z)\widehat{T}_{4,1}(-z),\qquad T_{2,4}(z)=\frac{x_{a}}{x_{1}}\lambda_{0}(z)\widehat{T}_{4,2}(-z),\\
T_{1,3}(z) & =\frac{x_{2}}{x_{1}}\lambda_{0}(z)\widehat{T}_{3,1}(-z),\qquad T_{1,4}(z)=-\frac{x_{a}}{x_{1}}\lambda_{0}(z)\widehat{T}_{3,2}(-z),
\end{split}
\end{equation}
and the action formula (\ref{eq:actTw})
\begin{equation}
\widehat{T}_{i,j}(-z)\mathbb{B}(\bar{t})=(-1)^{i-j}\alpha_{2}(z)\sum_{\mathrm{part}(\bar{w})}\mathbb{B}(\bar{\omega}_{\textsc{ii}})\alpha_{2}(\bar{\omega}_{\textsc{iii}}^{2})(\dots),
\end{equation}
for $j\leq2$, we obtain the following recursion for the overlap
\begin{equation}
\langle\Psi|\mathbb{B}(\bar{t}^{1},\{z,\bar{t}^{2}\},\bar{t}^{3})=\sum_{\mathrm{part}}\langle\Psi|\mathbb{B}(\bar{\omega}_{\textsc{ii}})\alpha_{2}(z)\alpha_{2}(\bar{\omega}_{\textsc{iii}}^{2})(\dots),\label{eq:sumTWtemp}
\end{equation}
where $\bar{\omega}_{\textsc{ii}}^{\nu},\bar{\omega}_{\textsc{iii}}^{\nu}\subset\{-z-\nu c,\bar{t}^{\nu}\}$
where $\bar{\omega}_{\textsc{ii}}^{2}=\#\bar{t}^{2}-1$ and $\bar{\omega}_{\textsc{ii}}^{\nu}\cap\bar{\omega}_{\textsc{iii}}^{\nu}=\emptyset$.
We can prove (\ref{eq:sumTWW}) for $N=4$ by induction on $r_{2}=\#\bar{t}^{2}$.
Let us start with $\#\bar{t}^{2}=2$ for which
\begin{equation}
\langle\Psi|\mathbb{B}(\bar{t}^{1},\{z,t^{2}\},\bar{t}^{3})=\sum_{\mathrm{part}}\langle\Psi|\mathbb{B}(\bar{\omega}_{\textsc{ii}}^{1},\emptyset,\bar{\omega}_{\textsc{ii}}^{3})\alpha_{2}(z)\alpha_{2}(\bar{\omega}_{\textsc{iii}}^{2})(\dots).
\end{equation}
Since the overlap with $r_{2}=0$ factorize as
\begin{equation}
\langle\Psi|\mathbb{B}(\bar{\omega}_{\textsc{ii}}^{1},\emptyset,\bar{\omega}_{\textsc{ii}}^{3})=S_{\alpha_{1}}^{(1)}(\bar{\omega}_{\textsc{ii}}^{1})S_{\alpha_{3}}^{(3)}(\bar{\omega}_{\textsc{ii}}^{3}),
\end{equation}
which does not depend on $\alpha_{2}$ we obtain that
\begin{equation}
\langle\Psi|\mathbb{B}(\bar{t}^{1},\{z,t^{2}\},\bar{t}^{3})=\sum_{\mathrm{part}}\alpha_{2}(z)\alpha_{2}(\bar{\omega}_{\textsc{iii}}^{2})(\dots).
\end{equation}
We have two possibilities: if $\bar{\omega}_{\textsc{iii}}^{2}=\{-z-2c\}$
then the r.h.s is $\alpha_{2}(z)\alpha_{2}(-z-2c)=1$ or if $\bar{\omega}_{\textsc{iii}}^{2}=\{t^{2}\}$
then the r.h.s is $\alpha_{2}(z)\alpha_{2}(t^{2})$ therefore
\begin{equation}
\langle\Psi|\mathbb{B}(\bar{t}^{1},\{z,t^{2}\},\bar{t}^{3})=\sum_{\mathrm{part}}\alpha_{2}(z)\alpha_{2}(t^{2})(\dots)+(\dots),
\end{equation}
i.e., we just proved that
\begin{equation}
\langle\Psi|\mathbb{B}(\bar{t}^{1},\bar{t}^{2},\bar{t}^{3})=\sum_{\mathrm{part}}\alpha_{2}(\bar{t}_{\textsc{i}}^{2})(\dots),\label{eq:Sum_TW_4}
\end{equation}
for $r_{2}=2$. Now let assume that (\ref{eq:Sum_TW_4}) is true for
$\#\bar{t}^{2}=r_{2}-2$. Let us apply the relation (\ref{eq:sumTWtemp})
for $\#\bar{t}^{2}=r_{2}-1$. In this case the number of the second
Bethe roots is $r_{2}-2$ therefore we can use the induction hypothesis
(\ref{eq:Sum_TW_4}):
\begin{equation}
\langle\Psi|\mathbb{B}(\bar{t}^{1},\{z,\bar{t}^{2}\},\bar{t}^{3})=\sum_{\mathrm{part}}\alpha_{2}(\bar{\omega}_{\textsc{i}}^{2})\alpha_{2}(z)\alpha_{2}(\bar{\omega}_{\textsc{iii}}^{2})(\dots),
\end{equation}
where $\bar{\omega}_{\textsc{i}}^{\nu}\cap\bar{\omega}_{\textsc{iii}}^{\nu}=\emptyset$.
We have three possibilities
\begin{equation}
\langle\Psi|\mathbb{B}(\bar{t}^{1},\{z,\bar{t}^{2}\},\bar{t}^{3})=\begin{cases}
\sum_{\mathrm{part}}\alpha_{2}(\bar{t}_{\textsc{i}}^{2})(\dots), & -z-2c\in\bar{\omega}_{\textsc{i}}^{2},\\
\sum_{\mathrm{part}}\alpha_{2}(\bar{t}_{\textsc{i}}^{2})(\dots), & \{-z-2c\}=\bar{\omega}_{\textsc{iii}}^{2},\\
\sum_{\mathrm{part}}\alpha_{2}(z)\alpha_{2}(\bar{t}_{\textsc{i}}^{2})(\dots), & \{-z-2c\}\notin\bar{\omega}_{\textsc{i}}^{2},\bar{\omega}_{\textsc{iii}}^{2},
\end{cases}
\end{equation}
therefore we obtain that
\begin{equation}
\langle\Psi|\mathbb{B}(\bar{t}^{1},\{z,\bar{t}^{2}\},\bar{t}^{3})=\sum_{\mathrm{part}}\alpha_{2}(z)\alpha_{2}(\bar{t}_{\textsc{i}}^{2})(\dots)+\sum_{\mathrm{part}}\alpha_{2}(\bar{t}_{\textsc{i}}^{2})(\dots),
\end{equation}
which proves (\ref{eq:Sum_TW_4}) for any even $r_{2}$, i.e we proved
(\ref{eq:sumTWW}) for $N=4$.

Let us continue with general $N$. Let us assume that the $\mathfrak{gl}(N-2)$
overlaps have the sum formula (\ref{eq:sumTWW}). Let us turn to the
type $\mathfrak{gl}(N)$ overlaps. We can use the recurrence formula
(\ref{eq:rect2})
\begin{align}
\mathbb{B}(\bar{t}^{1},\{z,\bar{t}^{2}\},\{\bar{t}^{s}\}_{s=3}^{N-1}) & =\sum_{a=2}^{N/2}\sum_{\mathrm{part}(\bar{t})}\frac{T_{2,2a-1}(z)}{\lambda_{3}(z)}\mathbb{B}(\bar{t}^{1},\bar{t}^{2},\{\bar{t}_{\textsc{ii}}^{s}\}_{s=3}^{2a-2},\{\bar{t}^{s}\}_{s=2a-1}^{N-1})\prod_{b=2}^{a-1}\alpha_{2b}(\bar{t}_{\textsc{i}}^{2b})(\dots)\nonumber \\
 & +\sum_{a=2}^{N/2}\sum_{\mathrm{part}(\bar{t})}\frac{T_{2,2a}(z)}{\lambda_{3}(z)}\mathbb{B}(\bar{t}^{1},\bar{t}^{2},\{\bar{t}_{\textsc{ii}}^{s}\}_{s=3}^{2a-1},\{\bar{t}^{s}\}_{s=2a}^{N-1})\prod_{b=2}^{a-1}\alpha_{2b}(\bar{t}_{\textsc{i}}^{2b})(\dots)\nonumber \\
 & +\sum_{a=2}^{N/2}\sum_{\mathrm{part}(\bar{t})}\frac{T_{1,2a-1}(z)}{\lambda_{3}(z)}\mathbb{B}(\bar{t}_{\textsc{ii}}^{1},\bar{t}^{2},\{\bar{t}_{\textsc{ii}}^{s}\}_{s=3}^{2a-2},\{\bar{t}^{s}\}_{s=2a-1}^{N-1})\prod_{b=2}^{a-1}\alpha_{2b}(\bar{t}_{\textsc{i}}^{2b})(\dots)\label{eq:rect2-1}\\
 & +\sum_{a=2}^{N/2}\sum_{\mathrm{part}(\bar{t})}\frac{T_{1,2a}(z)}{\lambda_{3}(z)}\mathbb{B}(\bar{t}_{\textsc{ii}}^{1},\bar{t}^{2},\{\bar{t}_{\textsc{ii}}^{s}\}_{s=3}^{2a-1},\{\bar{t}^{s}\}_{s=2a}^{N-1})\prod_{b=2}^{a-1}\alpha_{2b}(\bar{t}_{\textsc{i}}^{2b})(\dots),\nonumber 
\end{align}
and the $KT$-relation
\begin{equation}
\begin{split}T_{2,2a-1}(z) & =-\frac{x_{a}}{x_{1}}\lambda_{0}(z)\widehat{T}_{N,N+1-2a}(-z),\qquad T_{2,2a}(z)=\frac{x_{a}}{x_{1}}\lambda_{0}(z)\widehat{T}_{N,N+2-2a}(-z),\\
T_{1,2a-1}(z) & =\frac{x_{a}}{x_{1}}\lambda_{0}(z)\widehat{T}_{N-1,N+1-2a}(-z),\qquad T_{1,2a}(z)=-\frac{x_{a}}{x_{1}}\lambda_{0}(z)\widehat{T}_{N-1,N+2-2a}(-z),
\end{split}
\end{equation}
and the action formula (\ref{eq:actTw})
\begin{equation}
\widehat{T}_{i,N-j}(-z)\mathbb{B}(\bar{t})=\alpha_{2}(z)\sum_{\mathrm{part}(\bar{w})}\mathbb{B}(\bar{\omega}_{\textsc{ii}})\prod_{b=1}^{j/2}\alpha_{2b}(\bar{\omega}_{\textsc{iii}}^{2b})(\dots).
\end{equation}
We can see that in these formulas the $\alpha$-dependent terms are
$\alpha_{2b}(t_{k}^{2b})$ and $\alpha_{2b}(z)$ therefore combining
the equations above, we obtain the following recursion for the overlap
\begin{equation}
\langle\Psi|\mathbb{B}(\bar{t}^{1},\{z,\bar{t}^{2}\},\{\bar{t}^{s}\}_{s=3}^{N-1})=\sum_{\mathrm{part}}\langle\Psi|\mathbb{B}(\bar{\omega}_{\textsc{ii}})\prod_{b=1}^{\frac{N}{2}-1}\alpha_{2b}(\bar{t}_{\textsc{i}}^{2b})(\alpha_{2b}(z),\dots),\label{eq:sumTWtemp-1}
\end{equation}
where $\bar{\omega}_{\textsc{ii}}^{\nu}\subset\{-z-\nu c,\bar{t}_{\textsc{ii}}^{\nu}\}$
and $\bar{t}=\bar{t}_{\textsc{i}}\cup\bar{t}_{\textsc{ii}}$. Applying
this recursion for the overlap we can eliminate the Bethe roots $\bar{t}^{2}$
as
\begin{equation}
\langle\Psi|\mathbb{B}(\bar{t})=\sum_{\mathrm{part}}\langle\Psi|\mathbb{B}(\bar{\omega}_{\textsc{ii}}^{1},\emptyset,\left\{ \bar{\omega}_{\textsc{ii}}^{k}\right\} _{k=3}^{N-1})\prod_{b=2}^{\frac{N}{2}-1}\alpha_{2b}(\bar{t}_{\textsc{i}}^{2b})\times(\alpha_{2b}(t_{k}^{2}),\dots),
\end{equation}
where $\bar{\omega}_{\textsc{ii}}^{\nu}\subset\{-\bar{t}^{1}-\nu c,\bar{t}_{\textsc{ii}}^{\nu}\}$.
The overlaps in the rhs factorize to $\mathfrak{gl}(2)$ and $\mathfrak{gl}(N-2)$
overlaps as
\begin{equation}
\langle\Psi|\mathbb{B}(\bar{\omega}_{\textsc{ii}}^{1},\emptyset,\left\{ \bar{\omega}_{\textsc{ii}}^{k}\right\} _{k=3}^{N-1})=S_{\alpha_{1}}^{(1)}(\bar{\omega}_{\textsc{ii}}^{1})\langle\Psi|\mathbb{B}(\emptyset,\emptyset,\left\{ \bar{\omega}_{\textsc{ii}}^{k}\right\} _{k=3}^{N-1}).
\end{equation}
The $\mathfrak{gl}(2)$ overlap does not depend on $\alpha_{2b}$
and $\mathfrak{gl}(N-2)$ overlap satisfies the induction hypothesis
therefore we obtain that
\begin{equation}
\mathcal{S}_{\bar{\alpha}}(\bar{t})=\prod_{b=2}^{\frac{N}{2}-1}\alpha_{2b}(\bar{t}_{\textsc{i}}^{2b})\times(\alpha_{2b}(t_{k}^{2}),\dots).\label{eq:prop1-1-1}
\end{equation}

We can repeat the previous calculation using the alternative recurrence
formula (\ref{eq:rect2-2})
\begin{align}
\mathbb{B}(\{\bar{t}^{s}\}_{s=1}^{N-3},\{z,\bar{t}^{N-2}\},\bar{t}^{N-2}) & =\sum_{a=2}^{N/2}\sum_{\mathrm{part}(\bar{t})}\frac{T_{2a-1,N-1}(z)}{\lambda_{N-1}(z)}\mathbb{B}(\{\bar{t}^{s}\}_{s=1}^{2a-2},\{\bar{t}_{\textsc{ii}}^{s}\}_{s=2a-1}^{N-3},\bar{t}^{N-2},\bar{t}^{N-1})(\dots)\nonumber \\
 & +\sum_{a=2}^{N/2}\sum_{\mathrm{part}(\bar{t})}\frac{T_{2a,N-1}(z)}{\lambda_{N-1}(z)}\mathbb{B}(\{\bar{t}^{s}\}_{s=1}^{2a-1},\{\bar{t}_{\textsc{ii}}^{s}\}_{s=2a}^{N-3},\bar{t}^{N-2},\bar{t}^{N-1})(\dots)\nonumber \\
 & +\sum_{a=2}^{N/2}\sum_{\mathrm{part}(\bar{t})}\frac{T_{2a-1,2N}(z)}{\lambda_{N-1}(z)}\mathbb{B}(\{\bar{t}^{s}\}_{s=1}^{2a-2},\{\bar{t}_{\textsc{ii}}^{s}\}_{s=2a-1}^{N-3},\bar{t}^{N-2},\bar{t}_{\textsc{ii}}^{N-1})(\dots)\label{eq:rect2-1-1}\\
 & +\sum_{a=2}^{N/2}\sum_{\mathrm{part}(\bar{t})}\frac{T_{2a,2N}(z)}{\lambda_{N-1}(z)}\mathbb{B}(\{\bar{t}^{s}\}_{s=1}^{2a-1},\{\bar{t}_{\textsc{ii}}^{s}\}_{s=2a}^{N-3},\bar{t}^{N-2},\bar{t}_{\textsc{ii}}^{N-1})(\dots),\nonumber 
\end{align}
and the $KT$-relation
\begin{equation}
\begin{split}T_{2a-1,N-1}(z) & =\frac{x_{\frac{N}{2}}}{x_{a}}\lambda_{0}(z)\widehat{T}_{N-2a+1,1}(-z),\qquad T_{2a,N-1}(z)=-\frac{x_{\frac{N}{2}}}{x_{a}}\lambda_{0}(z)\widehat{T}_{N-2a+2,1}(-z),\\
T_{2a-1,N}(z) & =-\frac{x_{\frac{N}{2}}}{x_{a}}\lambda_{0}(z)\widehat{T}_{N-2a+1,2}(-z),\qquad T_{2a,N}(z)=\frac{x_{\frac{N}{2}}}{x_{a}}\lambda_{0}(z)\widehat{T}_{N-2a+2,2}(-z),
\end{split}
\end{equation}
and the action formula (\ref{eq:actTw}). We can see that in these
formulas the $\alpha$-dependent terms are $\alpha_{2b}(t_{k}^{2b})$
and $\alpha_{2b}(z)$ therefore combining them, we obtain the following
recursion for the overlap
\begin{equation}
\langle\Psi|(\{\bar{t}^{s}\}_{s=1}^{N-3},\{z,\bar{t}^{N-2}\},\bar{t}^{N-2})=\sum_{\mathrm{part}}\langle\Psi|\mathbb{B}(\bar{\omega}_{\textsc{ii}})\prod_{b=1}^{\frac{N}{2}-1}\alpha_{2b}(\bar{t}_{\textsc{i}}^{2b})(\alpha_{2b}(z),\dots),\label{eq:sumTWtemp-1-1}
\end{equation}
where $\bar{\omega}_{\textsc{ii}}^{\nu}\subset\{-z-\nu c,\bar{t}_{\textsc{ii}}^{\nu}\}$
and $\bar{t}=\bar{t}_{\textsc{i}}\cup\bar{t}_{\textsc{ii}}$. Applying
this recursion for the overlap we can eliminate the Bethe roots $\bar{t}^{N-2}$
as
\begin{equation}
\langle\Psi|\mathbb{B}(\bar{t})=\sum_{\mathrm{part}}\langle\Psi|\mathbb{B}(\left\{ \bar{\omega}_{\textsc{ii}}^{k}\right\} _{k=1}^{N-3},\emptyset,\bar{\omega}_{\textsc{ii}}^{N-1})\prod_{b=1}^{\frac{N}{2}-2}\alpha_{2b}(\bar{t}_{\textsc{i}}^{2b})\times(\alpha_{2b}(t_{k}^{N-2}),\dots),
\end{equation}
where $\bar{\omega}_{\textsc{ii}}^{\nu}\subset\{-\bar{t}^{N-1}-\nu c,\bar{t}_{\textsc{ii}}^{\nu}\}$.
The overlaps in the rhs factorize to $\mathfrak{gl}(2)$ and $\mathfrak{gl}(N-2)$
overlaps as
\begin{equation}
\langle\Psi|\mathbb{B}(\left\{ \bar{\omega}_{\textsc{ii}}^{k}\right\} _{k=1}^{N-3},\emptyset,\bar{\omega}_{\textsc{ii}}^{N-1})=\langle\Psi|\mathbb{B}(\left\{ \bar{\omega}_{\textsc{ii}}^{k}\right\} _{k=1}^{N-3},\emptyset,\emptyset)S_{\alpha_{1}}^{(1)}(\bar{\omega}_{\textsc{ii}}^{1}).
\end{equation}
The $\mathfrak{gl}(2)$ overlap does not depend on $\alpha_{2b}$
and $\mathfrak{gl}(N-2)$ overlap satisfies the induction hypothesis
therefore we obtain that
\begin{equation}
\mathcal{S}_{\bar{\alpha}}(\bar{t})=\sum_{\mathrm{part}}\prod_{b=1}^{\frac{N}{2}-2}\alpha_{2b}(\bar{t}_{\textsc{i}}^{2b})\times(\alpha_{2b}(t_{k}^{N-2}),\dots).\label{eq:prop1-1-1-1}
\end{equation}
Combining the two properties (\ref{eq:prop1-1-1}) and (\ref{eq:prop1-1-1-1})
we obtain that
\begin{equation}
\mathcal{S}_{\bar{\alpha}}(\bar{t})=\sum_{\mathrm{part}}\prod_{b=1}^{\frac{N}{2}-1}\alpha_{2b}(\bar{t}_{\textsc{i}}^{2b})\times(\dots),\label{eq:finalform1-1-1}
\end{equation}
which proves (\ref{eq:sumTWW}) for general $N$. 

\subsubsection{Proof of (\ref{eq:sumTW})}

Now we derive some useful identities for the undetermined coefficients
\begin{equation}
\mathcal{W}_{\{\alpha_{2a-1}\}_{a=1}^{N/2}}(\{\bar{t}_{\textsc{i}}^{2a}\}_{a=1}^{N/2-1}|\{\bar{t}_{\textsc{ii}}^{2a}\}_{a=1}^{N/2-1}|\{\bar{t}^{2a-1}\}_{a=1}^{N/2}),
\end{equation}
of the sum formula (\ref{eq:sumTWW}). 

Let us start with the overlap where $r_{2b}=0$ for $b=1,\dots,\frac{N}{2}-1$.
Since the corresponding Bethe states are generated by commuting $Y(2)$
subalgebras, these overlaps are factorized as
\begin{equation}
\langle\Psi|\mathbb{B}(\bar{t}^{1},\emptyset,\bar{t}^{3},\emptyset,\dots,\emptyset,\bar{t}^{N-1})=\prod_{b=1}^{N/2}S_{\alpha_{2b-1}}^{(2b-1)}(\bar{t}^{2b-1}),
\end{equation}
where
\begin{equation}
S_{\alpha_{2b-1}}^{(2b-1)}(\bar{t}^{2b-1}):=\langle\Psi|\mathbb{B}(\emptyset^{\times2b-2},\bar{t}^{2b-1},\emptyset^{\times N-2b}).
\end{equation}
For $N=2$ the untwisted and twisted $KT$-relations are equivalent
but there are different conventions for the monodromy matrices (there
is a shift in the spectral parameter) and the $K$-matrices. In the
untwisted convention the K-matrix is the type $(2,0)$ for which the
overlap is given by (\ref{eq:offov2}),(\ref{eq:HC2}):
\begin{equation}
S_{\alpha}(\bar{t})=\sum f(\bar{t}_{\textsc{ii}}^{\nu},\bar{t}_{\textsc{i}}^{\nu})Z^{0}(\bar{t}_{\textsc{i}})Z^{0}(-\bar{t}_{\textsc{ii}})\alpha(\bar{t}_{\textsc{i}}),
\end{equation}
where
\begin{equation}
Z^{0}(\bar{t})=\kappa(\bar{t})\prod_{k<l}f(-t_{k},t_{l}),\qquad\kappa(z)=\frac{1}{z}.
\end{equation}
Introducing the proper shifts (\ref{eq:gl2Equiv}), the overlaps in
the twisted convention reads as
\begin{equation}
S_{\alpha}^{(s)}(\bar{t})=S_{\alpha}(\bar{t}+c\frac{s}{2})\Biggr|_{\alpha_{2a-1}(z)\to\alpha_{2a-1}(z-c\frac{s}{2})},
\end{equation}
therefore
\begin{equation}
\mathcal{S}_{\bar{\alpha}}(\bar{t}^{1},\emptyset,\bar{t}^{3},\emptyset\dots,\emptyset,\bar{t}^{N-1})=\prod_{a=1}^{N/2}S_{\alpha_{2a-1}}^{(2a-1)}(\bar{t}_{\mathrm{i}}^{2a-1}).
\end{equation}

Let us renormalize the general overlap formula as
\begin{equation}
\tilde{\mathcal{S}}_{\bar{\alpha}}(\bar{t})=\mathcal{S}_{\bar{\alpha}}(\bar{t})\prod_{s=1}^{N-1}\lambda_{s+1}(\bar{t}^{s}).
\end{equation}
For the renormalized overlap the sum formula reads as
\begin{multline}
\tilde{\mathcal{S}}_{\bar{\alpha}}(\bar{t})=\sum_{\mathrm{part}(\{\bar{t}^{2a}\}_{a=1}^{N/2-1})}\mathcal{W}_{\{\alpha_{2a-1}\}_{a=1}^{N/2}}(\{\bar{t}_{\textsc{i}}^{2a}\}_{a=1}^{N/2-1}|\{\bar{t}_{\textsc{ii}}^{2a}\}_{a=1}^{N/2-1}|\{\bar{t}^{2a-1}\}_{a=1}^{N/2})\times\\
\prod_{a=1}^{N/2-1}\lambda_{2a}(\bar{t}_{\textsc{i}}^{2a})\lambda_{2a+1}(\bar{t}_{\textsc{ii}}^{2a})\prod_{a=1}^{N/2}\lambda_{2a}(\bar{t}^{2a-1}).\label{eq:sumt-1}
\end{multline}
We can use the co-product formula (\ref{eq:coproduct}):
\begin{equation}
\tilde{\mathcal{S}}_{\bar{\alpha}}(\bar{t})=\sum_{\mathrm{part}(\bar{t})}\frac{\prod_{\nu=1}^{N-1}\lambda_{\nu}^{(2)}(\bar{t}_{\mathrm{i}}^{\nu})\lambda_{\nu+1}^{(1)}(\bar{t}_{\mathrm{ii}}^{\nu})f(\bar{t}_{\mathrm{ii}}^{\nu},\bar{t}_{\mathrm{i}}^{\nu})}{\prod_{\nu=1}^{N-2}f(\bar{t}_{\mathrm{ii}}^{\nu+1},\bar{t}_{\mathrm{i}}^{\nu})}\tilde{\mathcal{S}}_{\bar{\alpha}^{(1)}}(\bar{t}_{\mathrm{i}})\tilde{\mathcal{S}}_{\bar{\alpha}^{(2)}}(\bar{t}_{\mathrm{ii}}).\label{eq:cp}
\end{equation}
Let us fix a partition $\bar{t}^{2a}=\bar{t}_{\mathrm{i}}^{2a}\cup\bar{t}_{\mathrm{ii}}^{2a}$
for every $a=1,\dots,\frac{N}{2}-1$. Let us choose the vacuum eigenvalues
$\lambda_{\nu}(z)$ as
\begin{equation}
\begin{split}\lambda_{2a}^{(2)}(t_{k}^{2a}) & =0,\quad\text{where }t_{k}^{2a}\in\bar{t}_{\mathrm{ii}}^{2a},\text{ for }a=1,\dots,\frac{N}{2}-1,\\
\lambda_{2a+1}^{(1)}(t_{k}^{2a}) & =0,\quad\text{where }t_{k}^{2a}\in\bar{t}_{\mathrm{i}}^{2a},\text{ for }a=1,\dots,\frac{N}{2}-1.
\end{split}
\end{equation}
Using these conditions in the sum formula (\ref{eq:sumt-1}) we obtain
that
\begin{equation}
\tilde{\mathcal{S}}_{\bar{\alpha}}(\bar{t})=\mathcal{W}_{\{\alpha_{2a-1}\}_{a=1}^{N/2}}(\{\bar{t}_{\mathrm{i}}^{2a}\}_{a=1}^{N/2-1}|\{\bar{t}_{\mathrm{ii}}^{2a}\}_{a=1}^{N/2-1}|\{\bar{t}^{2a-1}\}_{a=1}^{N/2})\prod_{a=1}^{N/2-1}\lambda_{2a}(\bar{t}_{\mathrm{i}}^{2a})\lambda_{2a+1}(\bar{t}_{\mathrm{ii}}^{2a})\prod_{a=1}^{N/2}\lambda_{2a}(\bar{t}^{2a-1}).
\end{equation}
Substituting to the co-product formula (\ref{eq:cp}) we obtain that
\begin{equation}
\tilde{\mathcal{S}}_{\bar{\alpha}}(\bar{t})=\sum_{\mathrm{part}(\{\bar{t}^{2a-1}\}_{a=1}^{N/2})}\frac{\prod_{\nu=1}^{N-1}f(\bar{t}_{\mathrm{ii}}^{\nu},\bar{t}_{\mathrm{i}}^{\nu})}{\prod_{\nu=1}^{N-2}f(\bar{t}_{\mathrm{ii}}^{\nu+1},\bar{t}_{\mathrm{i}}^{\nu})}\tilde{\mathcal{S}}_{\bar{\alpha}^{(1)}}(\bar{t}_{\mathrm{i}})\tilde{\mathcal{S}}_{\bar{\alpha}^{(2)}}(\bar{t}_{\mathrm{ii}})\prod_{\nu=1}^{N-1}\lambda_{\nu}^{(2)}(\bar{t}_{\mathrm{i}}^{\nu})\lambda_{\nu+1}^{(1)}(\bar{t}_{\mathrm{ii}}^{\nu}),
\end{equation}
where the sum goes through to the partitions $\bar{t}^{2a-1}=\bar{t}_{\mathrm{i}}^{2a-1}\cup\bar{t}_{\mathrm{ii}}^{2a-1}$
for $a=1,\dots,N/2$ and the partitions of $\bar{t}^{2a}$ are fixed.
The overlaps are 
\begin{align}
\tilde{\mathcal{S}}_{\bar{\alpha}^{(1)}}(\bar{t}_{\mathrm{i}}) & =\mathcal{W}_{\{\alpha_{2a-1}^{(1)}\}_{a=1}^{N/2}}(\{\bar{t}_{\mathrm{i}}^{2a}\}_{a=1}^{N/2-1}|\emptyset|\{\bar{t}_{\mathrm{i}}^{2a-1}\}_{a=1}^{N/2})\prod_{a=1}^{N/2-1}\lambda_{2a}^{(1)}(\bar{t}_{\mathrm{i}}^{2a})\prod_{a=1}^{N/2}\lambda_{2a}^{(1)}(\bar{t}_{\mathrm{i}}^{2a-1}),\\
\tilde{\mathcal{S}}_{\bar{\alpha}^{(2)}}(\bar{t}_{\mathrm{ii}}) & =\mathcal{W}_{\{\alpha_{2a-1}^{(2)}\}_{a=1}^{N/2}}(\emptyset|\{\bar{t}_{\mathrm{ii}}^{2a}\}_{a=1}^{N/2-1}|\{\bar{t}_{\mathrm{ii}}^{2a-1}\}_{a=1}^{N/2})\prod_{a=1}^{N/2-1}\lambda_{2a+1}^{(2)}(\bar{t}_{\mathrm{ii}}^{2a})\prod_{a=1}^{N/2}\lambda_{2a}^{(2)}(\bar{t}_{\mathrm{ii}}^{2a-1}).
\end{align}
Substituting back we obtain that
\begin{multline}
\mathcal{W}_{\{\alpha_{2a-1}\}_{a=1}^{N/2}}(\{\bar{t}_{\mathrm{i}}^{2a}\}_{a=1}^{N/2-1}|\{\bar{t}_{\mathrm{ii}}^{2a}\}_{a=1}^{N/2-1}|\{\bar{t}^{2a-1}\}_{a=1}^{N/2})=\sum_{\mathrm{part}(\{\bar{t}^{2a-1}\}_{a=1}^{N/2})}\frac{\prod_{\nu=1}^{N-1}f(\bar{t}_{\mathrm{ii}}^{\nu},\bar{t}_{\mathrm{i}}^{\nu})}{\prod_{\nu=1}^{N-2}f(\bar{t}_{\mathrm{ii}}^{\nu+1},\bar{t}_{\mathrm{i}}^{\nu})}\times\\
\mathcal{W}_{\{\alpha_{2a-1}^{(1)}\}_{a=1}^{N/2}}(\{\bar{t}_{\mathrm{i}}^{2a}\}_{a=1}^{N/2-1}|\emptyset|\{\bar{t}_{\mathrm{i}}^{2a-1}\}_{a=1}^{N/2})\mathcal{W}_{\{\alpha_{2a-1}^{(2)}\}_{a=1}^{N/2}}(\emptyset|\{\bar{t}_{\mathrm{ii}}^{2a}\}_{a=1}^{N/2-1}|\{\bar{t}_{\mathrm{ii}}^{2a-1}\}_{a=1}^{N/2})\prod_{a=1}^{N/2}\alpha_{2a-1}^{(2)}(\bar{t}_{\mathrm{i}}^{2a-1}).
\end{multline}
Let us choose the quantum space $\mathcal{H}^{(1)}$ as $\alpha_{2a-1}^{(1)}(z)=1$
for which the formula above simplifies as
\begin{multline}
\mathcal{W}_{\{\alpha_{2a-1}\}_{a=1}^{N/2}}(\{\bar{t}_{\mathrm{i}}^{2a}\}_{a=1}^{N/2-1}|\{\bar{t}_{\mathrm{ii}}^{2a}\}_{a=1}^{N/2-1}|\{\bar{t}^{2a-1}\}_{a=1}^{N/2})=\\
\sum_{\mathrm{part}(\{\bar{t}^{2a-1}\}_{a=1}^{N/2})}\frac{\prod_{\nu=1}^{N-1}f(\bar{t}_{\mathrm{ii}}^{\nu},\bar{t}_{\mathrm{i}}^{\nu})}{\prod_{\nu=1}^{N-2}f(\bar{t}_{\mathrm{ii}}^{\nu+1},\bar{t}_{\mathrm{i}}^{\nu})}\mathcal{Z}(\bar{t}_{\mathrm{i}})\mathcal{W}_{\{\alpha_{2a-1}\}_{a=1}^{N/2}}(\emptyset|\{\bar{t}_{\mathrm{ii}}^{2a}\}_{a=1}^{N/2-1}|\{\bar{t}_{\mathrm{ii}}^{2a-1}\}_{a=1}^{N/2})\prod_{a=1}^{N/2}\alpha_{2a-1}(\bar{t}_{\mathrm{i}}^{2a-1}),\label{eq:WW1-1}
\end{multline}
where we defined the HC as
\begin{equation}
\mathcal{Z}(\bar{t}):=\mathcal{W}_{\{\alpha_{2a-1}\}_{a=1}^{N/2}}(\{\bar{t}^{2a}\}_{a=1}^{N/2-1}|\emptyset|\{\bar{t}^{2a-1}\}_{a=1}^{N/2})\Biggr|_{\alpha_{2a-1}(z)=1}.
\end{equation}
We can also choose the other quantum space $\mathcal{H}^{(2)}$ as
$\alpha_{2a-1}^{(2)}(z)=1$ for which we have
\begin{multline}
\mathcal{W}_{\{\alpha_{2a-1}\}_{a=1}^{N/2}}(\{\bar{t}_{\mathrm{i}}^{2a}\}_{a=1}^{N/2-1}|\{\bar{t}_{\mathrm{ii}}^{2a}\}_{a=1}^{N/2-1}|\{\bar{t}^{2a-1}\}_{a=1}^{N/2})=\\
\sum_{\mathrm{part}(\{\bar{t}^{2a-1}\}_{a=1}^{N/2})}\frac{\prod_{\nu=1}^{N-1}f(\bar{t}_{\mathrm{ii}}^{\nu},\bar{t}_{\mathrm{i}}^{\nu})}{\prod_{\nu=1}^{N-2}f(\bar{t}_{\mathrm{ii}}^{\nu+1},\bar{t}_{\mathrm{i}}^{\nu})}\mathcal{W}_{\{\alpha_{2a-1}\}_{a=1}^{N/2}}(\{\bar{t}_{\mathrm{i}}^{2a}\}_{a=1}^{N/2-1}|\emptyset|\{\bar{t}_{\mathrm{i}}^{2a-1}\}_{a=1}^{N/2})\bar{\mathcal{Z}}(\bar{t}_{\mathrm{ii}}),\label{eq:secff}
\end{multline}
where we defined the other HC as
\begin{equation}
\bar{\mathcal{Z}}(\bar{t}):=\mathcal{W}_{\{\alpha_{2a-1}\}_{a=1}^{N/2}}(\emptyset|\{\bar{t}^{2a}\}_{a=1}^{N/2-1}|\{\bar{t}^{2a-1}\}_{a=1}^{N/2})\Biggr|_{\alpha_{2a-1}(z)=1}.
\end{equation}
Substituting $\bar{t}_{\mathrm{i}}^{2a}=\emptyset$ for $a=1,\dots,\frac{N}{2}-1$
to the second formula (\ref{eq:secff}) we obtain that
\begin{equation}
\mathcal{W}_{\{\alpha_{2a-1}\}_{a=1}^{N/2}}(\emptyset|\{\bar{t}_{\mathrm{ii}}^{2a}\}_{a=1}^{N/2-1}|\{\bar{t}^{2a-1}\}_{a=1}^{N/2})=\sum_{\mathrm{part}(\{\bar{t}^{2a-1}\}_{a=1}^{N/2})}\frac{\prod_{a=1}^{N/2}f(\bar{t}_{\mathrm{ii}}^{2a-1},\bar{t}_{\mathrm{i}}^{2a-1})}{\prod_{\nu=a}^{N/2-1}f(\bar{t}_{\mathrm{ii}}^{2a},\bar{t}_{\mathrm{i}}^{2a-1})}\prod_{a=1}^{N/2}S_{\alpha_{2a-1}}^{(2a-1)}(\bar{t}_{\mathrm{i}}^{2a-1})\bar{\mathcal{Z}}(\bar{t}_{\mathrm{ii}}),
\end{equation}
where we used that 
\begin{equation}
\mathcal{W}_{\{\alpha_{2a-1}\}_{a=1}^{N/2}}(\emptyset|\emptyset|\{\bar{t}_{\mathrm{i}}^{2a-1}\}_{a=1}^{N/2})=\mathcal{S}_{\bar{\alpha}}(\bar{t}^{1},\emptyset,\bar{t}^{3},\emptyset\dots,\emptyset,\bar{t}^{N-1})=\prod_{a=1}^{N/2}S_{\alpha_{2a-1}}^{(2a-1)}(\bar{t}_{\mathrm{i}}^{2a-1}).
\end{equation}
We can substitute to (\ref{eq:WW1-1}):
\begin{multline}
\mathcal{W}_{\{\alpha_{2a-1}\}_{a=1}^{N/2}}(\{\bar{t}_{\mathrm{i}}^{2a}\}_{a=1}^{N/2-1}|\{\bar{t}_{\mathrm{ii}}^{2a}\}_{a=1}^{N/2-1}|\{\bar{t}^{2a-1}\}_{a=1}^{N/2})=\\
\sum_{\mathrm{part}(\{\bar{t}^{2a-1}\}_{a=1}^{N/2})}\frac{\prod_{\nu=1}^{N-1}f(\bar{t}_{\mathrm{ii}}^{\nu},\bar{t}_{\mathrm{i}}^{\nu})}{\prod_{\nu=1}^{N-2}f(\bar{t}_{\mathrm{ii}}^{\nu+1},\bar{t}_{\mathrm{i}}^{\nu})}\frac{\prod_{a=1}^{N/2}f(\bar{t}_{\mathrm{iii}}^{2a-1},\bar{t}_{\mathrm{i}}^{2a-1})f(\bar{t}_{\mathrm{ii}}^{2a-1},\bar{t}_{\mathrm{iii}}^{2a-1})}{\prod_{a=1}^{N/2-1}f(\bar{t}_{\mathrm{iii}}^{2a+1},\bar{t}_{\mathrm{i}}^{2a})f(\bar{t}_{\mathrm{ii}}^{2a},\bar{t}_{\mathrm{iii}}^{2a-1})}\times\\
\mathcal{Z}(\bar{t}_{\mathrm{i}})\bar{\mathcal{Z}}(\bar{t}_{\mathrm{ii}})\prod_{a=1}^{N/2}S_{\alpha_{2a-1}}^{(2a-1)}(\bar{t}_{\mathrm{iii}}^{2a-1})\prod_{a=1}^{N/2}\alpha_{2a-1}(\bar{t}_{\mathrm{i}}^{2a-1}).
\end{multline}
Substituting back to the original sum formula we obtain that
\begin{align}
\mathcal{S}_{\bar{\alpha}}(\bar{t}) & =\sum_{\mathrm{part}(\bar{t})}\frac{\prod_{\nu=1}^{N-1}f(\bar{t}_{\mathrm{ii}}^{\nu},\bar{t}_{\mathrm{i}}^{\nu})}{\prod_{\nu=1}^{N-2}f(\bar{t}_{\mathrm{ii}}^{\nu+1},\bar{t}_{\mathrm{i}}^{\nu})}\frac{\prod_{a=1}^{N/2}f(\bar{t}_{\mathrm{iii}}^{2a-1},\bar{t}_{\mathrm{i}}^{2a-1})f(\bar{t}_{\mathrm{ii}}^{2a-1},\bar{t}_{\mathrm{iii}}^{2a-1})}{\prod_{a=1}^{N/2-1}f(\bar{t}_{\mathrm{iii}}^{2a+1},\bar{t}_{\mathrm{i}}^{2a})f(\bar{t}_{\mathrm{ii}}^{2a},\bar{t}_{\mathrm{iii}}^{2a-1})}\times\nonumber \\
 & \times\mathcal{Z}(\bar{t}_{\mathrm{i}})\bar{\mathcal{Z}}(\bar{t}_{\mathrm{ii}})\prod_{a=1}^{N/2}S_{\alpha_{2a-1}}^{(2a-1)}(\bar{t}_{\mathrm{iii}}^{2a-1})\prod_{s=1}^{N-1}\alpha_{s}(\bar{t}_{\mathrm{i}}^{s}),\label{eq:sumFormulaTw}
\end{align}
which is what we wanted to prove (\ref{eq:sumTW}).

\section{Elementary overlaps}

In this section we derive some elementary overlaps for two- and one-site
states.

\subsection*{Untwisted case}

For the untwisted case we use the type $(N,M)$ regular form $K$-matrices.
We derive the elementary overlaps where the quantum space is the tensor
product of the representations $(1,0,\dots,0)$ and $(0,\dots,0,-1)$
for which the monodromy matrix is
\begin{equation}
T_{0}(z)=\bar{L}_{0,2}(z+\theta)L_{0,1}(z-\theta).\label{eq:H1}
\end{equation}
In this convention the basis vectors $e_{j}\otimes e_{k}$ have weights
$\Lambda_{i}^{(1)}=\delta_{i,j}-\delta_{i,N+1-k}$. The pseudo-vacuum
has weight $(1,0,\dots,0,-1)$. The following Bethe vectors span the
whole quantum space
\begin{align}
 & \mathbb{B}(\{t^{1}\},\dots,\{t^{k-1}\},\emptyset\dots,\emptyset,\{t^{N+1-l}\},\dots,\{t^{N-1}\})\sim e_{k}\otimes e_{l},\text{ for }k+l\leq N,\nonumber \\
 & \mathbb{B}(\{t^{1}\},\dots,\{t^{N-1}\})\in\mathrm{span}(\{e_{k}\otimes e_{N+1-k}\}_{k=1}^{N}),\\
 & \mathbb{B}(\{t^{1}\},\dots,\{t^{k-1}\},\{t_{1}^{k},t_{2}^{k}\},\dots,\{t_{1}^{N-l},t_{2}^{N-l}\},\{t^{N+1-l}\},\dots,\{t^{N-1}\})\sim e_{N+1-l}\otimes e_{N+1-k},\text{ for }k+l\leq N.\nonumber 
\end{align}
We already saw that the corresponding two-site state $\langle\psi(\theta)|$
has the form (\ref{eq:elemTwoSite})
\begin{equation}
\langle\psi(\theta)|=\sum_{i,j=1}^{N}K_{N+1-j,i}(\theta)\left(e_{i}\right)^{t}\otimes\left(e_{j}\right)^{t}=\sum_{i=1}^{M}\mathfrak{b}_{i}\left(e_{i}\right)^{t}\otimes\left(e_{i}\right)^{t}+\sum_{i=1}^{N}K_{i,i}\left(e_{i}\right)^{t}\otimes\left(e_{N+1-i}\right)^{t},
\end{equation}
therefore the non-vanishing overlaps are
\begin{equation}
S_{\bar{\alpha}^{(1)}}(\{t^{s}\}_{s=1}^{k-1},\emptyset^{\times N-2k+1},\{t^{s}\}_{s=N+1-k}^{N-1})=\langle\psi(\theta)|\mathbb{B}(\{t^{s}\}_{s=1}^{k-1},\emptyset^{\times N-2k+1},\{t^{s}\}_{s=N+1-k}^{N-1}),
\end{equation}
for $k=1,\dots,M$ and 
\begin{equation}
S_{\bar{\alpha}^{(1)}}(\{t^{1}\},\dots,\{t^{N-1}\})=\langle\psi(\theta)|\mathbb{B}(\{t^{1}\},\dots,\{t^{N-1}\}).\label{eq:elemUTW2}
\end{equation}

Let us derive these ''elementary'' overlaps. We can us the recursion
equation (\ref{eq:rec1}):
\begin{multline}
\langle\psi(\theta)|\mathbb{B}(\{t^{s}\}_{s=1}^{k-1},\emptyset^{\times N-2k+1},\{t^{s}\}_{s=N+1-k}^{N-1})=\\
\frac{1}{\lambda_{2}(t^{1})}\langle\psi(\theta)|T_{1,k}(t^{1})\mathbb{B}(\emptyset,\dots,\emptyset,\left\{ t^{s}\right\} _{s=N+1-k}^{N-1})\frac{1}{\prod_{\nu=2}^{k-1}h(t^{\nu},t^{\nu-1})}.
\end{multline}
We also need the $KT$-relation
\begin{equation}
\langle\psi(\theta)|T_{1,k}(z)=\frac{\mathfrak{b}_{k}}{\mathfrak{b}_{1}}\langle\psi(\theta)|T_{N,N+1-k}(-z)+\frac{K_{k,k}(z)}{\mathfrak{b}_{1}}\langle\psi(\theta)|T_{N,k}(-z)-\frac{K_{N,N}(z)}{\mathfrak{b}_{1}}\langle\psi(\theta)|T_{N,k}(z).
\end{equation}
Since
\begin{equation}
T_{N,k}(z)\mathbb{B}(\emptyset,\dots,\emptyset,\left\{ t^{s}\right\} _{s=N+1-k}^{N-1})=0,
\end{equation}
for $k\leq M$, we have
\begin{multline}
\langle\psi(\theta)|\mathbb{B}(\{t^{s}\}_{s=1}^{k-1},\emptyset^{\times N-2k+1},\{t^{s}\}_{s=N+1-k}^{N-1})=\\
\frac{\mathfrak{b}_{k}}{\mathfrak{b}_{1}}\frac{1}{\lambda_{2}(t^{1})}\langle\psi(\theta)|T_{N,N+1-k}(-t^{1})\mathbb{B}(\emptyset,\dots,\emptyset,\left\{ t^{s}\right\} _{s=N+1-k}^{N-1})\frac{1}{\prod_{\nu=2}^{k-1}h(t^{\nu},t^{\nu-1})},
\end{multline}
we can also use the other recursion formula (\ref{eq:rec2-1}):
\begin{equation}
\mathbb{B}(\emptyset,\dots,\emptyset,\left\{ t^{s}\right\} _{s=N+1-k}^{N-1})=\frac{1}{\lambda_{N}(t^{N-1})}T_{N+1-k,N}(t^{N-1})|0\rangle\frac{1}{\prod_{\nu=N+2-k}^{N-1}h(t^{\nu},t^{\nu-1})},
\end{equation}
therefore
\begin{multline}
\langle\psi(\theta)|\mathbb{B}(\{t^{s}\}_{s=1}^{k-1},\emptyset^{\times N-2k+1},\{t^{s}\}_{s=N+1-k}^{N-1})=\frac{\mathfrak{b}_{k}}{\mathfrak{b}_{1}}\frac{1}{\lambda_{2}(t^{1})}\frac{1}{\lambda_{N}(t^{N-1})}\times\\
\times\langle\psi(\theta)|T_{N,N+1-k}(-t^{1})T_{N+1-k,N}(t^{N-1})|0\rangle\frac{1}{\prod_{\nu=2}^{k-1}h(t^{\nu},t^{\nu-1})}\frac{1}{\prod_{\nu=N+2-k}^{N-1}h(t^{\nu},t^{\nu-1})}.\label{eq:ovtemp}
\end{multline}
We can use the $RTT$-relation (\ref{eq:RTT}):
\begin{multline}
\left[T_{N,N+1-k}(-t^{1}),T_{N+1-k,N}(t^{N-1})\right]=\\
g(-t^{1},t^{N-1})\left(T_{N+1-k,N+1-k}(t^{N-1})T_{N,N}(-t^{1})-T_{N+1-k,N+1-k}(-t^{1})T_{N,N}(t^{N-1})\right),
\end{multline}
therefore
\begin{equation}
\frac{1}{\lambda_{2}(t^{1})}T_{N,N+1-k}(-t^{1})T_{N+1-k,N}(t^{N-1})|0\rangle=g(-t^{1},t^{N-1})\left(\alpha_{1}(t^{1})\alpha_{N-1}(t^{N-1})-1\right)|0\rangle.
\end{equation}
Substituting back to the overlap formula (\ref{eq:ovtemp}), we obtain
that
\begin{align}
\langle\psi(\theta)|\mathbb{B}(\{t^{s}\}_{s=1}^{k-1},\emptyset^{\times N-2k+1},\{t^{s}\}_{s=N+1-k}^{N-1} & =\frac{\mathfrak{b}_{k}}{\mathfrak{b}_{1}}g(-t^{1},t^{N-1})\left(\alpha_{1}(t^{1})\alpha_{N-1}(t^{N-1})-1\right)\times\nonumber \\
 & \times\frac{1}{\prod_{\nu=2}^{k-1}h(t^{\nu},t^{\nu-1})}\frac{1}{\prod_{\nu=N+2-k}^{N-1}h(t^{\nu},t^{\nu-1})}.\label{eq:elemOvUtw1}
\end{align}

Let us continue with the other elementary overlap (\ref{eq:elemUTW2}).
We can us the recursion equation (\ref{eq:rec1}):
\begin{align}
\langle\psi(\theta)|\mathbb{B}(\{t^{s}\}_{s=1}^{N-1}) & =\frac{1}{\lambda_{2}(t^{1})}\sum_{j=2}^{N-1}\langle\psi(\theta)|T_{1,j}(t^{1})\mathbb{B}(\emptyset,\dots,\emptyset,\left\{ t^{k}\right\} _{k=j}^{N-1})\frac{1}{\prod_{\nu=2}^{j-1}h(t^{\nu},t^{\nu-1})}\frac{1}{f(t^{j},t^{j-1})}+\nonumber \\
 & +\frac{1}{\lambda_{2}(t^{1})}\alpha_{N-1}(\bar{t}^{N-1})\langle\psi(\theta)|T_{1,N}(t^{1})\mathbb{B}(\emptyset)\frac{1}{\prod_{\nu=2}^{N-1}h(t^{\nu},t^{\nu-1})}.
\end{align}
Let us use the other recursion (\ref{eq:rec2-1}):
\[
\mathbb{B}(\emptyset,\dots,\emptyset,\left\{ t^{k}\right\} _{k=j}^{N-1})=\frac{1}{\lambda_{N}(t^{N-1})}T_{j,N}(t^{N-1})|0\rangle\frac{1}{\prod_{\nu=j+1}^{N-1}h(t^{\nu},t^{\nu-1})},
\]
therefore the elementary overlap simplifies as
\begin{align}
\langle\psi(\theta)|\mathbb{B}(\{t^{s}\}_{s=1}^{N-1}) & =\frac{1}{\lambda_{2}(t^{1})}\frac{1}{\lambda_{N}(t^{N-1})}\sum_{j=2}^{N-1}\langle\psi(\theta)|T_{1,j}(t^{1})T_{j,N}(t^{N-1})|0\rangle\frac{1}{\prod_{\nu=2}^{N-1}h(t^{\nu},t^{\nu-1})}\frac{1}{g(t^{j},t^{j-1})}+\nonumber \\
 & +\frac{1}{\lambda_{2}(t^{1})}\alpha_{N-1}(\bar{t}^{N-1})\langle\psi(\theta)|T_{1,N}(t^{1})|0\rangle\frac{1}{\prod_{\nu=2}^{N-1}h(t^{\nu},t^{\nu-1})}.
\end{align}
We can also use the KT-relation
\begin{equation}
\langle\psi|T_{1,k}(z)=\frac{\mathfrak{b}_{k}}{\mathfrak{b}_{1}}\langle\psi|T_{N,N+1-k}(-z)+\frac{K_{k,k}(z)}{\mathfrak{b}_{1}}\langle\psi|T_{N,k}(-z)-\frac{K_{N,N}(z)}{\mathfrak{b}_{1}}\langle\psi|T_{N,k}(z),
\end{equation}
for $k\leq M$ and
\begin{equation}
\langle\psi|T_{1,k}(z)=\frac{K_{k,k}(z)}{\mathfrak{b}_{1}}\langle\psi|T_{N,k}(-z)-\frac{K_{N,N}(z)}{\mathfrak{b}_{1}}\langle\psi|T_{N,k}(z),
\end{equation}
for $k>M$. Since 
\begin{equation}
T_{N,N+1-j}(-z)T_{j,N}(t^{N-1})|0\rangle=0,
\end{equation}
for $j\leq M$, we have 
\begin{align}
\langle\psi(\theta)|\mathbb{B}(\{t^{s}\}_{s=1}^{N-1}) & =\frac{K_{j,j}(t^{1})}{\mathfrak{b}_{1}}\sum_{j=2}^{N-1}\langle\psi(\theta)|\frac{T_{N,j}(-t^{1})T_{j,N}(t^{N-1})}{\lambda_{2}(t^{1})\lambda_{N}(t^{N-1})}|0\rangle\frac{1}{\prod_{\nu=2}^{N-1}h(t^{\nu},t^{\nu-1})}\frac{1}{g(t^{j},t^{j-1})}\nonumber \\
 & -\frac{K_{N,N}(t^{1})}{\mathfrak{b}_{1}}\sum_{j=2}^{N-1}\langle\psi(\theta)|\frac{T_{N,j}(t^{1})T_{j,N}(t^{N-1})}{\lambda_{2}(t^{1})\lambda_{N}(t^{N-1})}|0\rangle\frac{1}{\prod_{\nu=2}^{N-1}h(t^{\nu},t^{\nu-1})}\frac{1}{g(t^{j},t^{j-1})}\label{eq:ovttt}\\
 & +\frac{K_{N,N}(t^{1})}{\mathfrak{b}_{1}}\left(\alpha_{1}(t^{1})\alpha_{N-1}(\bar{t}^{N-1})-\frac{\alpha_{N-1}(\bar{t}^{N-1})}{\alpha_{N-1}(t^{1})}\right)\frac{1}{\prod_{\nu=2}^{N-1}h(t^{\nu},t^{\nu-1})}.\nonumber 
\end{align}
We can use the $RTT$-relation (\ref{eq:RTT}):
\begin{equation}
\left[T_{N,j}(z),T_{j,N}(t^{N-1})\right]=g(z,t^{N-1})\left(T_{j,j}(t^{N-1})T_{N,N}(z)-T_{j,j}(z)T_{N,N}(t^{N-1})\right),
\end{equation}
therefore
\begin{align}
\frac{1}{\lambda_{2}(t^{1})}\frac{1}{\lambda_{N}(t^{N-1})}\langle\psi(\theta)|T_{N,j}(-t^{1})T_{j,N}(t^{N-1})\mathbb{B}(\emptyset) & =g(-t^{1},t^{N-1})\left(\alpha_{1}(t^{1})\alpha_{N-1}(t^{N-1})-1\right),\\
\frac{1}{\lambda_{2}(t^{1})}\frac{1}{\lambda_{N}(t^{N-1})}\langle\psi(\theta)|T_{N,j}(t^{1})T_{j,N}(t^{N-1})\mathbb{B}(\emptyset) & =g(t^{1},t^{N-1})\left(\frac{\alpha_{N-1}(t^{N-1})}{\alpha_{N-1}(t^{1})}-1\right).
\end{align}
Substituting back to the overlap formula (\ref{eq:ovttt}) we obtain
that
\begin{align}
\langle\psi(\theta)|\mathbb{B}(\{t^{s}\}_{s=1}^{N-1}) & =g(-t^{1},t^{N-1})\left(\alpha_{1}(t^{1})\alpha_{N-1}(t^{N-1})-1\right)\frac{1}{\prod_{\nu=2}^{N-1}h(t^{\nu},t^{\nu-1})}\sum_{j=2}^{N-1}\frac{K_{j,j}(t^{1})}{\mathfrak{b}_{1}}\frac{1}{g(t^{j},t^{j-1})}\nonumber \\
 & -\frac{K_{N,N}(t^{1})}{\mathfrak{b}_{1}}g(t^{1},t^{N-1})\left(\frac{\alpha_{N-1}(t^{N-1})}{\alpha_{N-1}(t^{1})}-1\right)\frac{1}{\prod_{\nu=2}^{N-1}h(t^{\nu},t^{\nu-1})}\sum_{j=2}^{N-1}\frac{1}{g(t^{j},t^{j-1})}\\
 & +\frac{K_{N,N}(t^{1})}{\mathfrak{b}_{1}}\left(\alpha_{1}(t^{1})\alpha_{N-1}(\bar{t}^{N-1})-\frac{\alpha_{N-1}(\bar{t}^{N-1})}{\alpha_{N-1}(t^{1})}\right)\frac{1}{\prod_{\nu=2}^{N-1}h(t^{\nu},t^{\nu-1})}.\nonumber 
\end{align}
We have the identity
\begin{equation}
\sum_{j=2}^{N-1}\frac{1}{g(t^{j},t^{j-1})}=\frac{1}{g(t^{N-1},t^{1})},
\end{equation}
therefore the overlap formula simplifies as
\begin{align}
\langle\psi(\theta)|\mathbb{B}(\{t^{s}\}_{s=1}^{N-1}) & =g(-t^{1},t^{N-1})\left(\alpha_{1}(t^{1})\alpha_{N-1}(t^{N-1})-1\right)\frac{1}{\prod_{\nu=2}^{N-1}h(t^{\nu},t^{\nu-1})}\sum_{j=2}^{N-1}\frac{K_{j,j}(t^{1})}{\mathfrak{b}_{1}}\frac{1}{g(t^{j},t^{j-1})}\nonumber \\
 & +\frac{K_{N,N}(t^{1})}{\mathfrak{b}_{1}}\left(\alpha_{1}(t^{1})\alpha_{N-1}(\bar{t}^{N-1})-1\right)\frac{1}{\prod_{\nu=2}^{N-1}h(t^{\nu},t^{\nu-1})},
\end{align}
i.e.
\begin{align}
\langle\psi(\theta)|\mathbb{B}(\{t^{s}\}_{s=1}^{N-1}) & =\left(\alpha_{1}(t^{1})\alpha_{N-1}(t^{N-1})-1\right)\frac{1}{\prod_{\nu=2}^{N-1}h(t^{\nu},t^{\nu-1})}\times\nonumber \\
 & \times\left(g(-t^{1},t^{N-1})\sum_{j=2}^{N-1}\frac{K_{j,j}(t^{1})}{\mathfrak{b}_{1}}\frac{1}{g(t^{j},t^{j-1})}+\frac{K_{N,N}(t^{1})}{\mathfrak{b}_{1}}\right).
\end{align}

We concentrate on two types of K-matrices.
\begin{enumerate}
\item For the singular $K$-matrix where $K_{N,1}=\mathfrak{b}_{1}$, $K_{1,1}(z)=\dots=K_{N,N}(z)=\frac{1}{u}$
we have
\begin{equation}
\sum_{j=2}^{N-1}\frac{K_{j,j}(t^{1})}{K_{N,1}(t^{1})}\frac{1}{g(t^{j},t^{j-1})}=\frac{K_{N,N}(t^{1})}{K_{N,1}(t^{1})}\frac{1}{g(t^{N-1},t^{1})},
\end{equation}
for which the overlap reads as
\begin{align}
\langle\psi|\mathbb{B}(\left\{ t^{k}\right\} _{k=1}^{N-1}) & =\frac{K_{N,N}(t^{1})}{K_{N,1}(t^{1})}\left(\alpha_{1}(t^{1})\alpha_{N-1}(t^{N-1})-1\right)\prod_{s=1}^{N-2}\frac{1}{h(t^{s+1},t^{s})}\left(\frac{g(-t^{1},t^{N-1})}{g(t^{N-1},t^{1})}+1\right).
\end{align}
We also have the identity
\begin{equation}
\frac{g(-t^{1},t^{N-1})}{g(t^{N-1},t^{1})}+1=g(-t^{1},t^{N-1})\left(\frac{1}{g(t^{N-1},t^{1})}+\frac{1}{g(-t^{1},t^{N-1})}\right)=\frac{g(-t^{1},t^{N-1})}{g(-t^{1},t^{1})},
\end{equation}
therefore the overlap simplifies as
\begin{multline}
S_{0}^{s}(\left\{ t^{k}\right\} _{k=1}^{N-1}):=\langle\psi|\mathbb{B}(\left\{ t^{k}\right\} _{k=1}^{N-1})=\\
\left(\frac{1}{g(-t^{1},t^{1})}\frac{K_{N,N}(t^{1})}{K_{N,1}(t^{1})}\right)g(-t^{1},t^{N-1})\left(\alpha_{1}(t^{1})\alpha_{N-1}(t^{N-1})-1\right)\prod_{s=1}^{N-2}\frac{1}{h(t^{s+1},t^{s})},
\end{multline}
or equivalently
\begin{align}
S_{0}^{s}(\left\{ t^{k}\right\} _{k=1}^{N-1}) & =\left(\frac{1}{g(-t^{M},t^{M})}\frac{K_{N,N}(t^{M})}{K_{N,1}(t^{M})}\right)g(-t^{1},t^{N-1})\left(\alpha_{1}(t^{1})\alpha_{N-1}(t^{N-1})-1\right)\prod_{s=1}^{N-2}\frac{1}{h(t^{s+1},t^{s})}.
\end{align}
\item For the $K$-matrix where $K_{N,1}=\mathfrak{b}$, $K_{1,1}(z)=K_{2,2}(z)=\dots=K_{M,M}(z)=\frac{\mathfrak{a}-u}{u}$,
$K_{M+1,M+1}(z)=\dots=K_{N,N}(z)=\frac{\mathfrak{a}+u}{u}$ we have
\begin{multline}
\langle\psi|\mathbb{B}(\left\{ t^{k}\right\} _{k=1}^{N-1})=\left(\frac{K_{1,1}(t^{1})}{K_{N,1}(t^{1})}\frac{g(-t^{1},t^{N-1})}{g(t^{M},t^{1})}+\frac{K_{N,N}(t^{1})}{K_{N,1}(t^{1})}\left(\frac{g(-t^{1},t^{N-1})}{g(t^{N-1},t^{M})}+1\right)\right)\times\\
\times\prod_{s=1}^{N-2}\frac{1}{h(t^{s+1},t^{s})}\left(\alpha_{1}(t^{1})\alpha_{N-1}(t^{N-1})-1\right).
\end{multline}
We can use the identity
\begin{equation}
\frac{g(-t^{1},t^{N-1})}{g(t^{N-1},t^{M})}+1=\frac{g(-t^{1},t^{N-1})}{g(-t^{1},t^{M})}
\end{equation}
in the following expression
\begin{equation}
\frac{K_{1,1}(t^{1})}{K_{N,1}(t^{1})}\frac{g(-t^{1},t^{N-1})}{g(t^{M},t^{1})}+\frac{K_{N,N}(t^{1})}{K_{N,1}(t^{1})}\left(\frac{g(-t^{1},t^{N-1})}{g(t^{N-1},t^{M})}+1\right)=\frac{K_{N,N}(t^{M})}{K_{N,1}(t^{M})}\frac{1}{g(-t^{M},t^{M})}g(-t^{1},t^{N-1}),
\end{equation}
therefore the overlap simplifies as
\begin{align}
S_{0}^{M}(\left\{ t^{k}\right\} _{k=1}^{N-1}) & :=\langle\Psi|\mathbb{B}(\left\{ t^{k}\right\} _{k=1}^{N-1})\nonumber \\
 & =\frac{1}{g(-t^{M},t^{M})}\frac{K_{N,N}(t^{M})}{K_{N,1}(t^{M})}g(-t^{1},t^{N-1})\left(\alpha_{1}(t^{1})\alpha_{N-1}(t^{N-1})-1\right)\prod_{s=1}^{N-2}\frac{1}{h(t^{s+1},t^{s})}\nonumber \\
 & =(t^{M}+\mathfrak{a})S_{0}^{s}(\{t^{1}\},\left\{ t^{k}\right\} _{k=2}^{N-1}).
\end{align}
\end{enumerate}
We can see that there is a common form for the two types of the boundary
states which is
\begin{align}
 & \langle\Psi|\mathbb{B}(\left\{ t^{k}\right\} _{k=1}^{N-1})\label{eq:elemOvUtw2}\\
 & =\frac{1}{g(-t^{M},t^{M})}\frac{K_{N,N}(t^{M})}{K_{N,1}(t^{M})}g(-t^{1},t^{N-1})\left(\alpha_{1}(t^{1})\alpha_{N-1}(t^{N-1})-1\right)\prod_{s=1}^{N-2}\frac{1}{h(t^{s+1},t^{s})}.\nonumber 
\end{align}

\subsection*{Twisted case}

In this section we derive the elementary overlaps where the quantum
space is the rectangular representation where $s=1,a=2$ for which
the monodromy matrix is
\begin{equation}
T_{0}(z)=L_{0,1}^{(1,2)}(z+c/2),\label{eq:elemMonTw}
\end{equation}
for which
\begin{align}
\alpha_{1}(z) & =1,\nonumber \\
\alpha_{2}(z) & =\frac{z+3c/2}{z+c/2},\label{eq:explA}\\
\alpha_{k}(z) & =1,\qquad\qquad\text{for }k>2.\nonumber 
\end{align}
We fix the $K$-matrix to its regular form (\ref{eq:regTwK}). We
saw that the corresponding one-site state is (\ref{eq:one-site})
\begin{equation}
\langle\psi|=\sum_{a=1}^{N/2}x_{a}e_{(2a-1,2a)}^{t},
\end{equation}
where $e_{(i,j)}$ are the basis in the quantum space for $1\leq i<j\leq N$.
The following Bethe vectors span the quantum space
\begin{align}
 & \mathbb{B}(\emptyset,\emptyset,\emptyset,\dots,\emptyset)=e_{(1,2)},\nonumber \\
 & \mathbb{B}(\emptyset,\{t^{2}\}\dots,\{t^{k-1}\},\emptyset\dots,\emptyset)\sim e_{(1,k)},\text{ for }3\leq k\leq N,\label{eq:nonvan}\\
 & \mathbb{B}(\{t^{1}\},\{t^{2}\}\dots,\{t^{k-1}\},\emptyset\dots,\emptyset)\sim e_{(2,k)},\text{ for }k+l\leq N,\nonumber \\
 & \mathbb{B}(\{t^{1}\},\{t_{1}^{2},t_{2}^{2}\},\dots,\{t_{1}^{k-1},t_{2}^{k-1}\},\{t^{k-1}\},\dots,\{t^{l-1}\},\emptyset\dots,\emptyset)\sim e_{(k,l)},\text{ for }3\leq k<l\leq N.\nonumber 
\end{align}
The non-vanishing overlaps are
\begin{equation}
S_{\bar{\alpha}}(\{t^{1}\},\bar{t}^{2},\dots,\bar{t}^{2a-2},\{t^{2a-1}\},\emptyset,\dots\emptyset)=\langle\psi|\mathbb{B}(\{t^{1}\},\bar{t}^{2},\dots,\bar{t}^{2a-2},\{t^{2a-1}\},\emptyset,\dots\emptyset),
\end{equation}
for $a=2,\dots,N/2$ and $\#\bar{t}^{s}=2$ for $s=2,\dots,2a-2$.

Let us derive these ''elementary'' overlaps. We can use the recursion
(\ref{eq:rect2}). For this representation the non-vanishing Bethe
vectors are listed in (\ref{eq:nonvan}) therefore the recursion of
the off-shell Bethe vector simplifies as
\begin{align}
\mathbb{B}(\{t^{1}\},\{z,t^{2}\} & ,\bar{t}^{3},\dots,\bar{t}^{2a-2},\{t^{2a-1}\},\emptyset,\dots\emptyset)=\nonumber \\
=\sum_{\mathrm{part}(\bar{t})}\Biggl[ & \frac{T_{2,2a-1}(z)}{\lambda_{3}(z)}\mathbb{B}(\{t^{1}\},\{t^{2}\},\{\bar{t}_{\textsc{ii}}^{s}\}_{s=3}^{2a-2},\{t^{2a-1}\},\emptyset,\dots\emptyset)\frac{1}{f(z,t^{1})}\frac{1}{f(t^{2a-1},\bar{t}_{\textsc{i}}^{2a-2})}\nonumber \\
+ & \frac{T_{1,2a-1}(z)}{\lambda_{3}(z)}\mathbb{B}(\emptyset,\{t^{2}\},\{\bar{t}_{\textsc{ii}}^{s}\}_{s=3}^{2a-2},\{t^{2a-1}\},\emptyset,\dots\emptyset)\frac{1}{h(z,t^{1})}\frac{1}{f(t^{2a-1},\bar{t}_{\textsc{i}}^{2a-2})}\nonumber \\
+ & \frac{T_{2,2a}(z)}{\lambda_{3}(z)}\mathbb{B}(\{t^{1}\},\{t^{2}\},\{\bar{t}_{\textsc{ii}}^{s}\}_{s=3}^{2a-2},\emptyset,\dots\emptyset)\frac{1}{f(z,t^{1})}\frac{1}{h(t^{2a-1},\bar{t}_{\textsc{i}}^{2a-2})}\\
+ & \frac{T_{1,2a}(z)}{\lambda_{3}(z)}\mathbb{B}(\emptyset,\{t^{2}\},\{\bar{t}_{\textsc{ii}}^{s}\}_{s=3}^{2a-2},\emptyset,\dots\emptyset)\frac{1}{h(z,\bar{t}^{1})}\frac{1}{h(t^{2a-1},\bar{t}_{\textsc{i}}^{2a-2})}\Biggr]\prod_{s=3}^{2a-2}\frac{g(\bar{t}_{\textsc{i}}^{s},\bar{t}_{\textsc{i}}^{s-1})f(\bar{t}_{\textsc{ii}}^{s},\bar{t}_{\textsc{i}}^{s})}{f(\bar{t}^{s},\bar{t}_{\textsc{i}}^{s-1})}.\nonumber 
\end{align}
We can use the $KT$-relation
\begin{equation}
\begin{split}\langle\Psi|T_{2,2b-1}(z) & =-\lambda_{0}(z)\frac{x_{b}}{x_{1}}\langle\Psi|\widehat{T}_{N,N+1-2b}(-z),\qquad\langle\Psi|T_{2,2b}(z)=\lambda_{0}(z)\frac{x_{b}}{x_{1}}\langle\Psi|\widehat{T}_{N,N+2-2b}(-z),\\
\langle\Psi|T_{1,2b-1}(z) & =\lambda_{0}(z)\frac{x_{b}}{x_{1}}\langle\Psi|\widehat{T}_{N-1,N+1-2b}(-z),\qquad\langle\Psi|T_{1,2b}(z)=-\lambda_{0}(z)\frac{x_{b}}{x_{1}}\langle\Psi|\widehat{T}_{N-1,N+2-2b}(-z),
\end{split}
\end{equation}
therefore we obtain that
\begin{align}
 & \langle\Psi|\mathbb{B}(\{t^{1}\},\{z,t^{2}\},\bar{t}^{3},\dots,\bar{t}^{2a-2},\{t^{2a-1}\},\emptyset,\dots\emptyset)=\frac{\lambda_{0}(z)}{\lambda_{3}(z)}\frac{x_{a}}{x_{1}}\sum_{\mathrm{part}(\bar{t})}\Biggl[\nonumber \\
 & -\langle\Psi|\widehat{T}_{N,N+1-2a}(-z)\mathbb{B}(\{t^{1}\},\{t^{2}\},\{\bar{t}_{\textsc{ii}}^{s}\}_{s=3}^{2a-2},\{t^{2a-1}\},\emptyset,\dots\emptyset)\frac{1}{f(z,t^{1})}\frac{1}{f(t^{2a-1},\bar{t}_{\textsc{i}}^{2a-2})}\nonumber \\
 & +\langle\Psi|\widehat{T}_{N-1,N+1-2a}(-z)\mathbb{B}(\emptyset,\{t^{2}\},\{\bar{t}_{\textsc{ii}}^{s}\}_{s=3}^{2a-2},\{t^{2a-1}\},\emptyset,\dots\emptyset)\frac{1}{h(z,t^{1})}\frac{1}{f(t^{2a-1},\bar{t}_{\textsc{i}}^{2a-2})}\nonumber \\
 & +\langle\Psi|\widehat{T}_{N,N+2-2a}(-z)\mathbb{B}(\{t^{1}\},\{t^{2}\},\{\bar{t}_{\textsc{ii}}^{s}\}_{s=3}^{2a-2},\emptyset,\dots\emptyset)\frac{1}{f(z,t^{1})}\frac{1}{h(t^{2a-1},\bar{t}_{\textsc{i}}^{2a-2})}\label{eq:elovtt}\\
 & -\langle\Psi|\widehat{T}_{N-1,N+2-2a}(-z)\mathbb{B}(\emptyset,\{t^{2}\},\{\bar{t}_{\textsc{ii}}^{s}\}_{s=3}^{2a-2},\emptyset,\dots\emptyset)\frac{1}{h(z,t^{1})}\frac{1}{h(t^{2a-1},\bar{t}_{\textsc{i}}^{2a-2})}\Biggr]\prod_{s=3}^{2a-2}\frac{g(\bar{t}_{\textsc{i}}^{s},\bar{t}_{\textsc{i}}^{s-1})f(\bar{t}_{\textsc{ii}}^{s},\bar{t}_{\textsc{i}}^{s})}{f(\bar{t}^{s},\bar{t}_{\textsc{i}}^{s-1})}.\nonumber 
\end{align}
Let us use the identity (\ref{eq:connBhatB}) for the Bethe states
of the rhs:
\begin{align}
\mathbb{B}(\{t^{s}\}_{s=1}^{j-1},\emptyset,\dots,\emptyset) & =(-1)^{j-1}\prod_{s=N+1-j}^{N-2}f(w^{s+1},w^{s})\hat{\mathbb{B}}(\emptyset,\dots,\emptyset,\{w^{s}\}_{s=N+1-j}^{N-1}),\\
\mathbb{B}(\emptyset,\{t^{s}\}_{s=2}^{j-1},\emptyset,\dots,\emptyset) & =(-1)^{j}\prod_{s=N+1-j}^{N-2}f(w^{s+1},w^{s})\hat{\mathbb{B}}(\emptyset,\dots,\emptyset,\{w^{s}\}_{s=N+1-j}^{N-2},\emptyset),
\end{align}
where $w^{s}=t^{N-s}+(N-s)c$ and the Bethe states $\hat{\mathbb{B}}$
are built from the twisted monodromy matrix entries $\widehat{T}_{i,j}$.
We can use the recursion (\ref{eq:rect2-2})
\begin{align}
\hat{\mathbb{B}}(\emptyset,\dots,\emptyset,\{w^{s}\}_{s=N+1-j}^{N-1}) & =\frac{\widehat{T}_{N+1-j,N}(w^{N-2})}{\hat{\lambda}_{N-1}(w^{N-2})}|0\rangle\frac{1}{\prod_{\nu=N+1-j}^{N-2}h(w^{\nu+1},w^{\nu})},\\
\hat{\mathbb{B}}(\emptyset,\dots,\emptyset,\{w^{s}\}_{s=N+1-j}^{N-2},\emptyset) & =\frac{\widehat{T}_{N+1-j,N-1}(w^{N-2})}{\hat{\lambda}_{N-1}(w^{N-2})}|0\rangle\frac{1}{\prod_{\nu=N+1-j}^{N-3}h(w^{\nu+1},w^{\nu})}.
\end{align}
Substituting back to (\ref{eq:elovtt}) we obtain that
\begin{align}
 & \langle\Psi|\mathbb{B}(\{t^{1}\},\{z,t^{2}\},\bar{t}^{3},\dots,\bar{t}^{2a-2},\{t^{2a-1}\},\emptyset,\dots\emptyset)=\frac{\lambda_{0}(z)}{\lambda_{3}(z)}\frac{1}{\hat{\lambda}_{N-1}(t^{2}+2c)}\frac{x_{a}}{x_{1}}\sum_{\mathrm{part}(\bar{t})}\Biggl[\nonumber \\
 & +\langle\Psi|\widehat{T}_{N,N+1-2a}(-z)\widehat{T}_{N+1-2a,N}(t^{2}+2c)|0\rangle\frac{1}{f(z,t^{1})}\frac{1}{h(t^{2},t^{1})}\frac{1}{f(t^{2a-1},\bar{t}_{\textsc{i}}^{2a-2})}\frac{1}{h(t^{2a-1},t_{\textsc{ii}}^{2a-2})}\nonumber \\
 & -\langle\Psi|\widehat{T}_{N-1,N+1-2a}(-z)\widehat{T}_{N+1-2a,N-1}(t^{2}+2c)|0\rangle\frac{1}{h(z,t^{1})}\frac{1}{f(t^{2a-1},\bar{t}_{\textsc{i}}^{2a-2})}\frac{1}{h(t^{2a-1},t_{\textsc{ii}}^{2a-2})}\nonumber \\
 & -\langle\Psi|\widehat{T}_{N,N+2-2a}(-z)\widehat{T}_{N+2-2a,N}(t^{2}+2c)|0\rangle\frac{1}{f(z,t^{1})}\frac{1}{h(t^{2},t^{1})}\frac{1}{h(t^{2a-1},\bar{t}_{\textsc{i}}^{2a-2})}\label{eq:elovtt-1}\\
 & +\langle\Psi|\widehat{T}_{N-1,N+2-2a}(-z)\widehat{T}_{N+2-2a,N-1}(t^{2}+2c)|0\rangle\frac{1}{h(z,t^{1})}\frac{1}{h(t^{2a-1},\bar{t}_{\textsc{i}}^{2a-2})}\Biggr]\prod_{s=3}^{2a-2}\frac{g(\bar{t}_{\textsc{i}}^{s},\bar{t}_{\textsc{i}}^{s-1})f(\bar{t}_{\textsc{ii}}^{s},\bar{t}_{\textsc{i}}^{s})}{f(\bar{t}^{s},\bar{t}_{\textsc{i}}^{s-1})h(t_{\textsc{ii}}^{s},t_{\textsc{ii}}^{s-1})},\nonumber 
\end{align}
where we used the identity
\begin{equation}
\frac{f(w^{\nu+1},w^{\nu})}{h(w^{\nu+1},w^{\nu})}=g(t^{N-\nu-1}-c,t^{N-\nu})=-\frac{1}{h(t^{N-\nu},t^{N-\nu-1})}.
\end{equation}
Now we can use the $RTT$-relation (\ref{eq:RTT})
\begin{equation}
\left[\widehat{T}_{i,j}(-z),\widehat{T}_{j,i}(t^{2}+2c)\right]=g(-z,t^{2}+2c)\left(\widehat{T}_{j,j}(t^{2}+2c)\widehat{T}_{i,i}(-z)-\widehat{T}_{j,j}(-z)\widehat{T}_{i,i}(t^{2}+2c)\right),
\end{equation}
therefore we have
\begin{align}
\frac{\lambda_{0}(z)}{\lambda_{3}(z)}\frac{1}{\hat{\lambda}_{N-1}(t^{2}+2c)}\langle\Psi|\widehat{T}_{N,N+1-2a}(-z)\widehat{T}_{N+1-2a,N}(t^{2}+2c)|0\rangle & =\nonumber \\
\frac{\lambda_{0}(z)}{\lambda_{3}(z)}\frac{1}{\hat{\lambda}_{N-1}(t^{2}+2c)}\langle\Psi|\widehat{T}_{N-1,N+1-2a}(-z)\widehat{T}_{N+1-2a,N-1}(t^{2}+2c)|0\rangle & =\\
\frac{\lambda_{0}(z)}{\lambda_{3}(z)}\frac{1}{\hat{\lambda}_{N-1}(t^{2}+2c)}\langle\Psi|\widehat{T}_{N,N+2-2a}(-z)\widehat{T}_{N+2-2a,N}(t^{2}+2c)|0\rangle & =\nonumber \\
\frac{\lambda_{0}(z)}{\lambda_{3}(z)}\frac{1}{\hat{\lambda}_{N-1}(t^{2}+2c)}\langle\Psi|\widehat{T}_{N-1,N+2-2a}(-z)\widehat{T}_{N+2-2a,N-1}(t^{2}+2c)|0\rangle & =g(-z,t^{2}+2c)\left(\alpha_{2}(t^{2})\alpha_{2}(z)-1\right),\nonumber 
\end{align}
where we used the identity (\ref{eq:idAlph}), the symmetry property
(\ref{eq:twistedalpha}) and the explicit forms (\ref{eq:explA}).

Substituting back to (\ref{eq:elovtt-1}) the overlap simplifies as
\begin{align}
 & \langle\Psi|\mathbb{B}(\{t^{1}\},\{z,t^{2}\},\bar{t}^{3},\dots,\bar{t}^{2a-2},\{t^{2a-1}\},\emptyset,\dots\emptyset)=\nonumber \\
 & =\frac{x_{a}}{x_{1}}g(-z-2c,t^{2})(\alpha_{2}(z)\alpha_{2}(t^{2})-1)\left(\frac{1}{h(z,t^{1})}-\frac{1}{f(z,t^{1})}\frac{1}{h(t^{2},t^{1})}\right)\times\\
 & \times\sum_{\mathrm{part}(\bar{t})}\frac{\prod_{s=3}^{2a-2}g(\bar{t}_{\textsc{i}}^{s},\bar{t}_{\textsc{i}}^{s-1})f(\bar{t}_{\textsc{ii}}^{s},\bar{t}_{\textsc{i}}^{s})}{\prod_{s=3}^{2a-2}f(\bar{t}^{s},\bar{t}_{\textsc{i}}^{s-1})h(\bar{t}_{\textsc{ii}}^{s},\bar{t}_{\textsc{ii}}^{s-1})}\left(\frac{1}{h(t^{2a-1},\bar{t}_{\textsc{i}}^{2a-2})}-\frac{1}{f(t^{2a-1},\bar{t}_{\textsc{i}}^{2a-2})}\frac{1}{h(t^{2a-1},t_{\textsc{ii}}^{2a-2})}\right).\nonumber 
\end{align}
Using the identities
\begin{equation}
\begin{split}\frac{1}{h(u,v_{1})}-\frac{1}{f(u,v_{1})}\frac{1}{h(u,v_{2})} & =\frac{h(v_{1},v_{2})}{h(u,v_{1}),h(u,v_{2})},\\
\frac{1}{h(v_{1},u)}-\frac{1}{f(v_{1},u)}\frac{1}{h(v_{2},u)} & =\frac{h(v_{2},v_{1})}{h(v_{1},u),h(v_{2},u)},
\end{split}
\end{equation}
we obtain that
\begin{align}
 & \langle\Psi|\mathbb{B}(\{t^{1}\},\{z,t^{2}\},\bar{t}^{3},\dots,\bar{t}^{2a-2},\{t^{2a-1}\},\emptyset,\dots\emptyset)=\frac{x_{a}}{x_{1}}g(-z-2c,t^{2})(\alpha_{2}(z)\alpha_{2}(t^{2})-1)\nonumber \\
 & \times\frac{h(t^{2},z)}{h(z,t^{1})h(t^{2},t^{1})}\frac{1}{h(t^{2a-1},\bar{t}^{2a-2})}\sum_{\mathrm{part}(\bar{t})}\frac{\prod_{s=3}^{2a-2}g(\bar{t}_{\textsc{i}}^{s},\bar{t}_{\textsc{i}}^{s-1})f(\bar{t}_{\textsc{ii}}^{s},\bar{t}_{\textsc{i}}^{s})}{\prod_{s=3}^{2a-2}f(\bar{t}^{s},\bar{t}_{\textsc{i}}^{s-1})h(\bar{t}_{\textsc{ii}}^{s},\bar{t}_{\textsc{ii}}^{s-1})}h(\bar{t}_{\textsc{i}}^{2a-2},\bar{t}_{\textsc{ii}}^{2a-2}).\label{eq:ovTWtemp}
\end{align}
Let us try to simplify the sum. At first we try to do the summation
for the partition on $\bar{t}^{2a-2}$ as
\begin{align}
 & \sum_{\mathrm{part}(\bar{t})}\frac{\prod_{s=3}^{2a-2}g(\bar{t}_{\textsc{i}}^{s},\bar{t}_{\textsc{i}}^{s-1})f(\bar{t}_{\textsc{ii}}^{s},\bar{t}_{\textsc{i}}^{s})}{\prod_{s=3}^{2a-2}f(\bar{t}^{s},\bar{t}_{\textsc{i}}^{s-1})h(\bar{t}_{\textsc{ii}}^{s},\bar{t}_{\textsc{ii}}^{s-1})}h(\bar{t}_{\textsc{i}}^{2a-2},\bar{t}_{\textsc{ii}}^{2a-2})=\label{eq:ttt-1}\\
 & \sum_{\mathrm{part}(\{\bar{t}^{s}\}_{s=3}^{2a-3})}\frac{\prod_{s=3}^{2a-3}g(\bar{t}_{\textsc{i}}^{s},\bar{t}_{\textsc{i}}^{s-1})f(\bar{t}_{\textsc{ii}}^{s},\bar{t}_{\textsc{i}}^{s})}{\prod_{s=3}^{2a-3}f(\bar{t}^{s},\bar{t}_{\textsc{i}}^{s-1})h(\bar{t}_{\textsc{ii}}^{s},\bar{t}_{\textsc{ii}}^{s-1})}\left(\sum_{\mathrm{part}(\bar{t}^{2a-2})}\frac{g(\bar{t}_{\textsc{i}}^{2a-2},\bar{t}_{\textsc{i}}^{2a-3})f(\bar{t}_{\textsc{ii}}^{2a-2},\bar{t}_{\textsc{i}}^{2a-2})}{f(\bar{t}^{2a-2},\bar{t}_{\textsc{i}}^{2a-3})h(\bar{t}_{\textsc{ii}}^{2a-2},\bar{t}_{\textsc{ii}}^{2a-3})}h(\bar{t}_{\textsc{i}}^{2a-2},\bar{t}_{\textsc{ii}}^{2a-2})\right).\nonumber 
\end{align}
We can do the second summation as
\begin{multline}
\sum_{\mathrm{part}(\bar{t}^{2a-2})}\frac{g(\bar{t}_{\textsc{i}}^{2a-2},\bar{t}_{\textsc{i}}^{2a-3})f(\bar{t}_{\textsc{ii}}^{2a-2},\bar{t}_{\textsc{i}}^{2a-2})}{f(\bar{t}^{2a-2},\bar{t}_{\textsc{i}}^{2a-3})h(\bar{t}_{\textsc{ii}}^{2a-2},\bar{t}_{\textsc{ii}}^{2a-3})}h(t_{\textsc{i}}^{2a-2},t_{\textsc{ii}}^{2a-2})=\\
\frac{g(t_{1}^{2a-2},\bar{t}_{\textsc{i}}^{2a-3})f(t_{2}^{2a-2},t_{1}^{2a-2})}{f(\bar{t}^{2a-2},\bar{t}_{\textsc{i}}^{2a-3})h(t_{2}^{2a-2},\bar{t}_{\textsc{ii}}^{2a-3})}h(t_{1}^{2a-2},t_{2}^{2a-2})+\frac{g(t_{2}^{2a-2},\bar{t}_{\textsc{i}}^{2a-3})f(t_{1}^{2a-2},t_{2}^{2a-2})}{f(\bar{t}^{2a-2},\bar{t}_{\textsc{i}}^{2a-3})h(t_{1}^{2a-2},\bar{t}_{\textsc{ii}}^{2a-3})}h(t_{2}^{2a-2},t_{1}^{2a-2})=\\
\frac{h(t_{1}^{2a-2},t_{2}^{2a-2})h(t_{2}^{2a-2},t_{1}^{2a-2})}{h(\bar{t}^{2a-2},\bar{t}^{2a-3})}h(t_{\textsc{i}}^{2a-3},t_{\textsc{ii}}^{2a-3}).
\end{multline}
Substituting back to (\ref{eq:ttt-1}) we obtain that
\begin{multline}
\sum_{\mathrm{part}(\{\bar{t}^{s}\}_{s=3}^{2a-2})}\frac{\prod_{s=3}^{2a-2}g(\bar{t}_{\textsc{i}}^{s},\bar{t}_{\textsc{i}}^{s-1})f(\bar{t}_{\textsc{ii}}^{s},\bar{t}_{\textsc{i}}^{s})}{\prod_{s=3}^{2a-2}f(\bar{t}^{s},\bar{t}_{\textsc{i}}^{s-1})h(\bar{t}_{\textsc{ii}}^{s},\bar{t}_{\textsc{ii}}^{s-1})}h(\bar{t}_{\textsc{i}}^{2a-2},\bar{t}_{\textsc{ii}}^{2a-2})=\\
\frac{h(t_{1}^{2a-2},t_{2}^{2a-2})h(t_{2}^{2a-2},t_{1}^{2a-2})}{h(\bar{t}^{2a-2},\bar{t}^{2a-3})}\sum_{\mathrm{part}(\{\bar{t}^{s}\}_{s=3}^{2a-3})}\frac{\prod_{s=3}^{2a-3}g(\bar{t}_{\textsc{i}}^{s},\bar{t}_{\textsc{i}}^{s-1})f(\bar{t}_{\textsc{ii}}^{s},\bar{t}_{\textsc{i}}^{s})}{\prod_{s=3}^{2a-3}f(\bar{t}^{s},\bar{t}_{\textsc{i}}^{s-1})h(\bar{t}_{\textsc{ii}}^{s},\bar{t}_{\textsc{ii}}^{s-1})}h(t_{\textsc{i}}^{2a-3},t_{\textsc{ii}}^{2a-3}).
\end{multline}
In the second line we obtained the original sum with less partitions
therefore we can finish the summation by iteration and the result
is
\begin{equation}
\sum_{\mathrm{part}(\{\bar{t}^{s}\}_{s=3}^{2a-2})}\frac{\prod_{s=3}^{2a-2}g(\bar{t}_{\textsc{i}}^{s},\bar{t}_{\textsc{i}}^{s-1})f(\bar{t}_{\textsc{ii}}^{s},\bar{t}_{\textsc{i}}^{s})}{\prod_{s=3}^{2a-2}f(\bar{t}^{s},\bar{t}_{\textsc{i}}^{s-1})h(\bar{t}_{\textsc{ii}}^{s},\bar{t}_{\textsc{ii}}^{s-1})}h(\bar{t}_{\textsc{i}}^{2a-2},\bar{t}_{\textsc{ii}}^{2a-2})=\prod_{s=3}^{2a-2}\frac{h(t_{1}^{s},t_{2}^{s})h(t_{2}^{s},t_{1}^{s})}{h(\bar{t}^{s},\bar{t}^{s-1})}h(t_{\textsc{i}}^{2},t_{\textsc{ii}}^{2}).
\end{equation}
Substituting back to (\ref{eq:ovTWtemp}) we obtain that
\begin{align}
 & \langle\Psi|\mathbb{B}(\{t^{1}\},\{z,t^{2}\},\bar{t}^{3},\dots,\bar{t}^{2a-2},\{t^{2a-1}\},\emptyset,\dots\emptyset)=\nonumber \\
 & \frac{x_{a}}{x_{1}}g(-z-2c,t^{2})(\alpha_{2}(z)\alpha_{2}(t^{2})-1)\frac{h(t^{2},z)h(z,t^{2})}{h(z,t^{1})h(t^{2},t^{1})}\frac{\prod_{s=3}^{2a-2}h(t_{1}^{s},t_{2}^{s})h(t_{2}^{s},t_{1}^{s})}{\prod_{s=3}^{2a-1}h(\bar{t}^{s},\bar{t}^{s-1})},
\end{align}
i.e.
\begin{equation}
\langle\Psi|\mathbb{B}(\{t^{1}\},\{\bar{t}^{2}\}_{s=2}^{2a-2},\{t^{2a-1}\},\emptyset,\dots\emptyset)=\frac{x_{a}}{x_{1}}g(-t_{1}^{2}-2c,t_{2}^{2})(\alpha_{2}(\bar{t}^{2})-1)\frac{\prod_{s=2}^{2a-2}h(t_{1}^{s},t_{2}^{s})h(t_{2}^{s},t_{1}^{s})}{\prod_{s=1}^{2a-2}h(\bar{t}^{s+1},\bar{t}^{s})}.
\end{equation}

\section{Recursion for the HC-s $\mathcal{Z}$ and $\bar{\mathcal{Z}}$\label{sec:Reqursion-for-the}}

\subsection*{Untwisted case}

Let us define the tensor product Hilbert space as $\mathcal{H}^{mod}=\mathcal{H}^{(1)}\otimes\mathcal{H}$
where the $\mathcal{H}^{(1)}$ quantum space is defined by the monodromy
matrix (\ref{eq:H1}) for which the $\alpha$-functions are
\begin{equation}
\begin{split}\alpha_{1}^{(1)}(z) & =f(z,\theta),\quad\alpha_{N-1}^{(1)}(z)=\frac{1}{f(-z,\theta)},\\
\alpha_{s}^{(1)}(z) & =1,\quad\text{for }s=2,\dots,N-2.
\end{split}
\end{equation}
On the tensor product quantum space the $\alpha$-functions are
\begin{equation}
\alpha_{s}^{mod}(z)=\alpha_{s}^{(1)}(z)\alpha_{s}(z),\quad\text{for }s=1,\dots,N-1.
\end{equation}
Since the HC $\mathcal{Z}$ is independent on the $\alpha$-functions,
we can derive a recursion for the HC $\mathcal{Z}$ from the co-product
formula:
\begin{equation}
\mathcal{S}_{\bar{\alpha}^{mod}}(\bar{t})=\sum_{\mathrm{part}(\bar{t})}\frac{\prod_{\nu=1}^{N-1}\alpha_{\nu}(\bar{t}_{\mathrm{i}}^{\nu})f(\bar{t}_{\mathrm{ii}}^{\nu},\bar{t}_{\mathrm{i}}^{\nu})}{\prod_{\nu=1}^{N-2}f(\bar{t}_{\mathrm{ii}}^{\nu+1},\bar{t}_{\mathrm{i}}^{\nu})}\mathcal{S}_{\bar{\alpha}^{(1)}}(\bar{t}_{\mathrm{i}})\mathcal{S}_{\bar{\alpha}}(\bar{t}_{\mathrm{ii}}).\label{eq:coprodd}
\end{equation}
We fix some of the $\alpha$-functions as $\alpha_{s}(z)=1$ for $s\in\mathfrak{s}^{-}$
for which the sum formula simplifies as
\begin{equation}
\mathcal{S}_{\bar{\alpha}}(\bar{t})=\sum_{\mathrm{part}(\bar{t})}\frac{\prod_{\nu=1}^{N-1}f(\bar{t}_{\mathrm{ii}}^{\nu},\bar{t}_{\mathrm{i}}^{\nu})}{\prod_{\nu=1}^{N-2}f(\bar{t}_{\mathrm{ii}}^{\nu+1},\bar{t}_{\mathrm{i}}^{\nu})}\mathcal{Z}(\bar{t}_{\mathrm{i}})\bar{\mathcal{Z}}(\bar{t}_{\mathrm{ii}})\prod_{s\in\mathfrak{s}^{+}}\alpha_{s}(\bar{t}_{\mathrm{i}}^{s}).
\end{equation}
Substituting to the co-product formula (\ref{eq:coprodd}) we obtain
that
\begin{align}
 & \sum_{\mathrm{part}(\bar{t})}\frac{\prod_{\nu=1}^{N-1}f(\bar{t}_{\mathrm{ii}}^{\nu},\bar{t}_{\mathrm{i}}^{\nu})}{\prod_{\nu=1}^{N-2}f(\bar{t}_{\mathrm{ii}}^{\nu+1},\bar{t}_{\mathrm{i}}^{\nu})}\times\mathcal{Z}(\bar{t}_{\mathrm{i}})\bar{\mathcal{Z}}(\bar{t}_{\mathrm{ii}})\alpha_{1}^{(1)}(\bar{t}_{\mathrm{i}}^{1})\alpha_{N-1}^{(1)}(\bar{t}_{\mathrm{i}}^{N-1})\prod_{s\in\mathfrak{s}^{+}}\alpha_{s}(\bar{t}_{\mathrm{i}}^{s})=\\
 & \sum_{\mathrm{part}(\bar{t})}\frac{\prod_{\nu=1}^{N-1}f(\bar{t}_{\mathrm{ii}}^{\nu},\bar{t}_{\mathrm{i}}^{\nu})f(\bar{t}_{\mathrm{iii}}^{\nu},\bar{t}_{\mathrm{i}}^{\nu})}{\prod_{\nu=1}^{N-2}f(\bar{t}_{\mathrm{ii}}^{\nu+1},\bar{t}_{\mathrm{i}}^{\nu})f(\bar{t}_{\mathrm{iii}}^{\nu+1},\bar{t}_{\mathrm{i}}^{\nu})}\mathcal{S}_{\bar{\alpha}^{(1)}}(\bar{t}_{\mathrm{i}})\left[\frac{\prod_{\nu=1}^{N-1}f(\bar{t}_{\mathrm{iii}}^{\nu},\bar{t}_{\mathrm{ii}}^{\nu})}{\prod_{\nu=1}^{N-2}f(\bar{t}_{\mathrm{iii}}^{\nu+1},\bar{t}_{\mathrm{ii}}^{\nu})}\mathcal{Z}(\bar{t}_{\mathrm{ii}})\bar{\mathcal{Z}}(\bar{t}_{\mathrm{iii}})\prod_{s\in\mathfrak{s}^{+}}\alpha_{s}(\bar{t}_{\mathrm{ii}}^{s})\right]\prod_{s\in\mathfrak{s}^{+}}\alpha_{s}(\bar{t}_{\mathrm{i}}^{s}).\nonumber 
\end{align}
In the lhs and the rhs the sum goes through to the partitions with
condition $\#\bar{t}_{\mathrm{i}}^{s}=\#\bar{t}_{\mathrm{i}}^{M},$
$\#\bar{t}_{\mathrm{ii}}^{s}=\#\bar{t}_{\mathrm{ii}}^{M}$ and $\#\bar{t}_{\mathrm{iii}}^{s}=\#\bar{t}_{\mathrm{iii}}^{M}$
for $s\in\mathfrak{s}^{-}$. 

Let us get the coefficient for the term $\prod_{s\in\mathfrak{s}^{+}}\alpha_{s}(\bar{t}^{s})$.
In the lhs site we have to take $\bar{t}_{\mathrm{ii}}^{s}=\emptyset$
for $s\in\mathfrak{s}^{+}$. We also have the condition $\#\bar{t}_{\mathrm{ii}}^{s}=\#\bar{t}_{\mathrm{ii}}^{M}=\emptyset$
for the remaining $s\in\mathfrak{s}^{-}$. Analogous way we have to
take $\bar{t}_{\mathrm{iii}}^{s}=\emptyset$ in the rhs. We obtain
the condition
\begin{equation}
\mathcal{Z}(\bar{t})\alpha_{1}^{(1)}(\bar{t}^{1})\alpha_{N-1}^{(1)}(\bar{t}^{N-1})=\sum_{\mathrm{part}(\bar{t})}\frac{\prod_{\nu=1}^{N-1}f(\bar{t}_{\mathrm{ii}}^{\nu},\bar{t}_{\mathrm{i}}^{\nu})}{\prod_{\nu=1}^{N-2}f(\bar{t}_{\mathrm{ii}}^{\nu+1},\bar{t}_{\mathrm{i}}^{\nu})}\mathcal{S}_{\bar{\alpha}^{(1)}}(\bar{t}_{\mathrm{i}})\mathcal{Z}(\bar{t}_{\mathrm{ii}}),
\end{equation}
where the sum goes through all the subsets for which the overlap $\mathcal{S}_{\bar{\alpha}^{(1)}}(\bar{t}_{\mathrm{i}})$
is non-vanishing. In the previous section we derived these elementary
overlaps. We obtain that the non-vanish overlaps are
\begin{equation}
\mathcal{S}_{\bar{\alpha}^{(1)}}(\emptyset,\dots,\emptyset)=1,
\end{equation}
and (\ref{eq:elemOvUtw1})
\begin{align}
\mathcal{S}_{\bar{\alpha}^{(1)}}(\{t^{s}\}_{s=1}^{k-1},\emptyset^{\times N-2k+1},\{t^{s}\}_{s=N+1-k}^{N-1}) & =\frac{\mathfrak{b}_{k}}{\mathfrak{b}_{1}}g(-t^{1},t^{N-1})\left(\alpha_{1}^{(1)}(t^{1})\alpha_{N-1}^{(1)}(t^{N-1})-1\right)\times\nonumber \\
 & \times\frac{1}{\prod_{\nu=2}^{k-1}h(t^{\nu},t^{\nu-1})}\frac{1}{\prod_{\nu=N+2-k}^{N-1}h(t^{\nu},t^{\nu-1})},
\end{align}
for $k=2,\dots,M$ and
\begin{equation}
\mathcal{S}_{\bar{\alpha}^{(1)}}(\{t^{s}\}_{s=1}^{N-1})=\frac{1}{g(-t^{M},t^{M})}\frac{K_{N,N}(t^{M})}{K_{N,1}(t^{M})}g(-t^{1},t^{N-1})\left(\alpha_{1}^{(1)}(t^{1})\alpha_{N-1}^{(1)}(t^{N-1})-1\right)\prod_{s=1}^{N-2}\frac{1}{h(t^{s+1},t^{s})}.
\end{equation}
Using this result we obtain the following formula
\begin{align}
\mathcal{Z}(\bar{t}) & =\frac{1}{\alpha_{1}^{(1)}(\bar{t}^{1})\alpha_{N-1}^{(1)}(\bar{t}^{N-1})-1}\times\\
 & \Biggl(\sum_{k=2}^{M}\frac{\mathfrak{b}_{k}}{\mathfrak{b}_{1}}\sum_{\mathrm{part}_{k}(\bar{t})}g(-\bar{t}_{\mathrm{i}}^{1},\bar{t}_{\mathrm{i}}^{N-1})\left(\alpha_{1}^{(1)}(\bar{t}_{\mathrm{i}}^{1})\alpha_{N-1}^{(1)}(\bar{t}_{\mathrm{i}}^{N-1})-1\right)\frac{\prod_{\nu=1}^{N-1}f(\bar{t}_{\mathrm{ii}}^{\nu},\bar{t}_{\mathrm{i}}^{\nu})}{\prod_{\nu=1}^{N-2}f(\bar{t}_{\mathrm{ii}}^{\nu+1},\bar{t}_{\mathrm{i}}^{\nu})}\prod_{s=1}^{N-2}\frac{1}{h(\bar{t}_{\mathrm{i}}^{s+1},\bar{t}_{\mathrm{i}}^{s})}\mathcal{Z}(\bar{t}_{\mathrm{ii}})\nonumber \\
+ & \sum_{\mathrm{part}(\bar{t})}\mathcal{G}(\bar{t}_{\mathrm{i}}^{M})g(-\bar{t}_{\mathrm{i}}^{1},\bar{t}_{\mathrm{i}}^{N-1})\left(\alpha_{1}^{(1)}(\bar{t}_{\mathrm{i}}^{1})\alpha_{N-1}^{(1)}(\bar{t}_{\mathrm{i}}^{N-1})-1\right)\frac{\prod_{\nu=1}^{N-1}f(\bar{t}_{\mathrm{ii}}^{\nu},\bar{t}_{\mathrm{i}}^{\nu})}{\prod_{\nu=1}^{N-2}f(\bar{t}_{\mathrm{ii}}^{\nu+1},\bar{t}_{\mathrm{i}}^{\nu})}\prod_{s=1}^{N-2}\frac{1}{h(\bar{t}_{\mathrm{i}}^{s+1},\bar{t}_{\mathrm{i}}^{s})}\mathcal{Z}(\bar{t}_{\mathrm{ii}})\Biggr),\nonumber 
\end{align}
where in the first line the $\mathrm{part}_{k}(\bar{t})$ denotes
the partitions $\bar{t}^{s}=\bar{t}_{\mathrm{i}}^{s}\cup\bar{t}_{\mathrm{ii}}^{s}$
where $\#\bar{t}_{\mathrm{i}}^{s}=1$ for $s<k$ or $s>N-k$ and $\#\bar{t}_{\mathrm{i}}^{s}=0$
for $k\leq s\leq N-k$. In the second line the summation goes through
the partitions where $\#\bar{t}_{\mathrm{i}}^{s}=1$ for $s=1,\dots,N-1$.
We also used the notation
\begin{equation}
\mathcal{G}(z)=\frac{1}{g(-z,z)}\frac{K_{N,N}(z)}{K_{N,1}(z)}.
\end{equation}

This recursion can be used to eliminate all $t_{k}^{1}$-s to obtain
the HC for the type $(N-2,M-1)$ and continuing the recursion the
HC is completely defined. It is interesting that the rhs of the recursive
definition contain a parameter $\theta$, but the HC is independent
on $\theta$. We can choose special value to obtain $\theta$ independent
formula, e.g. choosing $\theta\to t_{k}^{1}$. However, the present
form of the formula is fully adequate from now on.

Let us renormalize the HC as
\begin{equation}
\mathcal{Z}^{0}(\bar{t})=\frac{1}{\mathcal{G}(\bar{t}^{M})}\mathcal{Z}(\bar{t}).
\end{equation}
For the renormalized HC we have the following recursion
\begin{align}
\mathcal{Z}^{0}(\bar{t}) & =\frac{1}{\alpha_{1}^{(1)}(\bar{t}^{1})\alpha_{N-1}^{(1)}(\bar{t}^{N-1})-1}\times\label{eq:reqZ0}\\
 & \Biggl(\sum_{k=2}^{M}\frac{\mathfrak{b}_{k}}{\mathfrak{b}_{1}}\sum_{\mathrm{part}_{k}(\bar{t})}g(-\bar{t}_{\mathrm{i}}^{1},\bar{t}_{\mathrm{i}}^{N-1})\left(\alpha_{1}^{(1)}(\bar{t}_{\mathrm{i}}^{1})\alpha_{N-1}^{(1)}(\bar{t}_{\mathrm{i}}^{N-1})-1\right)\frac{\prod_{\nu=1}^{N-1}f(\bar{t}_{\mathrm{ii}}^{\nu},\bar{t}_{\mathrm{i}}^{\nu})}{\prod_{\nu=1}^{N-2}f(\bar{t}_{\mathrm{ii}}^{\nu+1},\bar{t}_{\mathrm{i}}^{\nu})}\prod_{s=1}^{N-2}\frac{1}{h(\bar{t}_{\mathrm{i}}^{s+1},\bar{t}_{\mathrm{i}}^{s})}\mathcal{Z}^{0}(\bar{t}_{\mathrm{ii}})\nonumber \\
+ & \sum_{\mathrm{part}(\bar{t})}g(-\bar{t}_{\mathrm{i}}^{1},\bar{t}_{\mathrm{i}}^{N-1})\left(\alpha_{1}^{(1)}(\bar{t}_{\mathrm{i}}^{1})\alpha_{N-1}^{(1)}(\bar{t}_{\mathrm{i}}^{N-1})-1\right)\frac{\prod_{\nu=1}^{N-1}f(\bar{t}_{\mathrm{ii}}^{\nu},\bar{t}_{\mathrm{i}}^{\nu})}{\prod_{\nu=1}^{N-2}f(\bar{t}_{\mathrm{ii}}^{\nu+1},\bar{t}_{\mathrm{i}}^{\nu})}\prod_{s=1}^{N-2}\frac{1}{h(\bar{t}_{\mathrm{i}}^{s+1},\bar{t}_{\mathrm{i}}^{s})}\mathcal{Z}^{0}(\bar{t}_{\mathrm{ii}})\Biggr),\nonumber 
\end{align}
which is independent of the diagonal part of the $K$-matrix.

We can repeat this calculation for the other HC $\bar{\mathcal{Z}}$.
Interchanging the the $\alpha_{s}$-s with $\alpha_{s}^{(1)}$ in
the co-product formula (\ref{eq:coprodd}) we obtain that 
\begin{equation}
\mathcal{S}_{\bar{\alpha}^{mod}}(\bar{t})=\sum_{\mathrm{part}(\bar{t})}\frac{\prod_{\nu=1}^{N-1}f(\bar{t}_{\mathrm{ii}}^{\nu},\bar{t}_{\mathrm{i}}^{\nu})}{\prod_{\nu=1}^{N-2}f(\bar{t}_{\mathrm{ii}}^{\nu+1},\bar{t}_{\mathrm{i}}^{\nu})}\mathcal{S}_{\bar{\alpha}}(\bar{t}_{\mathrm{i}})\mathcal{S}_{\bar{\alpha}^{(1)}}(\bar{t}_{\mathrm{ii}})\alpha_{1}^{(1)}(\bar{t}_{\mathrm{i}}^{1})\alpha_{N-1}^{(1)}(\bar{t}_{\mathrm{i}}^{N-1}).
\end{equation}
Using the sum formulas where $\alpha_{s}(z)=1$ for $s=M+1,\dots,N-M-1$:
\begin{multline}
\sum_{\mathrm{part}(\bar{t})}\frac{\prod_{\nu=1}^{N-1}f(\bar{t}_{\mathrm{ii}}^{\nu},\bar{t}_{\mathrm{i}}^{\nu})}{\prod_{\nu=1}^{N-2}f(\bar{t}_{\mathrm{ii}}^{\nu+1},\bar{t}_{\mathrm{i}}^{\nu})}\mathcal{Z}(\bar{t}_{\mathrm{i}})\bar{\mathcal{Z}}(\bar{t}_{\mathrm{ii}})\alpha_{1}^{(1)}(\bar{t}_{\mathrm{i}}^{1})\alpha_{N-1}^{(1)}(\bar{t}_{\mathrm{i}}^{N-1})\prod_{s\in\mathfrak{s}^{+}}\alpha_{s}(\bar{t}_{\mathrm{i}}^{s})=\\
\sum_{\mathrm{part}(\bar{t})}\frac{\prod_{\nu=1}^{N-1}f(\bar{t}_{\mathrm{iii}}^{\nu},\bar{t}_{\mathrm{i}}^{\nu})f(\bar{t}_{\mathrm{iii}}^{\nu},\bar{t}_{\mathrm{ii}}^{\nu})}{\prod_{\nu=1}^{N-2}f(\bar{t}_{\mathrm{iii}}^{\nu+1},\bar{t}_{\mathrm{i}}^{\nu})f(\bar{t}_{\mathrm{iii}}^{\nu+1},\bar{t}_{\mathrm{ii}}^{\nu})}\left[\frac{\prod_{\nu=1}^{N-1}f(\bar{t}_{\mathrm{ii}}^{\nu},\bar{t}_{\mathrm{i}}^{\nu})}{\prod_{\nu=1}^{N-2}f(\bar{t}_{\mathrm{ii}}^{\nu+1},\bar{t}_{\mathrm{i}}^{\nu})}\mathcal{Z}(\bar{t}_{\mathrm{i}})\bar{\mathcal{Z}}(\bar{t}_{\mathrm{ii}})\prod_{s\in\mathfrak{s}^{+}}\alpha_{s}(\bar{t}_{\mathrm{i}}^{s})\right]\times\\
\times\mathcal{S}_{\bar{\alpha}^{(1)}}(\bar{t}_{\mathrm{iii}})\alpha_{1}^{(1)}(\bar{t}_{\mathrm{i}}^{1})\alpha_{1}^{(1)}(\bar{t}_{\mathrm{ii}}^{1})\alpha_{N-1}^{(1)}(\bar{t}_{\mathrm{i}}^{N-1})\alpha_{N-1}^{(1)}(\bar{t}_{\mathrm{ii}}^{N-1}).
\end{multline}
In the lhs and the rhs the sums go through to the partitions with
condition $\bar{t}_{\mathrm{i}}^{s}=\bar{t}_{\mathrm{i}}^{M},$ $\bar{t}_{\mathrm{ii}}^{s}=\bar{t}_{\mathrm{ii}}^{M}$
and $\bar{t}_{\mathrm{iii}}^{s}=\bar{t}_{\mathrm{iii}}^{M}$ for $s\in\mathfrak{s}^{-}$. 

Let us get the $\alpha$ independent terms. In the lhs site we have
to take $\bar{t}_{\mathrm{i}}^{s}=\emptyset$ for $s\in\mathfrak{s}^{+}$.
We have the condition $\bar{t}_{\mathrm{i}}^{s}=\bar{t}_{\mathrm{i}}^{M}=\emptyset$
for the reaming $s\in\mathfrak{s}^{-}$. Analogous way we have to
take $\bar{t}_{\mathrm{i}}^{s}=\emptyset$ in the rhs. We obtain the
condition
\begin{equation}
\bar{\mathcal{Z}}(\bar{t})=\sum_{\mathrm{part}(\bar{t})}\frac{\prod_{\nu=1}^{N-1}f(\bar{t}_{\mathrm{ii}}^{\nu},\bar{t}_{\mathrm{i}}^{\nu})}{\prod_{\nu=1}^{N-2}f(\bar{t}_{\mathrm{ii}}^{\nu+1},\bar{t}_{\mathrm{i}}^{\nu})}\alpha_{1}^{(1)}(\bar{t}_{\mathrm{i}}^{1})\alpha_{N-1}^{(1)}(\bar{t}_{\mathrm{i}}^{N-1})\bar{\mathcal{Z}}(\bar{t}_{\mathrm{i}})\mathcal{S}_{\bar{\alpha}^{(1)}}(\bar{t}_{\mathrm{ii}}),
\end{equation}
where the sum goes through all the subsets for which the overlap $\mathcal{S}_{\bar{\alpha}^{(1)}}(\bar{t}_{\mathrm{ii}})$
is non-vanishing. Using the explicit overlaps we obtain the following
formula
\begin{align}
 & \bar{\mathcal{Z}}(\bar{t})=\frac{\alpha_{1}^{(1)}(\bar{t}^{1})\alpha_{N-1}^{(1)}(\bar{t}^{N-1})}{1-\alpha_{1}^{(1)}(\bar{t}^{1})\alpha_{N-1}^{(1)}(\bar{t}^{N-1})}\times\\
 & \Biggl(\sum_{k=2}^{M}\frac{\mathfrak{b}_{k}}{\mathfrak{b}_{1}}\sum_{\mathrm{part}_{k}(\bar{t})}g(-\bar{t}_{\mathrm{ii}}^{1},\bar{t}_{\mathrm{ii}}^{N-1})\frac{\alpha_{1}^{(1)}(\bar{t}_{\mathrm{ii}}^{1})\alpha_{N-1}^{(1)}(\bar{t}_{\mathrm{ii}}^{N-1})-1}{\alpha_{1}^{(1)}(\bar{t}_{\mathrm{ii}}^{1})\alpha_{N-1}^{(1)}(\bar{t}_{\mathrm{ii}}^{N-1})}\frac{\prod_{\nu=1}^{N-1}f(\bar{t}_{\mathrm{ii}}^{\nu},\bar{t}_{\mathrm{i}}^{\nu})}{\prod_{\nu=1}^{N-2}f(\bar{t}_{\mathrm{ii}}^{\nu+1},\bar{t}_{\mathrm{i}}^{\nu})h(\bar{t}_{\mathrm{ii}}^{\nu+1},t_{\mathrm{ii}}^{\nu})}\bar{\mathcal{Z}}(\bar{t}_{\mathrm{i}})+\nonumber \\
 & +\sum_{\mathrm{part}(\bar{t})}\mathcal{G}(\bar{t}_{\mathrm{ii}}^{M})g(-\bar{t}_{\mathrm{ii}}^{1},\bar{t}_{\mathrm{ii}}^{N-1})\frac{\alpha_{1}^{(1)}(\bar{t}_{\mathrm{ii}}^{1})\alpha_{N-1}^{(1)}(\bar{t}_{\mathrm{ii}}^{N-1})-1}{\alpha_{1}^{(1)}(\bar{t}_{\mathrm{ii}}^{1})\alpha_{N-1}^{(1)}(\bar{t}_{\mathrm{ii}}^{N-1})}\frac{\prod_{\nu=1}^{N-1}f(\bar{t}_{\mathrm{ii}}^{\nu},\bar{t}_{\mathrm{i}}^{\nu})}{\prod_{\nu=1}^{N-2}f(\bar{t}_{\mathrm{ii}}^{\nu+1},\bar{t}_{\mathrm{i}}^{\nu})h(\bar{t}_{\mathrm{ii}}^{\nu+1},t_{\mathrm{ii}}^{\nu})}\bar{\mathcal{Z}}(\bar{t}_{\mathrm{i}}),\nonumber 
\end{align}
where in the first line the $\mathrm{part}_{k}(\bar{t})$ denotes
the partitions $\bar{t}^{s}=\bar{t}_{\mathrm{i}}^{s}\cup\bar{t}_{\mathrm{ii}}^{s}$
where $\#\bar{t}_{\mathrm{ii}}^{s}=1$ for $s<k$ or $s>N-k$ and
$\#\bar{t}_{\mathrm{ii}}^{s}=0$ for $k\leq s\leq N-k$. In the second
line the summation goes through the partitions where $\#\bar{t}_{\mathrm{ii}}^{s}=1$
for $s=1,\dots,N-1$. This recursion completely defines the HC and
it is depend on a parameter $\theta$, but the HC is independent on
it. 

Let us renormalize the HC as
\begin{equation}
\bar{\mathcal{Z}}^{0}(\bar{t})=\frac{1}{\mathcal{G}(\bar{t}^{M})}\bar{\mathcal{Z}}(\bar{t}).
\end{equation}
For the renormalized HC we have the following recursion
\begin{align}
\bar{\mathcal{Z}}^{0}(\bar{t}) & =\frac{\alpha_{1}^{(1)}(\bar{t}^{1})\alpha_{N-1}^{(1)}(\bar{t}^{N-1})}{1-\alpha_{1}^{(1)}(\bar{t}^{1})\alpha_{N-1}^{(1)}(\bar{t}^{N-1})}\times\nonumber \\
 & \Biggl(\sum_{k=2}^{M}\frac{\mathfrak{b}_{k}}{\mathfrak{b}_{1}}\sum_{\mathrm{part}_{k}(\bar{t})}g(-\bar{t}_{\mathrm{ii}}^{1},\bar{t}_{\mathrm{ii}}^{N-1})\frac{\alpha_{1}^{(1)}(\bar{t}_{\mathrm{ii}}^{1})\alpha_{N-1}^{(1)}(\bar{t}_{\mathrm{ii}}^{N-1})-1}{\alpha_{1}^{(1)}(\bar{t}_{\mathrm{ii}}^{1})\alpha_{N-1}^{(1)}(\bar{t}_{\mathrm{ii}}^{N-1})}\frac{\prod_{\nu=1}^{N-1}f(\bar{t}_{\mathrm{ii}}^{\nu},\bar{t}_{\mathrm{i}}^{\nu})}{\prod_{\nu=1}^{N-2}f(\bar{t}_{\mathrm{ii}}^{\nu+1},\bar{t}_{\mathrm{i}}^{\nu})h(\bar{t}_{\mathrm{ii}}^{\nu+1},t_{\mathrm{ii}}^{\nu})}\bar{\mathcal{Z}}^{0}(\bar{t}_{\mathrm{i}})+\nonumber \\
 & +\sum_{\mathrm{part}(\bar{t})}g(-\bar{t}_{\mathrm{ii}}^{1},\bar{t}_{\mathrm{ii}}^{N-1})\frac{\alpha_{1}^{(1)}(\bar{t}_{\mathrm{ii}}^{1})\alpha_{N-1}^{(1)}(\bar{t}_{\mathrm{ii}}^{N-1})-1}{\alpha_{1}^{(1)}(\bar{t}_{\mathrm{ii}}^{1})\alpha_{N-1}^{(1)}(\bar{t}_{\mathrm{ii}}^{N-1})}\frac{\prod_{\nu=1}^{N-1}f(\bar{t}_{\mathrm{ii}}^{\nu},\bar{t}_{\mathrm{i}}^{\nu})}{\prod_{\nu=1}^{N-2}f(\bar{t}_{\mathrm{ii}}^{\nu+1},\bar{t}_{\mathrm{i}}^{\nu})h(\bar{t}_{\mathrm{ii}}^{\nu+1},t_{\mathrm{ii}}^{\nu})}\bar{\mathcal{Z}}^{0}(\bar{t}_{\mathrm{i}})\Biggr),\label{eq:reqZb0}
\end{align}
which is independent of the diagonal part of the $K$-matrix.

\subsection*{Twisted case\label{subsec:Recursion-Twisted-case}}

Let us define the tensor product Hilbert space as $\mathcal{H}^{mod}=\mathcal{H}^{(1)}\otimes\mathcal{H}$
where the $\mathcal{H}^{(1)}$ quantum space is defined by the monodromy
matrix (\ref{eq:elemMonTw}) for which the $\alpha$-functions are
\begin{equation}
\begin{split}\alpha_{2}^{(1)}(z) & =\frac{f(z,\theta)}{f(-z-2c,\theta)},\quad\alpha_{1}^{(1)}(z)=1,\\
\alpha_{s}^{(1)}(z) & =1,\quad\text{for }s=3,\dots,N-1.
\end{split}
\end{equation}
Since the HC $\mathcal{Z}$ is independent on the $\alpha$-functions,
we can derive a recursion for the HC $\mathcal{Z}$ from the co-product
formula (\ref{eq:coprodd}). Fixing some of the $\alpha$-functions
as $\alpha_{2a-1}(z)=1$ for $a=1,\dots,N/2$, the sum formula simplifies
as
\begin{equation}
\mathcal{S}_{\bar{\alpha}}(\bar{t})=\sum_{\mathrm{part}(\bar{t})}\frac{\prod_{\nu=1}^{N-1}f(\bar{t}_{\mathrm{ii}}^{\nu},\bar{t}_{\mathrm{i}}^{\nu})}{\prod_{\nu=1}^{N-2}f(\bar{t}_{\mathrm{ii}}^{\nu+1},\bar{t}_{\mathrm{i}}^{\nu})}\mathcal{Z}(\bar{t}_{\mathrm{i}})\bar{\mathcal{Z}}(\bar{t}_{\mathrm{ii}})\prod_{a=1}^{N/2-1}\alpha_{2a}(\bar{t}_{\mathrm{i}}^{2a}).
\end{equation}
Substituting back to the co-product formula (\ref{eq:coprodd}) we
obtain that
\begin{multline}
\sum_{\mathrm{part}(\bar{t})}\frac{\prod_{\nu=1}^{N-1}f(\bar{t}_{\mathrm{ii}}^{\nu},\bar{t}_{\mathrm{i}}^{\nu})}{\prod_{\nu=1}^{N-2}f(\bar{t}_{\mathrm{ii}}^{\nu+1},\bar{t}_{\mathrm{i}}^{\nu})}\times\mathcal{Z}(\bar{t}_{\mathrm{i}})\bar{\mathcal{Z}}(\bar{t}_{\mathrm{ii}})\alpha_{2}^{(1)}(\bar{t}_{\mathrm{i}}^{2})\prod_{a=1}^{N/2-1}\alpha_{2a}(\bar{t}_{\mathrm{i}}^{2a})=\sum_{\mathrm{part}(\bar{t})}\frac{\prod_{\nu=1}^{N-1}f(\bar{t}_{\mathrm{ii}}^{\nu},\bar{t}_{\mathrm{i}}^{\nu})f(\bar{t}_{\mathrm{iii}}^{\nu},\bar{t}_{\mathrm{i}}^{\nu})}{\prod_{\nu=1}^{N-2}f(\bar{t}_{\mathrm{ii}}^{\nu+1},\bar{t}_{\mathrm{i}}^{\nu})f(\bar{t}_{\mathrm{iii}}^{\nu+1},\bar{t}_{\mathrm{i}}^{\nu})}\times\\
\times\mathcal{S}_{\bar{\alpha}^{(1)}}(\bar{t}_{\mathrm{i}})\left[\frac{\prod_{\nu=1}^{N-1}f(\bar{t}_{\mathrm{iii}}^{\nu},\bar{t}_{\mathrm{ii}}^{\nu})}{\prod_{\nu=1}^{N-2}f(\bar{t}_{\mathrm{iii}}^{\nu+1},\bar{t}_{\mathrm{ii}}^{\nu})}\mathcal{Z}(\bar{t}_{\mathrm{ii}})\bar{\mathcal{Z}}(\bar{t}_{\mathrm{iii}}^{s})\prod_{a=1}^{N/2-1}\alpha_{2a}(\bar{t}_{\mathrm{ii}}^{2a})\right]\prod_{a=1}^{N/2-1}\alpha_{2a}(\bar{t}_{\mathrm{i}}^{2a}).
\end{multline}
In the lhs and the rhs the sum goes through to the partitions with
condition $\#\bar{t}_{\mathrm{i}}^{2a-1}=\frac{\#\bar{t}_{\mathrm{i}}^{2a-2}+\#\bar{t}_{\mathrm{i}}^{2a}}{2},$
$\#\bar{t}_{\mathrm{ii}}^{2a-1}=\frac{\#\bar{t}_{\mathrm{ii}}^{2a-2}+\#\bar{t}_{\mathrm{ii}}^{2a}}{2}$
and $\#\bar{t}_{\mathrm{iii}}^{2a-1}=\frac{\#\bar{t}_{\mathrm{iii}}^{2a-2}+\#\bar{t}_{\mathrm{iii}}^{2a}}{2}$
for $a=1,\dots,N/2$. 

Let us get the coefficient of the term $\prod_{a=1}^{N/2-1}\alpha_{2a}(\bar{t}^{2a})$.
In the lhs site we have to take $\bar{t}_{\mathrm{ii}}^{2a}=\emptyset$
for $a=1,\dots,N/2$ therefore $\#\bar{t}_{\mathrm{ii}}^{2a-1}=\frac{\#\bar{t}_{\mathrm{ii}}^{2a-2}+\#\bar{t}_{\mathrm{ii}}^{2a}}{2}=0$
for the reaming sets. Analogous way we have to take $\bar{t}_{\mathrm{iii}}^{s}=\emptyset$
in the rhs. We obtain the condition
\begin{equation}
\mathcal{Z}(\bar{t})\alpha_{2}^{(1)}(\bar{t}^{2})=\sum_{\mathrm{part}(\bar{t})}\frac{\prod_{\nu=1}^{N-1}f(\bar{t}_{\mathrm{ii}}^{\nu},\bar{t}_{\mathrm{i}}^{\nu})}{\prod_{\nu=1}^{N-2}f(\bar{t}_{\mathrm{ii}}^{\nu+1},\bar{t}_{\mathrm{i}}^{\nu})}\mathcal{S}_{\bar{\alpha}^{(1)}}(\bar{t}_{\mathrm{i}})\mathcal{Z}(\bar{t}_{\mathrm{ii}}),
\end{equation}
where the sum goes through all the subsets for which the overlap $\mathcal{S}_{\bar{\alpha}^{(1)}}(\bar{t}_{\mathrm{i}})$
is non-vanishing. In the previous section we derived these elementary
overlaps. We obtain that the non-vanish overlaps are
\begin{equation}
\mathcal{S}_{\bar{\alpha}^{(1)}}(\emptyset,\dots,\emptyset)=1,
\end{equation}
and
\begin{equation}
\mathcal{S}_{\bar{\alpha}^{(1)}}(\{t^{1}\},\{\bar{t}^{s}\}_{s=2}^{2a-2},\{t^{2a-1}\},\emptyset,\dots\emptyset)=\frac{x_{a}}{x_{1}}g(-t_{1}^{2}-2c,t_{2}^{2})(\alpha_{2}(\bar{t}^{2})-1)\frac{\prod_{s=2}^{2a-2}h(t_{1}^{s},t_{2}^{s})h(t_{2}^{s},t_{1}^{s})}{\prod_{s=1}^{2a-2}h(\bar{t}^{s+1},\bar{t}^{s})},
\end{equation}
for $a=2,\dots,N/2$. Using this result we obtain the following formula
\begin{equation}
\mathcal{Z}(\bar{t})=\frac{1}{\alpha_{2}^{(1)}(\bar{t}^{2})-1}\sum_{a=2}^{M}\sum_{\mathrm{part}_{a}(\bar{t})}\frac{\prod_{\nu=1}^{N-1}f(\bar{t}_{\mathrm{ii}}^{\nu},\bar{t}_{\mathrm{i}}^{\nu})}{\prod_{\nu=1}^{N-2}f(\bar{t}_{\mathrm{ii}}^{\nu+1},\bar{t}_{\mathrm{i}}^{\nu})}\mathcal{S}_{\bar{\alpha}^{(1)}}(\bar{t}_{\mathrm{i}})\mathcal{Z}(\bar{t}_{\mathrm{ii}}),\label{eq:recZTw}
\end{equation}
where the $\mathrm{part}_{a}(\bar{t})$ denotes the partitions $\bar{t}^{b}=\bar{t}_{\mathrm{i}}^{b}\cup\bar{t}_{\mathrm{ii}}^{b}$
where $\#\bar{t}_{\mathrm{i}}^{1}=\#\bar{t}_{\mathrm{i}}^{2a-1}=1$,
$\#\bar{t}_{\mathrm{i}}^{b}=2$ for $2\leq b\leq2a-2$ and $\#\bar{t}_{\mathrm{i}}^{b}=0$
for $2a-1<b$.

We can repeat this calculation for the other HC $\bar{\mathcal{Z}}$.
Interchanging the $\alpha_{s}$-s with $\alpha_{s}^{(1)}$ in the
co-product formula (\ref{eq:coprodd}) we obtain that 
\begin{equation}
\mathcal{S}_{\bar{\alpha}^{mod}}(\bar{t})=\sum_{\mathrm{part}(\bar{t})}\frac{\prod_{\nu=1}^{N-1}f(\bar{t}_{\mathrm{ii}}^{\nu},\bar{t}_{\mathrm{i}}^{\nu})}{\prod_{\nu=1}^{N-2}f(\bar{t}_{\mathrm{ii}}^{\nu+1},\bar{t}_{\mathrm{i}}^{\nu})}\mathcal{S}_{\bar{\alpha}}(\bar{t}_{\mathrm{i}})\mathcal{S}_{\bar{\alpha}^{(1)}}(\bar{t}_{\mathrm{ii}})\alpha_{2}^{(1)}(\bar{t}_{\mathrm{i}}^{2}).
\end{equation}
Using the sum formulas where $\alpha_{2a-1}(z)=1$:
\begin{multline}
\sum_{\mathrm{part}(\bar{t})}\frac{\prod_{\nu=1}^{N-1}f(\bar{t}_{\mathrm{ii}}^{\nu},\bar{t}_{\mathrm{i}}^{\nu})}{\prod_{\nu=1}^{N-2}f(\bar{t}_{\mathrm{ii}}^{\nu+1},\bar{t}_{\mathrm{i}}^{\nu})}\times\mathcal{Z}(\bar{t}_{\mathrm{i}})\bar{\mathcal{Z}}(\bar{t}_{\mathrm{ii}})\alpha_{2}^{(1)}(\bar{t}_{\mathrm{i}}^{2})\prod_{a=1}^{N/2-1}\alpha_{2a}(\bar{t}_{\mathrm{i}}^{2a})=\sum_{\mathrm{part}(\bar{t})}\frac{\prod_{\nu=1}^{N-1}f(\bar{t}_{\mathrm{iii}}^{\nu},\bar{t}_{\mathrm{i}}^{\nu})f(\bar{t}_{\mathrm{iii}}^{\nu},\bar{t}_{\mathrm{ii}}^{\nu})}{\prod_{\nu=1}^{N-2}f(\bar{t}_{\mathrm{iii}}^{\nu+1},\bar{t}_{\mathrm{i}}^{\nu})f(\bar{t}_{\mathrm{iii}}^{\nu+1},\bar{t}_{\mathrm{ii}}^{\nu})}\times\\
\times\left[\frac{\prod_{\nu=1}^{N-1}f(\bar{t}_{\mathrm{ii}}^{\nu},\bar{t}_{\mathrm{i}}^{\nu})}{\prod_{\nu=1}^{N-2}f(\bar{t}_{\mathrm{ii}}^{\nu+1},\bar{t}_{\mathrm{i}}^{\nu})}\times\mathcal{Z}(\bar{t}_{\mathrm{i}})\bar{\mathcal{Z}}(\bar{t}_{\mathrm{ii}})\prod_{a=1}^{N/2-1}\alpha_{2a}(\bar{t}_{\mathrm{i}}^{2a})\right]\mathcal{S}_{\bar{\alpha}^{(1)}}(\bar{t}_{\mathrm{iii}})\alpha_{2}^{(1)}(\bar{t}_{\mathrm{i}}^{2})\alpha_{2}^{(1)}(\bar{t}_{\mathrm{ii}}^{2}).
\end{multline}
In the lhs and the rhs the sums go through to the partitions with
condition $\#\bar{t}_{\mathrm{i}}^{2a-1}=\frac{\#\bar{t}_{\mathrm{i}}^{2a-2}+\#\bar{t}_{\mathrm{i}}^{2a}}{2},$
$\#\bar{t}_{\mathrm{ii}}^{2a-1}=\frac{\#\bar{t}_{\mathrm{ii}}^{2a-2}+\#\bar{t}_{\mathrm{ii}}^{2a}}{2}$
and $\#\bar{t}_{\mathrm{iii}}^{2a-1}=\frac{\#\bar{t}_{\mathrm{iii}}^{2a-2}+\#\bar{t}_{\mathrm{iii}}^{2a}}{2}$
for $a=1,\dots,N/2$. 

Let us get the $\alpha$ independent terms. In the lhs site we have
to take $\bar{t}_{\mathrm{i}}^{2a}=\emptyset$ for $a=1,\dots,N/2$
therefore $\#\bar{t}_{\mathrm{i}}^{2a-1}=\frac{\#\bar{t}_{\mathrm{i}}^{2a-2}+\#\bar{t}_{\mathrm{i}}^{2a}}{2}=0$
for the reaming sets. Analogous way we have to take $\bar{t}_{\mathrm{i}}^{s}=\emptyset$
in the rhs. We obtain the condition
\begin{equation}
\bar{\mathcal{Z}}(\bar{t})=\sum_{\mathrm{part}(\bar{t})}\frac{\prod_{\nu=1}^{N-1}f(\bar{t}_{\mathrm{ii}}^{\nu},\bar{t}_{\mathrm{i}}^{\nu})}{\prod_{\nu=1}^{N-2}f(\bar{t}_{\mathrm{ii}}^{\nu+1},\bar{t}_{\mathrm{i}}^{\nu})}\alpha_{2}^{(1)}(\bar{t}_{\mathrm{i}}^{2})\bar{\mathcal{Z}}(\bar{t}_{\mathrm{i}})\mathcal{S}_{\bar{\alpha}^{(1)}}(\bar{t}_{\mathrm{ii}}),
\end{equation}
where the sum goes through all the subsets for which the overlap $\mathcal{S}_{\bar{\alpha}^{(1)}}(\bar{t}_{\mathrm{ii}})$
is non-vanishing therefore
\begin{equation}
\bar{\mathcal{Z}}(\bar{t})=\frac{1}{1-\alpha_{2}^{(1)}(\bar{t}^{2})}\sum_{a=2}^{N/2}\sum_{\mathrm{part}_{a}(\bar{t})}\alpha_{2}^{(1)}(\bar{t}_{\mathrm{i}}^{2})\frac{\prod_{\nu=1}^{N-1}f(\bar{t}_{\mathrm{ii}}^{\nu},\bar{t}_{\mathrm{i}}^{\nu})}{\prod_{\nu=1}^{N-2}f(\bar{t}_{\mathrm{ii}}^{\nu+1},\bar{t}_{\mathrm{i}}^{\nu})}\bar{\mathcal{Z}}(\bar{t}_{\mathrm{i}})\mathcal{S}_{\bar{\alpha}^{(1)}}(\bar{t}_{\mathrm{ii}}),\label{eq:recZbTw}
\end{equation}
where $\mathrm{part}_{a}(\bar{t})$ denotes the partitions $\bar{t}^{b}=\bar{t}_{\mathrm{i}}^{b}\cup\bar{t}_{\mathrm{ii}}^{b}$
where $\#\bar{t}_{\mathrm{ii}}^{1}=\#\bar{t}_{\mathrm{ii}}^{2a-1}=1$,
$\#\bar{t}_{\mathrm{ii}}^{b}=2$ for $2\leq b\leq2a-2$ and $\#\bar{t}_{\mathrm{ii}}^{b}=0$
for $2a-1<b$.

We close this section by the proof of the identity
\begin{equation}
\mathcal{Z}(\bar{t})=\frac{1}{\prod_{s=1}^{N-2}f(\bar{t}^{s+1},\bar{t}^{s})}\bar{\mathcal{Z}}(\pi^{c}(\bar{t})).
\end{equation}
Let us apply the the recursion for
\begin{equation}
\bar{\mathcal{Z}}(\pi^{c}(\bar{t}))=\frac{\alpha_{2}^{(1)}(\bar{t}^{2})}{\alpha_{2}^{(1)}(\bar{t}^{2})-1}\sum_{a=2}^{N/2}\sum_{\mathrm{part}_{a}(\bar{t})}\frac{\prod_{\nu=1}^{N-1}f(\bar{t}_{\mathrm{i}}^{\nu},\bar{t}_{\mathrm{ii}}^{\nu})\prod_{\nu=1}^{N-2}f(\bar{t}_{\mathrm{ii}}^{\nu+1},\bar{t}_{\mathrm{i}}^{\nu})}{\alpha_{2}^{(1)}(\bar{t}_{\mathrm{i}}^{2})}\bar{\mathcal{Z}}(\pi^{c}(\bar{t}_{\mathrm{i}}))\mathcal{S}_{\bar{\alpha}^{(1)}}(\pi^{c}(\bar{t}_{\mathrm{ii}})).
\end{equation}
For the elementary overlap we have 
\begin{equation}
\mathcal{S}_{\bar{\alpha}^{(1)}}(\pi^{c}(\bar{t}))=\frac{1}{\alpha_{2}(\bar{t}^{2})}\prod_{s=1}^{2a-2}f(\bar{t}^{s+1},\bar{t}^{s})\mathcal{S}_{\bar{\alpha}^{(1)}}(\bar{t}).
\end{equation}
Substituting back we obtain that
\begin{equation}
\bar{\mathcal{Z}}(\pi^{c}(\bar{t}))=\prod_{\nu=1}^{N-2}f(\bar{t}^{\nu+1},\bar{t}^{\nu})\frac{1}{\alpha_{2}^{(1)}(\bar{t}^{2})-1}\sum_{a=2}^{N/2}\sum_{\mathrm{part}_{a}(\bar{t})}\frac{\prod_{\nu=1}^{N-1}f(\bar{t}_{\mathrm{i}}^{\nu},\bar{t}_{\mathrm{ii}}^{\nu})}{\prod_{\nu=1}^{N-2}f(\bar{t}_{\mathrm{i}}^{\nu+1},\bar{t}_{\mathrm{ii}}^{\nu})}\mathcal{Z}(\bar{t}_{\mathrm{i}})\mathcal{S}_{\bar{\alpha}^{(1)}}(\bar{t}_{\mathrm{ii}}).
\end{equation}
Interchanging the notation we obtain
\begin{align}
\bar{\mathcal{Z}}(\pi^{c}(\bar{t})) & =\prod_{\nu=1}^{N-2}f(\bar{t}^{\nu+1},\bar{t}^{\nu})\frac{1}{\alpha_{2}^{(1)}(\bar{t}^{2})-1}\sum_{a=2}^{N/2}\sum_{\mathrm{part}_{a}(\bar{t})}\frac{\prod_{\nu=1}^{N-1}f(\bar{t}_{\mathrm{ii}}^{\nu},\bar{t}_{\mathrm{i}}^{\nu})}{\prod_{\nu=1}^{N-2}f(\bar{t}_{\mathrm{ii}}^{\nu+1},\bar{t}_{\mathrm{i}}^{\nu})}\mathcal{S}_{\bar{\alpha}^{(1)}}(\bar{t}_{\mathrm{i}})\mathcal{Z}(\bar{t}_{\mathrm{ii}})\nonumber \\
 & =\prod_{\nu=1}^{N-2}f(\bar{t}^{\nu+1},\bar{t}^{\nu})\mathcal{Z}(\bar{t}),
\end{align}
and that is we wanted to prove.

\section{Poles of the twisted HC-s\label{sec:Poles-of-the}}

In this section we derive the residues of the poles at the pair structure
limit $t_{k}^{2a}+t_{l}^{2a}+2ac\to0$ in the twisted HC-s $\bar{\mathcal{Z}}$
\begin{align}
\bar{\mathcal{Z}}(\bar{t}) & \to g(t_{l}^{2a},-t_{k}^{2a}-2ac)\left[\frac{x_{a+1}}{x_{a}}h(t_{k}^{2a},t_{l}^{2a})h(t_{l}^{2a},t_{k}^{2a})\right]\frac{f(\bar{\tau}^{2a},t_{k}^{2a})f(\bar{\tau}^{2a},t_{l}^{2a})}{f(\bar{\tau}^{2a+1},t_{k}^{2a})f(\bar{\tau}^{2a+1},t_{l}^{2a})}\nonumber \\
 & \sum_{\mathrm{part}(\bar{\tau}^{2a-1},\bar{\tau}^{2a+1})}\bar{\mathcal{Z}}(\bar{\tau}_{\mathrm{i}})\frac{f(\bar{\tau}_{\mathrm{i}}^{2a-1},\bar{\tau}_{\mathrm{iii}}^{2a-1})}{f(\bar{\tau}^{2a},\bar{\tau}_{\mathrm{iii}}^{2a-1})}\frac{f(\bar{\tau}_{\mathrm{i}}^{2a+1},\bar{\tau}_{\mathrm{iii}}^{2a+1})}{f(\bar{\tau}^{2a+2},\bar{\tau}_{\mathrm{iii}}^{2a+1})}\frac{g(\bar{\tau}_{\mathrm{iii}}^{2a+1},t_{k}^{2a})g(\bar{\tau}_{\mathrm{iii}}^{2a+1},t_{l}^{2a})}{h(t_{k}^{2a},\bar{\tau}_{\mathrm{iii}}^{2a-1})h(t_{l}^{2a},\bar{\tau}_{\mathrm{iii}}^{2a-1})}+reg.,\label{eq:poleZb}
\end{align}
where $\bar{\tau}=\bar{t}\backslash\{t_{k}^{2},t_{l}^{2}\}$. The
summation goes thought the partitions $\bar{\tau}^{s}=\bar{\tau}_{\mathrm{i}}^{s}\cup\bar{\tau}_{\mathrm{iii}}^{s}$
where $\#\bar{\tau}_{\mathrm{iii}}^{s}=\delta_{s,2a-1}+\delta_{s,2a+1}$. 

At first we derive the formula for $a=1$. We start with the $r_{2}=2$
case. We start with the recursion for the off-shell Bethe state (\ref{eq:rect2})
for the spin chains where $\alpha_{2a-1}(z)=1$ which is equivalent
to the selection rules $\#\bar{t}^{2a-1}=\frac{\#\bar{t}^{2a-2}+\#\bar{t}^{2a}}{2}$:
\begin{align}
 & \mathbb{B}(\{t^{1}\},\{z,t^{2}\},\{\bar{t}^{s}\}_{s=3}^{N-1})=\\
 & =\sum_{j=3}^{N}\sum_{\mathrm{part}(\bar{t})}\frac{T_{2,j}(z)}{\lambda_{3}(z)}\mathbb{B}(\{t^{1}\},\{t^{2}\},\{\bar{t}_{\textsc{ii}}^{s}\}_{s=3}^{j-1},\{\bar{t}^{s}\}_{s=j}^{N-1})\prod_{a=2}^{\frac{j-1}{2}}\alpha_{2a}(\bar{t}_{\textsc{i}}^{2a})\frac{1}{f(z,t^{1})}\frac{\prod_{s=3}^{j-1}g(\bar{t}_{\textsc{i}}^{s},\bar{t}_{\textsc{i}}^{s-1})f(\bar{t}_{\textsc{ii}}^{s},\bar{t}_{\textsc{i}}^{s})}{\prod_{s=2}^{j-1}f(\bar{t}^{s+1},\bar{t}_{\textsc{i}}^{s})}\nonumber \\
 & +\sum_{j=3}^{N}\sum_{\mathrm{part}(\bar{t})}\frac{T_{1,j}(z)}{\lambda_{3}(z)}\mathbb{B}(\emptyset,\{t^{2}\},\{\bar{t}_{\textsc{ii}}^{s}\}_{s=3}^{j-1},\{\bar{t}^{s}\}_{s=j}^{N-1})\prod_{a=2}^{\frac{j-1}{2}}\alpha_{2a}(\bar{t}_{\textsc{i}}^{2a})\frac{1}{h(z,t^{1})}\frac{\prod_{s=3}^{j-1}g(\bar{t}_{\textsc{i}}^{s},\bar{t}_{\textsc{i}}^{s-1})f(\bar{t}_{\textsc{ii}}^{s},\bar{t}_{\textsc{i}}^{s})}{\prod_{s=2}^{j-1}f(\bar{t}^{s+1},\bar{t}_{\textsc{i}}^{s})},\nonumber 
\end{align}
where $\bar{t}_{\textsc{i}}^{2}=\{z\}$. From the sum formula (\ref{eq:sumFormulaTw})
we can see that the $\alpha_{2a}$ independent term in the overlap
is just the HC $\bar{\mathcal{Z}}(\bar{t})$. In this calculation
we want to derive the pole structure for these quantities therefore
we use the notation $\cong$ for the equality up to $\alpha$-dependent
terms i.e.
\begin{equation}
\langle\Psi|\mathbb{B}(\bar{t})\cong\bar{\mathcal{Z}}(\bar{t}).
\end{equation}
Dropping the $\alpha_{2a}(t_{k}^{2a})$ dependent terms we obtain
that
\begin{align}
\mathbb{B}(\{t^{1}\},\{z,t^{2}\},\{\bar{t}^{s}\}_{s=3}^{N-1}) & \cong\frac{1}{\lambda_{3}(z)}T_{2,3}(z)\mathbb{B}(\{t^{1}\},\{t^{2}\},\{\bar{t}^{s}\}_{s=3}^{N-1})\frac{1}{f(z,t^{1})}\frac{1}{f(\bar{t}^{3},z)}\nonumber \\
 & +\frac{1}{\lambda_{3}(z)}\sum_{\mathrm{part}(\bar{t}^{3})}T_{2,4}(z)\mathbb{B}(\{t^{1}\},\{t^{2}\},\bar{t}_{\textsc{ii}}^{3},\{\bar{t}^{s}\}_{s=4}^{N-1})\frac{1}{f(z,t^{1})}\frac{g(\bar{t}_{\textsc{i}}^{3},\bar{t}_{\textsc{i}}^{2})f(\bar{t}_{\textsc{ii}}^{3},\bar{t}_{\textsc{i}}^{3})}{f(\bar{t}^{3},\bar{t}_{\textsc{i}}^{2})f(\bar{t}^{4},\bar{t}_{\textsc{i}}^{3})}\nonumber \\
 & +\frac{1}{\lambda_{3}(z)}T_{1,3}(z)\mathbb{B}(\emptyset,\{t^{2}\},\{\bar{t}^{s}\}_{s=3}^{N-1})\frac{1}{h(z,t^{1})}\frac{1}{f(\bar{t}^{3},z)}\\
 & +\frac{1}{\lambda_{3}(z)}\sum_{\mathrm{part}(\bar{t}^{3})}T_{1,4}(z)\mathbb{B}(\emptyset,\{t^{2}\},\bar{t}_{\textsc{ii}}^{3},\{\bar{t}^{s}\}_{s=4}^{N-1})\frac{1}{h(z,t^{1})}\frac{g(\bar{t}_{\textsc{i}}^{3},\bar{t}_{\textsc{i}}^{2})f(\bar{t}_{\textsc{ii}}^{3},\bar{t}_{\textsc{i}}^{3})}{f(\bar{t}^{3},\bar{t}_{\textsc{i}}^{2})f(\bar{t}^{4},\bar{t}_{\textsc{i}}^{3})}.\nonumber 
\end{align}
We can us the KT-relation
\begin{equation}
\begin{split}\langle\Psi|T_{2,3}(z) & =-\lambda_{0}(z)\frac{x_{2}}{x_{1}}\langle\Psi|\widehat{T}_{N,N-3}(-z),\qquad\langle\Psi|T_{1,3}(z)=\lambda_{0}(z)\frac{x_{2}}{x_{1}}\langle\Psi|\widehat{T}_{N-1,N-3}(-z),\\
\langle\Psi|T_{2,4}(z) & =\lambda_{0}(z)\frac{x_{2}}{x_{1}}\langle\Psi|\widehat{T}_{N,N-2}(-z),\qquad\langle\Psi|T_{1,4}(z)=-\lambda_{0}(z)\frac{x_{2}}{x_{1}}\langle\Psi|\widehat{T}_{N-1,N-2}(-z).
\end{split}
\end{equation}
Therefore
\begin{align}
 & \langle\Psi|\mathbb{B}(\{t^{1}\},\{z,t^{2}\},\{\bar{t}^{s}\}_{s=3}^{N-1})\cong\nonumber \\
 & -\frac{\lambda_{0}(z)}{\lambda_{3}(z)}\frac{x_{2}}{x_{1}}\langle\Psi|\widehat{T}_{N,N-3}(-z)\mathbb{B}(\{t^{1}\},\{t^{2}\},\{\bar{t}^{s}\}_{s=3}^{N-1})\frac{1}{f(z,t^{1})}\frac{1}{f(\bar{t}^{3},z)}\nonumber \\
 & +\frac{\lambda_{0}(z)}{\lambda_{3}(z)}\frac{x_{2}}{x_{1}}\sum_{\mathrm{part}(\bar{t})}\langle\Psi|\widehat{T}_{N,N-2}(-z)\mathbb{B}(\{t^{1}\},\{t^{2}\},\bar{t}_{\textsc{ii}}^{3}\{\bar{t}^{s}\}_{s=4}^{N-1})\frac{1}{f(z,t^{1})}\frac{g(\bar{t}_{\textsc{i}}^{3},\bar{t}_{\textsc{i}}^{2})f(\bar{t}_{\textsc{ii}}^{3},\bar{t}_{\textsc{i}}^{3})}{f(\bar{t}^{3},\bar{t}_{\textsc{i}}^{2})f(\bar{t}^{4},\bar{t}_{\textsc{i}}^{3})}\\
 & +\frac{\lambda_{0}(z)}{\lambda_{3}(z)}\frac{x_{2}}{x_{1}}\langle\Psi|\widehat{T}_{N-1,N-3}(-z)\mathbb{B}(\emptyset,\{t^{2}\},\{\bar{t}^{s}\}_{s=3}^{N-1})\frac{1}{h(z,t^{1})}\frac{1}{f(\bar{t}^{3},z)}\nonumber \\
 & -\frac{\lambda_{0}(z)}{\lambda_{3}(z)}\frac{x_{2}}{x_{1}}\sum_{\mathrm{part}(\bar{t})}\langle\Psi|\widehat{T}_{N-1,N-2}(-z)\mathbb{B}(\emptyset,\{t^{2}\},\bar{t}_{\textsc{ii}}^{3}\{\bar{t}^{s}\}_{s=4}^{N-1})\frac{1}{h(z,t^{1})}\frac{g(\bar{t}_{\textsc{i}}^{3},\bar{t}_{\textsc{i}}^{2})f(\bar{t}_{\textsc{ii}}^{3},\bar{t}_{\textsc{i}}^{3})}{f(\bar{t}^{3},\bar{t}_{\textsc{i}}^{2})f(\bar{t}^{4},\bar{t}_{\textsc{i}}^{3})}.\nonumber 
\end{align}
We can apply the action formula (\ref{eq:actTw})
\begin{align}
 & \frac{\lambda_{0}(z)}{\lambda_{3}(z)}\widehat{T}_{N,N-3}(-z)\mathbb{B}(\{t^{1}\},\{t^{2}\},\{\bar{t}^{s}\}_{s=3}^{N-1})=\nonumber \\
 & -\alpha_{2}(z)\sum_{\mathrm{part}(\bar{w})}\alpha_{2}(\bar{w}_{\textsc{iii}}^{2})\mathbb{B}(\emptyset,\emptyset,\{\bar{w}_{\textsc{ii}}^{s}\}_{s=3}^{N-1})\frac{\prod_{s=2}^{N-1}f(\bar{t}^{s-1}-c,\bar{t}^{s})}{\prod_{s=4}^{N-1}f(\bar{w}_{\textsc{ii}}^{s-1}-c,\bar{w}_{\textsc{ii}}^{s})}\frac{\prod_{s=1}^{3}f(\bar{w}_{\textsc{i}}^{s},\bar{w}_{\textsc{iii}}^{s})}{\prod_{s=2}^{3}f(\bar{w}_{\textsc{i}}^{s-1}-c,\bar{w}_{\textsc{iii}}^{s})}\times\nonumber \\
 & \times\frac{1}{h(\bar{w}_{\textsc{i}}^{1},\bar{w}_{\textsc{i}}^{2}+c)}\frac{1}{h(\bar{w}_{\textsc{i}}^{2},\bar{w}_{\textsc{i}}^{3}+c)f(\bar{w}_{\textsc{i}}^{2},\bar{w}_{\textsc{ii}}^{3}+c)}\prod_{s=3}^{N-1}\frac{f(\bar{w}_{\textsc{i}}^{s},\bar{w}_{\textsc{ii}}^{s})}{h(\bar{w}_{\textsc{i}}^{s},\bar{w}_{\textsc{i}}^{s+1}+c)f(\bar{w}_{\textsc{i}}^{s},\bar{w}_{\textsc{ii}}^{s+1}+c)}\times\\
 & \times\frac{1}{h(-z-c,\bar{w}_{\textsc{iii}}^{1})}\frac{1}{h(\bar{w}_{\textsc{iii}}^{1}-c,\bar{w}_{\textsc{iii}}^{2})}\frac{f(\bar{w}_{\textsc{ii}}^{3},\bar{w}_{\textsc{iii}}^{3})}{h(\bar{w}_{\textsc{iii}}^{2}-c,\bar{w}_{\textsc{iii}}^{3})}.\nonumber 
\end{align}
The $\alpha_{2}$ independent terms are where $\bar{w}_{\textsc{iii}}^{2}=\{-z-2c\}$,
i.e.,
\begin{align}
 & \frac{\lambda_{0}(z)}{\lambda_{3}(z)}\widehat{T}_{N,N-3}(-z)\mathbb{B}(\{t^{1}\},\{t^{2}\},\{\bar{t}^{s}\}_{s=3}^{N-1})\cong\nonumber \\
 & -\frac{f(t^{1}-c,t^{2})}{h(t^{1}-c,t^{2})}\sum_{\mathrm{part}(\bar{w})}\mathbb{B}(\emptyset,\emptyset,\{\bar{w}_{\textsc{ii}}^{s}\}_{s=3}^{N-1})\frac{\prod_{s=4}^{N-1}f(\bar{t}^{s-1}-c,\bar{t}^{s})}{\prod_{s=4}^{N-1}f(\bar{w}_{\textsc{ii}}^{s-1}-c,\bar{w}_{\textsc{ii}}^{s})}\times\\
 & \times g(t^{2}-c,\bar{w}_{\textsc{i}}^{3})\frac{f(\bar{w}_{\textsc{i}}^{3},\bar{w}_{\textsc{iii}}^{3})f(\bar{w}_{\textsc{ii}}^{3},\bar{w}_{\textsc{iii}}^{3})}{h(-z-3c,\bar{w}_{\textsc{iii}}^{3})}\prod_{s=3}^{N-1}\frac{f(\bar{w}_{\textsc{i}}^{s},\bar{w}_{\textsc{ii}}^{s})}{h(\bar{w}_{\textsc{i}}^{s},\bar{w}_{\textsc{i}}^{s+1}+c)f(\bar{w}_{\textsc{i}}^{s},\bar{w}_{\textsc{ii}}^{s+1}+c)}.\nonumber 
\end{align}
The pole at $t^{2}\to-z-2c$ appears only if $\bar{w}_{\textsc{i}}^{3}=\{-z-3c\}$
therefore
\begin{align}
 & \frac{\lambda_{0}(z)}{\lambda_{3}(z)}\widehat{T}_{N,N-3}(-z)\mathbb{B}(\{t^{1}\},\{t^{2}\},\{\bar{t}^{s}\}_{s=3}^{N-1})\cong\nonumber \\
 & -g(t^{2},-z-2c)\frac{f(t^{1}-c,t^{2})}{h(t^{1}-c,t^{2})}\sum_{\mathrm{part}(\bar{w})}\mathbb{B}(\emptyset,\emptyset,\bar{t}_{\textsc{ii}}^{3},\{\bar{w}_{\textsc{ii}}^{s}\}_{s=4}^{N-1})\frac{\prod_{s=4}^{N-1}f(\bar{t}^{s-1}-c,\bar{t}^{s})}{f(\bar{t}_{\textsc{ii}}^{3}-c,\bar{w}_{\textsc{ii}}^{4})\prod_{s=5}^{N-1}f(\bar{w}_{\textsc{ii}}^{s-1}-c,\bar{w}_{\textsc{ii}}^{s})}\times\nonumber \\
 & \times\frac{f(-z-3c,\bar{t}_{\textsc{ii}}^{3})}{h(-z-4c,\bar{w}_{\textsc{i}}^{4})f(-z-4c,\bar{w}_{\textsc{ii}}^{4})}\prod_{s=4}^{N-1}\frac{f(\bar{w}_{\textsc{i}}^{s},\bar{w}_{\textsc{ii}}^{s})}{h(\bar{w}_{\textsc{i}}^{s},\bar{w}_{\textsc{i}}^{s+1}+c)f(\bar{w}_{\textsc{i}}^{s},\bar{w}_{\textsc{ii}}^{s+1}+c)}\times\\
 & \times\frac{f(-z-3c,\bar{t}_{\textsc{iii}}^{3})f(\bar{t}_{\textsc{ii}}^{3},\bar{t}_{\textsc{iii}}^{3})}{h(-z-3c,\bar{t}_{\textsc{iii}}^{3})}+reg.\nonumber 
\end{align}
Since there is a $f(-z-4c,\bar{w}_{\textsc{ii}}^{4})$ term in the
denominator, the residue is non-zero only if $-z-4c\notin\bar{w}_{\textsc{ii}}^{4}$
therefore $\bar{w}_{\textsc{i}}^{4}=-z-4c$. Since there are also
$f(\bar{w}_{\textsc{i}}^{s},\bar{w}_{\textsc{ii}}^{s+1}+c)$ terms
in the denominator, repeating the previous argument, only the partitions
$\bar{w}_{\textsc{i}}^{s}=-z-sc$, $\bar{w}_{\textsc{ii}}^{s}=\bar{t}^{s}$
are relevant for the residue:
\begin{multline}
\frac{\lambda_{0}(z)}{\lambda_{3}(z)}\widehat{T}_{N,N-3}(-z)\mathbb{B}(\{t^{1}\},\{t^{2}\},\{\bar{t}^{s}\}_{s=3}^{N-1})\cong\\
\frac{g(t^{2},-z-2c)}{h(t^{2},t^{1})}\sum_{\mathrm{part}(\bar{t}^{3})}\mathbb{B}(\emptyset,\emptyset,\bar{t}_{\textsc{ii}}^{3},\{\bar{t}^{s}\}_{s=4}^{N-1})\frac{f(-z-3c,\bar{t}^{3})f(\bar{t}_{\textsc{ii}}^{3},\bar{t}_{\textsc{iii}}^{3})f(\bar{t}_{\textsc{iii}}^{3}-c,\bar{t}^{4})}{h(-z-3c,\bar{t}_{\textsc{iii}}^{3})}+reg.
\end{multline}
Analogous way we can also derive that
\begin{align}
\frac{\lambda_{0}(z)}{\lambda_{3}(z)}\widehat{T}_{N,N-2}(-z)\mathbb{B}(\{t^{1}\},\{t^{2}\},\{\bar{t}^{s}\}_{s=3}^{N-1}) & \cong-\frac{g(t^{2},-z-2c)}{h(t^{2},t^{1})}f(-z-3c,\bar{t}^{3})\mathbb{B}(\emptyset,\emptyset,\{\bar{t}^{s}\}_{s=3}^{N-1})+reg.,\nonumber \\
\frac{\lambda_{0}(z)}{\lambda_{3}(z)}\widehat{T}_{N-1,N-3}(-z)\mathbb{B}(\emptyset,\{t^{2}\},\{\bar{t}^{s}\}_{s=3}^{N-1}) & \cong g(t^{2},-z-2c)f(-z-3c,\bar{t}^{3})\times\\
 & \times\sum_{\mathrm{part}(\bar{t}^{3})}\mathbb{B}(\emptyset,\emptyset,\bar{t}_{\textsc{ii}}^{3},\{\bar{t}^{s}\}_{s=4}^{N-1})\frac{f(\bar{t}_{\textsc{ii}}^{3},\bar{t}_{\textsc{iii}}^{3})f(\bar{t}_{\textsc{i}}^{3}-c,\bar{t}^{4})}{h(-z-3c,\bar{t}_{\textsc{iii}}^{3})}+reg.,\nonumber \\
\frac{\lambda_{0}(z)}{\lambda_{3}(z)}\widehat{T}_{N-1,N-2}(-z)\mathbb{B}(\emptyset,\{t^{2}\},\{\bar{t}^{s}\}_{s=3}^{N-1}) & \cong-g(t^{2},-z-2c)f(-z-3c,\bar{t}^{3})\mathbb{B}(\emptyset,\emptyset,\{\bar{t}^{s}\}_{s=3}^{N-1})+reg.\nonumber 
\end{align}
Substituting back we obtain that
\begin{align}
 & \bar{\mathcal{Z}}(\{t^{1}\},\{z,t^{2}\},\{\bar{t}^{s}\}_{s=3}^{N-1})\cong\nonumber \\
 & g(t^{2},-z-2c)\frac{x_{2}}{x_{1}}\sum_{\mathrm{part}(\bar{t}^{3})}\bar{\mathcal{Z}}(\emptyset,\emptyset,\bar{t}_{\textsc{ii}}^{3},\{\bar{t}^{s}\}_{s=4}^{N-1})\left(\frac{1}{h(z,t^{1})}-\frac{1}{h(t^{2},t^{1})f(z,t^{1})}\right)\times\\
 & \times\frac{f(\bar{t}_{\textsc{ii}}^{3},\bar{t}_{\textsc{i}}^{3})}{f(\bar{t}^{4},\bar{t}_{\textsc{i}}^{3})f(-z-3c,\bar{t}_{\textsc{i}}^{3})}\frac{f(-z-3c,\bar{t}^{3})}{f(\bar{t}^{3},z)}\left(g(-z-3c,\bar{t}_{\textsc{i}}^{3})+g(\bar{t}_{\textsc{i}}^{3},z)\right)+reg.\nonumber 
\end{align}
We can use the following identities
\begin{align}
g(-z-3c,\bar{t}_{\textsc{i}}^{3})+g(\bar{t}_{\textsc{i}}^{3},z) & =\frac{g(-z-3c,\bar{t}_{\textsc{i}}^{3})g(\bar{t}_{\textsc{i}}^{3},z)}{g(-z-3c,z)},\\
\frac{1}{h(z,t^{1})}-\frac{1}{h(t^{2},t^{1})f(z,t^{1})} & =\frac{h(t^{2},z)}{h(t^{2},t^{1})h(z,t^{1})},
\end{align}
therefore we have
\begin{align}
 & \bar{\mathcal{Z}}(\{t^{1}\},\{z,t^{2}\},\{\bar{t}^{s}\}_{s=3}^{N-1})=\nonumber \\
 & =g(t^{2},-z-2c)\left[\frac{x_{2}}{x_{1}}h(t^{2},z)h(z,t^{2})\right]\frac{1}{f(\bar{t}^{3},t^{2})f(\bar{t}^{3},z)}\times\\
 & \times\sum_{\mathrm{part}(\bar{t}^{3})}\frac{f(\bar{t}_{\textsc{ii}}^{3},\bar{t}_{\textsc{i}}^{3})}{f(\bar{t}^{4},\bar{t}_{\textsc{i}}^{3})}\bar{\mathcal{Z}}(\emptyset,\emptyset,\bar{t}_{\textsc{ii}}^{3},\{\bar{t}^{s}\}_{s=4}^{N-1})\times\left[\frac{1}{h(t^{2},t^{1})h(z,t^{1})}\right]\left[g(\bar{t}_{\textsc{i}}^{3},t^{2})g(\bar{t}_{\textsc{i}}^{3},z)\right]+reg.\nonumber 
\end{align}
In summary
\begin{align}
\bar{\mathcal{Z}}(\{t^{1}\},\{t_{1}^{2},t_{2}^{2}\},\{\bar{t}^{s}\}_{s=3}^{N-1}) & \to g(t_{2}^{2},-t_{1}^{2}-2c)\left[\frac{x_{2}}{x_{1}}h(t_{2}^{2},t_{1}^{2})h(t_{1}^{2},t_{2}^{2})\right]\frac{1}{f(\bar{t}^{3},t_{2}^{2})f(\bar{t}^{3},t_{1}^{2})}\times\\
 & \sum_{\mathrm{part}(\bar{t}^{3})}\frac{f(\bar{t}_{\textsc{i}}^{3},\bar{t}_{\textsc{iii}}^{3})}{f(\bar{t}^{4},\bar{t}_{\textsc{iii}}^{3})}\bar{\mathcal{Z}}(\emptyset,\emptyset,\bar{t}_{\textsc{i}}^{3},\{\bar{t}^{s}\}_{s=4}^{N-1})\frac{g(\bar{t}_{\textsc{iii}}^{3},t_{2}^{2})g(\bar{t}_{\textsc{iii}}^{3},t_{1}^{2})}{h(t_{2}^{2},t^{1})h(t_{1}^{2},t^{1})}+reg.\nonumber 
\end{align}
We prove the general formula (\ref{eq:poleZb}) by induction in $r_{2}$.
Let us assume that the formula (\ref{eq:poleZb}) is true for $\#\bar{t}^{2}=r_{2}-2$.
For $\#\bar{t}^{2}=r_{2}$ we can use the recursion (\ref{eq:recZbTw})
\begin{equation}
\bar{\mathcal{Z}}(\bar{t})=\frac{1}{1-\alpha_{2}^{(1)}(\bar{t}^{2})}\sum_{a=2}^{N/2}\sum_{\mathrm{part}_{a}(\bar{t})}\alpha_{2}^{(1)}(\bar{t}_{\mathrm{i}}^{2})\frac{\prod_{\nu=1}^{N-1}f(\bar{t}_{\mathrm{ii}}^{\nu},\bar{t}_{\mathrm{i}}^{\nu})}{\prod_{\nu=1}^{N-2}f(\bar{t}_{\mathrm{ii}}^{\nu+1},\bar{t}_{\mathrm{i}}^{\nu})}\bar{\mathcal{Z}}(\bar{t}_{\mathrm{i}})\mathcal{S}_{\bar{\alpha}^{(1)}}(\bar{t}_{\mathrm{ii}}).
\end{equation}
Now we can take the $t_{k}^{2}+t_{l}^{2}+2c\to0$ limit. The elementary
overlaps $\mathcal{S}_{\bar{\alpha}^{(1)}}(\bar{t}_{\mathrm{ii}})$
have no poles in this limit, the poles only appears in $\bar{\mathcal{Z}}(\bar{t}_{\mathrm{i}})$
where $\#\bar{t}_{\mathrm{i}}^{2}=r_{2}-2$ therefore we can use the
induction assumption in the rhs:
\begin{align}
\bar{\mathcal{Z}}(\bar{t}) & \to g(t_{l}^{2},-t_{k}^{2}-2c)\left[\frac{x_{2}}{x_{1}}h(t_{k}^{2},t_{l}^{2})h(t_{l}^{2},t_{k}^{2})\right]\frac{f(\bar{\tau}^{2},t_{k}^{2})f(\bar{\tau}^{2},t_{l}^{2})}{f(\bar{\tau}^{3},t_{k}^{2})f(\bar{\tau}^{3},t_{l}^{2})}\times\nonumber \\
 & \times\left[\frac{1}{1-\alpha_{2}^{(1)}(\bar{\tau}^{2})}\sum_{a=2}^{\frac{N}{2}}\sum_{\mathrm{part}_{a}(\bar{t})'}\frac{\prod_{\nu=1}^{2a-1}f(\bar{\tau}_{\mathrm{ii}}^{\nu},\bar{\tau}_{\mathrm{i}}^{\nu})}{\prod_{\nu=1}^{2a-2}f(\bar{\tau}_{\mathrm{ii}}^{\nu+1},\bar{\tau}_{\mathrm{i}}^{\nu})}\alpha_{2}^{(1)}(\bar{\tau}_{\mathrm{i}}^{2})\bar{\mathcal{Z}}(\bar{\tau}_{\mathrm{i}})\mathcal{S}_{\bar{\alpha}^{(1)}}(\bar{\tau}_{\mathrm{ii}})\right]\\
 & \frac{f(\bar{\tau}_{\mathrm{i}}^{1},\bar{\tau}_{\mathrm{iii}}^{1})f(\bar{\tau}_{\mathrm{ii}}^{1},\bar{\tau}_{\mathrm{iii}}^{1})}{f(\bar{\tau}^{2},\bar{\tau}_{\mathrm{iii}}^{1})}\frac{f(\bar{\tau}_{\mathrm{i}}^{3},\bar{\tau}_{\mathrm{iii}}^{3})f(\bar{\tau}_{\mathrm{ii}}^{3},\bar{\tau}_{\mathrm{iii}}^{3})}{f(\bar{\tau}^{4},\bar{\tau}_{\mathrm{iii}}^{3})}\frac{g(\bar{\tau}_{\mathrm{iii}}^{3},t_{k}^{2})g(\bar{\tau}_{\mathrm{iii}}^{3},t_{l}^{2})}{h(t_{k}^{2},\bar{\tau}_{\mathrm{iii}}^{1})h(t_{l}^{2},\bar{\tau}_{\mathrm{iii}}^{1})}.\nonumber 
\end{align}
Now we can use the recursion (\ref{eq:recZbTw}) again.
\begin{align}
\bar{\mathcal{Z}}(\bar{t}) & \to g(t_{l}^{2},-t_{k}^{2}-2c)\left[\frac{x_{2}}{x_{1}}h(t_{k}^{2},t_{l}^{2})h(t_{l}^{2},t_{k}^{2})\right]\frac{f(\bar{\tau}^{2},t_{k}^{2})f(\bar{\tau}^{2},t_{l}^{2})}{f(\bar{\tau}^{3},t_{k}^{2})f(\bar{\tau}^{3},t_{l}^{2})}\times\nonumber \\
 & \times\bar{\mathcal{Z}}(\bar{\tau}_{\mathrm{i}})\frac{f(\bar{\tau}_{\mathrm{i}}^{1},\bar{\tau}_{\mathrm{iii}}^{1})}{f(\bar{\tau}^{2},\bar{\tau}_{\mathrm{iii}}^{1})}\frac{f(\bar{\tau}_{\mathrm{i}}^{3},\bar{\tau}_{\mathrm{iii}}^{3})}{f(\bar{\tau}^{4},\bar{\tau}_{\mathrm{iii}}^{3})}\frac{g(\bar{\tau}_{\mathrm{iii}}^{3},t_{k}^{2})g(\bar{\tau}_{\mathrm{iii}}^{3},t_{l}^{2})}{h(t_{k}^{2},\bar{\tau}_{\mathrm{iii}}^{1})h(t_{l}^{2},\bar{\tau}_{\mathrm{iii}}^{1})},
\end{align}
which proves (\ref{eq:poleZb}) for $a=1$. One can also do an induction
on $N$ and combining it with the previous derivation we obtain the
formula (\ref{eq:poleZb}) for general $a$.

Now we can derive the pole for the other HC simply from the identity
\begin{equation}
\mathcal{Z}(\bar{t})=\frac{1}{\prod_{s=1}^{N-2}f(\bar{t}^{s+1},\bar{t}^{s})}\bar{\mathcal{Z}}(\pi^{c}(\bar{t}))).\label{eq:idZZb}
\end{equation}
Taking the $t_{k}^{2a}+t_{l}^{2a}+2ac$ limit we can use the pole
formula (\ref{eq:poleZb}) on the rhs 
\begin{align}
\mathcal{Z}(\bar{t}) & \to g(-t_{l}^{2a}-2c,t_{k}^{2a})\left[\frac{x_{a+1}}{x_{a}}h(t_{k}^{2a},t_{l}^{2a})h(t_{l}^{2a},t_{k}^{2a})\right]\frac{f(t_{k}^{2a},\bar{\tau}^{2a})f(t_{l}^{2a},\bar{\tau}^{2a})}{f(t_{k}^{2a},\bar{\tau}^{2a-1})f(t_{l}^{2a},\bar{\tau}^{2a-1})}\times\nonumber \\
 & \sum_{\mathrm{part}(\bar{\tau}^{2a-1},\bar{\tau}^{2a+1})}\bar{\mathcal{Z}}(\pi^{c}(\bar{\tau}_{\mathrm{i}}))\frac{1}{\prod_{s=1}^{N-2}f(\bar{\tau}_{\mathrm{i}}^{s+1},\bar{\tau}_{\mathrm{i}}^{s})}\times\\
 & \frac{f(\bar{\tau}_{\mathrm{iii}}^{2a-1},\bar{\tau}_{\mathrm{i}}^{2a-1})}{f(\bar{\tau}_{\mathrm{iii}}^{2a-1},\bar{\tau}^{2a-2})}\frac{f(\bar{\tau}_{\mathrm{iii}}^{2a+1},\bar{\tau}_{\mathrm{i}}^{2a+1})}{f(\bar{\tau}_{\mathrm{iii}}^{2a+1},\bar{\tau}^{2a})}\left[\frac{g(\bar{\tau}_{\mathrm{iii}}^{2a-1},t_{k}^{2a})g(\bar{\tau}_{\mathrm{iii}}^{2a-1},t_{l}^{2a})}{h(\bar{\tau}_{\mathrm{iii}}^{2a+1},t_{k}^{2a})h(\bar{\tau}_{\mathrm{iii}}^{2a+1},t_{l}^{2a})}\right]+reg.,\nonumber 
\end{align}
using the identity (\ref{eq:idZZb}) again we obtain that
\begin{align}
\mathcal{Z}(\bar{t}) & \to g(-t_{l}^{2a}-2c,t_{k}^{2a})\left[\frac{x_{a+1}}{x_{a}}h(t_{k}^{2a},t_{l}^{2a})h(t_{l}^{2a},t_{k}^{2a})\right]\frac{f(t_{k}^{2a},\bar{\tau}^{2a})f(t_{l}^{2a},\bar{\tau}^{2a})}{f(t_{k}^{2a},\bar{\tau}^{2a-1})f(t_{l}^{2a},\bar{\tau}^{2a-1})}\times\\
 & \sum_{\mathrm{part}(\bar{\tau}^{2a-1},\bar{\tau}^{2a+1})}\mathcal{Z}(\bar{\tau}_{\mathrm{i}}^{1})\frac{f(\bar{\tau}_{\mathrm{iii}}^{2a-1},\bar{\tau}_{\mathrm{i}}^{2a-1})}{f(\bar{\tau}_{\mathrm{iii}}^{2a-1},\bar{\tau}^{2a-2})}\frac{f(\bar{\tau}_{\mathrm{iii}}^{2a+1},\bar{\tau}_{\mathrm{i}}^{2a+1})}{f(\bar{\tau}_{\mathrm{iii}}^{2a+1},\bar{\tau}^{2a})}\frac{g(\bar{\tau}_{\mathrm{iii}}^{2a-1},t_{k}^{2a})g(\bar{\tau}_{\mathrm{iii}}^{2a-1},t_{l}^{2a})}{h(\bar{\tau}_{\mathrm{iii}}^{2a+1},t_{k}^{2a})h(\bar{\tau}_{\mathrm{iii}}^{2a+1},t_{l}^{2a})}+reg.\nonumber 
\end{align}

\section{Pair structure limit of the twisted overlaps\label{sec:Pair-structure-limit}}

In this section we derive the pair structure limit $t_{l}^{s}\to-t_{k}^{s}-sc$
of the off-shell overlap. In section \ref{subsec:Sum-formulaTW} we
saw that there is a significant difference between the even and odd
$s$. For $s=2b$ the there is no extra selection rule for the $\alpha$-functions
$\alpha_{2b}(z)$ i.e. it can contain arbitrary number of parameters
(inhomogeneities).  On the other hand we have extra selection rules
for $s=2b-1$, see section \ref{subsec:Sum-formulaTW} for the details.

In the following we calculate the pair structure limit $t_{l}^{2b}\to-t_{k}^{2b}-2bc$
of the off-shell overlap (\ref{eq:sumTW}): 
\begin{align}
\mathcal{S}_{\bar{\alpha}}(\bar{t}) & =\sum_{\mathrm{part}(\bar{t})}\frac{\prod_{\nu=1}^{N-1}f(\bar{t}_{\mathrm{ii}}^{\nu},\bar{t}_{\mathrm{i}}^{\nu})}{\prod_{\nu=1}^{N-2}f(\bar{t}_{\mathrm{ii}}^{\nu+1},\bar{t}_{\mathrm{i}}^{\nu})}\frac{\prod_{a=1}^{N/2}f(\bar{t}_{\mathrm{iii}}^{2a-1},\bar{t}_{\mathrm{i}}^{2a-1})f(\bar{t}_{\mathrm{ii}}^{2a-1},\bar{t}_{\mathrm{iii}}^{2a-1})}{\prod_{a=1}^{N/2-1}f(\bar{t}_{\mathrm{iii}}^{2a+1},\bar{t}_{\mathrm{i}}^{2a})f(\bar{t}_{\mathrm{ii}}^{2a},\bar{t}_{\mathrm{iii}}^{2a-1})}\nonumber \\
 & \times\mathcal{Z}(\bar{t}_{\mathrm{i}})\bar{\mathcal{Z}}(\bar{t}_{\mathrm{ii}})\prod_{a=1}^{N/2}S_{\alpha_{2a-1}}^{(2a-1)}(\bar{\tau}_{\mathrm{iii}}^{2a-1})\prod_{s=1}^{N-1}\alpha_{s}(\bar{t}_{\mathrm{i}}^{s}).\label{eq:sumFormulaTw-1}
\end{align}
We also take limits $t_{l}^{2b-1}\to-t_{k}^{2b-1}-(2b-1)c$, $t_{k}^{2b-1}\to-(b-1/2)c$
when $n_{2b-1}\geq2$.

\subsection*{Pair structure limit $t_{l}^{2b}\to-t_{k}^{2b}-2bc$}

Let us calculate the limit $t_{l}^{2b}\to-t_{k}^{2b}-2bc$ of the
overlap formula (\ref{eq:sumFormulaTw-1}). Formal poles appear in
the HC in this limit for the partition where $\{t_{k}^{2b},t_{l}^{2b}\}\in\bar{t}_{\mathrm{i}}^{2b}$
or $\{t_{k}^{2b},t_{l}^{2b}\}\in\bar{t}_{\mathrm{ii}}^{2b}$. Let
$\bar{\tau}=\bar{t}\backslash\{t_{k}^{2b},t_{l}^{2b}\}$ and let us
get the partitions in (\ref{eq:sumFormulaTw-1}) where $\bar{t}_{\mathrm{ii}}^{2b}=\{t_{k}^{2b},t_{l}^{2b}\}\cup\bar{\tau}_{\mathrm{ii}}$.
We can take the limit $t_{l}^{2b}\to-t_{k}^{2b}-2bc$ as
\begin{align}
 & g(t_{l}^{2b},-t_{k}^{2b}-2bc)\left[\frac{x_{b+1}}{x_{b}}h(t_{k}^{2b},t_{l}^{2b})h(t_{l}^{2b},t_{k}^{2b})\right]\sum_{\mathrm{part}(\bar{\tau})}\frac{f(\bar{\tau}_{\mathrm{ii}}^{2b},t_{k}^{2b})f(\bar{\tau}_{\mathrm{ii}}^{2b},t_{l}^{2b})}{f(\bar{\tau}_{\mathrm{ii}}^{2b+1},t_{k}^{2b})f(\bar{\tau}_{\mathrm{ii}}^{2b+1},t_{l}^{2b})f(\bar{\tau}_{\mathrm{iv}}^{2b+1},t_{k}^{2b})f(\bar{\tau}_{\mathrm{iv}}^{2b+1},t_{l}^{2b})}\times\nonumber \\
 & \times\frac{f(t_{k}^{2b},\bar{\tau}_{\mathrm{i}}^{2b})f(t_{l}^{2b},\bar{\tau}_{\mathrm{i}}^{2b})}{f(t_{k}^{2b},\bar{\tau}_{\mathrm{i}}^{2b-1})f(t_{l}^{2b},\bar{\tau}_{\mathrm{i}}^{2b-1})f(t_{k}^{2b},\bar{\tau}_{\mathrm{iii}}^{2b-1})f(t_{l}^{2b},\bar{\tau}_{\mathrm{iii}}^{2b-1})}\times\nonumber \\
 & \times\frac{f(\bar{\tau}_{\mathrm{iv}}^{2b-1},\bar{\tau}_{\mathrm{i}}^{2b-1})f(\bar{\tau}_{\mathrm{iv}}^{2b+1},\bar{\tau}_{\mathrm{i}}^{2b+1})f(\bar{\tau}_{\mathrm{iv}}^{2b-1},\bar{\tau}_{\mathrm{iii}}^{2b-1})f(\bar{\tau}_{\mathrm{iv}}^{2b+1},\bar{\tau}_{\mathrm{iii}}^{2b+1})}{f(\bar{\tau}_{\mathrm{iv}}^{2b-1},\bar{\tau}_{\mathrm{i}}^{2b-2})f(\bar{\tau}_{\mathrm{iv}}^{2b+1},\bar{\tau}_{\mathrm{i}}^{2b})}\times\nonumber \\
 & \times\frac{\prod_{\nu=1}^{N-1}f(\bar{\tau}_{\mathrm{ii}}^{\nu},\bar{\tau}_{\mathrm{i}}^{\nu})}{\prod_{\nu=1}^{N-2}f(\bar{\tau}_{\mathrm{ii}}^{\nu+1},\bar{\tau}_{\mathrm{i}}^{\nu})}\frac{\prod_{a=1}^{N/2}f(\bar{\tau}_{\mathrm{ii}}^{2a-1},\bar{\tau}_{\mathrm{iii}}^{2a-1})}{\prod_{a=1}^{N/2-1}f(\bar{\tau}_{\mathrm{ii}}^{2a},\bar{\tau}_{\mathrm{iii}}^{2a-1})}\frac{\prod_{a=1}^{N/2}f(\bar{\tau}_{\mathrm{iii}}^{2a-1},\bar{\tau}_{\mathrm{i}}^{2a-1})}{\prod_{a=1}^{N/2-1}f(\bar{\tau}_{\mathrm{iii}}^{2a+1},\bar{\tau}_{\mathrm{i}}^{2a})}\times\\
 & \times\mathcal{Z}(\bar{\tau}_{\mathrm{i}})\bar{\mathcal{Z}}(\bar{\tau}_{\mathrm{ii}})\prod_{a=1}^{N/2}S_{\alpha_{2b-1}}(\bar{\tau}_{\mathrm{iii}}^{2a-1})\prod_{s=1}^{N-1}\alpha_{s}(\bar{\tau}_{\mathrm{i}}^{s})\times\nonumber \\
 & \times\frac{f(\bar{\tau}_{\mathrm{ii}}^{2b-1},\bar{\tau}_{\mathrm{iv}}^{2b-1})}{f(\bar{\tau}_{\mathrm{ii}}^{2b},\bar{\tau}_{\mathrm{iv}}^{2b-1})}\frac{f(\bar{\tau}_{\mathrm{ii}}^{2b+1},\bar{\tau}_{\mathrm{iv}}^{2b+1})}{f(\bar{\tau}_{\mathrm{ii}}^{2b+2},\bar{\tau}_{\mathrm{iv}}^{2b+1})}\left[\frac{1}{h(t_{k}^{2b},\bar{\tau}_{\mathrm{iv}}^{2b-1})h(t_{l}^{2b},\bar{\tau}_{\mathrm{iv}}^{2b-1})}\right]\left[g(\bar{\tau}_{\mathrm{iv}}^{2b+1},t_{k}^{2b})g(\bar{\tau}_{\mathrm{iv}}^{2b+1},t_{l}^{2b})\right],\nonumber 
\end{align}
where we used the pole formula (\ref{eq:poleZb}) and the sum goes
through the partitions $\bar{\tau}=\bar{\tau}_{\mathrm{i}}\cup\bar{\tau}_{\mathrm{ii}}\cup\bar{\tau}_{\mathrm{iii}}\cup\bar{\tau}_{\mathrm{iv}}$
where $\#\bar{\tau}_{\mathrm{i}}^{2a-1}=\frac{\#\bar{\tau}_{\mathrm{i}}^{2a-2}+\#\bar{\tau}_{\mathrm{i}}^{2a}}{2}$,
$\#\bar{\tau}_{\mathrm{ii}}^{2a-1}=\frac{\#\bar{\tau}_{\mathrm{ii}}^{2a-2}+\#\bar{\tau}_{\mathrm{ii}}^{2a}}{2}$
for $a=1,\dots,\frac{N}{2}$ and $\#\bar{\tau}_{\mathrm{iii}}^{2a}=\#\bar{\tau}_{\mathrm{iv}}^{2a}=0$
and $\#\bar{\tau}_{\mathrm{iv}}^{2b-1}=\#\bar{\tau}_{\mathrm{iv}}^{2b+1}=1$
and $\#\bar{\tau}_{\mathrm{iv}}^{2a-1}=0$ for $a\neq b,b+1$. Collecting
the terms we obtain that
\begin{align}
 & g(t_{l}^{2b},-t_{k}^{2b}-2bc)\left[\frac{x_{b+1}}{x_{b}}h(t_{k}^{2b},t_{l}^{2b})h(t_{l}^{2b},t_{k}^{2b})\right]\frac{f(\bar{\tau}^{2b},t_{k}^{2b})f(\bar{\tau}^{2b}t_{l}^{2b})}{f(\bar{\tau}^{2b+1},t_{k}^{2b})f(\bar{\tau}^{2b+1},t_{l}^{2b})}\times\nonumber \\
 & \times\sum_{\mathrm{part}(\bar{\tau})}\frac{\prod_{\nu=1}^{N-1}f(\bar{\tau}_{\mathrm{ii}}^{\nu},\bar{\tau}_{\mathrm{i}}^{\nu})}{\prod_{\nu=1}^{N-2}f(\bar{\tau}_{\mathrm{ii}}^{\nu+1},\bar{\tau}_{\mathrm{i}}^{\nu})}\left[\frac{\prod_{a=1}^{N/2}f(\bar{\tau}_{\mathrm{ii}}^{2a-1},\bar{\tau}_{\mathrm{iii}}^{2a-1})}{\prod_{a=1}^{N/2-1}f(\bar{\tau}_{\mathrm{ii}}^{2a},\bar{\tau}_{\mathrm{iii}}^{2a-1})}\frac{f(\bar{\tau}_{\mathrm{ii}}^{2b-1},\bar{\tau}_{\mathrm{iv}}^{2b-1})}{f(\bar{\tau}_{\mathrm{ii}}^{2b},\bar{\tau}_{\mathrm{iv}}^{2b-1})}\frac{f(\bar{\tau}_{\mathrm{ii}}^{2b+1},\bar{\tau}_{\mathrm{iv}}^{2b+1})}{f(\bar{\tau}_{\mathrm{ii}}^{2b+2},\bar{\tau}_{\mathrm{iv}}^{2b+1})}\right]\times\nonumber \\
 & \times\left[\frac{\prod_{a=1}^{N/2}f(\bar{\tau}_{\mathrm{iii}}^{2a-1},\bar{\tau}_{\mathrm{i}}^{2a-1})}{\prod_{a=1}^{N/2-1}f(\bar{\tau}_{\mathrm{iii}}^{2a+1},\bar{\tau}_{\mathrm{i}}^{2a})}\frac{f(\bar{\tau}_{\mathrm{iv}}^{2b-1},\bar{\tau}_{\mathrm{i}}^{2b-1})f(\bar{\tau}_{\mathrm{iv}}^{2b+1},\bar{\tau}_{\mathrm{i}}^{2b+1})}{f(\bar{\tau}_{\mathrm{iv}}^{2b-1},\bar{\tau}_{\mathrm{i}}^{2b-2})f(\bar{\tau}_{\mathrm{iv}}^{2b+1},\bar{\tau}_{\mathrm{i}}^{2b})}\right]\times\nonumber \\
 & \times\mathcal{Z}(\bar{\tau}_{\mathrm{i}})\bar{\mathcal{Z}}(\bar{\tau}_{\mathrm{ii}})\left[\prod_{s=1}^{N-1}\alpha_{s}(\bar{\tau}_{\mathrm{i}}^{s})\frac{f(t_{k}^{2b},\bar{\tau}_{\mathrm{i}}^{2b})f(t_{l}^{2b},\bar{\tau}_{\mathrm{i}}^{2b})}{f(\bar{\tau}_{\mathrm{i}}^{2b},t_{k}^{2b})f(\bar{\tau}_{\mathrm{i}}^{2b}t_{l}^{2b})}\frac{f(\bar{\tau}_{\mathrm{i}}^{2b+1},t_{k}^{2b})f(\bar{\tau}_{\mathrm{i}}^{2b+1},t_{l}^{2b})}{f(t_{k}^{2b},\bar{\tau}_{\mathrm{i}}^{2b-1})f(t_{l}^{2b},\bar{\tau}_{\mathrm{i}}^{2b-1})}\right]\times\label{eq:temp1-1}\\
 & \times f(\bar{\tau}_{\mathrm{iv}}^{2b-1},\bar{\tau}_{\mathrm{iii}}^{2b-1})f(\bar{\tau}_{\mathrm{iv}}^{2b+1},\bar{\tau}_{\mathrm{iii}}^{2b+1})\prod_{a=1}^{N/2}S_{\alpha_{2a-1}}^{(2a-1)}(\bar{\tau}_{\mathrm{iii}}^{2a-1})\frac{f(\bar{\tau}_{\mathrm{iii}}^{2b+1},t_{k}^{2b})f(\bar{\tau}_{\mathrm{iii}}^{2b+1},t_{l}^{2b})}{f(t_{k}^{2b},\bar{\tau}_{\mathrm{iii}}^{2b-1})f(t_{l}^{2b},\bar{\tau}_{\mathrm{iii}}^{2b-1})}\frac{g(\bar{\tau}_{\mathrm{iv}}^{2b+1},t_{k}^{2b})g(\bar{\tau}_{\mathrm{iv}}^{2b+1},t_{l}^{2b})}{h(t_{k}^{2b},\bar{\tau}_{\mathrm{iv}}^{2b-1})h(t_{l}^{2b},\bar{\tau}_{\mathrm{iv}}^{2b-1})}.\nonumber 
\end{align}
Now let us define a spin chain where the quantum space $\mathcal{H}^{(2)}$
is a tensor product two rectangular representation $(1,2b-1)$:
\begin{equation}
T_{0}^{(2)}(z)=L_{0,2}^{(1,2b-1)}(z+t_{k}^{2b}-c(1-2b))L_{0,1}^{(1,2b-1)}(z-t_{k}^{2b}-c),
\end{equation}
for which the $\alpha$-functions are
\begin{equation}
\begin{split}\alpha_{2b-1}^{(2)}(z) & =\frac{1}{f(t_{k}^{2b},z)f(-t_{k}^{2b}-2bc,z)}=\frac{1}{f(t_{k}^{2b},z)f(t_{l}^{2b},z)},\\
\alpha_{k}^{(2)}(z) & =1,\quad\text{for }k\neq2b-1.
\end{split}
\end{equation}
Defining the tensor product quantum space as $\mathcal{H}^{mod}=\mathcal{H}\otimes\mathcal{H}^{(2)}$,
we can use the co-product formula (\ref{eq:coproduct})
\begin{equation}
S_{\alpha_{2b-1}^{mod}}^{(2b-1)}(\bar{t}^{2b-1})=\sum_{\mathrm{part}(\bar{t}^{2b-1})}\alpha_{2b-1}^{(2)}(\bar{t}_{\mathrm{i}}^{2b-1})f(\bar{t}_{\mathrm{ii}}^{2b-1},\bar{t}_{\mathrm{i}}^{2b-1})S_{\alpha_{2b-1}}^{(2b-1)}(\bar{t}_{\mathrm{i}}^{2b-1})S_{\alpha_{2b-1}^{(2)}}^{(2b-1)}(\bar{t}_{\mathrm{ii}}^{2b-1}),\label{eq:cp2}
\end{equation}
where the elementary overlap is 
\begin{align}
S_{\alpha_{2b-1}^{(2)}}^{(2b-1)}(\{u\}) & =S_{\alpha_{2b-1}^{(2)}(z-c\frac{2b-1}{2})}(\{u+c\frac{2b-1}{2}\})=\frac{1}{u+c\frac{2b-1}{2}}(\alpha_{2b-1}^{(2)}(u)-1)\nonumber \\
 & =\frac{2}{c}\frac{1}{h(t_{k}^{b},u)h(t_{l}^{b},u)}.
\end{align}
Substituting back to the co-product formula (\ref{eq:coproduct})
we obtain that
\begin{equation}
S_{\alpha_{2b-1}^{mod}}^{(2b-1)}(\bar{t}^{2b-1})=\frac{2}{c}\sum_{\mathrm{part}(\bar{t})}f(\bar{t}_{\mathrm{ii}}^{2b-1},\bar{t}_{\mathrm{i}}^{2b-1})S_{\alpha_{2b-1}}^{(2b-1)}(\bar{t}_{\mathrm{i}}^{2b-1})\frac{1}{f(t_{k}^{2b},\bar{t}_{\mathrm{i}}^{2b-1})f(t_{l}^{2b},\bar{t}_{\mathrm{i}}^{2b-1})}\frac{1}{h(t_{k}^{b},\bar{t}_{\mathrm{ii}}^{2b-1})h(t_{l}^{b},\bar{t}_{\mathrm{ii}}^{2b-1})}.\label{eq:2coprod1}
\end{equation}
Repeating the calculation for the representation $(1,2b+1)$, we
obtain the following co-product formula
\begin{equation}
S_{\alpha_{2b+1}^{mod}}^{(2b+1)}(\bar{t}^{2b+1})=\frac{2}{c}\sum_{\mathrm{part}(\bar{t})}f(\bar{t}_{\mathrm{ii}}^{2b+1},\bar{t}_{\mathrm{i}}^{2b+1})S_{\alpha_{2b+1}}^{(2b+1)}(\bar{t}_{\mathrm{i}}^{2b+1})f(\bar{t}_{\mathrm{i}}^{2b+1},t_{k}^{2b})f(\bar{t}_{\mathrm{i}}^{2b+1},t_{l}^{2b})g(t_{k}^{2b},\bar{t}_{\mathrm{ii}}^{2b+1})g(t_{l}^{2b},\bar{t}_{\mathrm{ii}}^{2b+1}).\label{eq:2coprod2}
\end{equation}
Substituting back the co-product formulas (\ref{eq:2coprod1}) and
(\ref{eq:2coprod2}) to (\ref{eq:temp1-1}) we obtain that
\begin{align}
 & g(t_{l}^{2b},-t_{k}^{2b}-2bc)\left[\frac{x_{b+1}}{x_{b}}\frac{c^{2}}{4}h(t_{k}^{2b},t_{l}^{2b})h(t_{l}^{2b},t_{k}^{2b})\right]\times\frac{f(\bar{\tau}^{2b},t_{k}^{2b})f(\bar{\tau}^{2b}t_{l}^{2b})}{f(\bar{\tau}^{2b+1},t_{k}^{2b})f(\bar{\tau}^{2b+1},t_{l}^{2b})}\nonumber \\
 & \sum_{\mathrm{part}(\bar{\tau})}\frac{\prod_{\nu=1}^{N-1}f(\bar{\tau}_{\mathrm{ii}}^{\nu},\bar{\tau}_{\mathrm{i}}^{\nu})}{\prod_{\nu=1}^{N-2}f(\bar{\tau}_{\mathrm{ii}}^{\nu+1},\bar{\tau}_{\mathrm{i}}^{\nu})}\frac{\prod_{a=1}^{N/2}f(\bar{\tau}_{\mathrm{ii}}^{2a-1},\bar{\tau}_{\mathrm{iii}}^{2a-1})}{\prod_{a=1}^{N/2-1}f(\bar{\tau}_{\mathrm{ii}}^{2a},\bar{\tau}_{\mathrm{iii}}^{2a-1})}\frac{\prod_{a=1}^{N/2}f(\bar{\tau}_{\mathrm{iii}}^{2a-1},\bar{\tau}_{\mathrm{i}}^{2a-1})}{\prod_{a=1}^{N/2-1}f(\bar{\tau}_{\mathrm{iii}}^{2a+1},\bar{\tau}_{\mathrm{i}}^{2a})}\nonumber \\
 & \times\mathcal{Z}(\bar{\tau}_{\mathrm{i}})\bar{\mathcal{Z}}(\bar{\tau}_{\mathrm{ii}})\left[\prod_{s=1}^{N-1}\alpha_{s}(\bar{\tau}_{\mathrm{i}}^{s})\frac{f(t_{k}^{2b},\bar{\tau}_{\mathrm{i}}^{2b})f(t_{l}^{2b},\bar{\tau}_{\mathrm{i}}^{2b})}{f(\bar{\tau}_{\mathrm{i}}^{2b},t_{k}^{2b})f(\bar{\tau}_{\mathrm{i}}^{2b}t_{l}^{2b})}\frac{f(\bar{\tau}_{\mathrm{i}}^{2b+1},t_{k}^{2b})f(\bar{\tau}_{\mathrm{i}}^{2b+1},t_{l}^{2b})}{f(t_{k}^{2b},\bar{\tau}_{\mathrm{i}}^{2b-1})f(t_{l}^{2b},\bar{\tau}_{\mathrm{i}}^{2b-1})}\right]\label{eq:temp1-1-1}\\
 & \prod_{a=1}^{N/2}S_{\alpha_{2a-1}^{mod}}^{(2a-1)}(\bar{\tau}_{\mathrm{iii}}^{2a-1}),\nonumber 
\end{align}
where the sum goes through the partitions $\bar{\tau}=\bar{\tau}_{\mathrm{i}}\cup\bar{\tau}_{\mathrm{ii}}\cup\bar{\tau}_{\mathrm{iii}}$
where $\#\bar{\tau}_{\mathrm{i}}^{2a-1}=\frac{\#\bar{\tau}_{\mathrm{i}}^{2a-2}+\#\bar{\tau}_{\mathrm{i}}^{2a}}{2}$,
$\#\bar{\tau}_{\mathrm{ii}}^{2a-1}=\frac{\#\bar{\tau}_{\mathrm{ii}}^{2a-2}+\#\bar{\tau}_{\mathrm{ii}}^{2a}}{2}$
for $a=1,\dots,\frac{N}{2}$ and $\#\bar{\tau}_{\mathrm{iii}}^{2a}=0$.
We also defined the modified $\alpha$-s as
\begin{align}
\alpha_{2b-1}^{mod}(z) & =\alpha_{2b-1}(z)\frac{1}{f(t_{k}^{2b},z)f(t_{l}^{2b},z)},\nonumber \\
\alpha_{2b+1}^{mod}(z) & =\alpha_{2b+1}(z)f(z,t_{k}^{2b})f(z,t_{l}^{2b}),\\
\alpha_{2a-1}^{mod}(z) & =\alpha_{2a-1}(z),\quad\text{for }a\neq b,b+1.\nonumber 
\end{align}
Defining the remaining $\alpha^{mod}$-s as
\begin{equation}
\begin{split}\alpha_{2b}^{mod}(z) & =\alpha_{2b}(z)\frac{f(t_{k}^{2b},z)f(t_{l}^{2b},z)}{f(z,t_{k}^{2b})f(z,t_{l}^{2b})},\\
\alpha_{2a}^{mod}(z) & =\alpha_{2a}(z),\quad\text{for }a\neq b,
\end{split}
\end{equation}
we obtain that
\begin{align}
 & g(t_{l}^{2b},-t_{k}^{2b}-2bc)\left[\frac{x_{b+1}}{x_{b}}\frac{c^{2}}{4}h(t_{k}^{2b},t_{l}^{2b})h(t_{l}^{2b},t_{k}^{2b})\right]\times\frac{f(\bar{\tau}^{2b},t_{k}^{2b})f(\bar{\tau}^{2b}t_{l}^{2b})}{f(\bar{\tau}^{2b+1},t_{k}^{2b})f(\bar{\tau}^{2b+1},t_{l}^{2b})}\nonumber \\
 & \sum_{\mathrm{part}(\bar{\tau})}\frac{\prod_{\nu=1}^{N-1}f(\bar{\tau}_{\mathrm{ii}}^{\nu},\bar{\tau}_{\mathrm{i}}^{\nu})}{\prod_{\nu=1}^{N-2}f(\bar{\tau}_{\mathrm{ii}}^{\nu+1},\bar{\tau}_{\mathrm{i}}^{\nu})}\frac{\prod_{a=1}^{N/2}f(\bar{\tau}_{\mathrm{ii}}^{2a-1},\bar{\tau}_{\mathrm{iii}}^{2a-1})f(\bar{\tau}_{\mathrm{iii}}^{2a-1},\bar{\tau}_{\mathrm{i}}^{2a-1})}{\prod_{a=1}^{N/2-1}f(\bar{\tau}_{\mathrm{ii}}^{2a},\bar{\tau}_{\mathrm{iii}}^{2a-1})f(\bar{\tau}_{\mathrm{iii}}^{2a+1},\bar{\tau}_{\mathrm{i}}^{2a})}\label{eq:ppoleZb}\\
 & \times\mathcal{Z}(\bar{\tau}_{\mathrm{i}})\bar{\mathcal{Z}}(\bar{\tau}_{\mathrm{ii}})\prod_{a=1}^{N/2}S_{\alpha_{2a-1}^{mod}}^{(2a-1)}(\bar{\tau}_{\mathrm{iii}}^{2a-1})\left[\prod_{s=1}^{N-1}\alpha_{s}^{mod}(\bar{\tau}_{\mathrm{i}}^{s})\right].\nonumber 
\end{align}

Repeating the previous calculation for the partition $\bar{t}_{\mathrm{i}}^{2b}=\{t_{k}^{2b},t_{l}^{2b}\}\cup\bar{\tau}_{\mathrm{i}}$
we obtain the following pole in the $t_{k}^{2b}+t_{l}^{2b}+2bc\to0$
limit
\begin{align}
 & g(-t_{k}^{2b}-2bc,t_{l}^{2b})\alpha_{2b}(t_{k}^{2b})\alpha_{2b}(t_{l}^{2b})\left[\frac{x_{b+1}}{x_{b}}\frac{c^{2}}{4}h(t_{k}^{2b},t_{l}^{2b})h(t_{l}^{2b},t_{k}^{2b})\right]\times\frac{f(\bar{\tau}^{2b},t_{k}^{2b})f(\bar{\tau}^{2b}t_{l}^{2b})}{f(\bar{\tau}^{2b+1},t_{k}^{2b})f(\bar{\tau}^{2b+1},t_{l}^{2b})}\nonumber \\
 & \sum_{\mathrm{part}(\bar{\tau})}\frac{\prod_{\nu=1}^{N-1}f(\bar{\tau}_{\mathrm{ii}}^{\nu},\bar{\tau}_{\mathrm{i}}^{\nu})}{\prod_{\nu=1}^{N-2}f(\bar{\tau}_{\mathrm{ii}}^{\nu+1},\bar{\tau}_{\mathrm{i}}^{\nu})}\frac{\prod_{a=1}^{N/2}f(\bar{\tau}_{\mathrm{ii}}^{2a-1},\bar{\tau}_{\mathrm{iii}}^{2a-1})f(\bar{\tau}_{\mathrm{iii}}^{2a-1},\bar{\tau}_{\mathrm{i}}^{2a-1})}{\prod_{a=1}^{N/2-1}f(\bar{\tau}_{\mathrm{ii}}^{2a},\bar{\tau}_{\mathrm{iii}}^{2a-1})f(\bar{\tau}_{\mathrm{iii}}^{2a+1},\bar{\tau}_{\mathrm{i}}^{2a})}\label{eq:ppoleZ}\\
 & \times\mathcal{Z}(\bar{\tau}_{\mathrm{i}})\bar{\mathcal{Z}}(\bar{\tau}_{\mathrm{ii}})\prod_{a=1}^{N/2}S_{\alpha_{2a-1}^{mod}}^{(2a-1)}(\bar{\tau}_{\mathrm{iii}}^{2a-1})\prod_{s=1}^{N-1}\alpha_{s}^{mod}(\bar{\tau}_{\mathrm{i}}^{s}).\nonumber 
\end{align}
Applying the formulas (\ref{eq:ppoleZb}) and (\ref{eq:ppoleZ}) we
obtain that the formal pole at the limit $t_{k}^{2b}+t_{l}^{2b}+2bc\to0$
reads as 
\begin{align}
S_{\bar{\alpha}}(\bar{t}) & \to g(-t_{k}^{2b}-2bc,t_{l}^{2b})\left(\alpha_{2b}(t_{k}^{2b})\alpha_{2b}(t_{l}^{2b})-1\right)\times\nonumber \\
 & \times\left[\frac{x_{b+1}}{x_{b}}\frac{c^{2}}{4}h(t_{k}^{2b},t_{l}^{2b})h(t_{l}^{2b},t_{k}^{2b})\right]\frac{f(\bar{\tau}^{2b},t_{k}^{2b})f(\bar{\tau}^{2b}t_{l}^{2b})}{f(\bar{\tau}^{2b+1},t_{k}^{2b})f(\bar{\tau}^{2b+1},t_{l}^{2b})}\times\nonumber \\
 & \times\sum_{\mathrm{part}(\bar{\tau})}\frac{\prod_{\nu=1}^{N-1}f(\bar{\tau}_{\mathrm{ii}}^{\nu},\bar{\tau}_{\mathrm{i}}^{\nu})}{\prod_{\nu=1}^{N-2}f(\bar{\tau}_{\mathrm{ii}}^{\nu+1},\bar{\tau}_{\mathrm{i}}^{\nu})}\frac{\prod_{a=1}^{N/2}f(\bar{\tau}_{\mathrm{ii}}^{2a-1},\bar{\tau}_{\mathrm{iii}}^{2a-1})f(\bar{\tau}_{\mathrm{iii}}^{2a-1},\bar{\tau}_{\mathrm{i}}^{2a-1})}{\prod_{a=1}^{N/2-1}f(\bar{\tau}_{\mathrm{ii}}^{2a},\bar{\tau}_{\mathrm{iii}}^{2a-1})f(\bar{\tau}_{\mathrm{iii}}^{2a+1},\bar{\tau}_{\mathrm{i}}^{2a})}\times\\
 & \times\mathcal{Z}(\bar{\tau}_{\mathrm{i}})\bar{\mathcal{Z}}(\bar{\tau}_{\mathrm{ii}})\prod_{a=1}^{N/2}S_{\alpha_{2a-1}^{mod}}^{(2a-1)}(\bar{\tau}_{\mathrm{iii}}^{2a-1})\prod_{s=1}^{N-1}\alpha_{s}^{mod}(\bar{\tau}_{\mathrm{i}}^{s})+reg.\nonumber 
\end{align}
Since $\alpha_{2b}(t_{k}^{2b})\alpha_{2b}(-t_{k}^{2b}-2bc)=1$, the
limit $t_{k}^{2b}+t_{l}^{2b}+2bc\to0$ of the overlap is finite and
the derivative of the $\alpha_{2b}$ appears as
\begin{equation}
\lim_{t_{l}^{2b}\to-t_{k}^{2b}-2bc}g(-t_{k}^{2b}-2bc,t_{l}^{2b})\left(\alpha_{2b}(t_{k}^{2b})\alpha_{2b}(t_{l}^{2b})-1\right)=-c\frac{\alpha_{2b}'(t_{k}^{2b})}{\alpha_{2b}(t_{k}^{2b})}=X_{k}^{2b}.
\end{equation}
 Using the sum formula (\ref{eq:sumFormulaTw-1}) again we obtain
that
\begin{equation}
\lim_{t_{l}^{2b}\to-t_{k}^{2b}-2bc}S_{\bar{\alpha}}(\bar{t})=X_{k}^{2b}\times F^{(2b)}(t_{k}^{2b})\frac{f(\bar{\tau}^{2b},t_{k}^{2b})f(\bar{\tau}^{2b}t_{l}^{2b})}{f(\bar{\tau}^{2b-1},t_{k}^{2b})f(\bar{\tau}^{2b-1},t_{l}^{2b})}S_{\bar{\alpha}^{mod}}(\bar{\tau})+\tilde{S},
\end{equation}
where $\tilde{S}$ is independent from $X_{k}^{2b}$ and
\begin{equation}
F^{(2b)}(z)=\frac{x_{b+1}}{x_{b}}\frac{c^{2}}{4}h(z,-z-2bc)h(-z-2bc,z)
\end{equation}

\subsection*{Pair structure limit $t_{l}^{2b-1}\to-t_{k}^{2b-1}-(2b-1)c$}

Let us calculate the limit $t_{l}^{2b-1}\to-t_{k}^{2b-1}-(2b-1)c$
of the overlap formula (\ref{eq:sumFormulaTw-1}) when $n_{2b-1}\geq2$.
Let us notice that $n_{2a-1}=\#\bar{t}_{\mathrm{iii}}^{2a-1}$ for
$a=1,\dots,\frac{N}{2}$, therefore $\#\bar{t}_{\mathrm{iii}}^{2a-1}\geq2$.
We know that HC-s $\mathcal{Z},\bar{\mathcal{Z}}$ are regular in
the limit $t_{l}^{2b-1}\to-t_{k}^{2b-1}-(2b-1)c$ and the $t_{l}^{2b-1}\to-t_{k}^{2b-1}-(2b-1)c$
limit of the $\mathfrak{gl}(2)$ overlaps are (\ref{eq:XdepGL2})
\begin{equation}
S_{\alpha_{2b-1}}^{(2b-1)}(\bar{t}^{2b-1})\to X_{k}^{2b-1}\times F^{(2b-1)}(t_{k}^{2b-1})f(\bar{\tau}^{2b-1},t_{k}^{2b-1})f(\bar{\tau}^{2b-1},t_{l}^{2b-1})S_{\alpha_{2b-1}^{mod}}^{(2b-1)}(\bar{\tau})+\tilde{S},
\end{equation}
where $\tilde{S}$ is independent from $X_{k}^{2b-1}$, the modified
$\alpha$ is defined as
\begin{equation}
\alpha_{2b-1}^{mod}(z)=\alpha_{2b-1}(z)\frac{f(t_{k}^{2b-1},z)f(t_{l}^{2b-1},z)}{f(z,t_{k}^{2b-1})f(z,t_{l}^{2b-1})},
\end{equation}
 and
\begin{equation}
F^{(2b-1)}(z)=-\left(\frac{1}{z+c\frac{2b-1}{2}}\right)^{2}.
\end{equation}
We can see that the $X_{k}^{2b-1}$ term appears only for the partitions
where $\{t_{k}^{2b-1},t_{l}^{2b-1}\}\in\bar{t}_{\mathrm{iii}}^{2b-1}$.
Let $\bar{\tau}=\bar{t}\backslash t_{k}^{2b-1},t_{l}^{2b-1}\}$ and
let us get the partitions in (\ref{eq:sumFormulaTw-1}) where $\bar{t}_{\mathrm{iii}}^{2b-1}=\{t_{k}^{2b-1},t_{l}^{2b-1}\}\cup\bar{\tau}_{\mathrm{iii}}^{2b-1}$.
\begin{align}
\mathcal{S}_{\bar{\alpha}}(\bar{t}) & \to X_{k}^{2b-1}\times F^{(2b-1)}(t_{k}^{2b-1})\sum_{\mathrm{part}(\bar{\tau})}\Biggl[f(\bar{\tau}_{\mathrm{iii}}^{2b-1},t_{k}^{2b-1})f(\bar{\tau}_{\mathrm{iii}}^{2b-1},t_{l}^{2b-1})\nonumber \\
 & \frac{f(t_{k}^{2b-1},\bar{\tau}_{\mathrm{i}}^{2b-1})f(t_{l}^{2b-1},\bar{\tau}_{\mathrm{i}}^{2b-1})}{f(t_{k}^{2b-1},\bar{\tau}_{\mathrm{i}}^{2b-2})f(t_{l}^{2b-1},\bar{\tau}_{\mathrm{i}}^{2b-2})}\frac{f(\bar{\tau}_{\mathrm{ii}}^{2b-1},t_{k}^{2b-1})f(\bar{\tau}_{\mathrm{ii}}^{2b-1},t_{l}^{2b-1})}{f(\bar{\tau}_{\mathrm{ii}}^{2b},t_{k}^{2b-1})f(\bar{\tau}_{\mathrm{ii}}^{2b},t_{l}^{2b-1})}\nonumber \\
 & \frac{\prod_{\nu=1}^{N-1}f(\bar{\tau}_{\mathrm{ii}}^{\nu},\bar{\tau}_{\mathrm{i}}^{\nu})}{\prod_{\nu=1}^{N-2}f(\bar{\tau}_{\mathrm{ii}}^{\nu+1},\bar{\tau}_{\mathrm{i}}^{\nu})}\frac{\prod_{a=1}^{N/2}f(\bar{\tau}_{\mathrm{iii}}^{2a-1},\bar{\tau}_{\mathrm{i}}^{2a-1})f(\bar{\tau}_{\mathrm{ii}}^{2a-1},\bar{\tau}_{\mathrm{iii}}^{2a-1})}{\prod_{a=1}^{N/2-1}f(\bar{\tau}_{\mathrm{iii}}^{2a+1},\bar{\tau}_{\mathrm{i}}^{2a})f(\bar{\tau}_{\mathrm{ii}}^{2a},\bar{\tau}_{\mathrm{iii}}^{2a-1})}\\
 & \times\mathcal{Z}(\bar{\tau}_{\mathrm{i}})\bar{\mathcal{Z}}(\bar{\tau}_{\mathrm{ii}})\prod_{a=1}^{N/2}S_{\alpha_{2a-1}}^{(2a-1)}(\bar{\tau}_{\mathrm{iii}}^{2a-1})\prod_{s=1}^{N-1}\alpha_{s}(\bar{\tau}_{\mathrm{i}}^{s})\Biggr]+\tilde{S},\nonumber 
\end{align}
where $\tilde{S}$ is independent from $X_{k}^{2b-1}$. Collecting
the terms we obtain that
\begin{align}
\mathcal{S}_{\bar{\alpha}}(\bar{t}) & \to X_{k}^{2b-1}\times F^{(2b-1)}(t_{k}^{2b-1})\sum_{\mathrm{part}(\bar{\tau})}\Biggl[\frac{f(\bar{\tau}^{2b-1},t_{k}^{2b-1})f(\bar{\tau}^{2b-1},t_{l}^{2b-1})}{f(\bar{\tau}^{2b},t_{k}^{2b-1})f(\bar{\tau}^{2b},t_{l}^{2b-1})}\\
 & \times\frac{\prod_{\nu=1}^{N-1}f(\bar{\tau}_{\mathrm{ii}}^{\nu},\bar{\tau}_{\mathrm{i}}^{\nu})}{\prod_{\nu=1}^{N-2}f(\bar{\tau}_{\mathrm{ii}}^{\nu+1},\bar{\tau}_{\mathrm{i}}^{\nu})}\frac{\prod_{a=1}^{N/2}f(\bar{\tau}_{\mathrm{iii}}^{2a-1},\bar{\tau}_{\mathrm{i}}^{2a-1})f(\bar{\tau}_{\mathrm{ii}}^{2a-1},\bar{\tau}_{\mathrm{iii}}^{2a-1})}{\prod_{a=1}^{N/2-1}f(\bar{\tau}_{\mathrm{iii}}^{2a+1},\bar{\tau}_{\mathrm{i}}^{2a})f(\bar{\tau}_{\mathrm{ii}}^{2a},\bar{\tau}_{\mathrm{iii}}^{2a-1})}\mathcal{Z}(\bar{\tau}_{\mathrm{i}})\bar{\mathcal{Z}}(\bar{\tau}_{\mathrm{ii}})\prod_{a=1}^{N/2}S_{\alpha_{2a-1}^{mod}}^{(2a-1)}(\bar{\tau}_{\mathrm{iii}}^{2a-1})\nonumber \\
 & \times\left[\prod_{s=1}^{N-1}\alpha_{s}(\bar{\tau}_{\mathrm{i}}^{s})\frac{f(t_{k}^{2b-1},\bar{\tau}_{\mathrm{i}}^{2b-1})f(t_{l}^{2b-1},\bar{\tau}_{\mathrm{i}}^{2b-1})}{f(\bar{\tau}_{\mathrm{i}}^{2b-1},t_{k}^{2b-1})f(\bar{\tau}_{\mathrm{i}}^{2b-1},t_{l}^{2b-1})}\frac{f(\bar{\tau}_{\mathrm{i}}^{2b},t_{k}^{2b-1})f(\bar{\tau}_{\mathrm{i}}^{2b},t_{l}^{2b-1})}{f(t_{k}^{2b-1},\bar{\tau}_{\mathrm{i}}^{2b-2})f(t_{l}^{2b-1},\bar{\tau}_{\mathrm{i}}^{2b-2})}\right]\Biggr]+\tilde{S}.\nonumber 
\end{align}
Defining the remaining $\alpha^{mod}$-s as
\begin{align}
\alpha_{2b-2}^{mod}(z) & =\alpha_{2b-2}(z)\frac{1}{f(t_{k}^{2b-1},z)f(t_{l}^{2b-1},z)},\nonumber \\
\alpha_{2b}^{mod}(z) & =\alpha_{2b}(z)f(z,t_{k}^{2b-1})f(z,t_{l}^{2b-1}),\\
\alpha_{s}^{mod}(z) & =\alpha_{s}(z),\quad\text{for }s\neq2b-2,2b-1,2b,\nonumber 
\end{align}
the overlap formula simplifies as
\begin{align}
\mathcal{S}_{\bar{\alpha}}(\bar{t}) & \to X_{k}^{2b-1}\times F^{(2b-1)}(t_{k}^{2b-1})\Biggl[\frac{f(\bar{\tau}^{2b-1},t_{k}^{2b-1})f(\bar{\tau}^{2b-1},t_{l}^{2b-1})}{f(\bar{\tau}^{2b},t_{k}^{2b-1})f(\bar{\tau}^{2b},t_{l}^{2b-1})}\nonumber \\
 & \times\sum_{\mathrm{part}(\bar{\tau})}\frac{\prod_{\nu=1}^{N-1}f(\bar{\tau}_{\mathrm{ii}}^{\nu},\bar{\tau}_{\mathrm{i}}^{\nu})}{\prod_{\nu=1}^{N-2}f(\bar{\tau}_{\mathrm{ii}}^{\nu+1},\bar{\tau}_{\mathrm{i}}^{\nu})}\frac{\prod_{a=1}^{N/2}f(\bar{\tau}_{\mathrm{iii}}^{2a-1},\bar{\tau}_{\mathrm{i}}^{2a-1})f(\bar{\tau}_{\mathrm{ii}}^{2a-1},\bar{\tau}_{\mathrm{iii}}^{2a-1})}{\prod_{a=1}^{N/2-1}f(\bar{\tau}_{\mathrm{iii}}^{2a+1},\bar{\tau}_{\mathrm{i}}^{2a})f(\bar{\tau}_{\mathrm{ii}}^{2a},\bar{\tau}_{\mathrm{iii}}^{2a-1})}\\
 & \times\mathcal{Z}(\bar{\tau}_{\mathrm{i}})\bar{\mathcal{Z}}(\bar{\tau}_{\mathrm{ii}})\prod_{a=1}^{N/2}S_{\alpha_{2a-1}^{mod}}^{(2a-1)}(\bar{\tau}_{\mathrm{iii}}^{2a-1})\prod_{s=1}^{N-1}\alpha_{s}^{mod}(\bar{\tau}_{\mathrm{i}}^{s})\Biggr]+\tilde{S}.\nonumber 
\end{align}
Applying the sum formula again we obtain that
\begin{equation}
\mathcal{S}_{\bar{\alpha}}(\bar{t})\to X_{k}^{2b-1}\times F^{(2b-1)}(t_{k}^{2b-1})\frac{f(\bar{\tau}^{2b-1},t_{k}^{2b-1})f(\bar{\tau}^{2b-1},t_{l}^{2b-1})}{f(\bar{\tau}^{2b},t_{k}^{2b-1})f(\bar{\tau}^{2b},t_{l}^{2b-1})}\mathcal{S}_{\bar{\alpha}^{mod}}(\bar{\tau})+\tilde{S}.
\end{equation}

\subsection*{Pair structure limit $t_{k}^{2b-1}\to-(b-1/2)c$\label{subsec:Pair-structure-limit0}}

Let us calculate the limit $t_{k}^{2b-1}\to-(b-1/2)c$ of the overlap
formula (\ref{eq:sumFormulaTw-1}) when $n_{2b-1}\geq1$ or $\#\bar{t}_{\mathrm{iii}}^{2b-1}\geq1$.
We know that HC-s $\mathcal{Z},\bar{\mathcal{Z}}$ are regular in
the limit $t_{k}^{2b-1}\to-(b-1/2)c$ and the $t_{k}^{2b-1}\to-(b-1/2)c$
limit of the $\mathfrak{gl}(2)$ overlaps are (\ref{eq:XdepGL2-1})
\begin{equation}
S_{\alpha_{2b-1}}^{(2b-1)}(\bar{t}^{2b-1})\to X^{0,2b-1}\left(-\frac{2}{c}\right)f(\bar{\tau}^{2b-1},-(b-1/2)c)S_{\alpha_{2b-1}^{mod}}^{(2b-1)}(\bar{\tau})+\tilde{S},
\end{equation}
where $\tilde{S}$ is independent from $X^{0,2b-1}$, the modified
$\alpha$ is defined as
\begin{equation}
\alpha_{2b-1}^{mod}(z)=\alpha_{2b-1}(z)\frac{f(-(b-1/2)c,z)}{f(z,-(b-1/2)c)}.
\end{equation}
We can see that the $X^{0,2b-1}$ term appears only for the partitions
where $\{t_{k}^{2b-1}\}\in\bar{t}_{\mathrm{iii}}^{2b-1}$. Let $\bar{\tau}=\bar{t}\backslash t_{k}^{2b-1}$
and let us get the partitions in (\ref{eq:sumFormulaTw-1}) where
$\bar{t}_{\mathrm{iii}}^{2b-1}=\{t_{k}^{2b-1}\}\cup\bar{\tau}_{\mathrm{iii}}^{2b-1}$.
\begin{align}
\mathcal{S}_{\bar{\alpha}}(\bar{t}) & \to X^{0,2b-1}\left(-\frac{2}{c}\right)\times\sum_{\mathrm{part}(\bar{\tau})}\Biggl[f(\bar{\tau}_{\mathrm{iii}}^{2b-1},-(b-1/2)c)\nonumber \\
 & \frac{f(-(b-1/2)c,\bar{\tau}_{\mathrm{i}}^{2b-1})}{f(-(b-1/2)c,\bar{\tau}_{\mathrm{i}}^{2b-2})}\frac{f(\bar{\tau}_{\mathrm{ii}}^{2b-1},-(b-1/2)c)}{f(\bar{\tau}_{\mathrm{ii}}^{2b},-(b-1/2)c)}\\
 & \frac{\prod_{\nu=1}^{N-1}f(\bar{\tau}_{\mathrm{ii}}^{\nu},\bar{\tau}_{\mathrm{i}}^{\nu})}{\prod_{\nu=1}^{N-2}f(\bar{\tau}_{\mathrm{ii}}^{\nu+1},\bar{\tau}_{\mathrm{i}}^{\nu})}\frac{\prod_{a=1}^{N/2}f(\bar{\tau}_{\mathrm{iii}}^{2a-1},\bar{\tau}_{\mathrm{i}}^{2a-1})f(\bar{\tau}_{\mathrm{ii}}^{2a-1},\bar{\tau}_{\mathrm{iii}}^{2a-1})}{\prod_{a=1}^{N/2-1}f(\bar{\tau}_{\mathrm{iii}}^{2a+1},\bar{\tau}_{\mathrm{i}}^{2a})f(\bar{\tau}_{\mathrm{ii}}^{2a},\bar{\tau}_{\mathrm{iii}}^{2a-1})}\nonumber \\
 & \times\mathcal{Z}(\bar{\tau}_{\mathrm{i}})\bar{\mathcal{Z}}(\bar{\tau}_{\mathrm{ii}})\prod_{a=1}^{N/2}S_{\alpha_{2a-1}^{mod}}^{(2a-1)}(\bar{\tau}_{\mathrm{iii}}^{2a-1})\prod_{s=1}^{N-1}\alpha_{s}(\bar{\tau}_{\mathrm{i}}^{s})\Biggr]+\tilde{S},\nonumber 
\end{align}
where $\tilde{S}$ is independent from $X^{0,2b-1}$. Collecting the
terms we obtain that
\begin{align}
\mathcal{S}_{\bar{\alpha}}(\bar{t}) & \to X^{0,2b-1}\left(-\frac{2}{c}\right)\times\sum_{\mathrm{part}(\bar{\tau})}\Biggl[\frac{f(\bar{\tau}^{2b-1},-(b-1/2)c)}{f(\bar{\tau}^{2b},-(b-1/2)c)}\\
 & \times\frac{\prod_{\nu=1}^{N-1}f(\bar{\tau}_{\mathrm{ii}}^{\nu},\bar{\tau}_{\mathrm{i}}^{\nu})}{\prod_{\nu=1}^{N-2}f(\bar{\tau}_{\mathrm{ii}}^{\nu+1},\bar{\tau}_{\mathrm{i}}^{\nu})}\frac{\prod_{a=1}^{N/2}f(\bar{\tau}_{\mathrm{iii}}^{2a-1},\bar{\tau}_{\mathrm{i}}^{2a-1})f(\bar{\tau}_{\mathrm{ii}}^{2a-1},\bar{\tau}_{\mathrm{iii}}^{2a-1})}{\prod_{a=1}^{N/2-1}f(\bar{\tau}_{\mathrm{iii}}^{2a+1},\bar{\tau}_{\mathrm{i}}^{2a})f(\bar{\tau}_{\mathrm{ii}}^{2a},\bar{\tau}_{\mathrm{iii}}^{2a-1})}\mathcal{Z}(\bar{\tau}_{\mathrm{i}})\bar{\mathcal{Z}}(\bar{\tau}_{\mathrm{ii}})\prod_{a=1}^{N/2}S_{\alpha_{2a-1}^{mod}}^{(2a-1)}(\bar{\tau}_{\mathrm{iii}}^{2a-1})\nonumber \\
 & \times\left[\prod_{s=1}^{N-1}\alpha_{s}(\bar{\tau}_{\mathrm{i}}^{s})\frac{f(-(b-1/2)c,\bar{\tau}_{\mathrm{i}}^{2b-1})}{f(\bar{\tau}_{\mathrm{i}}^{2b-1},-(b-1/2)c)}\frac{f(\bar{\tau}_{\mathrm{i}}^{2b},-(b-1/2)c)}{f(-(b-1/2)c,\bar{\tau}_{\mathrm{i}}^{2b-2})}\right]\Biggr]+\tilde{S}.\nonumber 
\end{align}
Defining the remaining $\alpha^{mod}$-s as
\begin{align}
\alpha_{2b-2}^{mod}(z) & =\alpha_{2b-2}(z)\frac{1}{f(-(b-1/2)c,z)},\nonumber \\
\alpha_{2b}^{mod}(z) & =\alpha_{2b}(z)f(z,-(b-1/2)c),\\
\alpha_{s}^{mod}(z) & =\alpha_{s}(z),\quad\text{for }s\neq2b-2,2b-1,2b,\nonumber 
\end{align}
the overlap formula simplifies as
\begin{align}
\mathcal{S}_{\bar{\alpha}}(\bar{t}) & \to X^{0,2b-1}\left(-\frac{2}{c}\right)\times\Biggl[\frac{f(\bar{\tau}^{2b-1},-(b-1/2)c)}{f(\bar{\tau}^{2b},-(b-1/2)c)}\nonumber \\
 & \times\sum_{\mathrm{part}(\bar{\tau})}\frac{\prod_{\nu=1}^{N-1}f(\bar{\tau}_{\mathrm{ii}}^{\nu},\bar{\tau}_{\mathrm{i}}^{\nu})}{\prod_{\nu=1}^{N-2}f(\bar{\tau}_{\mathrm{ii}}^{\nu+1},\bar{\tau}_{\mathrm{i}}^{\nu})}\frac{\prod_{a=1}^{N/2}f(\bar{\tau}_{\mathrm{iii}}^{2a-1},\bar{\tau}_{\mathrm{i}}^{2a-1})f(\bar{\tau}_{\mathrm{ii}}^{2a-1},\bar{\tau}_{\mathrm{iii}}^{2a-1})}{\prod_{a=1}^{N/2-1}f(\bar{\tau}_{\mathrm{iii}}^{2a+1},\bar{\tau}_{\mathrm{i}}^{2a})f(\bar{\tau}_{\mathrm{ii}}^{2a},\bar{\tau}_{\mathrm{iii}}^{2a-1})}\\
 & \times\mathcal{Z}(\bar{\tau}_{\mathrm{i}})\bar{\mathcal{Z}}(\bar{\tau}_{\mathrm{ii}})\prod_{a=1}^{N/2}S_{\alpha_{2a-1}^{mod}}^{(2a-1)}(\bar{\tau}_{\mathrm{iii}}^{2a-1})\prod_{s=1}^{N-1}\alpha_{s}^{mod}(\bar{\tau}_{\mathrm{i}}^{s})\Biggr]+\tilde{S}.\nonumber 
\end{align}
Applying the sum formula again we obtain that
\begin{equation}
\mathcal{S}_{\bar{\alpha}}(\bar{t})\to X^{0,2b-1}\left(-\frac{2}{c}\right)\times\frac{f(\bar{\tau}^{2b-1},-(b-1/2)c)}{f(\bar{\tau}^{2b},-(b-1/2)c)}\mathcal{S}_{\bar{\alpha}^{mod}}(\bar{\tau})+\tilde{S}.
\end{equation}

\bibliographystyle{elsarticle-num}
\bibliography{ref}

\end{document}